\documentclass[reqno,a4paper,10pt]{book}
\usepackage[pdftex,unicode,implicit]{hyperref}
\hypersetup{%
  pdftitle    = {Black Holes in Supergravity with Applications to String Theory},
  pdfkeywords = {Supergravity,String Theory,Black Holes,microscopic entropy,hidden conformal symmetry, Special Kahler Geometry}, 
  pdfsubject  = {hep-th, gr-qc, math.dg},
  pdfauthor   = {C.S.Shahbazi},
  pdfcreator  = {pdf\LaTeXe\ with package \flqq hyperref\frqq},
  pdffitwindow= false,  % Open the pdf file without the sidebars open.
  bookmarksopen   = false,
  bookmarksopenlevel = 1,
  bookmarksnumbered = true,
  unicode     = true,
  plainpages  = false,
  colorlinks  = true,  % enable to avoid having coloured squares around hyperlinks
  citecolor   = black,
  urlcolor    = black,
  linkcolor   = black,
}

\usepackage{amssymb,amstext,amsthm,amsfonts,mathrsfs,amsbsy,amsmath,array,bbm,multirow}

\usepackage{slashed,fancyhdr,hojadeestilotesisdoctoral,bbding}

\usepackage{a4wide} % para hacer mas ancho la pagina
\usepackage[english,greek]{babel}
\usepackage[usenames,dvipsnames,svgnames,table]{xcolor}
\usepackage[pdftex]{graphicx}
\usepackage{xparse}
\usepackage{collref} % serves to collect multiple \bibitem references (in the same sequence) into a single \bibitem block the problem with 'collref' is that it blocks the references automatically, and it makes them look less important against ones that were mentioned earlier in the text and thus have are a unique ref
\usepackage{tikz}

\usepackage[a4paper,bindingoffset=0.2in,left=0.6in,right=0.6in,top=1in,bottom=1in]{geometry}

\usepackage{concmath}
\usepackage[T1]{fontenc}

\theoremstyle{plain}
\newtheorem{thm}{Theorem}[section]

\newtheorem{prop}[thm]{Proposition}
\newtheorem{lemma}[thm]{Lemma}

\theoremstyle{definition}
\newtheorem{definition}[thm]{Definition}

\theoremstyle{remark}

\NewDocumentCommand\mycite{mgggg}{\IfNoValueTF{#5}{\IfNoValueTF{#3}{\IfNoValueTF{#2}{\singlecite{#1}}{\singlecitedetail{#1}{#2}}}{\multicite{#1}{#2}{#3}}}{\multimulticite{#1}{#2}{#3}{#4}{#5}}} 
\usepackage[authoryear,round,square,colon,sort&compress,comma]{natbib}
\bibpunct{[}{]}{,}{n}{}{,}

\begin{document}
\setcounter{tocdepth}{1}
\selectlanguage{english}
%\renewcommand{\contentsname}{Contents/ Contenidos}
%%%%%%%%%%%%%%%%%%%%%%%%%%%%%%%%%%%%%%%%%%%%%%%%%%%%%%%%%%%%%%%%%%%%%%%
%%% PORTADA
%%%%%%%%%%%%%%%%%%%%%%%%%%%%%%%%%%%%%%%%%%%%%%%%%%%%%%%%%%%%%%%%%%%%%%%
\thispagestyle{empty}
\phantom{a}
\vspace{2cm}

\vspace*{-1.3cm}
\newcommand{\HRule}{\rule{\linewidth}{0.3mm}}
\setlength{\parindent}{1cm}
\setlength{\parskip}{1mm}
\noindent
\HRule
\begin{center}
%{\fontfamily{Calligra}\selectfont\textbf{Black Holes in Supergravity with Applications to String Theory}}

\setlength{\baselineskip}{3\baselineskip}
\textbf{Black Holes in Supergravity with Applications to String Theory}
\HRule
\vspace{4cm}

%{\sizea {\sizeA B}lack holes in{\sizeA S}upergravity and}\\
%{\sizea {\sizeA S}tring {\sizeA T}heory}}

\includegraphics[scale=0.5]{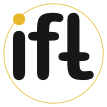}

\vspace{4cm}

\begin{minipage}{12cm}
\begin{center}
\bf{Carlos Shahbazi Alonso} \\ \vspace{0.2cm} \bf{Instituto de F\'isica Te\'orica} \\ \vspace{0.2cm} \bf{UAM-CSIC}
\end{center}
\end{minipage}

\vspace{2cm}

\begin{minipage}{12cm}
\begin{center}
\emph{A thesis submitted for the degree of} \\ \emph{Doctor of Philosophy} \\ \emph{June 2013}
\end{center}
\end{minipage}

\end{center}
\clearpage

%%%%%%%%%%%%%%%%%%%%%%%%%%%%%%%%%%%%%%%%%%%%%%%%%%%%%%%%%%%%%%%%%%%%%%%
%%% PAGINA VACIA
%%%%%%%%%%%%%%%%%%%%%%%%%%%%%%%%%%%%%%%%%%%%%%%%%%%%%%%%%%%%%%%%%%%%%%%
\thispagestyle{empty}
\phantom{1}

\clearpage

%%%%%%%%%%%%%%%%%%%%%%%%%%%%%%%%%%%%%%%%%%%%%%%%%%%%%%%%%%%%%%%%%%%%%%%
%%% CONTRAPORTADA
%%%%%%%%%%%%%%%%%%%%%%%%%%%%%%%%%%%%%%%%%%%%%%%%%%%%%%%%%%%%%%%%%%%%%%%
\thispagestyle{empty}
\phantom{a}
\begin{center}
{
\renewcommand{\tabcolsep}{2em}
\begin{tabular}{cc}  
\textbf{Universidad}& \textbf{Consejo Superior de} \\ 
\textbf{Aut\'onoma de Madrid} & \textbf{Investigaciones Cient\'ificas} \\  
%\hline
\includegraphics[scale=0.3]{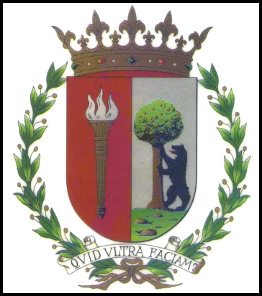} & \includegraphics[scale=0.3]{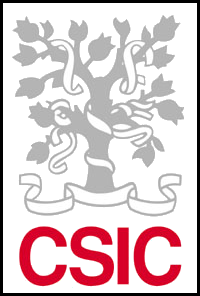} \\
Departamento de F\'isica Te\'orica & Instituto de F\'isica Te\'orica\\
Facultad de Ciencias\\
\end{tabular}} 
\end{center}
\vspace{2cm}

\vspace{\stretch{5}}
\begin{center}
\setlength{\baselineskip}{2\baselineskip}
\textbf{
{\sizea {\sizeA A}GUJEROS NEGROS EN SUPERGRAVEDAD}\\
{\sizea Y TEOR\'IA DE CUERDAS}}
	
%\vspace{\stretch{4}}
%\setlength{\baselineskip}{0.5\baselineskip}
%{Alberto R. Palomo-Lozano}
%\vspace{\stretch{6}}
\vspace{\stretch{1}}
\setlength{\baselineskip}{0.5\baselineskip}
\vspace{\stretch{6}}

\begin{minipage}{16cm}
\begin{center}
Memoria de Tesis Doctoral presentada ante el Departamento de F\'isica Te\'orica\\
de la Universidad Aut\'onoma de Madrid para la obtenci\'on del t\'itulo de Doctor en Ciencias F\'isicas
\end{center}
\end{minipage}
	
\vspace{\stretch{4}}
Tesis Doctoral dirigida por:
\\[1ex]
\textbf{Profesor D. Tom\'as Ort\'in Miguel}\\
Profesor de Investigaci\'on, Consejo Superior de Investigaciones Cient\'ificas
\\[5ex]
Junio, 2013
\end{center}

\newpage

%%%%%%%%%%%%%%%%%%%%%%%%%%%%%%%%%%%%%%%%%%%%%%%%%%%%%%%%%%%%%%%%%%%%%%%
%%% PAGINA VACIA
%%%%%%%%%%%%%%%%%%%%%%%%%%%%%%%%%%%%%%%%%%%%%%%%%%%%%%%%%%%%%%%%%%%%%%%
\thispagestyle{empty}
\phantom{1}

\clearpage

%%%%%%%%%%%%%%%%%%%%%%%%%%%%%%%%%%%%%%%%%%%%%%%%%%%%%%%%%%%%%%%%%%%%%%%
%%% AGRADECIMIENTOS
%%%%%%%%%%%%%%%%%%%%%%%%%%%%%%%%%%%%%%%%%%%%%%%%%%%%%%%%%%%%%%%%%%%%%%%
\section*{Agradecimientos}

%\thispagestyle{empty}
%\hspace{\stretch{3}}
%\begin{minipage}{25em}
%\begin{flushright}
%{\em `` "}\\
%\vspace{2ex}
%{\em }\phantom{"}
%\end{flushright}
%\end{minipage}
%\hspace{\stretch{1}}
%\vspace{8ex}

En esta tesis se recoge parcialmente el resultado de tres a\~nos de trabajo en el Instituto de F\'isica Te\'orica, perteneciente al Consejo Superior de Investigaciones Cient\'ificas, bajo la tutela del Profesor D. Tom\'as Ort\'in Miguel. No tengo palabras para expresar mi agradecimiento a Tom\'as: de \'el he aprendido b\'asicamente todo lo que s\'e sobre F\'isica y Matem\'aticas. Me ha ense\~nado a abordar problemas y a resolverlos de la manera m\'as efectiva, as\'i como la importancia de dominar los c\'alculos matem\'aticos y utilizar a la vez la intuici\'on; en definitiva, me ha ense\~nado lo que significa \emph{investigar} en F\'isica Te\'orica. Ha sido un tutor inmejorable y es sin duda uno de los f\'isicos m\'as brillantes que conozco. Gracias a \'el adem\'as, he tenido la oportunidad de viajar, conocer y colaborar con otros f\'isicos excepcionales, como Eric Bergshoeff, Renata Kallosh o Patrick Meessen, de los que tambi\'en he aprendido mucho, a los que estoy muy agradecido y con los que espero seguir en contacto en el futuro. 

Renata fue una anfitriona inmejorable de mi visita de medio a\~no al \emph{Stanford Insititute of Theoretical Physics}; me ayud\'o con todo desde el primer momento, tanto fuera como dentro del \'ambito acad\'emico, se preocup\'o siempre de que estuviera bien acomodado y me dio la oportunidad de colaborar con ella y su grupo de investigaci\'on, de lo que obtuve un conocimiento y experiencia que de otra manera no hubiera sido posible. %Fue una experiencia que trascendi\'o el plano meramente profesional o acad\'emico y que siempre atesorar\'e en mi recuerdo. 
Gracias Renata.

Una menci\'on de agradecimiento merecen todos los colaboradores con los que he tenido la oportunidad de trabajar durante este tiempo; todos me han aportado algo y de todos he aprendido. Me refiero a Antonio de Antonio Mart\'in, Eric Bergshoeff, Johannes Broedel, Pablo Bueno, Wissam Chemissany, Frederik Coomans, \'Alvaro de la Cruz-Dombriz, Rhys Davies, Pietro Galli, Mechthild Huebscher, Antonio L. Maroto, Patrick Meessen, Tom\'as Ort\'in, Jan Perz, Antoine Van Proeyen y Marco Zamb\'on. Gracias tambi\'en a \'Angel Uranga y Marco Zamb\'on por atenderme tan pacientemente con mis interminables dudas y preguntas. 

Debo agradecer al Consejo Superior de Investigaciones Cient\'ificas por la financiaci\'on recibida, en concepto tanto de mi salario como de la estancia realizada. Al personal administrativo, tanto del IFT, Isabel, Roxanna, Chabely, como del SITP, Karen y Julie, les doy las gracias: sin personas como ellas la investigaci\'on cient\'ifica no ser\'ia posible. 

Fuera del \'ambito acad\'emico hay muchas personas a las que tengo que agradecer el poder estar, en el momento de escribir estas l\'ineas, en disposici\'on de obtener el t\'itulo de doctor en F\'isica Te\'orica. Primero, a mis padres, Maria del Carmen y Mahmood; sin la eduaci\'on que me dieron y el posterior apoyo nunca hubiera llegado a donde hoy estoy, y por ello, entre otras cosas, estar\'e eternamente agradecido. Un lugar especial ocupa tambi\'en mi t\'ia Maria Jos\'e, que ya desde peque\~no  me pusiera en contacto con la ciencia y me mostrara las vicisitudes de la vida del cient\'ifico, siendo ella una gran cient\'ifica. Debo agradecer asimismo todo el apoyo recibido a mi t\'ia Teresa, siempre preocupada por m\'i, a mi hermana Marta, y a toda mi familia en Ir\'an; aunque la distancia es grande, siempre se han preocupado por m\'i como si estuvieran aqu\'i y me han deseado lo mejor: !`muchas gracias!, espero poder visitaros alg\'un d\'ia en vuestro tierra, Ir\'an, que tambi\'en es la m\'ia. Debo tambi\'en mencionar a mis abuelos espa\~noles, Generoso y Gloria, descansen en paz, que siempre quisieron que me dedicara a una ocupaci\'on intelectual.

Agradezco en especial el apoyo y cari\~no diario, as\'i como la comprensi\'on, a mi novia Sara, con ella los buenos momentos son a\'un mejores, y los malos momentos son menos amargos: !`gracias por todo!

Gracias tambi\'en a mis amigos, los que siempre hab\'eis estado ah\'i, \emph{dando el callo}: Santi, Charli, Iker, Roberto: !`qu\'e buenos ratos hemos pasado y los que a\'un nos quedan!

Finalmente, mencionar a Pablo Bueno y Alberto Palomo-Lozano, compa\~neros de despacho con los que he pasado buenos ratos que van m\'as all\'a de lo meramente profesional. Agradezco tambi\'en a Pablo Bueno y Diego Regalado por estimulantes discusiones sobre F\'isica y Matem\'aticas, as\'i como por leer y comentar las versiones iniciales de este manuscrito.\\

\vspace{1cm}
\hspace{\stretch{1}}Madrid, Junio 2013.
\clearpage

%%%%%%%%%%%%%%%%%%%%%%%%%%%%%%%%%%%%%%%%%%%%%%%%%%%%%%%%%%%%%%%%%%%%%%%
%%% TABLE OF CONTENTS
%%%%%%%%%%%%%%%%%%%%%%%%%%%%%%%%%%%%%%%%%%%%%%%%%%%%%%%%%%%%%%%%%%%%%%%
%\renewcommand{\thepage}{\Roman{page}}
\addtocontents{toc}{\protect\thispagestyle{empty}}
%\addtocontents{toc}{\vspace{-5ex}}
%\addtocontents{toc}{\contentsline {chapter}{\numberline {}}{}{}}
\tableofcontents
\thispagestyle{empty}
\clearpage

%%%%%%%%%%%%%%%%%%%%%%%%%%%%%%%%%%%%%%%%%%%%%%%%%%%%%%%%%%%%%%%%%%%%%%%
%%% PART I
%%%%%%%%%%%%%%%%%%%%%%%%%%%%%%%%%%%%%%%%%%%%%%%%%%%%%%%%%%%%%%%%%%%%%%%
\renewcommand{\thepage}{\arabic{page}}
%\addtocontents{toc}{\contentsline{chapter}{\numberline {\small{PART I}}}{}{}}
\pagestyle{fancy}
\renewcommand{\leftmark}{\MakeUppercase{Chapter \thechapter. Introduction}}
\renewcommand{\headrulewidth}{0.0pt}
\fancyhead[LE]{\thepage}
\fancyhead[RE]{\slshape \leftmark}
\fancyhead[LO]{\slshape \rightmark}
\fancyhead[RO]{\thepage}
\fancyfoot{}
%%%%%%%%%%%%%%%%%%%%%%%%%%%%%%%%%%%%%%%%%%%%%%%%%%%%%%%%%%%%%%%%%%%%%%%
%%% CHAPTER 1: INTRODUCTION
%%%%%%%%%%%%%%%%%%%%%%%%%%%%%%%%%%%%%%%%%%%%%%%%%%%%%%%%%%%%%%%%%%%%%%%
%\renewcommand{\chaptername}{Part I\\ Chapter}
%./chapters/Introduction

\renewcommand{\leftmark}{\MakeUppercase{Introduction}}
\chapter{Introduction}

The aim of this thesis is to study black holes in String Theory (ST) through their classical description as Supergravity solutions. ST \cite{Scherk:1974ca,Yoneya:1974jg} is a framework that attempts to offer a unified quantum description of all the known fundamental interactions and, in particular, of gravity. It would solve, if correct, the long-standing problem of Quantum Gravity. In the present day there is no experimental evidence of ST, and there is no hope in the ST community that such an evidence will soon be found. Despite the lack of experimental results, ST has passed several self-consistency checks that are expected to hold in the \emph{right} unifying theory of nature \cite{Green:1984sg,Gross:1985fr,Gross:1985rr,Strominger:1996sh,Ortin:2004ms,Ibanez:2012zz}, if it exists. As a consequence, an enormous effort has been devoted over the last decades to develop ST, leading to beautiful and important advances and insights in modern Theoretical Physics \cite{Strominger:1996sh,Maldacena:1997re,Bern:2007hh} and Mathematics \cite{deWit:1983rz,deWit:1984pk,Strominger:1990pd,Candelas:1990rm,Seiberg:1994rs,Witten:1998cd,Uranga:2000xp,2002math......9099H,2004math......1221G}. The ramifications of ST-inspired results are nowadays virtually everywhere in Theoretical Physics, and thus even if the \emph{right} theory of nature was not ST, we can expect that they will have some ingredients in common. Hundreds of thesis, books and reviews have been written over the years dealing with the principal aspects of ST, such as supersymmetry, perturbative ST, conformal field theory, D-branes, ST dualities, aDS/CFT correspondence, M-Theory... and hence we refer the interested reader to the existing literature, for instance \cite{gswst,Greene:1996cy,Ortin:2004ms,Kiritsis:2007zza,Ibanez:2012zz} and references therein.

I will focus instead (see chapters \ref{chapter:mathpreliminaries} and \ref{chapter:sugra}), for reasons that will become apparent later, on a different area, sometimes forgotten in ST applications: the precise mathematical structure of the ST \emph{effective actions}, \emph{i.e.} field theories that describe the dynamics of the massless modes of the ST spectrum and are used to make contact with four-dimensional low-energy physics. The mathematical structure of these effective actions is crucial in order to ensure the consistency of the theory and  it is the necessary background that we will need to pursue our goal, namely the study of black holes in ST. Let us see first how to go from the full-fledged ST to four-dimensional Supergravity. 

%%%%%%%%%%%%%%%%%%%%%%%%%%%%%%%%%%%%%%%%%%%%%%%%%%%%%%%%%%%%%%%%%%%%%%%%%%%%%%%%%%%%%%%%%%%%%%%%%%%
%%%%%%%%%%%%%%%%%%%%%%%%%%%%%%%%%%%%%%%%%%%%%%%%%%%%%%%%%%%%%%%%%%%%%%%%%%%%%%%%%%%%%%%%%%%%%%%%%%%

\section{From String Theory to Supergravity}

%%%%%%%%%%%%%%%%%%%%%%%%%%%%%%%%%%%%%%%%%%%%%%%%%%%%%%%%%%%%%%%%%%%%%%%%%%%%%%%%%%%%%%%%%%%%%%%%%%%
%%%%%%%%%%%%%%%%%%%%%%%%%%%%%%%%%%%%%%%%%%%%%%%%%%%%%%%%%%%%%%%%%%%%%%%%%%%%%%%%%%%%%%%%%%%%%%%%%%%

ST \emph{lives} in ten dimensions, that is, the mathematical object that represents the space-time in ST is a ten-dimensional manifold. Since experimentally, that is, at low energies, we observe four dimensions, some mechanism must be used to reconcile theory and experiment. The standard way to proceed is to assume that the space-time manifold $\mathcal{M}$\footnote{For more details about the precise properties of $\mathcal{M}$, see chapter \ref{chapter:mathpreliminaries}.} has the following fibre bundle structure

\begin{equation}
\mathcal{M}\xrightarrow{\pi}\mathcal{M}_{4}\, ,
\end{equation}

\noindent
where $\mathcal{M}_{4}$ is the base space manifold (which represents the space-time that we observe at low energies) and the fibre $\mathcal{M}_6\left(p\right)=\pi^{-1}(p)$ at each $p\in \mathcal{M}_{4}$ is a compact manifold, small enough to not be accessible in current high-energy experiments. So to say, $\mathcal{M}_6\left(p\right)$ is so small that we cannot \emph{see it} with the available technology. Notice that $\mathcal{M}_6\left(p\right)\cong \mathcal{M}_6\left(q\right)\, ,\forall ~p, q\in\mathcal{M}_{4}$, so we can denote the typical fibre simply by $\mathcal{M}_6$.

The general mechanism described above can be put in practice in several different ways, on which the precise size limits of $\mathcal{M}_6$ depend \cite{Duff:1986hr,Candelas:1985en,Kachru:2002he,Grana:2005jc}. Unfortunately, it is currently not known how to compactify ST on a non-trivial compact manifold $\mathcal{M}_6$, except for some particular cases. By non-trivial manifold here we mean a Riemannian manifold with a curved metric. The usual procedure to deal with this situation is to cook up an effective field theory action that encodes the dynamics of the massless modes present in the ST spectrum, which are the ones relevant to low energy physics. The reason is that the first massive states in the spectrum have masses of the order of the Planck mass, which is way out of reach for current particle accelerators.

As a consequence of ST having local space-time supersymmetry, the ST massless states action is a very special one: a Supergravity \cite{Freedman:1976xh,West:1986wua,Wess:1992cp,deWit:2002vz,Freedman:2012zz}. The key point is that, even though it is not know how to compactify ST, it is known how to compactify a field theory, and in particular a Supergravity. Therefore, although the procedure described above is sometimes called ``ST compactification'' what is actually compactified is a ten-dimensional Supergravity. We can add to it ST corrections, that is, modifications to the \emph{tree level} result as prescribed by ST, which is not a field theory but a theory of extended objects. The resulting corrected action is, again, a Supergravity but, tipically, it has higher order terms in the Lagrangian and modified couplings.

If we compactify now low energy ST, that is, ten-dimensional Supergravity, on a particular compact six-dimensional manifold $\mathcal{M}_6$\footnote{Notice that it is possible to compactify in  manifolds with dimension other than six, obtaining as effective actions Supergravities in dimensions other than four.}, we obtain a four-dimensional effective action which will also be a Supergravity, four-dimensional in this case, if some technical requirements are obeyed by the compact manifold $\mathcal{M}_{6}$. I will come back to this point later, but for the moment I refer the reader to \cite{Grana:2005jc} and references therein.

To summarize, we have started with ten-dimensional ST and we have ended up with four-dimensional Supergravity, which is a theory of gravity and matter that can be embedded in ST, and therefore seems to be the perfect starting point to study four-dimensional black holes in ST.

%%%%%%%%%%%%%%%%%%%%%%%%%%%%%%%%%%%%%%%%%%%%%%%%%%%%%%%%%%%%%%%%%%%%%%%%%%%%%%%%%%%%%%%%%%%%%%%%%%%
%%%%%%%%%%%%%%%%%%%%%%%%%%%%%%%%%%%%%%%%%%%%%%%%%%%%%%%%%%%%%%%%%%%%%%%%%%%%%%%%%%%%%%%%%%%%%%%%%%%

\section{Supergravity}

%%%%%%%%%%%%%%%%%%%%%%%%%%%%%%%%%%%%%%%%%%%%%%%%%%%%%%%%%%%%%%%%%%%%%%%%%%%%%%%%%%%%%%%%%%%%%%%%%%%
%%%%%%%%%%%%%%%%%%%%%%%%%%%%%%%%%%%%%%%%%%%%%%%%%%%%%%%%%%%%%%%%%%%%%%%%%%%%%%%%%%%%%%%%%%%%%%%%%%%

Supergravity is a locally supersymmetric theory of gravity, that is, a field theory invariant under local supersymmetry transformations. Supersymmetry is a not so new, and yet to be observed, hypothetical symmetry between bosons and fermions \cite{Wess:1974tw,Volkov:1973ix,Akulov:1974xz,Wess:1992cp}.

If we consider a field theory content with spin less or equal than two, there are seven different types of four-dimensional Supergravity $\mathcal{N}=1,\cdots,6,8$, depending on the amount $\mathcal{N}$ of supersymmetry of the theory. Supersymmetry transformations are generated by a set of spinors $\epsilon_{I}(p)\, , I=1,\cdots , \mathcal{N}$, where $\mathcal{N}$ is the \emph{number of supersymmetries} of the given Supergravity. In four dimensions, the minimal spinors $\epsilon_{I}(p)$ can be taken to be Weyl or Majorana, and therefore we have respectively $2\mathcal{N}$ complex or $4\mathcal{N}$ real associated \emph{charges}. Supersymmetry transformations can be schematically written as

\begin{equation}
\delta_{\epsilon}\phi_{b}\sim \bar{\epsilon}(p)\left(\phi_{b}+\bar{\phi}_{f}\phi_{f}\right)\phi_{f}\, , \qquad \delta_{\epsilon}\phi_{f}\sim \partial\epsilon(p) +\left(\phi_{b}+\bar{\phi}_{f}\phi_{f}\right) \epsilon(p)\phi_{f}\, ,
\end{equation}

\noindent
where $\phi_{b}$ denotes the bosonic fields and $\phi_{f}$ denotes the fermionic fields.

Since Supergravity is a supersymmetric theory, and supersymmetry relates bosonic a fermionic fields, every Supergravity contains bosons and fermions. Truncating the fermions is always consistent, thanks to the following $\mathbb{Z}_{2}$ symmetry, present in every Supergravity Lagrangian

\begin{equation}
\phi_{b}\rightarrow \phi_{b}\, ,\qquad \phi_{f}\rightarrow -\phi_{f}\, .
\end{equation}

\noindent
The bosonic sector of four-dimensional Supergravity is a particular instance of General Relativity, as formulated by Albert Einstein in 1915 \cite{Einstein}. That is, it is a \emph{metric theory} of gravity coupled to a particular matter content, which includes scalars and vector fields, and where the equation of motion for the metric ${\bf g}$ is given by

\begin{equation}
{\bf R}\left({\bf \nabla}\right)={\bf T}\, .
\end{equation} 

\noindent
Here ${\bf R}\left({\bf \nabla}\right)$ is the Ricci tensor of the Levi-Civita connection $\nabla$ associated to ${\bf g}$ on the space-time tangent bundle \cite{LeviCivita}, and ${\bf T}$ is the \emph{geometrized} energy-momentum tensor corresponding to the matter content of the theory. 

General Relativity cannot be, in principle, coupled to fermions, since it is formulated in a way on which only the diffeomorphisms group $\mathrm{Diff}\left(\mathcal{M}\right)$ acts naturally on the matter content. We need to make manifest the local action of the Lorentz group $\mathrm{SO}(1,3)$\footnote{Rather, the action of its double-cover, the spin group $\mathrm{Spin}(1,3)$, see below.} on the matter content of the theory, since fermions are associated to spinorial representations of $\mathrm{SO}(1,3)$.  Therefore, if we want to consider the complete Supergravity action, we have to change the set-up and use a more general formalism, which turns out to be the Cartan-Sciama-Kibble theory \cite{CartanSK1,CartanSK2,Hehl:1976kj,Ortin:2004ms}, a generalization of Einstein's General Relativity. Just as the bosonic sector of Supergravity is a particular case of General Relativity, the complete Supergravity theory is a particular case of the Cartan-Sciama-Kibble theory, which is an extension of General Relativity that can accommodate fermions. 

Before introducing the Cartan-Sciama-Kibble theory it is necessary to modify a bit the \emph{geometric} set-up, in order to geometrically introduce fermions. General Relativity coupled to matter can be described in terms of objects that transform as tensors under space-time diffeomorphisms (such as sections over the tensor products of $T\mathcal{M}$ and $T^{\ast}\mathcal{M}$ or connections on principal bundles over $\mathcal{M}$). For instance the metric, which \emph{describes} the gravitational interaction, is a non-degenerate section of $\mathrm{S}^{2}T^{\ast}\mathcal{M}$. The electromagnetic interaction, on the other hand, is described by a connection on a principal $\mathrm{U}(1)$ bundle over $\mathcal{M}$.

However, fermions are described in classical field theory as spinors, that is, representations of the spin group $\mathrm{Spin}(1,3)$, and they do not correspond to any section of the tangent or cotangent bundle. The spin group $\mathrm{Spin}(1,3)$ is the double cover of the Lorentz group $\mathrm{SO}(1,3)$ such that the following short exact sequence holds

\begin{equation}
 \mathbb{Z}_2 \rightarrow \mathrm{Spin}(1,3) \xrightarrow{\rho} \mathrm{SO}(1,3)\, .
\end{equation}

\noindent
In order to properly include fermions into the game, we have to consider first the bundle of frames $\mathrm{F}\left(T\mathcal{M}\right)$ instead of the tangent $T\mathcal{M}$ and the cotangent $T^{\ast}\mathcal{M}$ bundles. The frame bundle $\mathrm{F}\left(T\mathcal{M}\right)$ is the principal vector bundle associated to $T\mathcal{M}$, defined as follows

\definition{\label{def:fbundle} Let $\mathcal{M}$ be a differentiable manifold, in our case the space-time minfold. Define a manifold $\mathrm{F}\left(T\mathcal{M}\right)$ as

\begin{equation}
\mathrm{F}\left(T\mathcal{M}\right) = \left\{(p, {\bf e}_1,...,{\bf e}_{4}): p\in\mathcal{M}\, , ({\bf e}_1,...,{\bf e}_{4}) ~\mathrm{ordered~ basis~ of} ~ T_{p}\mathcal{M}\right\}\, .
\end{equation}

Define now a projection $\pi_{\mathrm{F}} : \mathrm{F}\left(T\mathcal{M}\right)\rightarrow\mathcal{M}$ by $\pi(p,{\bf e}_1,\cdots , {\bf e}_{4}) = p$ and define an action of $\mathrm{GL}(4,\mathbb{R})$ by ${\bf e}^{\prime}_{a}=A_{a}^{\, b} {\bf e}_{b}\, , a=1,\cdots , 4$, where $A\in\mathrm{GL}(4,\mathbb{R}) $. This makes $\mathrm{F}\left(T\mathcal{M}\right)$ into a principal bundle with fibre $\mathrm{GL}(4,\mathbb{R})$. 

}

\noindent
\newline Since the space-time manifold $\mathcal{M}$ is equipped with a Lorentzian metric ${\bf g}$, we can actually consider the oriented orthonormal frame bundle $\mathrm{F}_{\mathrm{SO}}\left(T\mathcal{M}\right)$, built only from the bases orthonormal respect to the metric ${\bf g}$ and positively oriented. Notice that this is possible because we are assuming from the onset that the space-time manifold is oriented (otherwise we could not write actions as integrals over the space-time). The structure group of $\mathrm{F}_{\mathrm{SO}}\left(T\mathcal{M}\right)$ is then reduced from $\mathrm{GL}(4,\mathbb{R})$ to $\mathrm{SO}(1,3)$.

\definition{\label{def:ofbundle} Let $\mathcal{M}$ be a differentiable manifold. Define a manifold $\mathrm{F}_{\mathrm{SO}}\left(T\mathcal{M}\right)$ as

\begin{equation}
\mathrm{F}_{\mathrm{SO}}\left(T\mathcal{M}\right) = \left\{(p, {\bf e}_1,...,{\bf e}_{4}): p\in\mathcal{M}\, , ({\bf e}_1,...,{\bf e}_{4}) ~T_{p}\mathcal{M}~\mathrm{ordered~basis} ~|~{\bf g}({\bf e}_{a},{\bf e}_{b}) = \eta_{ab}\, ,\mathrm{det}_{\partial}({\bf e})>0 \right\}\, .
\end{equation}

Define now a projection $\pi_{\mathrm{SO}} : \mathrm{F}_{\mathrm{SO}}\left(T\mathcal{M}\right)\rightarrow\mathcal{M}$ by $\pi(p,{\bf e}_1,\cdots , {\bf e}_{4}) = p$ and define an action of $\mathrm{SO}(1,3)$ by ${\bf e}^{\prime}_{a}=O_{a}^{\, b} {\bf e}_{b}\, , a=1,\cdots , 4$, where $O\in\mathrm{SO}(1,3) $. This makes $\mathrm{F}_{\mathrm{SO}}\left(T\mathcal{M}\right)$ into a principal bundle with fibre $\mathrm{SO}(1,3)$. 

}

\noindent
\newline Now, in order for the space-time manifold $\mathcal{M}$ to admit fermions, a technical requirement must be fulfilled: the bundle $\mathrm{F}_{\mathrm{SO}}\left(T\mathcal{M}\right)$ must admit an \emph{equivariant lift} with respect to $\mathrm{Spin}(1,3)$. In that case we can construct the spin bundle $\mathrm{P}\rightarrow \mathcal{M}$, which is a principal bundle with fibre $\mathrm{Spin}(1,3)$. The precise definition goes as follows

\definition{\label{def:spinstructure} A pair $(\mathrm{P},\Omega_{P})$ is a spin structure on the principal bundle $\mathrm{F}_{\mathrm{SO}} \to \mathcal{M}$ when

\begin{enumerate}

\item $\mathrm{P}\xrightarrow{\pi_{\mathrm{P}}}\mathcal{M}$ is a principal bundle with fibre $\mathrm{Spin}(1,3)$.

\item $\Omega_{\mathrm{P}}: \mathrm{P} \to \mathrm{F}_{\mathrm{SO}}$ is an equivariant two-fold covering map such that $\pi_{\mathrm{SO}}\circ \Omega_{\mathrm{P}}=\pi_{\mathrm{P}}$ and $\Omega_{\mathrm{P}}(p,O) = \Omega_{\mathrm{P}}(p)\rho(O)\, , \forall ~p\in\mathrm{P}\, , \forall ~O\in \mathrm{Spin}(1.3)$.

\end{enumerate}

}

The vector bundle $\mathrm{S}\rightarrow\mathcal{M}$ associated to $\mathrm{P}$ and the spin representation of $\mathrm{Spin}(1,3)$ is then the spin bundle, and spinors are thus section of $\mathrm{S}$. The vector bundle associated to a principal bundle can be defined in general. In our case the construction is given by

\definition{\label{def:spinbundle} Let $\kappa:\mathrm{Spin}(1,3)\to\mathrm{U}(V)$ a unitary representation of $\mathrm{Spin}(1,3)$ on a complex vector space $V$. Then $\mathrm{Spin}(1,3)$ acts on the product space $\mathrm{P}\times V$ by the principal bundle action on the first factor and $\kappa$ on the second. We define 

\begin{equation}
\mathrm{S}=\left(\mathrm{P}\times V\right)/\mathrm{Spin}(1,3)\, .
\end{equation}

\noindent
Since $\mathrm{P}/\mathrm{Spin}(1,3)=\mathcal{M}$, the obvious map $\pi_{\mathrm{S}}:\left(\mathrm{P}\times V\right)/\mathrm{Spin}(1,3)\to \mathrm{P}/\mathrm{Spin}(1,3)$ gives the projection on the base space $\mathcal{M}$. Since $\mathrm{Spin}(1,3)$ acts freely on $\mathrm{P}$, this projection has fibre $V$. Therefore $\mathrm{S}$ is a vector space with base $\mathcal{M}$ and fibre $V$. 

}

\noindent
\newline Sections of $\mathrm{S}$ are spinors. If we want fermions to exist in the space-time, we must require $\mathcal{M}$ to admit a spin bundle $\mathrm{S}$, since, as we have seen, spinor fields are sections of such bundle. The obstruction to construct a spin bundle over $\mathcal{M}$ is given by the  second Stiefel-Whitney class $w_2$, which should be zero. For more details, the reader is referred to \cite{Spin}.

Local sections ${\bf e}$ of $\mathrm{P}$ are called local spin frames or \emph{vierbiens}. In terms of them, the metric is given by

\begin{equation}
\label{eq:gee}
{\bf g} = \eta\left({\bf e},{\bf e}\right) \, ,
\end{equation}

\noindent
where $\eta = \mathrm{Diag}(+,-,-,-)$ is the Minkowski metric. The principal bundle $\mathrm{P}$ can now be equipped with the so-called \emph{spin connection} $\omega$, which makes it possible to construct derivatives covariant under local $\mathrm{Spin}(1,3)$ transformations. Using (\ref{eq:gee}) it is possible to rewrite General Relativity in terms of the vierbeins ${\bf e}$ instead of the metric. This is important because the choice of a vierbein makes each $\left( T_{p}\mathcal{M}\, , p\in\mathcal{M}\right)$ into Minkowski space, where we know how to write actions for spinor fields. In addition, the use of vierbeins makes explicit the invariance of the theory under local $\mathrm{Spin}(1,3)$ transformations. This is precisely what we need to include fermions in the theory, using the so-called \emph{first-order formalism} of the Cartan-Sciama-Kibble theory. It can be summarized as follows 

\begin{enumerate}

\item The dynamical field associated to gravity is taken to be ${\bf e}$ instead of the metric ${\bf g}$.

\item The Einstein-Hilbert term is now written in terms of the curvature $R_{a_{1} a_{2}}(\omega)$ of the spin connection $\omega$ as follows

\begin{equation}
\label{eq:CSKEH}
S\left[{\bf e},\omega\right] = \frac{1}{2}\int R_{a_{1} a_{2}}(\omega)\wedge {\bf e}_{a3}\wedge {\bf e}_{a4} \epsilon^{a_{1} a_{2} a_{3} a_{4}}\, ,
\end{equation}

\noindent
where $\epsilon^{a_{1} a_{2} a_{3} a_{4}}$ is the flat Levi-Civita tensor. The action (\ref{eq:CSKEH}) is equivalent to the first-order Einstein-Hilbert action for the metric and the affine connection $\Gamma$. See \cite{Ortin:2004ms} for a detailed explanation. 

\item The covariant derivatives are constructed using the connection $\omega$ which is considered an independent field of the theory. 

\item Kinetic terms for the spinors are introduced using the covariant derivative constructed from $\omega$. For instance, if $\psi$ is a Dirac spinor, then a kinetic term of the action would be of the form $\bar{\psi}\slashed{\mathcal{D}}\psi$, where $\mathcal{D}\psi \sim \partial\psi+\omega\psi$.

\item The equation of motion for the spin connection $\omega$ is an algebraic constraint which relates it with the other fields of the theory. In particular, it may have torsion. A metric compatible connection $\omega$ can be written as follows 

\begin{equation}
\omega = \omega_{\mathrm{LC}} + \mathrm{K}\, ,
\end{equation}

\noindent
where $\omega_{\mathrm{LC}}$ is the Levi-Civita connection and $\mathrm{K}$ is the \emph{contorsion} tensor \cite{Ortin:2004ms}. For Supergravity theories, $\mathrm{K}$ depends only on the fermionic fields of the theory, and therefore vanishes in a purely bosonic background.

\end{enumerate}

Therefore, using the Cartan-Sciama-Kibble theory we can extend the Lagrangian density $\mathcal{L}({\bf g},\phi_{b})$ of General Relativity to the Lagrangian density $\mathcal{L}({\bf e},\omega, \phi_{b},\phi_{f})$, which includes fermions and instead of the metric ${\bf g}$ has ${\bf e}$ as the dynamical field associated to gravity. Notice that, although every Lagrangian density of the form $\mathcal{L}({\bf g},\phi_{b})$ can be written as $\mathcal{L}({\bf e},\omega, \phi_{b},\phi_{f}=0)$, the converse is not true and, therefore, the Cartan-Sciama-Kibble theory is in general not equivalent to General Relativity.\\
As it happens in General Relativity, four-dimensional Supergravity assumes that the space-time can be described by a differentiable Pseudo-Riemannian manifold $\mathcal{M}$ modulo isometries of the Pseudo-Riemannian metric, that is, by an equivalence class of Pseudo-Riemannian isometric manifolds $\left[\mathcal{M}\right]$. Physicists are used to consider diffeomorphic space-times as equivalent. This can be easily accommodated in the definition of equivalence classes by isometries by equipping the image manifold with the induced metric. That is, if $f$ is the corresponding diffeomorphism, we would have $f:\left(\mathcal{M},{\bf g}\right)\rightarrow \left(\mathcal{M},{\bf \tilde{g}}=\left(f^{-1}\right)^{\ast} {\bf g}\right)$, which are isometric manifolds, and hence equivalent from a physical point of view.

Therefore Supergravity can be constructed explicitly in terms of geometric objects, that is, suitable sections of appropriate fibre bundle constructed over the space-time manifold, and thus it is a \emph{geometric theory}. In fact, as we will see in chapter \ref{chapter:sugra}, the structure of the Lagrangian itself can be determined in a geometric way by  using appropriate manifolds and sections therein. 

Supersymmetry imposes severe constraints on the field content and structure of the Lagrangian, making possible the study and classification of all the possible Supergravities. Initially developed independently from ST, it was realized in the eighties that the three different ten-dimensional Supergravities are the low-energy limit of the five, duality-related, ten-dimensional STs. Not only that: Supergravity includes crucial non-perturbative information about ST, through its BPS spectrum. That is, although Supergravity encodes the dynamics of only the massless states of the ST spectrum, it contains solitonic supersymmetric solutions with physical properties that are protected by supersymmetry from the ST corrections to the Supergravity action. Some of these supersymmetric solutions, like particular instances of supersymmetric black holes,  correspond to the long range fields created by bound states of non-perturbative ST objects, like D-branes. 

Its crucial relation with ST, together with the fact that Supergravity may be the right effective theory of nature (up to some scale) even if it is not embedded in ST, and the beautiful mathematical formulation that underlies the theory, has made Supergravity an extremely important research topic, which is still extensively studied, with never decreasing interest.

In addition, a remarkable and fascinating discovery that focused a lot of attention, and increased even more the interest in Supergravity, was made in 2007: explicit computations showed that $\mathcal{N}=8$ Supergravity was perturbatively UV finite up to three loops \cite{Bern:2007hh}! (Divergences were expected at one loop). Soon enough the computation was extended, with finite results, to four loops \cite{Bern:2009kd} and to three loops for $\mathcal{N}=4$ pure Supergravity \cite{Bern:2012cd}. Several powerful arguments have appeared in favor of $\mathcal{N}=8$ Supergravity and $\mathcal{N}=4$ pure Supergravity being perturbatively finite to all loops \cite{Kallosh:2009jb,Kallosh:2011dp,Kallosh:2011qt,Ferrara:2012ui}\footnote{For a pessimistic opinion, see \cite{Banks:2012dp}.}. Intense research is being performed to extend the explicit calculations to higher loops and to provide a final proof of the conjectured UV perturbative finitness of $\mathcal{N}=8$ and pure $\mathcal{N}=4$ Supergravities. %Interestingly enough, ST, which has always been assumed to contain a perturbatively finite theory of Quantum Gravity, has been explicitly checked to be finite only up to two loops \cite{Kiritsis:2007zza}. As in the Supergravity case, arguments exist in favor of finiteness to all orders. 

On a more mathematical side, Supergravity is suffering a quiet and elegant revolution: a new mathematical tool, called Generalized Complex Geometry \cite{2002math......9099H,2004math......1221G} and its \emph{extension}, Exceptional Generalized Geometry \cite{Hull:2007zu,Pacheco:2008ps,Grana:2009im,Grana:2011nb}, allow for a new and more geometric formulation of Supergravity, considering sections on a generalized bundle instead of sections of the tangent bundle (and its tensorial products) of the spacetime manifold to represent the bosonic fields of the theory. This new formulation may be relevant for the role that Supergravity plays in ST, since it allows a covariant and natural action of the ST duality groups on its massless content, through the \emph{extended structure group} that acts on the generalized tangent bundle. 

Generalized Complex Geometry considers the vector bundle $\mathbb{T}\mathbb{M} = T\mathcal{M}\oplus T^{\ast}\mathcal{M}$ over $\mathcal{M}$ instead of the tangent or contangent bundles. Here $\oplus$ stands for the \emph{Whitney sum} of vector bundles over the same base space $\mathcal{M}$. Elements of $\mathbb{T}\mathbb{M}$ are of the form $\mathbb{X}=X+\xi$, where $X\in T\mathcal{M}$ and $\xi\in  T^{\ast}\mathcal{M} $. $\mathbb{T}\mathbb{M}$, in contrast to the tangent bundle $T\mathcal{M}$, is naturally equipped with a canonical metric $\mathrm{G}$, defined as follows

\begin{equation}
\mathrm{G}\left(\mathbb{X},\mathbb{Y}\right) = \frac{1}{2}\left(\xi(Y)+\eta(X)\right)\, ,\qquad \forall ~\mathbb{X}=X+\xi\, , \mathbb{Y}=Y+\eta \in \mathbb{T}\mathbb{M}\, .
\end{equation}

\noindent
$\mathrm{G}$ has signature $(d,d)$, where $d$ is the dimension of $\mathcal{M}$. Interestingly enough, at every point $p\in\mathcal{M}$, $\mathrm{G}$ is invariant under the action of $\mathrm{SO}(d,d)$, the $\mathrm{T}$-duality symmetry group of ST compactified on a $d$-torus.  A generalized complex structure $\mathrm{J}$ on $\mathbb{T}\mathbb{M}$ is a bundle map

\begin{equation}
\mathrm{J}: T\mathcal{M}\oplus T^{\ast}\mathcal{M}\to T\mathcal{M}\oplus T^{\ast}\mathcal{M}\, ,
\end{equation}

\noindent
such that $\mathrm{J}^2=-1$ and $\mathrm{G}(\cdot,\cdot) = \mathrm{G}(\mathrm{J}\cdot,\mathrm{J}\cdot)$, where $\mathrm{G}$ is the canonical metric on $T\mathcal{M}\oplus T^{\ast}\mathcal{M}$. Remarkably enough, a complex structure $\mathcal{J}$ on $T\mathcal{M}$ and a symplectic structure $\omega$ on $\Lambda^2 T^{\ast}\mathcal{M}$ are special instances of $\mathrm{J}$, when their actions are suitable extended to $T\mathcal{M}\oplus T^{\ast}\mathcal{M}$. 

This is extremely important, and allows for Generalized Complex Geometry to give a unified description of the ten-dimensional supersymmetric Supergravity backgrounds. If we look for a supersymmetric solution of the equations of motion of ten-dimensional Supergravity of the form

\begin{equation}
\mathcal{M}=\mathcal{M}_{4}\times\mathcal{M}_{6}\, ,\qquad {\bf g}(x,y) ={\bf g}_{4}(x) + {\bf g}_{6}(y)\, ,
\end{equation}  

\noindent
where the metric is written in a patch ~$\mathcal{U}$ with coordinates $(x^{1},\dots,x^{4},y^{1},\dots,y^{6})$, then a possible solution is

\begin{equation}
{\bf g}(y) =\eta + {\bf g}_{CY}(y)
\end{equation}  

\noindent
where all the fluxes are taken to be zero and the dilaton is constant. This solution is phenomenologically relevant because, for Heterotic ST, it preserves four-dimensional $\mathcal{N}=1$ supersymmetry \cite{Candelas:1985en}. Here ${\bf g}_{CY}(y)$ stands for the metric of the Calabi-Yau internal space $\mathcal{M}_{6}$. However, supersymmetric backgrounds with non-trivial fluxes are far from being Calabi-Yau\footnote{Notice that fluxes cannot be turned on in compact spaces, unless negative tension sources, the so-called orientifold planes, are included in the vacuum structure.}. A Calabi-Yau manifold is in particular a symplectic manifold and a complex manifold in a compatible way, but more general supersymmetric backgrounds are in general not simultaneously complex and symplectic. It turns out that Generalized Complex Geometry gives a unified description of all internal spaces of supersymmetric flux backgrounds \cite{Grana:2005jc}. 

Generalized Complex Geometry can be also used to naturally describe the supersymmetric embedding of D-branes with world-volume fluxes into these backgrounds \cite{Cascales:2004qp,Martucci:2005ht,Koerber:2005qi,2007math......1740Z}. For more details and other applications of Generalized Complex Geometry, the reader is referred to the excellent reviews \cite{Grana:2005jc,Koerber:2010bx}.
 
Generalized Complex geometry and Exceptional Generalized Geometry could be the first step towards the correct mathematical formalism to describe ST.

%%%%%%%%%%%%%%%%%%%%%%%%%%%%%%%%%%%%%%%%%%%%%%%%%%%%%%%%%%%%%%%%%%%%%%%%%%%%%%%%%%%%%%%%%%%%%%%%%%%
%%%%%%%%%%%%%%%%%%%%%%%%%%%%%%%%%%%%%%%%%%%%%%%%%%%%%%%%%%%%%%%%%%%%%%%%%%%%%%%%%%%%%%%%%%%%%%%%%%%

\section{Remarks about Black Holes}

%%%%%%%%%%%%%%%%%%%%%%%%%%%%%%%%%%%%%%%%%%%%%%%%%%%%%%%%%%%%%%%%%%%%%%%%%%%%%%%%%%%%%%%%%%%%%%%%%%%
%%%%%%%%%%%%%%%%%%%%%%%%%%%%%%%%%%%%%%%%%%%%%%%%%%%%%%%%%%%%%%%%%%%%%%%%%%%%%%%%%%%%%%%%%%%%%%%%%%%

A black hole space-time \cite{Sch,RN1,RN2,Kerr:1963ud,Newman:1965tw,Misner:1974qy} is a particular kind of space-time that contains a region, called black hole, from which gravity prevents anything, including light, from escaping (we will give a precise definition in a moment). In this extreme situation, the classical laws of physics break down at the singularity and a quantum description of gravity seems to be needed. Since ST is one of the candidates to a theory of Quantum Gravity, it makes sense to use it to study situations where Quantum Gravity effects are expected to be important, as it happens for black holes, and see what we can learn from it, even if we don't know if ST actually describes the universe.

All the black holes that we will obtain and study in this thesis are asymptotically flat, static and spherically symmetric\footnote{With one exception: in chapter \ref{chapter:susysolutions} we will consider momentarily stationary space-times, which are anyway \emph{composed} of several \emph{static} black holes.}. Notice that we truncate the fermions, which is always a consistent truncation, because black-hole solutions describe classical macroscopic objects. Therefore, by an asymptotically flat, static and spherically symmetric black-hole solution we mean a solution of the equations of motion of the corresponding theory for all the bosonic fields in the Lagrangian, such that the space-time manifold $\mathcal{M}$ is asymptotically flat, static, spherically symmetric and contains a black hole in the following sense \cite{Garay}

\begin{itemize}

\item A space-time $\mathcal{M}$ is asymptotically simple if and only if it admits a conformal compactification and all the null geodesics of $\mathcal{M}$ start and end on $\partial \mathcal{M}$. 

\item A space-time $\mathcal{M}$ is asymptotically flat if and only if it has an open neighbourhood $\mathcal{U}$ isometric to an open neighbourhood of the boundary of the conformal compactification of an asymptotically simple space-time, and the Ricci tensor vanishes on $\mathcal{U}$.

\item A asymptotically flat space-time is stationary if and only if there exists a time-like Killing vector $\xi$ in a neighbourhood of the spatial infinity. 

\item A stationary space-time $\mathcal{M}$ is static if and only if there exists an space-like hypersurface orthogonal to the Killing vector $\xi$.

\item In a strongly asymptotically predictable space-time $\mathcal{M}$, we will call black hole, if it exists, the region $\mathcal{B}\subset\mathcal{M}$ which is not contained in the causal past of the infinite null future. 

\end{itemize}

Although the definition given above is completely precise, it is not very useful for practical purposes. A clearer characterization can be given using local coordinates $(t,r,\theta,\phi)$ adapted to an spherically symmetric and static space-time $\mathcal{M}$. Here $t$ is a time coordinate, $r$ is a radial coordinate, and $(\theta,\phi)$ are angular coordinates. The metric can thus be written as

\begin{equation}
\label{eq:metricin}
{\bf g} = g_{tt}(r) dt\otimes dt - g_{rr}(r) dr\otimes dr - r^2 \mathrm{h}_{S^2}\, ,
\end{equation}

\noindent 
where $\mathrm{h}_{S^2}=d\theta\otimes d\theta +\sin^2\theta d\phi\otimes d\phi$ is the round metric on the unit two-sphere. The event horizon typically lies at one of the the solutions of the equation $g_{tt}(r) = 0$. In general, the equation $g_{tt}(r) = 0$ has several solutions $r_{1}<,\cdots ,<r_{n}$, each one typically corresponding to the position of a particular \emph{horizon}.

In the cases considered in this thesis, the maximum number of solutions of the equation $g_{tt}(r) = 0$ is going to be two. Let us denote them by $r_{-}<r_{+}$. Then, at $r=r_{+}$ we have the event horizon and at $r=r_{-}$ we have the Cauchy horizon. When the physical parameters of the solution (such as the mass, the charge or the \emph{moduli}\footnote{In this context, the moduli is the arbitrary value at spatial infinity of the scalar fields present in the theory, which is not fixed by the equations of motion.}) are adjusted so $r_{-}=r_{+}$ we have an extremal black hole. Let's see how this works for the Reissner-Nordstr\"om black hole \cite{RN1,RN2} of mass $m$ and charge $q$, which is the simplest example of the kind of black holes that are considered in this thesis. By Birkhoff's theorem, the Reissner-Nordstr\"om solution is the only spherically symmetric solution of the Einstein-Maxwell theory, which in addition is also static and asymptotically flat. The metric is therefore of the form (\ref{eq:metricin}), with

\begin{equation}
g_{tt}(r) = 1-\frac{2m}{r} + \frac{q^2}{r^2}\, ,\qquad g_{rr}(r)=g^{-1}_{tt}(r)\, .
\end{equation}

\noindent
$g_{tt}(r)$ can be written as 

\begin{equation}
g_{tt}(r) = \frac{(r-r_{+})(r-r_{-})}{r^2}\, ,\qquad r_{\pm} = m\pm\sqrt{m^2-q^2}\, .
\end{equation}

\noindent
Hence, the equation $g_{tt}(r)=0$ has two solutions, given by $r_{\pm}$. The event horizon lies at $r=r_{+}$ and the Cauchy horizon lies at $r=r_{-}$. Therefore, the entropy and temperatures of the black hole are given by \cite{Bardeen:1973gs,Bekenstein:1973ur}

\begin{equation}
S=\frac{A}{4}= \pi r^2_{+}\, ,\qquad T = \frac{1}{2\pi}\left(\frac{r_0}{r_{+}}\right)^2\, ,
\end{equation}

\noindent
where $A$ is the area of the event horizon and $r_{0}=\mp M \pm r_{\pm}$ is the extremality parameter (see below). The black hole is extremal when the two horizons coincide, that is, when $m=q$ (and hence $r_{0} = 0$). Notice that in that case, the temperature $T$ of the black hole is always zero but the entropy $S$ is not, if the extremal black hole is regular. Extremal black holes cannot radiate \cite{Hawking:1974sw} and are thus stable. For more details the reader is referred to \cite{Ellis,Wald,Garay,Ortin:2004ms}.

%%%%%%%%%%%%%%%%%%%%%%%%%%%%%%%%%%%%%%%%%%%%%%%%%%%%%%%%%%%%%%%%%%%%%%%%%%%%%%%%%%%%%%%%%%%%%%%%%%%
%%%%%%%%%%%%%%%%%%%%%%%%%%%%%%%%%%%%%%%%%%%%%%%%%%%%%%%%%%%%%%%%%%%%%%%%%%%%%%%%%%%%%%%%%%%%%%%%%%%

\section{Supergravity solutions and the attractor mechanism}

%%%%%%%%%%%%%%%%%%%%%%%%%%%%%%%%%%%%%%%%%%%%%%%%%%%%%%%%%%%%%%%%%%%%%%%%%%%%%%%%%%%%%%%%%%%%%%%%%%%
%%%%%%%%%%%%%%%%%%%%%%%%%%%%%%%%%%%%%%%%%%%%%%%%%%%%%%%%%%%%%%%%%%%%%%%%%%%%%%%%%%%%%%%%%%%%%%%%%%%

Going back to the topic of this thesis, black holes in ST, we now have all the elements necessary to precisely define our goal: in order to study black holes in ST we are going to develop a mathematical procedure to obtain black-hole solutions of four-dimensional Supergravity theories. These classical solutions (they are classical simply because the theory is not quantized) play a relevant role in ST, since they are necessary, for instance, in order to check the match between the microscopic entropy, computed as an ensemble of D-branes and other ST objects, and the macroscopic entropy, given by the area of the classical solution. This match is mandatory if ST is to be the correct theory of nature and contains a consistent theory of Quantum Gravity. In fact, the match has been checked to some extent in several specific cases of extremal and near-extremal black holes \cite{Strominger:1996sh,Horowitz:1996fn,Breckenridge:1996is,Maldacena:1996gb,Breckenridge:1996sn,
Behrndt:1997gs,Sen:2007qy,Mandal:2010cj}. However, there are supersymmetric (and hence extremal and stable) simple black holes for which the microscopic interpretation of the entropy is not even known at the leading order \cite{Bueno:2012jc}. This means that even for the simplest kind of black holes the complete microscopic description of the entropy in ST has not always been achieved.

With the currently available tools, a necessary condition to obtain a match between the microscopic and the macroscopic computation is that the Supergravity solution depends only on quantized quantities, such as the charges of the black hole, but not, for example on the moduli, which are free parameters. Remarkably enough, for extremal black holes this is what generically happens: all the information about infinity is lost at the horizon\footnote{In the absence of flat directions.} and the entropy only depends on the quantized charges, \emph{e.i.}, the dependence on the moduli drops out. This is a consequence of the so-called \emph{attractor mechanism} \cite{Ferrara:1996dd,Ferrara:1997tw,Tripathy:2005qp,Sen:2005wa,Goldstein:2005hq,Bellucci:2007ds,Ferrara:2007qx,Ferrara:2007tu,Ceresole:2007wx,
Ferrara:2008hwa,Bellucci:2008cb,Bellucci:2008jq,Marrani:2010bn} for Supergravity black holes, which roughly speaking states that the scalar fields of a Supergravity extremal black-hole solution flow from a completely arbitrary value at spatial infinity to a completely fixed (in terms of the quantized charges of the black hole) value on the horizon, independent of the asymptotic value of the scalar at spatial infinity.

In fact, the attractor mechanism holds in a class of theories larger than the Supergravities. In particular, it holds in any theory of the form (\ref{eq:generalaction4}). As we will see in chapter \ref{chapter:sugrabhs}, the most general spherically symmetric and static metric solution of (\ref{eq:generalaction}) is given by

\begin{equation}
\label{eq:bhi}
{\bf g} 
 = 
e^{2U} dt\otimes dt - e^{-2U}\left[\frac{r^4_0}{\sinh^4 r_0\tau}d\tau\otimes d\tau +\frac{r^2_0}{\sinh^2 r_0\tau} \mathrm{h}_{S^2} \right]\, , 
\end{equation}

\noindent
where $\tau$ is the radial coordinate and $r_0$ is the non-extremality parameter when (\ref{eq:bhi}) represents a black hole. In that case, the exterior of the event horizon is covered by $\tau\in\left(-\infty,0\right)$, the event horizon being located at $\tau\rightarrow -\infty$
and the spatial infinity at $\tau \rightarrow 0^{-}$. The interior of the Cauchy
horizon (if any) is covered by $\tau\in\left(\tau_{S},\infty\right)$, the
inner horizon being located at $\tau\rightarrow +\infty$ while the singularity
is located at some finite, positive, value $\tau_S$ of the radial coordinate $\tau$ 
\cite{Galli:2011fq}. Since we are assuming that the space-time is spherically symmetric, all the fields of the theory, that is, the scalars and the vector fields, depend only on the radial coordinate $\tau$. In the background given by (\ref{eq:bhi}) the Maxwell equations can be explicitly integrated, giving the vector fields as functions of $\tau$ and the electric $q_{\Lambda}$ and magnetic $p^{\Lambda}$ charges. The other equations of motion form a system of second order ordinary differential equations for $(U(\tau),\phi(\tau))$, namely (\ref{eq:e1}), (\ref{eq:Vbh-r0-real}) and (\ref{eq:e3}).

The extremal limit $r_{0}\rightarrow 0$ of (\ref{eq:bhi}) is given by

\begin{equation}
\label{eq:bhie}
{\bf g} 
 = 
e^{2U} dt\otimes dt - e^{-2U}\left[\frac{1}{\tau^4}d\tau\otimes d\tau +\frac{1}{\tau^2} \mathrm{h}_{S^2} \right] = e^{2U} dt\otimes dt - e^{-2U}\left[\delta_{ij}dx^{i}\otimes dx^{j} \right]\, , 
\end{equation}

\noindent
where $x^{i}\, , i=1,\dots,3$ are three-dimensional \emph{cartesian coordinates} and $\delta_{ij}$ is the Kronecker delta. For an extremal regular black-hole solution of a theory with non-degenerate scalar metric $\mathcal{G}_{ij}$ and non-divergent scalars at the horizon, it can be proven that in the near horizon limit we have 

\begin{equation}
\label{eq:attractorass}
\lim_{\tau\to -\infty} e^{-2U} = \frac{A}{4\pi} \lim_{\tau\to -\infty}\tau^2\, ,\qquad \lim_{\tau\to -\infty} \tau\frac{d\phi^{i}}{d\tau} = 0\, ,i=1,\cdots , n_{v}\, ,
\end{equation}

\noindent
where $A$ is the area of the event horizon and $n_{v}$ is the number of scalars. Using Eq. (\ref{eq:attractorass}) we can obtain the near-horizon limit of the Eq. (\ref{eq:e3}), which reads

\begin{equation}
\lim_{\tau\to -\infty} \left[\frac{d^2\phi^{i}}{d\tau^2} + \frac{4\pi}{A} \mathcal{G}^{ij}(\phi_{{\rm h}})\partial_{j} V_{{\rm bh}}(\phi_{{\rm h}},\mathcal{Q})\frac{1}{\tau^2}\right] = 0\, ,
\end{equation}

\noindent
where $V_{{\rm bh}}(\phi,\mathcal{Q})$ is the \emph{black-hole potential} and $\mathcal{Q}=\left(p^{\Lambda},q_{\Lambda}\right)^{T}$ denotes the electric and magnetic charges of the black hole. The solution to the above differential equation is given by 

\begin{equation}
\lim_{\tau\to -\infty} \phi^{i} = \lim_{\tau\to -\infty}\left[-\frac{4\pi}{A} \mathcal{G}^{ij}(\phi_{{\rm h}})\partial_{j} V_{{\rm bh}}(\phi_{{\rm h}},\mathcal{Q})\log(-\tau)  +c^{i}_{1}\tau+c^{i}_{2}\right]\, ,
\end{equation}

\noindent
where $c^{i}_{1}$ and $c^{i}_{2}$ are arbitrary constants. Therefore, taking the $\tau$-derivative we obtain

\begin{equation}
\lim_{\tau\to -\infty} \frac{d\phi^{i}}{d\tau} =\lim_{\tau\to -\infty}\left[ -\frac{4\pi}{A} \mathcal{G}^{ij}(\phi_{{\rm h}})\partial_{j} V_{{\rm bh}}(\phi_{{\rm h}},\mathcal{Q})\frac{1}{\tau}  +c^{i}_{1}\right]\, .
\end{equation}

\noindent
Since Eq. (\ref{eq:attractorass}) must hold and, by assumption, the scalars do not diverge at the horizon, we conclude that 

\begin{equation}
c^{i}_{1} = 0\, , \qquad c^{i}_{2}=\phi^{i}_{{\rm h}}\, , \qquad\mathcal{G}^{ij}(\phi_{{\rm h}})\partial_{j} V_{{\rm bh}}(\phi_{{\rm h}},\mathcal{Q}) = 0 \, .
\end{equation}

\noindent
Finally, since the scalar metric is non-degenerate, the condition involving the black-hole potential can be rewritten as 

\begin{equation}
\label{eq:attractor}
\partial_{i} V_{{\rm bh}}(\phi_{{\rm h}},\mathcal{Q}) = 0 \, ,
\end{equation}

\noindent
which is the essence of the attractor mechanism \cite{Attractor}. The value of the scalars at the horizon $\phi^{i}_{{\rm h}}$ must be a critical point of the black hole potential $V_{{\rm bh}}(\phi,\mathcal{Q})$. If $V_{{\rm bh}}(\phi,\mathcal{Q})$ has no flat directions, that is, if (\ref{eq:attractor}) is a compatible system of $n_{v}$ independent equations, all the scalars are fixed at the horizon in terms of the charges of the black hole. From the near-horizon limit of Eq. (\ref{eq:Vbh-r0-real}) it can be easily obtained that, for extremal black holes, the entropy $S$ is given by

\begin{equation}
S = \pi V_{{\rm bh}}(\phi_{{\rm h}},\mathcal{Q})\, .
\end{equation}

\noindent
Therefore, we obtain a remarkable result: in the absence of flat directions the entropy $S$ and the value of the scalars at the horizon only depend on the charges of the black hole, that is 

\begin{equation}
\phi_{{\rm h}} = \phi_{{\rm h}}(\mathcal{Q})\, ,\qquad S = \pi V_{{\rm bh}}(\phi_{{\rm h}}(\mathcal{Q}),\mathcal{Q})\, .
\end{equation}

\noindent
When flat directions are present, the scalars are only partially fixed in terms of the charges, but the dependence of the entropy $S$ on the moduli still drops out. Supersymmetric attractors are always minima of the black hole potential. For non-supersymmetric attractors this issue must be studied on a case by case basis, although for homogeneous scalar manifolds it has been proven that the critical points of the black hole potential are always stable, but only up to possible flat directions \cite{Ferrara:2007pc,Ferrara:2007tu}. The attractor mechanism is only one example of the interesting features that Supergravity black holes, and other black objects such a strings or black $p$-branes display \cite{Stelle:1996tz,Stelle:1998xg,Chemissany:2011gr,Chemissany:2011sr,deAntonioMartin:2012bi}. 

The search of solutions of Supergravity theories started with the class of supersymmetric solutions, since they are easier to obtain and classify, thanks to the first order differential equations (the so-called \emph{Killing spinor equations}) that they obey.

A configuration of fields $(\phi_{b},\phi_{f})$ is supersymmetric invariant (also called B.P.S.) if and only if 

\begin{equation}
\delta_{\epsilon}\phi_{b} = 0\, ,\qquad \delta_{\epsilon}\phi_{f} = 0\, ,
\end{equation}

\noindent
for at least one spinor $\epsilon$. In that case, $\epsilon$ is called a \emph{Killing spinor}. If a given configuration is invariant under the maximum number of independent Killing spinors, given by $\mathcal{N}$, then it is said to be \emph{maximally supersymmetric}. Notice that a supersymmetric configuration in principle does not solve the equations of motion of the corresponding theory. However, a careful analysis of the Killing spinor equations and their integrability conditions, the so called \emph{Killing spinor identities}, shows that a supersymmetric configuration usually solves almost all the equations of motion of the theory, if not all \cite{Kallosh:1993wx,Bellorin:2005hy}.  

We are interested in black-hole solutions where the fermions are set to zero. For bosonic configurations, the equation $\delta_{\epsilon}\phi_{b} = 0$ is automatically solved and we have to solve only $\delta_{\epsilon}\phi_{f} = 0$ to obtain a supersymmetric configuration. These are the Killing spinor equations that must be solved in any supersymmetric theory to obtain a bosonic supersymmetric configuration. When a supersymmetric configuration also obeys the equations of motion of the theory it becomes a supersymmetric solution. Examples of supersymmetric solutions of Supergravity are the Reissner-Nordstr\"om extreme black hole, Minkowski space-time or the $\mathrm{aDS}$ space-time.

The supersymmetric solutions of Supergravity theories that describe vacua, black
holes or topological defects, play a fundamental role in the progress of
ST, since they represent non-perturbative stable states that can be trusted beyond the Supergravity approximation, and that, if supersymmetric enough, can be taken as exact states of ST \cite{Kallosh:1998qs,Meessen:2007ef}. Supersymmetric solutions have also played an important role in mathematics, and, in particular in differential geometry, since the classification of supersymmetric solutions is closely related to the classification of manifolds with special holonomy, given that a supersymmetric solution has at least a globally defined section of the corresponding spin bundle, and therefore has reduced \emph{generalized holonomy}. Therefore, great effort has been devoted in the last decades to study and classify as  many supersymmetric solutions as possible.

In his seminal work \cite{Tod:1983pm}, Tod showed that the Killing spinor equations, together with the corresponding integrability conditions, could be used to systematically classify all the supersymmetric solutions of a given Supergravity theory. He used the Newman-Penrose formalism and focused on pure $\mathcal{N}=2$ Supergravity, following \cite{Gibbons:1982fy}. Since the Newman-Penrose formalism can only be applied in four dimensions, new techniques had to be developed in order to deal with higher dimensional cases. One of the techniques developed was the spinor-bilinear method \cite{Gauntlett:2002nw}, which was used to classify all the supersymmetric solutions of minimal five-dimensiona Supergravity. This result was soon extended to the Abelian gauged case \cite{Gauntlett:2003fk}, to general matter contents and couplings \cite{Bellorin:2006yr} and to other Supergravities \cite{Caldarelli:2003pb,Gutowski:2003rg,Cariglia:2004qi,
Gauntlett:2002fz,Cacciatori:2004rt,
Bellorin:2005zc,Meessen:2006tu,Huebscher:2007hj,Gran:2008vx,Ortin:2008wj,Hubscher:2008yz,Klemm:2009uw,Klemm:2010mc,Deger:2010rb}.

Another approach, closer to the geometrical properties of having a spin manifold with global sections (it exploits the fact that a global spinor defines a \emph{G structure}) was developed in \cite{Gauntlett:2002nw,Gauntlett:2002fz,Gauntlett:2002sc}.  Finally, we can mention yet another
approach, that can be used to find black-hole solutions of four-dimensional
theories, which exploits the symmetries of the time-like dimensionally-reduced theories which become a non-linear $\sigma$-model coupled to 3-dimensional gravity \cite{Breitenlohner:1987dg,Bergshoeff:2008be,Bossard:2009at,Chemissany:2009hq,Chemissany:2010ay,Chemissany:2010zp,Chemissany:2012nb}.

%%%%%%%%%%%%%%%%%%%%%%%%%%%%%%%%%%%%%%%%%%%%%%%%%%%%%%%%%%%%%%%%%%%%%%%%%%%%%%%%%%%%%%%%%%%%%%%%%%%
%%%%%%%%%%%%%%%%%%%%%%%%%%%%%%%%%%%%%%%%%%%%%%%%%%%%%%%%%%%%%%%%%%%%%%%%%%%%%%%%%%%%%%%%%%%%%%%%%%%

\section{Outline of the thesis}

%%%%%%%%%%%%%%%%%%%%%%%%%%%%%%%%%%%%%%%%%%%%%%%%%%%%%%%%%%%%%%%%%%%%%%%%%%%%%%%%%%%%%%%%%%%%%%%%%%%
%%%%%%%%%%%%%%%%%%%%%%%%%%%%%%%%%%%%%%%%%%%%%%%%%%%%%%%%%%%%%%%%%%%%%%%%%%%%%%%%%%%%%%%%%%%%%%%%%%%

The outline of this thesis goes as follows: on chapter \ref{chapter:mathpreliminaries} we introduce the relevant mathematical background to formulate extended ungauged Supergravity\footnote{For the case of $\mathcal{N}=2$ Supergravity, in the absence of hypermultiplets.} in four-dimensions, that is, Special K\"ahler Geometry and homogeneous spaces. On chapter \ref{chapter:sugra} we briefly comment on the structure of four-dimensional extended ungauged Supergravity relevant for black-hole solutions. On chapter \ref{chapter:sugrabhs}, based on \cite{Ortin:2012mt}, we characterize the most general spherically symmetric and static black-hole solution of ungauged Supergravity, and use the result to study the hidden conformal symmetries of Supergravity black holes, obtaining the full Virasoro algebra of the dual conformal field theory. On chapter \ref{chapter:susysolutions}, based on \cite{Ortin:2012gg}, we obtain all the supersymmetric black-hole solutions of extended Supergravity by means of the algorithm provided in \cite{Meessen:2010fh}. On chapter \ref{chapter:HFGK}, based on \cite{Meessen:2011aa,Galli:2012pt}, we introduce the \emph{H-F.G.K. formalism}, which simplifies the construction of non-supersymmetric black-hole solutions in $\mathcal{N}=2$ Supergravity. On chapter \ref{chapter:quantumbhs}, based on \cite{Bueno:2012jc,Bueno:2013psa}, we apply the H-F.G.K. formalism to a class of theories corresponding to Type-IIA String Theory compactified on a Calabi-Yau (C.Y.) threefold, obtaining the so-called \emph{quantum} black holes, which only exist when certain quantum corrections (perturbative or non-perturbative, depending on the solution) are included in the prepotential. For the case of non-perturbative black holes we elaborate on the potential consequences of the appearance on the solution of multi-valued functions in relation to the no-hair theorem for four-dimensional black holes. This thesis is based on \cite{Ortin:2012mt,Meessen:2010fh,Ortin:2012gg,Galli:2012pt,Meessen:2011aa,
Bueno:2012jc,Bueno:2013psa}. Other works finished during my doctoral studies are \cite{Huebscher:2010ib,Galli:2011fq,deAntonioMartin:2012bi,Chemissany:2012pf,Bueno:2012sd,
Bueno:2012vx,Bergshoeff:2013pia,delaCruz-Dombriz:2013dha}.
%\cleardoublepage

%%%%%%%%%%%%%%%%%%%%%%%%%%%%%%%%%%%%%%%%%%%%%%%%%%%%%%%%%%%%%%%%%%%%%%%
%%% CHAPTER 2 MATHEMATICAL PRELIMINARIES
%%%%%%%%%%%%%%%%%%%%%%%%%%%%%%%%%%%%%%%%%%%%%%%%%%%%%%%%%%%%%%%%%%%%%%%
\renewcommand{\chaptername}{Chapter}

\renewcommand{\leftmark}{\MakeUppercase{Chapter \thechapter. Mathematical preliminaries}}
\chapter{Mathematical preliminaries}
\label{chapter:mathpreliminaries}

Extended four-dimensional Supergravity, as a classical field theory, admits an elegant \emph{geometric} formulation in the following sense: the lagrangian can be constructed from geometrical structures (such as sections) of particular manifolds that are naturally associated to the theory. Therefore, extended four-dimensional Supergravity can be completely specified by some \emph{geometrical data}, that is, particular fibre bundles and global sections thereof.

In fact, such geometric formulation is not just an elegant mathematical tool to describe Supergravity, it is also useful in a wide range of applications of Supergravity, for instance to relate Supergravity to ST compactifications or to provide the general formalism to deal with gaugings. In this thesis we will use it to simplify the task of obtaining black hole solutions: using the mathematical structure of the theory we can introduce a new set of variables which considerably eases the construction of non-supersymmetric black hole solutions. 

As far as Quantum maximally and half-maximally extended Supergravity \cite{Bern:2007hh,Bern:2009kd,Bern:2012cd} is concerned, the mathematical structure of the theory at the classical level is also of outermost importance: for example, in Refs. \cite{Kallosh:2011qt,Kallosh:2011dp} the symplectic action of the $E_{7(7)}$ group was key in order to explain the conjectured finiteness of $\mathcal{N}=8$ ungauged Supergravity at all orders in perturbation theory.

Before dealing with the geometric formulation of four-dimensional extended Supergravity\footnote{See chapter \ref{chapter:sugra}.}, it is therefore necessary to introduce the mathematical background that will play a relevant role in the construction of the theories, namely, Special K\"ahler geometry for $\mathcal{N}=2$ Supergravity (in the absence of Hypermultiplets) and \emph{Irreducible Riemannian Globally Symmetric} (I.R.G.S.) spaces for $\mathcal{N}>2$ Supergravity. That is the goal of this chapter.

%%%%%%%%%%%%%%%%%%%%%%%%%%%%%%%%%%%%%%%%%%%%%%%%%%%%%%%%%%%%%%%%%%%%%%
%%%%%%%%%%%%%%%%%%%%%%%%%%%%%%%%%%%%%%%%%%%%%%%%%%%%%%%%%%%%%%%%%%%%%%
%%%%%%%%%%%%%%%%%%%%%%%%%%%%%%%%%%%%%%%%%%%%%%%%%%%%%%%%%%%%%%%%%%%%%%
%%%%%%%%%%%%%%%%%%%%%%%%%%%%%%%%%%%%%%%%%%%%%%%%%%%%%%%%%%%%%%%%%%%%%%

\section{Special K\"ahler Geometry}

%%%%%%%%%%%%%%%%%%%%%%%%%%%%%%%%%%%%%%%%%%%%%%%%%%%%%%%%%%%%%%%%%%%%%%
%%%%%%%%%%%%%%%%%%%%%%%%%%%%%%%%%%%%%%%%%%%%%%%%%%%%%%%%%%%%%%%%%%%%%%
%%%%%%%%%%%%%%%%%%%%%%%%%%%%%%%%%%%%%%%%%%%%%%%%%%%%%%%%%%%%%%%%%%%%%%
%%%%%%%%%%%%%%%%%%%%%%%%%%%%%%%%%%%%%%%%%%%%%%%%%%%%%%%%%%%%%%%%%%%%%%

Some basic references for this section are \cite{Ceresole:1995ca,Ceresole:1995jg,Craps:1997gp,Freed:1997dp,Ana,KN,Moroianu}. See the appendices of \cite{Meessen:2006tu,Huebscher:2006mr,Meessen:2012sr}, for an extremely well written short review of Special K\"ahler Geometry and its relation to $\mathcal{N}=2$ Supergravity coupled to vector multiplets and its gaugings. The definition of Special K\"ahler manifold was made in ~\cite{Strominger:1990pd}, formalizing the original results of~\cite{deWit:1983rz}. We will follow \cite{Ana,Meessen:2006tu,Moroianu}.

A Special K\"ahler manifold is a particular instance of real differentiable manifold $\mathcal{M}$, that is, it is a differentiable manifold with some extra-structure defined on it. Therefore, we will proceed defining step-by-step all the necessary ingredients until we arrive to a Special K\"ahler manifold, By a real differentiable manifold $\mathcal{M}$ we mean a Hausdorff and second countable topological space equipped with a differentiable structure (therefore it is paracompact and metrizable).

\definition{\label{def:rmanifold} A differentiable Riemannian manifold $(\mathcal{M},\mathcal{G})$ is a differentiable real manifold equipped with a smooth,non-degenerate, point-wise positive definite, global section $g$ of $S^2T^{*}\mathcal{M} $.}

\definition{\label{def:smanifold}  A Symplectic manifold $(\mathcal{M},\omega)$ is a differentiable manifold real equipped with a smooth, non-degenerate , global section $\omega$ of $\Lambda^2T^{*}\mathcal{M} $.}

\definition{An almost-complex structure $\mathcal{J}:T\mathcal{M} \rightarrow T\mathcal{M}$ on a tangent bundle $T\mathcal{M}$ is a bundle endomorphism such that $\mathcal{J}^2=-1$.}

\definition{\label{def:acmanifold} A differentiable almost-complex manifold $(\mathcal{M},\mathcal{J})$ is a differentiable $2n$-dimensional real manifold equipped with an almost-complex structure.}

\noindent
\newline Let $(\mathcal{M},\mathcal{J})$ be an almost-complex manifold. The complexified tangent bundle of $\mathcal{M}$ is the bundle $T\mathcal{M}\otimes\mathbb{C}\rightarrow \mathcal{M}$ with fibre $\left(T\mathcal{M}\otimes\mathbb{C}\right)_{p} = T\mathcal{M}_{p}\otimes\mathbb{C}$ at each $p\in\mathcal{M}$. If $T_{p}\mathcal{M}$ is a real $2n$-dimensional vector space, then $T_{p}\mathcal{M}\otimes \mathbb{C}$ is a complex $2n$-dimensional vector space.

We can extend linearly $\mathcal{J}$ on $T\mathcal{M}\otimes\mathbb{C}$ as follows

\begin{equation}
\mathcal{J}( v\otimes c)=\mathcal{J}v\otimes c\, ,\qquad v\in T\mathcal{M},~c\in \mathbb{C}\, .
\end{equation}

\noindent
Since $\mathcal{J}^2=-1$, $\mathcal{J}_{p}$ acting on $T_{p}\mathcal{M}\otimes\mathbb{C}$ has eigenvalues $\pm i$. We define

\begin{eqnarray}
T_{(1,0)}\mathcal{M} &=&\left\{v\in T\mathcal{M}\otimes \mathbb{C} ~~|~~ \mathcal{J}v = iv\, ,\forall ~ v\in T\mathcal{M}\otimes\mathbb{C} \right\}\, ,\nonumber\\
T_{(0,1)}\mathcal{M} &=&\left\{v\in T\mathcal{M}\otimes \mathbb{C} ~~| ~~\mathcal{J}v = -iv\, ,\forall ~ v\in T\mathcal{M}\otimes\mathbb{C}\right\}\, .
\end{eqnarray}

\noindent
Since 

\begin{eqnarray}
\pi_{(1,0)}: T\mathcal{M}\otimes\mathbb{C} &\rightarrow & T_{(1,0)}\mathcal{M}\nonumber\\ 
                 v  &\rightarrow &\frac{1}{2}\left(v\otimes 1 -\mathcal{J} v\otimes i\right) \, , 
\end{eqnarray}

\begin{eqnarray}
\pi_{(0,1)}: T\mathcal{M}\otimes\mathbb{C} &\rightarrow & T_{(0,1)}\mathcal{M}\nonumber\\ 
                 v  &\rightarrow &\frac{1}{2}\left(v\otimes 1 +\mathcal{J} v\otimes i\right) \, , 
\end{eqnarray}

\noindent
are a real bundle isomorphisms such that $ \pi_{(1,0)}\circ \mathcal{J} = -i\pi_{(0,1)}$, we have

\begin{eqnarray}
& & T\mathcal{M}\cong T_{(1,0)}\mathcal{M} \cong \overline{T_{(0,1)}\mathcal{M}}\, ,\nonumber\\
& &\\ 
\left(\pi_{(1,0)},\pi_{(0,1)}\right)&:& T\mathcal{M}\otimes\mathbb{C} \xrightarrow{\cong} T_{(1,0)}\mathcal{M} \oplus T_{(0,1)}\mathcal{M}\, .
\end{eqnarray}

\noindent
Analogously, for the complexified cotangent bundle $T^{\ast}\mathcal{M}\otimes\mathbb{C}\rightarrow \mathcal{M}$ we can conclude

 \begin{eqnarray}
& & T^{\ast}\mathcal{M}\cong T^{(1,0)}\mathcal{M} \cong T^{(0,1)}\mathcal{M}\, ,\nonumber\\
& &\\ 
\left(\pi^{(1,0)},\pi^{(0,1)}\right)&:& T^{\ast}\mathcal{M}\otimes\mathbb{C} \xrightarrow{\cong} T^{(1,0)}\mathcal{M} \oplus T^{(0,1)}\mathcal{M}\, .
\end{eqnarray}

\noindent
where 

\begin{eqnarray}
T^{(1,0)}\mathcal{M} &=&\left\{\alpha\in T^{\ast}\mathcal{M}\otimes \mathbb{C} ~~|~~ \alpha\left(\mathcal{J}v\right) = i\alpha\left(v\right)\, , \forall ~ v\in T\mathcal{M}\otimes\mathbb{C} \right\}\, ,\nonumber\\
T^{(0,1)}\mathcal{M} &=&\left\{v\alpha\in T^{\ast}\mathcal{M}\otimes \mathbb{C} ~~|~~ \alpha\left(\mathcal{J}v\right) = -i\alpha\left(v\right)\, ,\forall ~ v\in T\mathcal{M}\otimes\mathbb{C}\right\}\, ,
\end{eqnarray}

\noindent
and we have defined the natural projections $\pi^{(1,0)}$ and $\pi^{(0.1)}$ of the complexified cotangent bundle as follows

\begin{eqnarray}
\pi^{(1,0)}: T^{\ast}\mathcal{M}\otimes\mathbb{C} &\rightarrow & T^{(1,0)}\mathcal{M}\nonumber\\ 
                 \alpha  &\rightarrow &\frac{1}{2}\left(\alpha\otimes 1 - \alpha\otimes i\circ\mathcal{J} \right) \, , 
\end{eqnarray}

\begin{eqnarray}
\pi^{(0,1)}: T^{\ast}\mathcal{M}\otimes\mathbb{C} &\rightarrow & T^{(0,1)}\mathcal{M}\nonumber\\ 
                 \alpha  &\rightarrow &\frac{1}{2}\left(\alpha\otimes 1 +\alpha\otimes i\circ\mathcal{J} \right) \, , 
\end{eqnarray}

\noindent
We are going to elucidate the structure of the space of forms on $(\mathcal{M},\mathcal{J})$, which, since $(\mathcal{M},\mathcal{J})$ is almost-complex, is going to be constructed from sections of $T^{\ast}\mathcal{M}\otimes\mathbb{C}$ and its exterior powers, and not from $T^{\ast}\mathcal{M}$. The main reason is that $\mathcal{J}$ can be diagonalized on $T^{\ast}\mathcal{M}\otimes\mathbb{C}$ but not on $T^{\ast}\mathcal{M}$. For an almost-complex manifold $\left(\mathcal{M},\mathcal{J}\right)$ let

\begin{equation}
\Omega^{k}\left(\mathcal{M},\mathbb{C}\right)\equiv\Gamma\left(\Lambda^{k}\left(T^{\ast}\mathcal{M}\otimes\mathbb{C}\right)\right)\, ,
\end{equation}

\noindent
where $\Gamma(\cdot)$ stands for the space of sections of the corresponding fibre bundle $\cdot$ and

\begin{equation}
\Lambda^{k}\left(T^{\ast}\mathcal{M}\otimes\mathbb{C}\right) =
\Lambda^{k}\left(T^{(0,1)}\mathcal{M}\oplus T^{(1,0)}\mathcal{M}\right) =
\bigoplus_{l+m=k}\Lambda^{l}\left(T^{(0,1)}\mathcal{M}\right)\wedge\Lambda^{m}\left(T^{(1,0)}\mathcal{M}\right)
\, .
\end{equation}

\noindent
\definition{\label{def:forms} The differential forms of type $(l,m)$ on an almost-complex manifold $(\mathcal{M},\mathcal{J})$ are the sections of $\bigoplus_{l+m=k}\Lambda^{l}\left(T^{(0,1)}\mathcal{M}\right)\wedge\Lambda^{m}\left(T^{(1,0)}\mathcal{M}\right)$
.}

\noindent
\newline It is convenient to define

\begin{equation}
\Omega^{(l,m)}\left(\mathcal{M},\mathbb{C}\right)=\Gamma\left( \Lambda^{l}\left(T^{(0,1)}\mathcal{M}\right)\wedge\Lambda^{m}\left(T^{(1,0)}\mathcal{M}\right)\right)
\, ,
\end{equation}

\noindent
and hence

\begin{equation}
\Omega^{k}\left(\mathcal{M},\mathbb{C}\right) = \bigoplus_{l+m=k}\Omega^{(l,m)}\left(\mathcal{M},\mathbb{C}\right)\, .
\end{equation}

\noindent
Therefore, when speaking of tensors on a complex manifold, it is generally referred to sections of the complexified tangent or cotangent bundle and its tensorial products. Let $\pi^{(l,m)}$ be the natural projection  $\pi^{(l,m)}:\Lambda^{k}\left(T^{\ast}\mathcal{M}\otimes\mathbb{C}\right) \rightarrow \Lambda^{l}\left(T^{(0,1)}\mathcal{M}\right)\wedge\Lambda^{m}\left(T^{(1,0)}\mathcal{M}\right)$. We define then

\begin{eqnarray}
&\partial & \equiv \pi^{(l+1,m)}\circ d : \Omega^{(l,m)}(\mathcal{M},\mathbb{C})\rightarrow \Omega^{(l+1,m)}(\mathcal{M},\mathbb{C})\, ,\nonumber\\
&\bar{\partial} & \equiv \pi^{(l,m+1)}\circ d : \Omega^{(l,m)}(\mathcal{M},\mathbb{C})\rightarrow \Omega^{(l,m+1)}(\mathcal{M},\mathbb{C})\, ,
\end{eqnarray} 

\noindent
which are differential operators that act on forms of type $(m,l)$. When considering complex manifolds, $\partial$ and $\bar{\partial}$ will become natural differential operators in terms of the complex coordinates of a given chart.

It is possible to define in a compatible, natural, way, an almost-complex structure $\mathcal{J}$ on a Symplectic manifold, which makes it also Riemannian.

\definition{\label{def:compatiblecomplex} Let $(\mathcal{M},\omega )$ be a Symplectic manifold. An almost-complex structure $\mathcal{J}$ is called compatible if $\mathcal{G}\left(\cdot,\cdot \right) =\omega\left(\cdot,\mathcal{J}\cdot\right)$ is a Riemannian metric on $\mathcal{M}$. The triple $\left(\omega,\mathcal{G},\mathcal{J} \right)$ is then called a compatible triple.}

For a compatible triple $\left(\omega,\mathcal{G},\mathcal{J} \right)$ we have that

\begin{equation}
\mathcal{G}(\mathcal{J}\cdot,\mathcal{J}\cdot) = \omega(\mathcal{J}\cdot,\mathcal{J}^2\cdot) = \omega(\cdot,\mathcal{J}\cdot) = \mathcal{G}(\cdot,\cdot)\, ,
\end{equation}

\noindent
that is, $\mathcal{J}$ preserves the Riemannian metric $G$.

\definition{\label{def:complexs} An almost complex structure $\mathcal{J}$ is integrable if and only if $N(u,v)\equiv [\mathcal{J}u,\mathcal{J}v] -\mathcal{J}[u,\mathcal{J}v] -\mathcal{J}[\mathcal{J}u,v]-[u,v]=0\, \forall u,v\in T\mathcal{M}$.}

\noindent
\newline $N(\cdot,\cdot)$ is the Nijenhuis tensor, and intuitively it parametrizes the obstruction to the possibility of defining holomorphic changes of coordinates in $\mathcal{M}$. When it vanishes, it is possible to construct an holomorphic atlas and therefore $\mathcal{M}$ is a complex manifold.

\definition{\label{def:cmanifold} A differentiable complex manifold $(\mathcal{M},\mathcal{J})$ is a differentiable $2n$-dimensional real manifold equipped with an integrable complex structure.}

\noindent
\newline Let $\mathcal{U}\subset\mathcal{M}$ a chart on a complex manifold $(\mathcal{M},\mathcal{J})$ with coordinates $z^{i}=x^{i}+iy^{i}$ and real coordinates $(x^{i},y^{i})$. At any $p\in\mathcal{M}$ we have

\begin{eqnarray}
T^{\ast}_{p}\mathcal{M} &=& \mathbb{R} {\rm -Span}\left[ dx^{i}, dy^{i}\right]_{p}\nonumber\, ,\\
T^{\ast}_{p}\mathcal{M}\otimes\mathbb{C} &=& \mathbb{C} {\rm -Span}\left[ dx^{i}, dy^{i}\right]_{p}\, .
\end{eqnarray}

\noindent
Then, defining $dz^{i} = dx^{i}+i dy^{i}$ and $d\bar{z}^{\bar{i}} = dx^{i}-i dy^{i}$ we can write

\begin{eqnarray}
T^{\ast}_{p}\mathcal{M}\otimes\mathbb{C} &=& \mathbb{C} {\rm -Span}\left[ dz^{i}\right]_{p} \oplus  \mathbb{C} {\rm -Span}\left[ d\bar{z}^{\bar{i}}\right]_{p}=T^{(1,0)}\mathcal{M} \oplus T^{(0,1)}\mathcal{M}\, ,
\end{eqnarray}

\noindent
since $dz^{i}\circ\mathcal{J}= idz^{i}$ and $d\bar{z}^{\bar{i}}\circ\mathcal{J}= -id\bar{z}^{\bar{i}}$. Therefore, a $(1,0)$-form $\alpha$ is written as $\alpha = \alpha_{i} dz^{i}$ and $(0,1)$-form $\beta$ is written as $\beta = \beta_{\bar{i}} d\bar{z}^{i}$. Since the manifold $(\mathcal{M},\mathcal{J})$ is complex, we can consistently define complex coordinates coordinates $\tilde{z}^{i}$ (this may fail in an almost-complex manifold) such that $d\tilde{z}^{i}=dz^{i}$. Of course, we will drop the tilde and call them $z^{i}$. An similar derivation allows to conclude that

\begin{equation}
\frac{\partial}{\partial z^{i}} = \frac{1}{2}\left(\frac{\partial}{\partial x^{i}} - i\frac{\partial}{\partial y^{i}} \right)\, ,\qquad \frac{\partial}{\partial \bar{z}^{\bar{i}}} = \frac{1}{2}\left(\frac{\partial}{\partial x^{i}} + i\frac{\partial}{\partial y^{i}} \right)\, ,
\end{equation}

\noindent
where $\frac{\partial}{\partial z^{i}}$ spans $T_{(1,0)}\mathcal{M}$ and $ \frac{\partial}{\partial \bar{z}^{\bar{i}}}$ spans $T_{(0,1)}\mathcal{M}$.

It can be also proven that the exterior derivative $d$ can be written in terms of $\partial$ and $\bar{\partial}$ as $d=\partial+\bar{\partial}$, where, given a function $f\in C^{\infty}(\mathcal{M},\mathbb{C})$ $\partial$ and $\bar{\partial}$ act as follows

\begin{equation}
\partial f = \frac{\partial f}{\partial z^{i}}dz^{i}\, ,\qquad \bar{\partial} f = \frac{\partial f}{\partial \bar{z}^{\bar{i}}} d\bar{z}^{\bar{i}}\, .
\end{equation}

\noindent
We are now ready to define K\"ahler manifolds:

\definition{\label{def:kmanifold}A K\"ahler manifold is a Symplectic manifold $(\mathcal{M},\omega)$, equipped with an integrable compatible almost complex structure $\mathcal{J}$. The symplectic form $\omega$ is then called the K\"ahler form.}

Since $\omega$ and $\mathcal{J}$ are compatible, they define a Riemannian metric: $\mathcal{G}(\cdot , \cdot)=\omega(\cdot , \mathcal{J} \cdot)$. Therefore, a K\"ahler manifold is a differentiable manifold which is Symplectic, Riemannian, and Complex in a compatible way, and therefore incorporates together the three basics types of geometry: Symplectic, Riemannian and Complex \footnote{K\"ahler manifolds are of outermost importance in Theoretical Physics: they appear for example in every supersymmetric non-linear $\sigma$ model, String Theory compactifications...}.

Since a K\"ahler manifold has an integrable complex structure, it immediately follows that a K\"ahler manifold is a complex manifold. As a consequence, tensors on a K\"ahler manifold are sections of the complexified tangent and cotangent bundles and their tensorial products. In addition we have the following propostion

\prop{\label{prop:Jsymp}Let $(\mathcal{M},\omega,\mathcal{J})$ a K\"ahler manifold, then $\mathcal{J}$ is a symplectomorphism, that is, $\mathcal{J}^{\ast}=\mathcal{J}$.}

\proof{
\begin{equation}
\mathcal{J}^{\ast}_{p}\omega_{p}(u,v) = \omega_{p}(\mathcal{J}_{p}u,\mathcal{J}_{p}v) = 
\mathcal{G}_{p}(v,\mathcal{J}_{p}u)= \omega_{p} (u,v)\, ,\qquad \forall p\in\mathcal{M}~~ {\rm and}~~ u,v\in T_{p}\mathcal{M}\, .
\end{equation}}

\noindent
In fact, propostion \ref{prop:Jsymp} holds also when $\omega$ is not closed, as long as $\left(\omega, \mathcal{G}, \mathcal{J}\right)$is a compatible triple. Since $\omega$ is a two-form, in principle $\omega\in\Omega^2\left(\mathcal{M},\mathbb{C}\right)$, that is, it is a section of the exterior product of two copies of the complexified cotangent bundle $T\mathcal{M}\otimes\mathbb{C}$. Therefore

\begin{equation}
\omega\in \Omega^2\left(\mathcal{M},\mathbb{C}\right) = \Omega^{(2,0)}\left(\mathcal{M},\mathbb{C}\right)\oplus\Omega^{(1,1)}\left(\mathcal{M},\mathbb{C}\right)\oplus\Omega^{(0,2)}\left(\mathcal{M},\mathbb{C}\right)\, .
\end{equation}

\noindent
However, imposing proposition \ref{prop:Jsymp} in a local chart $\left(\mathcal{U},z^{i}\right)$, one can see that 

\begin{equation}
\omega\in \Omega^{(1,1)}\left(\mathcal{M},\mathbb{C}\right)\, ,
\end{equation}

\noindent
\emph{i.e.} $\omega$ defines a Dolbeault $(1,1)$-cohomology class $[\omega]\in H^{(1,1)}_{{\rm Dolbeault}}(\mathcal{M})$. In other words, the local expression for $\omega$ does not contain terms of the form $dz^{i}\wedge dz^{j}$ or $d\bar{z}^{\bar{i}}\wedge d\bar{z}^{\bar{j}}$. Therefore, in a local chart $\left(\mathcal{U},z^{i}\right)$, we can write $\omega$ as follows

\begin{equation}
\omega= i h_{i\bar{j}} dz^{i}\wedge d\bar{z}^{\bar{j}}\, ,
\end{equation}

\noindent
where $h_{i\bar{j}}\in C^{\infty}\left(\mathcal{U},\mathbb{C}\right)$ and the $i$ has been introduced by convenience. Using that $\mathcal{G}(\cdot , \cdot)=\omega(\cdot , \mathcal{J} \cdot)$, we can write $\mathcal{G}$ in the same local chart as 

\begin{equation}
\mathcal{G} =  h_{i\bar{j}} \left( dz^{i}\otimes d\bar{z}^{\bar{j}}+ dz^{\bar{j}}\otimes d\bar{z}^{i}\right)\, .
\end{equation}

\noindent
Since a K\"ahler manifold is in particular Symplectic, the symplectic form $\omega$ must be real, that is $\omega=\bar{\omega}$, which is equivalent to $h_{ij}=\bar{h_{ji}}$. For the same reason,  $\omega$ is non-degenerate and its $n$-th exterior power defines a volume form on $\left(\mathcal{M},\omega,\mathcal{J}\right)$, making $\left(\mathcal{M},\omega,\mathcal{J}\right)$ orientable. To summarize, the K\"ahler form $\omega$ is a two-form compatible with the complex structure, closed, real valued and non-degenerate.  

\thm{\label{thm:K}Let $\omega$ be a closed real-valued $(1,1)$-form on a complex manifold $\mathcal{M}_{{\rm Complex}}$ and let $p\in\mathcal{M}_{{\rm Complex}}$. Then, $\exists ~\mathcal{U}(p) ~{\rm and}~ \mathcal{K}\in C^{\infty}\left(\mathcal{U},\mathbb{R}\right)$ such that, on  $\mathcal{U}(p)$, $\omega=i\partial\bar{\partial}\mathcal{K}$.}
\proof{See proposition 8.8 in \cite{Moroianu}.}

\noindent
If we apply theorem \ref{thm:K} to a K\"ahler manifold, which is in particular complex, we obtain

\begin{equation}
\label{eq:kahlermetric}
\mathcal{G} = \partial_{i}\partial_{\bar{j}}\mathcal{K}\left( dz^{i}\otimes d\bar{z}^{\bar{j}}+ dz^{\bar{j}}\otimes d\bar{z}^{i}\right)
\end{equation}

\noindent
and $\mathcal{K}$ is called the \emph{K\"ahler potential}. The metric therefore can be obtained from a single function, which allows to write simple expressions for the Levi-Civit\`a connection and the curvature tensors. Remarkably enough, the existence of normal holomorphic coordinates on a complex manifold around each point is equivalent to the metric being K\"ahler. The Levi-Civit\`a connection on a K\"ahler manifold is given by

\begin{equation}
\label{eq:KCCChrisSymb}
\Gamma_{jk}{}^{i} = 
\mathcal{G}^{i\bar{i}}\partial_{j}\mathcal{G}_{\bar{i}k}\, ,
\hspace{1cm}
\Gamma_{\bar{j}\bar{k}}{}^{\bar{i}}  = 
\mathcal{G}^{\bar{i}i}\partial_{\bar{j}}\mathcal{G}_{\bar{k}i} \, .\end{equation}

\noindent
The Riemann curvature tensor has as only non-vanishing components
$R_{i\bar{j}k\bar{l}}$, but we will not need their explicit expression.  The Ricci
tensor is given by

\begin{equation}
\label{eq:KCCRicci}
R_{i\bar{i}}  = \partial_{i}\partial_{\bar{i}}
\left(\textstyle{1\over 2}\log\det\mathcal{G}\right) \, .
\end{equation}

\noindent
The K\"ahler potential is not unique: it is defined only up to \textit{K\"ahler transformations} of the form

\begin{equation}
\label{eq:Kpotentialtransformation}
\mathcal{K}^{\prime}=\mathcal{K}+f+\bar{f}\, , 
\end{equation}

\noindent
where $f$ is any holomorphic function on $\left(\mathcal{M},\omega,\mathcal{J}\right)$. We need to introduce now the concepts of holomorphic vector bundle, line-bundel and \emph{K\"ahler-Hodge} manifold

\definition{\label{def:hvb} An holomorphic vector bundle is a complex vector bundle over a complex manifold $\mathcal{M}$ such that the total space $E$ is a complex manifold and the projection map $\pi:E\to \mathcal{M}$ is holomorphic.}

\noindent
In particular, the transition functions of an holomorphic vector bundle are holomorphic.

\definition{\label{def:lb} A line bundle $\mathcal{L}\xrightarrow{\pi}\mathcal{M}$ is a holomorphic vector bundle of rank $r=1$.}

\definition{\label{def:kh} A K\"ahler manifold $\left(\mathcal{M},\omega,\mathcal{J}\right)$ is Hodge if and only if there exists a line bundle $\mathcal{L}\xrightarrow{\pi}\mathcal{M}$ such that $c_1\left(\mathcal{L}\right)=\left[\omega\right]$, where $c_1\left(\mathcal{L}\right)$ denotes the first Chern class of $\mathcal{L}$.}

\noindent
\newline The condition on the Chern class is rather abstract, and we will not discuss it here. Intuitively, what is required in definition \ref{def:kh} is that the curvature of the line-bundle bundle is equal, up to an exact form, to the K\"ahler form $\omega$.

Let us define now, motivated by the structure of $\mathcal{N}=1$ Supergravity, a family of (complex) rank one bundles $\left\{\mathcal{L}^{(q,\bar{q})}\, , q\in\mathbb{R}\, , \bar{q}\in\mathbb{R}\right\}$ over a K\"ahler manifold $\left(\mathcal{M},\omega,\mathcal{J}\right)$ such that, given a section $W\in\Gamma\left(\mathcal{L}^{(q,\bar{q})} \right)$, on the overlap of two patches $\mathcal{U}_{(\alpha)}$ and $\mathcal{U}_{(\beta)}$, $W$ and $\mathcal{K}$ are related as follows

\begin{equation}
W_{(\alpha)} = e^{-(qf+\bar{q}\bar{f})/2} W_{(\beta)}\, ,\qquad \mathcal{K}_{(\alpha)} = \mathcal{K}_{(\alpha)}+f+\bar{f}\, ,
\end{equation}

\noindent
and define the \textit{K\"ahler (connection) 1-form}  $\mathcal{Q}$ as

\begin{equation}
\label{eq:K1form}
\mathcal{Q} \equiv (2i)^{-1}(dz^{i}\partial_{i}\mathcal{K} -
d\bar{z}^{\bar{i}}\partial_{\bar{i}}\mathcal{K})\, ,
\end{equation}

\noindent
such that the covariant derivative on sections of $\mathcal{L}^{(q,\bar{q})}$ is given by

\begin{equation}
\label{eq:Kcovariantderivative}
\mathfrak{D}_{i} \equiv \nabla_{i} +iq \mathcal{Q}_{i}\, ,
\hspace{1cm}
\mathfrak{D}_{\bar{i}} \equiv \nabla_{\bar{i}} -i\bar{q} \mathcal{Q}_{\bar{i}}\, ,
\end{equation}

\noindent
where $\nabla$ is the standard covariant derivative associated to the Levi-Civit\`a connection on $\mathcal{M}$. The K\"ahler connection one-form, on the overlap of two patches $\mathcal{U}_{(\alpha)}$ and $\mathcal{U}_{(\beta)}$, is related as follows

\begin{equation}
\label{eq:K1formtransformation}
\mathcal{Q}_{(\alpha)} =\mathcal{Q}_{(\beta)} 
-{\textstyle\frac{i}{2}}\partial f\, .
\end{equation}

\noindent    
Then, for $q=1,~\bar{q}=0$, $W\in\Gamma\left(\mathcal{L}^{(1,0)} \right)$ is the superpotential of $\mathcal{N}=1$ Supergravity and $\mathcal{L}^{(1,0)}$ is a line-bundle which is also K\"ahler-Hodge in the sense of definiton \ref{def:kh}. That is, the curvature of the line-bundle, obtaining by computing $[\mathfrak{D}_{i},\mathfrak{D}_{j}]$ is equal to the $\omega$ up to an exact form, which in this case is zero. Other objects of $\mathcal{N}=1$ Supergravity correspond to sections of $\mathcal{L}^{(q,\bar{q})}$ for other values of $q$ and $\bar{q}$. 

A K\"ahler-Hodge manifold provides the formal starting point for the definition of a Special K\"ahler manifold: Special K\"ahler Geometry appears in the scalar manifold corresponding to the vector multiplets of $\mathcal{N}=2$ Supergravity. Since $\mathcal{N}=2$ supersymmetry includes $\mathcal{N}=1$ supersymmetry, we expect Special K\"ahler manifolds to be also K\"ahler-Hodge, equipped with some extra structure, which is indeed the case. Before giving the definition of Special K\"ahler manifold, let us state that there are two kinds of Special K\"ahler geometry:  

\begin{enumerate}

\item Rigid Special K\"ahler Geometry $\rightarrow$ vector-multiplet scalar field sector of $\mathcal{N}=2$ Yang-Mills theories.

\item Local Special K\"ahler Geometry $\rightarrow$ vector-multiplet scalar field sector of $\mathcal{N}=2$ Supergravity theories.

\end{enumerate}
Here we will deal with the local case, since we are interested in Supergravity. In any case, the definition for the global case is similar to the local case, see \cite{Andrianopoli:1996cm}.

\definition{\label{def:SKG}A K\"ahler-Hodge manifold $\mathcal{L}\xrightarrow{\pi}\mathcal{M}$ is Special K\"ahler of the local type if there exists a bundle $\mathcal{S}\mathcal{V}=\mathcal{S}\mathcal{M}\otimes\mathcal{L}\xrightarrow{\pi}\mathcal{M}$ such that for some holomorphic section $\Omega \in\Gamma\left(\mathcal{S}\mathcal{V}\right)$ the K\"ahler 2-form is given by $\omega = i\partial\bar{\partial}\log \left(i\Omega_M\bar{\Omega}^M\right)$, where $\mathcal{S}\mathcal{M}\xrightarrow{\pi}\mathcal{M}$ is a flat $(2n_{v}+2)$-dimensional vector bundle with structure group
$\mathrm{Sp}(2n_{v}+2,\mathbb{R})$.}

\noindent
\newline $\Omega_M\bar{\Omega}^M = \langle \Omega \mid\bar{\Omega}\rangle $ denotes the hermitian and symplectic inner product of fibres.   ~$M,N,\cdots = 1\cdots 2n_{v}+2$ are called \emph{symplectic indices}, and can be \emph{decomposed} in two sets of indices $\Lambda,\Sigma\cdots=1,\cdots, n_{v}+1$ such that, for example, the section $\Omega$ would be written as $\Omega^M=(\mathcal{X}^{\Lambda},\mathcal{F}_{\Lambda})^{T}$, and therefore we have

\begin{equation}
\label{eq:SGDefFund2}
\Omega =
\left(
\begin{array}{c}
\mathcal{X}^{\Lambda}\\
\mathcal{F}_{\Sigma}\\
\end{array}
\right) 
\;\; \rightarrow \;\;
\left\{
\begin{array}{lcl}
\langle \Omega \mid\bar{\Omega}\rangle 
& \equiv & 
\bar{\mathcal{X}}^{\Lambda}\mathcal{F}_{\Lambda}  
-\mathcal{X}^{\Lambda}\bar{\mathcal{F}}_{\Lambda}
= -i\ e^{-\mathcal{K}}\, , \\
& & \\
\partial_{\bar{i}}\Omega & = & 0 \, ,\\
& & \\
\langle\partial_{i}\Omega\mid\Omega\rangle & = & 0 \; .
     \end{array}
  \right. 
\end{equation}

\noindent
$\Omega$ is everything we need to know in order to define $\mathcal{N}=2$ ungauged supergravity in the absence of hyper-multiplets: if we know $\Omega^M$ we can write the complete Lagrangian.

It is convenient to define a covariantly holomorphic symplectic section $\mathcal{V}=e^{\frac{\mathcal{K}}{2}}\Omega$\footnote{This is the section of a different vector bundle, which cannot be holomorphic since $\mathcal{V}$ is related in different patches through non-holomorphic transition functions.}, which therefore obeys

\begin{equation}
\label{eq:SGDefFund}
\mathcal{V} = 
\left(
\begin{array}{c}
\mathcal{L}^{\Lambda}\\
\mathcal{M}_{\Sigma}\\
\end{array}
\right) \;\; \rightarrow \;\;
\left\{
\begin{array}{lcl}
\langle \mathcal{V}\mid\bar{\mathcal{V}}\rangle 
& \equiv & 
\bar{\mathcal{L}}^{\Lambda}\mathcal{M}_{\Lambda} 
-\mathcal{L}^{\Lambda}\bar{\mathcal{M}}_{\Lambda}
= -i\, , \\
& & \\
\mathfrak{D}_{\bar{i}}\mathcal{V} & = & (\partial_{\bar{i}}+
{\textstyle\frac{1}{2}}\partial_{\bar{i}}\mathcal{K})\mathcal{V} =0 \, ,\\
& & \\
\langle\mathfrak{D}_{i}\mathcal{V}\mid\mathcal{V}\rangle & = & 0 \, .
\end{array}
\right. 
\end{equation}

\noindent
Notice if we define

\begin{equation}
\label{eq:SGDefU}
\mathcal{U}_{i}  \equiv \mathfrak{D}_{i}\mathcal{V}  
= 
\left(
\begin{array}{c}
f^{\Lambda}{}_{i}\\
h_{\scriptscriptstyle{\Sigma}\, i}
\end{array}
\right)\, ,\,\,\,\, 
\bar{\mathcal{U}}_{\bar{i}} = \overline{\mathcal{U}_{i}} \, ,
\end{equation}

\noindent
then it follows from the basic definitions that

\begin{equation}
\label{eq:SGProp1}
\begin{array}{rclrcl}
\mathfrak{D}_{\bar{i}}\ \mathcal{U}_{i} 
& = & 
\mathcal{G}_{i\bar{i}}\ \mathcal{V}\,\hspace{2cm} &
\langle\mathcal{U}_{i}\mid\bar{\mathcal{U}}_{\bar{i}}\rangle 
& = & 
i\mathcal{G}_{i\bar{i}} \, , \\
& & & & & \\
\langle\mathcal{U}_{i}\mid\bar{\mathcal{V}}\rangle &  = & 0\, , &
\langle\mathcal{U}_{i}\mid\mathcal{V}\rangle & = & 0 \, .\\
\end{array}
\end{equation}

Taking the covariant derivative of the last identity
$\langle\mathcal{U}_{i}\mid\mathcal{V}\rangle =0$ we find immediately that
$\langle\mathfrak{D}_{i}\mathcal{U}_{j}\mid\mathcal{V}\rangle = -\langle\ 
\mathcal{U}_{j}\mid \mathcal{U}_{i}\rangle$. It can be shown that the
r.h.s.~of this equation is antisymmetric while the l.h.s.~is symmetric, so that

\begin{equation}
\label{eq:UU}
\langle\mathfrak{D}_{i}\mathcal{U}_{j}\mid\mathcal{V}\rangle = \langle
\mathcal{U}_{j}\mid \mathcal{U}_{i}\rangle = 0\, .
\end{equation}

The importance of this last equation is that if we group together
$\mathcal{E}_{\Lambda} = (\mathcal{V},\mathcal{U}_{i})$, we can see that
$\langle \mathcal{E}_{\Sigma}\mid\bar{\mathcal{E}}_{\Lambda}\rangle$ is a
non-degenerate matrix. This then allows us to construct an identity operator
for the symplectic indices, such that for a given section of
$\mathcal{A}\ni\Gamma\left( E,\mathcal{M}\right)$ we have

\begin{equation}
\label{eq:SGSymplProj}
\mathcal{A} = i\langle\mathcal{A}\mid\bar{\mathcal{V}}\rangle \mathcal{V} 
-i\langle\mathcal{A}\mid\mathcal{V}\rangle\ \bar{\mathcal{V}}
+i\langle\mathcal{A}\mid\mathcal{U}_{i}\rangle\mathcal{G}^{i\bar{i}}\ 
\bar{\mathcal{U}}_{\bar{i}}
-i\langle\mathcal{A}\mid\bar{\mathcal{U}}_{\bar{i}}\rangle
\mathcal{G}^{i\bar{i}}\mathcal{U}_{i} \, .
\end{equation}

As we have seen $\mathfrak{D}_{i}\mathcal{U}_{j}$ is symmetric in $i$ and $j$,
but what more can be said about it: as one can easily see, the inner product
with $\bar{\mathcal{V}}$ and $\bar{\mathcal{U}}_{\bar{i}}$ vanishes due to the
basic properties. Let us then define the K\"ahler-weight 2 object

\begin{equation}
\label{eq:SGDefC}
\mathcal{C}_{ijk} \equiv 
\langle \mathfrak{D}_{i}\ \mathcal{U}_{j}\mid \mathcal{U}_{k}\rangle 
\;\; \rightarrow\;\; 
\mathfrak{D}_{i}\ \mathcal{U}_{j}  = 
i\mathcal{C}_{ijk}\mathcal{G}^{k\bar{l}}\bar{\mathcal{U}}_{\bar{l}} \, ,
\end{equation}

\noindent
where the last equation is a consequence of
Eq.~(\ref{eq:SGSymplProj}). Since the $\mathcal{U}$'s are orthogonal,
however, one can see that $\mathcal{C}$ is completely symmetric in its
3 indices. Furthermore one can show that

\begin{equation}
\label{eq:SGCProp}
\mathfrak{D}_{\bar{i}}\ \mathcal{C}_{jkl}  = 0\, ,\hspace{1cm}
\mathfrak{D}_{[i}\ \mathcal{C}_{j]kl} = 0\, .
\end{equation}

The period or monodromy matrix $\mathcal{N}$ is defined by the relations

\begin{equation}
\label{eq:SGDefN}
\mathcal{M}_{\Lambda}  = \mathcal{N}_{\Lambda\Sigma} \mathcal{L}^{\Sigma}\, ,
\hspace{1cm}
h_{\Lambda\, i}  = \bar{\mathcal{N}}_{\Lambda\Sigma} f^{\Sigma}{}_{i} \, .
\end{equation}

\noindent
The relation $\langle\mathcal{U}_{i}\mid\overline{\mathcal{V}}\rangle =0$
then implies that $\mathcal{N}$ is symmetric, which then also trivializes
$\langle\mathcal{U}_{i}\mid\mathcal{U}_{j}\rangle =0$.

From the other basic properties in (\ref{eq:SGProp1}) we find

\begin{eqnarray}
\mathcal{L}^{\Lambda} \Im{\rm m}\mathcal{N}_{\Lambda\Sigma}
\bar{\mathcal{L}}^{\Sigma} & = & -{\textstyle\frac{1}{2}}\, ,
\label{eq:esta}\\
& & \nonumber \\
\mathcal{L}^{\Lambda} \Im{\rm m}\mathcal{N}_{\Lambda\Sigma} f^{\Sigma}{}_{i} 
& = & 
\mathcal{L}^{\Lambda} \Im{\rm m}\mathcal{N}_{\Lambda\Sigma} 
\bar{f}^{\Sigma}{}_{\bar{i}} 
=0\, ,
\label{eq:esa}\\
& & \nonumber \\
f^{\Lambda}{}_{i}\ \Im{\rm m}\mathcal{N}_{\Lambda\Sigma}
\bar{f}^{\Sigma}{}_{\bar{i}} & = &   -\textstyle{1\over 2}\mathcal{G}_{i\bar{i}} \, .
\label{eq:esaotra}
\end{eqnarray}

Further identities that can be derived are

\begin{eqnarray}
\label{eq:SGColl1}
(\partial_{i}\mathcal{N}_{\Lambda\Sigma}) \mathcal{L}^{\Sigma} 
& = & 
-2i\Im{\rm m}(\mathcal{N})_{\Lambda\Sigma}\ f^{\Sigma}{}_{i} \, , \\
& & \nonumber \\
\label{eq:SGColl2}
\partial_{i}\bar{\mathcal{N}}_{\Lambda\Sigma}\ f^{\Sigma}{}_{j} 
& = & 
-2\mathcal{C}_{ijk}\mathcal{G}^{k\bar{k}}
\Im{\rm m}\mathcal{N}_{\Lambda\Sigma}
\bar{f}^{\Sigma}{}_{\bar{k}} \, ,\\
& & \nonumber \\
\label{eq:SGColl3}
\mathcal{C}_{ijk} 
& = & 
f^{\Lambda}{}_{i}f^{\Sigma}{}_{j} 
\partial_{k}\bar{\mathcal{N}}_{\Lambda\Sigma} \, ,\\
& & \nonumber \\
\label{eq:SGColl4}
\mathcal{L}^{\Sigma}\partial_{\bar{i}}\mathcal{N}_{\Lambda\Sigma} 
& = & 
0 \, , \\
& & \nonumber \\
\label{eq:SGColl5}
\partial_{\bar{i}}\bar{\mathcal{N}}_{\Lambda\Sigma}\ f^{\Sigma}{}_{i} 
& = & 
2i\mathcal{G}_{i\bar{i}}\Im{\rm m}\mathcal{N}_{\Lambda\Sigma}
 \mathcal{L}^{\Sigma} \, .
\end{eqnarray}

An important identity one can derive, and that will be used various times
in the main text, is given by

\begin{equation}
\label{eq:SGImpId}
U^{\Lambda\Sigma} \equiv  f^{\Lambda}{}_{i}\mathcal{G}^{i\bar{i}}
\bar{f}^{\Sigma}{}_{\bar{i}}
=
-\textstyle{1\over 2}\Im{\rm m}(\mathcal{N})^{-1|\Lambda\Sigma}
-\bar{\mathcal{L}}^{\Lambda}\mathcal{L}^{\Sigma} \; ,
\end{equation}

\noindent
whence $\bar{U}^{\Lambda\Sigma}=U^{\Sigma\Lambda}$.  

We can define the graviphoton and matter vector projectors

\begin{eqnarray}
\label{eq:projectorg}
\mathcal{T}_{\Lambda} & \equiv & 2i \mathcal{L}_{\Lambda}
=2i\mathcal{L}^{\Sigma}\Im{\rm m}\, 
\mathcal{N}_{\Sigma\Lambda}\, ,\\
& & \nonumber \\
\label{eq:projectorm}
\mathcal{T}^{i}{}_{\Lambda} & \equiv & -\bar{f}_{\Lambda}{}^{i}=
-\mathcal{G}^{i\bar{j}}
\bar{f}^{\Sigma}{}_{\bar{j}}\Im{\rm m}\, 
\mathcal{N}_{\Sigma\Lambda}\, .
\end{eqnarray}

Using these definitions and the above properties one can show the following
identities for the derivatives of the period matrix:

\begin{equation}
  \label{eq:dN}
  \begin{array}{rcl}
\partial_{i}\mathcal{N}_{\Lambda\Sigma} & = & 
4\mathcal{T}_{i(\Lambda}\mathcal{T}_{\Sigma)}\, ,\\
& & \\
\partial_{\bar{i}}\mathcal{N}_{\Lambda\Sigma} & = & 
4\bar{\mathcal{C}}_{\bar{i}\bar{j}\bar{k}}
\mathcal{T}^{\bar{i}}{}_{(\Lambda}\mathcal{T}^{\bar{j}}{}_{\Sigma)}\, .\\
  \end{array}
\end{equation}

\noindent
Observe that the first of Eqs.~(\ref{eq:SGDefFund2}) together with the
definition of the period matrix $\mathcal{N}$ imply the following
expression for the K\"ahler potential: 

\begin{equation}
\label{eq:kpotential}
e^{-\mathcal{K}}= -2\Im{\rm m}\mathcal{N}_{\Lambda\Sigma}
\mathcal{X}^{\Lambda}  \bar{\mathcal{X}}^{\Sigma}\, .
\end{equation}

If we now assume that $\mathcal{F}_{\Lambda}$ depends on $z^{i}$ through the
$\mathcal{X}$'s, then from the last equation we can derive that

\begin{equation}
\partial_{i}\mathcal{X}^{\Lambda}
\left[ 
2\mathcal{F}_{\Lambda}  
-\partial_{\Lambda}\left( \mathcal{X}^{\Sigma}\mathcal{F}_{\Sigma}\right)
\right] 
= 0 \; .
\end{equation}

If $\partial_{i}\mathcal{X}^{\Lambda}$ is invertible as an $n\times \bar{n}$
matrix, then we must conclude that
 
\begin{equation}
\label{eq:Prepot}
\mathcal{F}_{\Lambda}  = \partial_{\Lambda}\mathcal{F}(\mathcal{X}) \, ,
\end{equation}

\noindent
where $\mathcal{F}$ is a homogeneous function of degree 2, called the
\textit{prepotential}.

Making use of the prepotential and the definitions (\ref{eq:SGDefN}), we
can calculate

\begin{equation}
\label{eq:PrepotN}
\mathcal{N}_{\Lambda\Sigma} 
=\bar{\mathcal{F}}_{\Lambda\Sigma} 
+2i\frac{
\Im{\rm m}\mathcal{F}_{\Lambda\Lambda^{\prime}}
\mathcal{X}^{\Lambda^{\prime}}
\Im{\rm m}\mathcal{F}_{\Sigma\Sigma^{\prime}}\mathcal{X}^{\Sigma^{\prime}}
}
{                                          
\mathcal{X}^{\Omega}\Im{\rm m}\mathcal{F}_{\Omega\Omega^{\prime}}
\mathcal{X}^{\Omega^{\prime}}
}\, .
\end{equation}

Having the explicit form of $\mathcal{N}$, we can also derive an explicit
representation for $\mathcal{C}$ by applying Eq. (\ref{eq:SGColl4}). One finds

\begin{equation}
\label{eq:PrepC}
\mathcal{C}_{ijk} = 
e^{\mathcal{K}}\partial_{i}\mathcal{X}^{\Lambda}
 \partial_{j}\mathcal{X}^{\Sigma} \partial_{k}\mathcal{X}^{\Omega}
\mathcal{F}_{\Lambda\Sigma\Omega} \; ,
\end{equation}

\noindent
so that the prepotential really determines all structures in special geometry.

A last remark has to be made about the existence of a prepotential: clearly,
given a holomorphic section $\Omega$ a prepotential need not exist. It was
shown in Ref.~\cite{Craps:1997gp}, however, that one can always apply an
$Sp(2n_{v}+2,\mathbb{R})$ transformation such that a prepotential exists.

%%%%%%%%%%%%%%%%%%%%%%%%%%%%%%%%%%%%%%%%%%%%%%%%%%%%%%%%%%%%%%%%%%%%%%
%%%%%%%%%%%%%%%%%%%%%%%%%%%%%%%%%%%%%%%%%%%%%%%%%%%%%%%%%%%%%%%%%%%%%%
%%%%%%%%%%%%%%%%%%%%%%%%%%%%%%%%%%%%%%%%%%%%%%%%%%%%%%%%%%%%%%%%%%%%%%
%%%%%%%%%%%%%%%%%%%%%%%%%%%%%%%%%%%%%%%%%%%%%%%%%%%%%%%%%%%%%%%%%%%%%%

\section{Homogeneous spaces}
\label{sec:homspaces}

%%%%%%%%%%%%%%%%%%%%%%%%%%%%%%%%%%%%%%%%%%%%%%%%%%%%%%%%%%%%%%%%%%%%%%
%%%%%%%%%%%%%%%%%%%%%%%%%%%%%%%%%%%%%%%%%%%%%%%%%%%%%%%%%%%%%%%%%%%%%%
%%%%%%%%%%%%%%%%%%%%%%%%%%%%%%%%%%%%%%%%%%%%%%%%%%%%%%%%%%%%%%%%%%%%%%
%%%%%%%%%%%%%%%%%%%%%%%%%%%%%%%%%%%%%%%%%%%%%%%%%%%%%%%%%%%%%%%%%%%%%%

The relevant spaces in $\mathcal{N}>2$ Supergravity are \emph{Irreducible Riemannian Globally Symmetric} (I.R.G.S.) spaces, which are particular instances of homogeneous spaces (see \cite{Ferrara:2008de,Helgason,Warner,SS} and references therein). We will also closely follow the appendix of \cite{Meessen:2010fh}. We will start by defining the concept of homogeneous space and then we will move to the cases of interest in Supergravity, the symmetric spaces.

\thm{\label{thm:cosets} Let $H$ be a closed subgroup of a Lie group $G$, and let $G/H$ be the set $\left\{\sigma H:~\sigma\in G\right\}$ of left cosets modulo $H$. Let $\pi: G\rightarrow G/H$ denote the natural projection $\pi(\sigma)=\sigma H$. Then $G/H$ has a unique manifold structure such that

\begin{enumerate}

\item $\pi$ is $C^{\infty}$.

\item There exist local smooth sections of $G/H$ in $G$; that is, if $\sigma H\in G/H$, there is a
neighbourhood $\mathcal{U}\left(\sigma H\right)$ and a $C^{\infty}$ map $\tau : \mathcal{U}\left(\sigma H\right)\rightarrow G$ such that $\pi\circ\tau={\rm I}$.
\end{enumerate}

}

\definition{\label{def:homspaces} Manifolds of the form $G/H$ where $G$ is a Lie group, $H$ is a closed subgroup of $G$, and the manifold structure is the unique satisfying theorem \ref{thm:cosets}, are called \emph{homogeneous manifolds}. }

\definition{\label{def:action} Let $\eta:G\times \mathcal{M}\rightarrow\mathcal{M}$ be an action of $G$ on $\mathcal{M}$ on the left. We denote $\eta_{\sigma}(m) = \eta(\sigma,m)$. 

\begin{enumerate}

\item The action is called \emph{effective} if the identity $e$ of $G$ is the only element of $G$ for which $\eta_{e}$ is the identity map on $\mathcal{M}$.

\item The action is called \emph{transitive} if $\forall~p,q\in\mathcal{M}~~\exists ~ \sigma\in G~/~ \eta_{\sigma}(p)=q$.

\item Let $p_{0}\in\mathcal{M}$ and let $H=\left\{\sigma\in G : \eta_{\sigma}(p_{0})=p_{0}\right\}$. Then $H$ is a closed subgroup of $G$, the isotropy group at $m_{0}$.

\item The action $\eta$ restricted to $H$ gives an action of $H$ on $\mathcal{M}$ on the left with fixed point $m_{0}$. It can then be proven that $\alpha:H\rightarrow {\rm Aut}(T_{p_{0}}\mathcal{M})$, where $\alpha(\sigma)=d\eta_{\sigma} \left(T_{p_{0}}\mathcal{M}\right)$, is a representation of $H$, the \emph{linear isotropy group} at $p_{0}$.

\end{enumerate}

}

\thm{\label{thm:equiv} Let $\eta\times \mathcal{M}\rightarrow\mathcal{M}$ be a transitive action of the Lie group $G$ on the manifold $\mathcal{M}$ on the left. Let $p_{0}\in\mathcal{M}$, and let $H$ be the isotropy group at $p_{0}$. Then the map $\beta: G/H\rightarrow \mathcal{M}$ given by $\beta(\sigma H) = \eta_{\sigma}(p_{0})$ is a diffeomorphism.}
\proof{See \cite{Warner}, Theorem 3.62.}

If $G$ is a Lie group and $H$ a closed subgroup of $G$, then there is a natural transitive action $l$ of $G$ on the homogeneous manifold $G/H$ on the left, namely $l: G\times G/H\rightarrow G/H$ given by $l(\sigma,\tau H)=\sigma\tau H$.

A manifold $G/H$ of the kind defined in \ref{def:homspaces} is called homogeneous because it admits a transitive action of a group, in particular of $G$. Thanks to theorem \ref{thm:equiv} we know that the converse is also true, if a manifold $\mathcal{M}$ admits a transitive action of a group $G$, then it is diffeomorphic to $G/H$, where $H$ is the isotropy group of the action. 

\noindent
\thm{\label{them:metriccoset} If the isotropy group $H$ of a homogeneous space $G/H$ is compact, then it can be equipped with a $G$-invariant Riemannian metric. }

\proof{See \cite{Barut}, theorem 1, chapter 4.}

Notice that, under the assumptions of \ref{thm:cosets}, $G/H$ is a differentiable manifold, but it may not be a Lie group itself. The following theorem gives a sufficient condition for that to happen.

\thm{\label{thm:GHgroup} Let $G$ be a Lie group and $H$ a closed normal subgroup of $G$. Then the homogeneous manifold $G/H$ with its natural group structure is a Lie group.}

\proof{See \cite{Warner}, theorem 3.64.}

The non-linear $\sigma$-models appearing in extended Supergravity are homogeneous manifolds $G/H$, but $H$ is in general not a normal subgroup of $G$, and therefore we cannot use the tools available for group manifolds in order to study them. They are, however, of a particular kind: they are symmetric spaces, a specific type of homogeneous manifold, and in addition, its isotropy group $H$ is compact. Therefore, they can be endowed with a $G$-invariant Riemannian metric. We will thus consider Riemannian symmetric spaces, since they are the ones appearing in the non-linear $\sigma$-models of extended Supergravity.

A Riemannian symmetric space is a Riemannian manifold $\left(\mathcal{S},\mathcal{G}\right)$ with the property that the geodesic reflection at any point is an isometry of $\mathcal{S}$. That is, $\forall p\in\mathcal{M}\, ~\exists ~ s_{p}\in ~I_{{\rm Isometries}}\left(\mathcal{M}\right)$ with the properties

\begin{equation}
s_{p}(p)=p\, ,\qquad \left(ds_{p}\right)_{p}=-{\rm I}\, .
\end{equation}

\noindent
As a consequence of this definition, every symmetric space $\mathcal{S}$ is homogeneous: $\mathcal{S}$ can be shown to admit the transitive action of a Lie group $G$, which is indeed the isometry group of the Riemannian metric $\mathcal{G}$.

\definition{\label{def:symspace} A symmetric space $\mathcal{S}$ is precisely a homogeneous space with a symmetry $s_{p}$ at some point $p\in\mathcal{S}$.

\noindent
For our purposes it is therefore better to characterize a symmetric space in the form $G/H$ (which is possible since they are homogeneous), where $G$ is its isometry group and $H$ a closed subgroup of $G$, through the following result

\thm{\label{thm:symspace} Let $G$ be a connected Lie group with an involution $\sigma: G\rightarrow G$ and a left invariant metric $\mathcal{G}$, which is also right invariant under the closed subgroup $\tilde{K}=\left\{g\in G\, , g^{\sigma} = g\right\}$. Let $K$ be a closed subgroup of $G$ with $\tilde{K}^{0}\subset K\subset \tilde{K}$ where $\tilde{K}^{0}$ denotes the connected component (identity component) of $\tilde{K}$. Then $S=G/H$ is a symmetric space where the metric is induced from the given metric on $G$. Every symmetric space $S$ arises this way.}

\proof{See \cite{SS}, theorem 4.1.}

\noindent
\newline 
\lemma{\label{lemma:decomposition} A vector space decomposition $\mathfrak{g}=\mathfrak{t}\oplus \mathfrak{h}$ of a Lie algebra $\mathfrak{g}$ is the eigenspace decomposition of a order-two automorphism $\sigma$ of $\mathfrak{g}$ if and only if

\begin{equation}
[\mathfrak{h},\mathfrak{h}]\subset \mathfrak{h}\, , [\mathfrak{h},\mathfrak{t}]\subset\mathfrak{t} \, , [\mathfrak{t},\mathfrak{t}]\subset\mathfrak{h}\, ,
\end{equation}
}

\noindent
A decomposition of Lie algebra $\mathfrak{g} = \mathfrak{t}\oplus\mathfrak{h}$ obeying lemma \ref{lemma:decomposition} and such that ${\rm ad}\left(\mathfrak{h}\right)_{\mathfrak{t}}$ is the Lie algebra of a compact subgroup of $GL(\mathfrak{t})$ is called the \emph{Cartan decomposition}, and the corresponding involution $\sigma$ the \emph{Cartan involution}. We have the following result

\thm{\label{thm:symspaceII} Any symmetric space $\mathcal{S}$ determines a Cartan decomposition on the Lie algebra of Killing fields. Vice versa, to any Lie algebra $\mathfrak{g}$ with Cartan decomposition $\mathfrak{g} = \mathfrak{t}\oplus\mathfrak{h}$ there exists a unique simply connected symmetric space $\mathcal{S}=G/H$ where $G$ is the simply connected Lie group with Lie algebra $\mathfrak{g}$ and $H$ is the connected subgroup with Lie algebra $\mathfrak{h}$.}

\proof{See \cite{SS}, theorem 4.2.}

The extended Supergravity scalar manifolds are symmetric spaces: they are of the form $G/H$, where $G$ is the non-compact real form of a simple, finite-dimensional, Lie group and a $H$  is its maximal compact subgroup, which is the isotropy group of the manifold. It is equipped with a $G_{L}\times H_{R}$ invariant Riemannian metric, which has strictly negative definite signature.

\definition{\label{def:IRGS} An Irreducible Riemannian Globally Symmetric space is a symmetric space with strictly negative definite metric signature.}\footnote{In our conventions, the $\sigma$-model metric is positive definite. Therefore it is minus the metric of the corresponding I.R.G.S. space.}

\noindent
\newline We have arrived therefore to the specific kind of scalar manifolds appearing in the non-linear $\sigma$-models of extended four-dimensional Supergravity. In order to implement electromagnetic duality rotations in the theory, as it will be explained in section \ref{sec:duality}, it is needed to embed the group $G$ appearing in the Supergravity scalar manifold $G/H$ into the symplectic group $\mathrm{Sp}(2\bar{n},\mathbb{R})$, or, going to a complex basis, into $\mathrm{Usp}(\bar{n},\bar{n})$, a procedure that can be always performed in Supergravity. Therefore, all the scalar manifolds can be described by a $\mathrm{Usp}(\bar{n},\bar{n})$ matrix $U$ which is constructed in terms of the matrices\footnote{When we multiply these matrices we must include a factor $1/2$ for each contraction of pairs of antisymmetric indices $IJ$.}

\begin{equation}
f\equiv (f^{\Lambda}{}_{IJ}, f^{\Lambda}{}_{i})\, ,
\hspace{1cm}  
h\equiv (h_{\Lambda\, IJ}, h_{\Lambda\, i})\, ,
\end{equation}

\noindent 
which formally are sections of the following trivial flat, symplectic bundle

\begin{equation}
G\times_{H}\mathbb{R}^{2n}\rightarrow G/H
\end{equation}

\noindent
$I,J=1,\ldots \mathcal{N}$ are the graviton-supermultiplet, or equivalently
$\mathrm{U}(\mathcal{N})$, indices and $i(=1,\ldots n_{v})$ are indices labeling the vector
multiplets, and the embedding then imposes that  

\begin{equation}
\bar{n}=n+\frac{\mathcal{N}(\mathcal{N}-1)}{2}\, .
\end{equation}

\noindent
This information is represented in the following table:\footnote{ Observe that
  $\mathcal{N}=6$ has $n=1$, even though there are no vector supermultiplets in this
  case. }
\begin{table}[!h]
\begin{center}
\begin{tabular}{|c||c|c|c|c|c|}
\hline
$N$&3&4&5&6&8\\\hline\hline
%$\alpha$&1&1&0&1&0\\ \hline
$n$&$n$&$n$&0&1&0\\\hline
$\bar{n}$ & $n+3$ &$n+6$&10&16&28\\\hline
\end{tabular}
\end{center}
% \caption{x}
\label{tablealphabeta}
\end{table} 

\noindent
Using the above matrices one can then embed the generic scalar manifolds as

\begin{equation}
\label{eq:U}
U \equiv 
{\textstyle\frac{1}{\sqrt{2}}}
\left(
  \begin{array}{cc}
f+ih & \bar{f}+i\bar{h} \\
f-ih & \bar{f}-i\bar{h} \\
  \end{array}
\right)\, .  
\end{equation}

\noindent
The condition that  $U\in \mathrm{Usp}(\bar{n},\bar{n})$ 

\begin{equation}
  \begin{array}{rcl}
U^{-1} & = & \left( 
\begin{array}{cc}
1 & 0 \\ 0 & -1 \\ 
\end{array}
\right)
U^{\dagger}
\left( 
\begin{array}{cc}
1 & 0 \\ 0 & -1 \\ 
\end{array}
\right)
 = 
\left( 
\begin{array}{cc}
0 & 1 \\ -1 & 0 \\ 
\end{array}
\right)
U^{T}
\left( 
\begin{array}{cc}
0 & -1 \\ 1 & 0 \\ 
\end{array}
\right)
\\
& & \\
& = & 
{\textstyle\frac{1}{\sqrt{2}}}
\left(
  \begin{array}{cc}
f^{\dagger}-ih^{\dagger} &  -(f^{\dagger}+ih^{\dagger}) \\
-(f-ih ) &  f+ih \\
  \end{array}
\right)\, , \\
\end{array}
\end{equation}

\noindent
leads to the following conditions for $f$ and $h$:

\begin{equation}
\label{eq:fhnormalization}
i(f^{\dagger}h-h^{\dagger}f) = 1\, ,
\hspace{1cm}
f^{T}h-h^{T}f=0\, .
\end{equation}

\noindent
In terms of the symplectic sections

\begin{equation}\label{eq:symsec}
\mathcal{V}_{IJ}= 
\left(
  \begin{array}{c}
f^{\Lambda}{}_{IJ} \\ h_{\Lambda IJ} \\
  \end{array}
\right)\, ,  
\hspace{1cm}
\mathcal{V}_{i}= 
\left(
  \begin{array}{c}
f^{\Lambda}{}_{i} \\ h_{\Lambda\, i} \\
  \end{array}
\right)\, ,  
\end{equation}

\noindent
these constraints take the form\footnote{We use the convention 
\begin{equation}
\langle  \mathcal{A}\mid \mathcal{B}\rangle \equiv
  \mathcal{B}^{\Lambda}\mathcal{A}_{\Lambda} 
-\mathcal{B}_{\Lambda}\mathcal{A}^{\Lambda}\, .
\end{equation}
}

\begin{equation}
  \begin{array}{rcl}
\langle \mathcal{V}_{IJ}\mid\bar{\mathcal{V}}^{KL}\rangle 
& = &   
-2i\delta^{KL}{}_{IJ}\, , \\
& & \\
\langle \mathcal{V}_{i}\mid\bar{\mathcal{V}}^{j}\rangle 
&  = &   
-i\delta_{i}{}^{j}\, , \\
\end{array}
\end{equation}

\noindent
with the rest of the symplectic products vanishing.

The left-invariant Maurer-Cartan 1-form can be split into the Vielbeine $P$ and
the connection $\Omega$ as follows:

\begin{equation}
\Gamma \equiv U^{-1}dU = 
\left(
  \begin{array}{cc}
\Omega & \bar{P} \\
P & \bar{\Omega} \\
  \end{array}
\right)\, .  
\end{equation}

\noindent
Thus, the different components of the connection are

\begin{equation}
\label{eq:O}
\Omega =
\left(
  \begin{array}{cc}
\Omega^{KL}{}_{IJ} & \Omega^{j}{}_{IJ} \\
\Omega^{KL}{}_{i} & \Omega^{j}{}_{i} \\
  \end{array}
\right) 
=
\left(
  \begin{array}{cc}
i \langle d \mathcal{V}_{IJ}\mid\bar{\mathcal{V}}^{KL}\rangle & 
i \langle d \mathcal{V}_{IJ}\mid\bar{\mathcal{V}}^{j}\rangle \\
i \langle d \mathcal{V}_{i}\mid\bar{\mathcal{V}}^{KL}\rangle & 
i \langle d \mathcal{V}_{i}\mid\bar{\mathcal{V}}^{j}\rangle \\
  \end{array}
\right)\, ,
\end{equation}

\noindent
and those of the Vielbeine are 

\begin{equation}
\label{eq:P}
P =
\left(
  \begin{array}{cc}
P_{KLIJ} & P_{jIJ} \\
P_{KLi} & P_{ij} \\
  \end{array}
\right) 
=
\left(
  \begin{array}{cc}
-i \langle d \mathcal{V}_{IJ}\mid\mathcal{V}_{KL}\rangle & 
-i \langle d \mathcal{V}_{IJ}\mid\mathcal{V}_{j}\rangle \\
-i \langle d \mathcal{V}_{i}\mid\mathcal{V}_{KL}\rangle & 
-i \langle d \mathcal{V}_{i}\mid\mathcal{V}_{j}\rangle \\
  \end{array}
\right)\, .
\end{equation}

\noindent
The period matrix $\mathcal{N}_{\Lambda\Sigma}$ is defined by

\begin{equation}
\label{eq:periodmatrix}
\mathcal{N}=hf^{-1}=\mathcal{N}^{T}\, ,   
\end{equation}

\noindent
which implies properties which should be familiar from the $N=2$ case: for
instance

\begin{equation}
\label{eq:loweringindices}
\mathfrak{D}h_{\Lambda}=
\bar{\mathcal{N}}_{\Lambda\Sigma}\mathfrak{D}f^{\Lambda}\, ,
\hspace{1cm}
h_{\Lambda}=\mathcal{N}_{\Lambda\Sigma}f^{\Sigma}\, ,
\end{equation}

and 

\begin{equation} 
\label{eq:completeness}
-{\textstyle\frac{1}{2}}(\Im{\rm m}\mathcal{N})^{-1|\Lambda\Sigma}=
{\textstyle\frac{1}{2}}f^{\Lambda}{}_{IJ}\bar{f}^{\Sigma IJ}
+f^{\Lambda}{}_{i}\bar{f}^{\Sigma\, i}\, ,
\end{equation}

\noindent
which can be derived from the definition of $\mathcal{N}$ and
Eq.~(\ref{eq:fhnormalization}).

We also quote the completeness relation

\begin{equation}
\label{eq:completeness2}
{\textstyle\frac{1}{2}}
 \mid \mathcal{V}_{IJ} \rangle\langle\bar{\mathcal{V}}^{IJ}  \mid\,\,
-{\textstyle\frac{1}{2}}
 \mid \bar{\mathcal{V}}^{IJ} \rangle\langle\mathcal{V}_{IJ}  \mid\,\,
+ \mid \mathcal{V}_{i} \rangle\langle\bar{\mathcal{V}}^{i}  \mid\,\,
- \mid \bar{\mathcal{V}}^{i} \rangle\langle\mathcal{V}_{i}  \mid\,\,
=i\, .
\end{equation}

\noindent
Defining the $H_{Aut}\times H_{Matter}$ covariant derivative
according to

\begin{equation}
\label{eq:coder}
\mathfrak{D}\mathcal{V}=d\mathcal{V}-\mathcal{V}\Omega\, ,
\end{equation}  

\noindent
and using Eq.~(\ref{eq:loweringindices}) we obtain from (\ref{eq:O})

\begin{equation} 
\Omega^{KL}{}_{i}=\Omega^{j}{}_{IJ}=0\, ,
\end{equation}

\noindent
and from (\ref{eq:P})

\begin{eqnarray}
\label{eq:Paut}
P_{IJKL} & = & 
-2f^{\Lambda}{}_{IJ}\Im{\rm m}\mathcal{N}_{\Lambda\Sigma}\
\mathfrak{D}f^{\Sigma}{}_{KL}\, ,\\
& & \nonumber \\
\label{eq:Pmatter}
P_{iIJ} & = & -2f^{\Lambda}{}_{i}\Im{\rm m}\mathcal{N}_{\Lambda\Sigma}\
\mathfrak{D}f^{\Sigma}{}_{IJ}\, ,\\
& & \nonumber \\
\label{eq:Pij}
P_{ij} & = & -2f^{\Lambda}{}_{i}\Im{\rm m}\mathcal{N}_{\Lambda\Sigma}\
\mathfrak{D}f^{\Sigma}{}_{j}\, .
\end{eqnarray}

\noindent
The above equation can be inverted to give

\begin{eqnarray} 
\label{eq:dflij}
\mathfrak{D} f^{\Lambda}{}_{IJ} & = & 
\bar{f}^{\Lambda i} P_{iIJ}
+{\textstyle\frac{1}{2}}\bar{f}^{\Lambda KL}P_{IJKL}\, , \\
& & \nonumber \\
\label{eq:dfli}
\mathfrak{D} f^{\Lambda}{}_{i} & = & 
\bar{f}^{\Lambda j} P_{ij}
+{\textstyle\frac{1}{2}}\bar{f}^{\Lambda IJ}P_{iIJ}\, , 
\end{eqnarray}

\noindent
using Eq.~(\ref{eq:completeness}).

The definition of the covariant derivative leads to the identities

\begin{equation}
\langle\, \mathfrak{D}\mathcal{V}  \mid \bar{\mathcal{V}}\, \rangle  
= 0\, ,
\hspace{1cm}
\langle\, \mathfrak{D}\mathcal{V}  \mid \mathcal{V}\, \rangle  
=\langle\, d\mathcal{V}  \mid \mathcal{V}\, \rangle  
=i P\, .
\end{equation}

The inverse Vielbeine $\bar{P}^{IJKL},\bar{P}^{iIJ},\bar{P}^{ij}$, satisfy (here $A$
labels the physical fields)

\begin{equation}
\label{eq:inversevielbeine}
\bar{P}^{IJKL\, A}P_{MNOP\, A} = 4!\delta^{IJKL}{}_{MNOP}\, , 
\hspace{.5cm}   
\bar{P}^{iIJ\, A}P_{jKL\, A} = 2\delta^{i}{}_{j}\delta^{IJ}{}_{KL}\, . 
\end{equation}

\noindent 
Their crossed products vanish but their products with $P_{ij\, A}$ do not.

We find 

\begin{eqnarray}
\label{eq:partialcontractionvielbeine1}
\langle\,  \mathfrak{D}_{A}\mathcal{V}_{IJ} \mid 
\mathfrak{D}_{B}\bar{\mathcal{V}}^{KL}\, \rangle 
& = & 
{\textstyle\frac{i}{2}}P_{IJMN}{}_{A}\bar{P}^{KLMN}{}_{B}  
+iP_{i IJ}{}_{A}\bar{P}^{iKL}{}_{B}\, ,  
\\
& & \nonumber \\
\label{eq:partialcontractionvielbeine3}
\langle\,  \mathfrak{D}_{A}\mathcal{V}_{IJ} \mid 
\mathfrak{D}_{B}\bar{\mathcal{V}}^{i}\, \rangle 
& = & 
{\textstyle\frac{i}{2}}P_{IJKL}{}_{A}\bar{P}^{iKL}{}_{B}  
+iP_{jIJ}{}_{A}\bar{P}^{ij}{}_{B}\, ,  
\\
& & \nonumber \\
\label{eq:partialcontractionvielbeine2}
\langle\,  \mathfrak{D}_{A}\mathcal{V}_{i} \mid 
\mathfrak{D}_{B}\bar{\mathcal{V}}^{j}\, \rangle 
& = & 
{\textstyle\frac{i}{2}}P_{iIJ}{}_{A}\bar{P}^{iIJ}{}_{B}  
+iP_{ik}{}_{A}\bar{P}^{jk}{}_{B}\, ,  
\end{eqnarray}

\noindent
while $\langle\,  \mathfrak{D}_{A}\mathcal{V}_{IJ} \mid 
\mathfrak{D}_{B}\mathcal{V}_{KL}\, \rangle 
=
\langle\,  \mathfrak{D}_{A}\mathcal{V}_{IJ} \mid 
\mathfrak{D}_{B}\mathcal{V}_{i}\, \rangle 
=
\langle\,  \mathfrak{D}_{A}\mathcal{V}_{i} \mid 
\mathfrak{D}_{B}\mathcal{V}_{j}\, \rangle =0
$.

\noindent
Using the definition of the period matrix Eq.~(\ref{eq:periodmatrix}),
equation (\ref{eq:loweringindices}) and the first of
Eqs.~(\ref{eq:fhnormalization}) we get

\begin{equation}
d\mathcal{N} = 4i \Im{\rm m}\mathcal{N}\, \mathfrak{D}f f^{\dagger}\,
\Im{\rm m}\mathcal{N}\, .
\end{equation}

\noindent
This expression can be expanded in terms of the Vielbeine, using
Eqs.~(\ref{eq:dflij}) and (\ref{eq:dfli})

\begin{equation}
d\mathcal{N}_{\Lambda_{\Sigma}} = 
i \Im{\rm m}\mathcal{N}_{\Gamma(\Lambda} \Im{\rm m}\mathcal{N}_{\Sigma)\Omega}
\left[ P_{IJKL}\bar{f}^{\Gamma IJ}\bar{f}^{\Omega KL} 
+4 P_{iIJ}\bar{f}^{\Gamma i}\bar{f}^{\Omega IJ}
+4 P_{ij}\bar{f}^{\Gamma i}\bar{f}^{\Omega j}
\right]\, .
\end{equation}

\noindent
and, using Eqs.~(\ref{eq:inversevielbeine}) and taking into account that their
contraction with $P_{ij}$ does not necessarily vanish, implies

\begin{eqnarray}
\label{eq:PIJKLdN}
\bar{P}^{IJKL\, A}\frac{\partial}{\partial\phi^{A}}\mathcal{N}_{\Lambda\Sigma} 
& = & 
4!i \Im{\rm m}\mathcal{N}_{\Omega(\Lambda}   
\Im{\rm m}\mathcal{N}_{\Sigma) \Delta} 
\bar{f}^{\Omega [IJ|} \bar{f}^{\Delta |KL]}\, ,\\
& & \nonumber \\
\label{eq:PiIJdN}
\bar{P}^{iIJ\, A}\frac{\partial}{\partial\phi^{A}}\mathcal{N}_{\Lambda\Sigma} 
& = & 
8i \Im{\rm m}\mathcal{N}_{\Omega(\Lambda}   
\Im{\rm m}\mathcal{N}_{\Sigma) \Delta} 
\bar{f}^{\Omega i} \bar{f}^{\Delta IJ}\, .\\
& & \nonumber \\
\label{eq:P*IJKLdN}
\bar{P}^{IJKL\, A}\frac{\partial}{\partial\phi^{A}}\bar{\mathcal{N}}_{\Lambda\Sigma} 
& = & 
-4i \Im{\rm m}\mathcal{N}_{\Omega(\Lambda}   
\Im{\rm m}\mathcal{N}_{\Sigma) \Delta} 
\bar{P}^{IJKL\, A}\bar{P}^{ij}{}_{A}
f^{\Omega}{}_{i} f^{\Delta}{}_{j}\, ,\\
& & \nonumber \\
\label{eq:P*iIJdN}
\bar{P}^{iIJ\, A}\frac{\partial}{\partial\phi^{A}}\bar{\mathcal{N}}_{\Lambda\Sigma} 
& = & 
-4i \Im{\rm m}\mathcal{N}_{\Omega(\Lambda}   
\Im{\rm m}\mathcal{N}_{\Sigma) \Delta} 
\bar{P}^{iIJ\, A}\bar{P}^{jk}{}_{A}
f^{\Omega}{}_{i} f^{\Delta}{}_{j}\, .
\end{eqnarray}

\noindent
Using the Maurer-Cartan equations $d\Gamma+\Gamma\wedge \Gamma =0$ and direct
calculations we find that the curvatures of $\Omega^{KL}{}_{IJ}$ and
$\Omega^{j}{}_{i}$ are

\begin{eqnarray}
R^{KL}{}_{IJ} & = & 
d \Omega^{KL}{}_{IJ} +{\textstyle\frac{1}{2}}\Omega^{KL}{}_{MN}\wedge
\Omega^{MN}{}_{IJ} 
 \nonumber \\
& & \nonumber \\
& = &  
-{\textstyle\frac{1}{2}}\bar{P}^{KLMN}\wedge P_{MNIJ}
- \bar{P}^{i KL}\wedge P_{iIJ}
\label{eq:RO=PP}
\\
& & \nonumber \\
& = & 
-i\langle\, \mathfrak{D}\mathcal{V}_{IJ}\mid  
\mathfrak{D}\bar{\mathcal{V}}^{KL}\,
\rangle\, ,\\
& & \nonumber \\
R^{j}{}_{i} & = & d \Omega^{j}{}_{i} +\Omega^{j}{}_{k}\wedge
\Omega^{k}{}_{i}
= 
-{\textstyle\frac{1}{2}}\bar{P}^{jIJ}\wedge P_{iIJ}
- \bar{P}^{ik}\wedge P_{ik}
\\
& & \nonumber \\
& = &  
-i\langle\, \mathfrak{D}\mathcal{V}_{i} \mid  
\mathfrak{D}\bar{\mathcal{V}}^{j} \,
\rangle\, .
\end{eqnarray}

\noindent
The vanishing of the curvature of $\Omega^{i}{}_{IJ}$ leads to 

\begin{equation}
\label{eq:esaidentidad}
{\textstyle\frac{1}{2}}P_{IJKL}\wedge \bar{P}^{iKL}
+ P_{jIJ}\wedge \bar{P}^{ij}
=
-i\langle\,  \mathfrak{D}\mathcal{V}_{IJ} \mid 
\mathfrak{D}\bar{\mathcal{V}}^{i}\, \rangle =0\, .
\end{equation}

%\cleardoublepage

%%%%%%%%%%%%%%%%%%%%%%%%%%%%%%%%%%%%%%%%%%%%%%%%%%%%%%%%%%%%%%%%%%%%%%%
%%% CHAPTER 3 EXTENDED UNGAUGED SUPERGRAVITY IN FOUR DIMENSIONS
%%%%%%%%%%%%%%%%%%%%%%%%%%%%%%%%%%%%%%%%%%%%%%%%%%%%%%%%%%%%%%%%%%%%%%%
\renewcommand{\chaptername}{Chapter}

\renewcommand{\leftmark}{\MakeUppercase{Chapter \thechapter. Four-Dimensional Extended Supergravity}}
\chapter{Extended ungauged Supergravity in four dimensions}
\label{chapter:sugra}

After the mathematical background provided in chapter \ref{chapter:mathpreliminaries}, we are ready to summarize the \emph{geometric} formulation of extended four-dimensional ungauged Supergravity, following \cite{Andrianopoli:1996cm,Andrianopoli:1996ve}, where the interested reader will find a more detailed exposition. In the case of $\mathcal{N}=2$ Supergravity, we will focus only on the vector multiplet sector, given that black hole solutions with non-trivial hyper-scalars are believed to be singular since they would have \emph{scalar hair} and it is always consistent to set the hyper-scalars to a constant value. We start by reviewing electromagnetic duality in a class of gravity theories coupled to scalars and vector fields that includes the bosonic sector of any ungauged Supergravity.

%%%%%%%%%%%%%%%%%%%%%%%%%%%%%%%%%%%%%%%%%%%%%%%%%%%%%%%%%%%%%%%%%%%%%%
%%%%%%%%%%%%%%%%%%%%%%%%%%%%%%%%%%%%%%%%%%%%%%%%%%%%%%%%%%%%%%%%%%%%%%
%%%%%%%%%%%%%%%%%%%%%%%%%%%%%%%%%%%%%%%%%%%%%%%%%%%%%%%%%%%%%%%%%%%%%%
%%%%%%%%%%%%%%%%%%%%%%%%%%%%%%%%%%%%%%%%%%%%%%%%%%%%%%%%%%%%%%%%%%%%%%

\section{Extended electromagnetic duality}
\label{sec:duality}

%%%%%%%%%%%%%%%%%%%%%%%%%%%%%%%%%%%%%%%%%%%%%%%%%%%%%%%%%%%%%%%%%%%%%%
%%%%%%%%%%%%%%%%%%%%%%%%%%%%%%%%%%%%%%%%%%%%%%%%%%%%%%%%%%%%%%%%%%%%%%
%%%%%%%%%%%%%%%%%%%%%%%%%%%%%%%%%%%%%%%%%%%%%%%%%%%%%%%%%%%%%%%%%%%%%%
%%%%%%%%%%%%%%%%%%%%%%%%%%%%%%%%%%%%%%%%%%%%%%%%%%%%%%%%%%%%%%%%%%%%%%

For this section, the basic references are \cite{Gaillard:1981rj,Gibbons:1995cv,Ceresole:1995jg}. An excellent review and extension of these works can be found in \cite{Aschieri:2008ns}. We are going to consider four-dimensional theories of the general form
\begin{equation}
\label{eq:generalaction}
I
=
\int d^{4}x \sqrt{|g|}
\left\{
R +\mathcal{G}_{ij}(\phi)\partial_{\mu}\phi^{i}\partial^{\mu}\phi^{j}
+2 \Im{\rm m} \mathcal{N}_{\Lambda\Sigma}
F^{\Lambda}{}_{\mu\nu}F^{\Sigma\, \mu\nu}
-2 \Re{\rm e} \mathcal{N}_{\Lambda\Sigma}
F^{\Lambda}{}_{\mu\nu}\star F^{\Sigma\, \mu\nu}
\right\}\, ,   
\end{equation}

\noindent
which includes the bosonic sectors of all four-dimensional ungauged
supergravities for appropriate $\sigma$-model metrics $\mathcal{G}_{ij}(\phi)$
and (complex) kinetic matrix $\mathcal{N}_{\Lambda\Sigma}(\phi)$, with
negative-definite imaginary part (see sections \ref{sec:N2sugra} and \ref{sec:Nesugra}). The indices $i,j,\dots~= 1,\dots,n_{s}$ run over the scalar
fields and the indices $\Lambda, \Sigma,\dots~= 0,\dots,n_{v}$ over the 1-form fields. Their
numbers are related only for $\mathcal{N}\geq 2$ supergravity theories. 

We denote to the equations of motion corresponding to the action (\ref{eq:generalaction}) by 

\begin{equation}
\mathcal{E}_{a}{}^{\mu}\equiv 
-\frac{1}{2\sqrt{|g|}}\frac{\delta S}{\delta e^{a}{}_{\mu}}\, ,
\hspace{.5cm}
\mathcal{E}_{i} \equiv -\frac{1}{2\sqrt{|g|}}
\frac{\delta S}{\delta \phi^{i}}\, ,
\hspace{.5cm}
\mathcal{E}_{\Lambda}{}^{\mu}\equiv 
\frac{1}{8\sqrt{|g|}}\frac{\delta S}{\delta A^{\Lambda}{}_{\mu}}\, ,
\end{equation}

\noindent
and denote the Bianchi identities for the vector field strengths by

\begin{equation}
\label{eq:BL}
\mathcal{B}^{\Lambda\, \mu} \equiv \nabla_{\nu}{}^{\star}F^{\Lambda\,
  \nu\mu}\, .  
\end{equation}

\noindent
The explicit form of the equations of motion can be found to be

\begin{eqnarray}
\mathcal{E}_{\mu\nu} & = & 
G_{\mu\nu}
+\mathcal{G}_{ij}[\partial_{\mu}\phi^{i} \partial_{\nu}\phi^{j}
-{\textstyle\frac{1}{2}}g_{\mu\nu}
\partial_{\rho}\phi^{i}\partial^{\rho}\phi^{j}]\nonumber \\
& & \nonumber \\
& & 
+8\Im {\rm m}\mathcal{N}_{\Lambda\Sigma}
F^{\Lambda\, +}{}_{\mu}{}^{\rho}F^{\Sigma\, -}{}_{\nu\rho}\, ,
\label{eq:Emn}\\
& & \nonumber \\
\mathcal{E}_{i} & = & \nabla_{\mu}(\mathcal{G}_{ij}
\partial^{\mu}\phi^{i})
-\frac{1}{2}\partial_{i}\mathcal{G}_{jk}\partial_{\rho}\phi^{j}
\partial^{\rho}\phi^{k} 
+\partial_{i}[
\tilde{F}_{\Lambda}{}^{\mu\nu}{}^{\star}F^{\Lambda}{}_{\mu\nu}]\, ,
\label{eq:Ei}\\
& & \nonumber \\
\mathcal{E}_{\Lambda}{}^{\mu} & = & 
\nabla_{\nu}{}^{\star}\tilde{F}_{\Lambda}{}^{\nu\mu}\, ,
\label{eq:ERm}
\end{eqnarray}

\noindent
where we have defined the dual vector field strength $\tilde{F}_{\Lambda}$
by 

\begin{equation}
\label{eq:defFmag}
\tilde{F}_{\Lambda\, \mu\nu} \equiv   
-\frac{1}{4\sqrt{|g|}}\frac{\delta S}{\delta {}^{\star}F^{\Lambda}{}_{\mu\nu}}
= \Re {\rm e}\mathcal{N}_{\Lambda\Sigma}F^{\Sigma}{}_{\mu\nu}
+\Im {\rm m}\mathcal{N}_{\Lambda\Sigma}{}^{*}F^{\Sigma}{}_{\mu\nu}\, .
\end{equation}

\noindent
Let's focus our attention on the equations of motion for the vector fields (that is, the Maxwell identities) $\mathcal{E}_{\Lambda}{}^{\mu}$ together with the Bianchi identities $\mathcal{B}^{\Lambda\, \mu}$ and define the \emph{doublet}

\begin{equation}
\label{eq:maxwellvector}
\mathcal{E}^{M}_{\mu}\equiv \left(
  \begin{array}{c}
\mathcal{B}^{\Lambda}{}_{\mu}\\
\\
\mathcal{E}_{\Lambda~\mu}\\
  \end{array}
\right)\, ,
\end{equation}

\noindent
where $M=(\Lambda,\Lambda)$. The Maxwell and Bianchi identities can be now succinctly written as

\begin{equation}
\mathcal{E}^{M}_{\mu}=0\, ,
\end{equation}

\noindent
and therefore they admit as a symmetry an arbitrary $\mathrm{GL}(2n_{v}+2,\mathbb{R})$ rotation acting on $M$. That is

\begin{equation}
\mathcal{E}^{M}_{\mu}=0\Rightarrow \mathfrak{A}^{M}{}_{N}\mathcal{E}^{N}_{\mu}=0\, ,\qquad \mathfrak{A}\in \mathrm{G}L(2n_{v}+2,\mathbb{R})\, .
\end{equation}

\noindent
It is convenient to write $\mathfrak{A}$ in terms of $(n_{v}+1)\times(n_{v}+1)$ blocks

\begin{equation}
\label{eq:matrixA}
\mathfrak{A}=\left(
  \begin{array}{cc}
D & C \\
& \\
B & A \\
  \end{array}
\right)\, ,
\end{equation}

\noindent
These transformations act in the same form on the vector of $2n_{v}+2$ two-forms

\begin{equation}
\label{eq:2formvector}
F^M \equiv 
\left(
  \begin{array}{c}
F^{\Lambda}\\
\\
\tilde{F}_{\Lambda}\\
  \end{array}
\right)\, , \qquad F^{\prime\, M} = \mathfrak{A}^{M}{}_{N} F^{N}\, .
\end{equation}

\noindent
However, $\tilde{F}_{\Lambda}$ is not an independent set of fields, it is related to $F^{\Lambda}$ by Eq. (\ref{eq:defFmag}), and therefore we must require the same definition to hold for the transformed $\tilde{F}^{\prime}_{\Lambda}$ in terms of the transformed action $S^{\prime}$ and the transformed $F^{\prime\, \Lambda}$, that is, we require that

\begin{equation}
\label{eq:defFmagprime}
\tilde{F}^{\prime}_{\Lambda\, \mu\nu} \equiv   
-\frac{1}{4\sqrt{|g|}}\frac{\delta S^{\prime}}{\delta {}^{\star}F^{\prime\, \Lambda}{}_{\mu\nu}}\, .
\end{equation}

\noindent
In order to implement (\ref{eq:defFmagprime}) consistently, we have to consider simultaneously a transformation $\xi\in {\rm Diff}\left(\mathcal{M}_{{\rm scalar}}\right)$ on the scalar manifold $\mathcal{M}_{{\rm scalar}}$, since imposing (\ref{eq:defFmagprime}) requires the scalar matrix $\mathcal{N}_{\Lambda\Sigma}$ to transform in a prescribed way, which in turn has to be implemented through a transformation of the scalars. Therefore, we assume the existence of a group homomorphism  

\begin{equation}
\mathfrak{i}:~{\rm Diff}\left(\mathcal{M}_{{\rm scalar}}\right) \rightarrow \mathrm{GL}\left(2n_{v}+2,\mathbb{R}\right)\, ,
\end{equation}

\noindent
which maps every diffeomorphism $\xi\in {\rm Diff}\left(\mathcal{M}_{{\rm scalar}}\right)$ to a general linear transformation $\mathfrak{i}\left(\xi\right)\in \mathrm{GL}(2n_{v}+2,\mathbb{R})$.

Using the homomorphism $\mathfrak{i}$ we can define now a simultaneous action of $\xi$ in all of the fields of the theory. Writing schematically $\phi^{\prime}=\xi\left(\phi\right)$, we define the action of an arbitrary diffeomorphism $\xi\in {\rm Diff}\left(\mathcal{M}_{{\rm scalar}}\right)$ on the theory (\ref{eq:generalaction}) to be given by \cite{Andrianopoli:1996cm}

\begin{equation}
\left\{\phi,F^{M},\mathcal{N}_{\Lambda\Sigma}\right\}\xrightarrow{\xi}\left\{\phi^{\prime},\left(\mathfrak{i}\left(\xi\right)\right)^{M}{}_{N}F^{N},\mathcal{N}^{\prime}_{\Lambda\Sigma}\left(\phi^{\prime}\right)\right\}\, ,
\end{equation}

\noindent
and consistency with Eq. (\ref{eq:defFmagprime}) requires that

\begin{equation}
\label{eq:Ntransformation}
\mathcal{N}^{\prime}\left(\phi^{\prime}\right) = (A\mathcal{N}\left(\phi\right)+B)(C\mathcal{N}\left(\phi\right)+D)^{-1}\, .  
\end{equation}

\noindent
Furthermore, the transformations must preserve the symmetry of the period
matrix, which requires

\begin{equation}
A^{T}C=C^{T}A\, ,
\hspace{1cm}
D^{T}B=B^{T}D\, ,  
\hspace{1cm}
A^{T}D-C^{T}B=1\, ,
\end{equation}

\noindent
{\em i.e.\/} the transformations must belong to $\mathrm{Sp}(2n_{v}+2,\mathbb{R})$. Therefore, the homomorphism $\mathfrak{i}$ must be reduced to 

\begin{equation}
\label{eq:homi1}
\mathfrak{i}:~{\rm Diff}\left(\mathcal{M}_{{\rm scalar}}\right) \rightarrow \mathrm{Sp}\left(2n_{v}+2,\mathbb{R}\right)\, .
\end{equation}

\noindent
Notice that (\ref{eq:homi1}) can never be an isomorphism, since $\mathrm{Sp}\left(2n_{v}+2,\mathbb{R}\right)$ is a finite-dimensional Lie group and ${\rm Diff}\left(\mathcal{M}_{{\rm scalar}}\right)$ is infinite-dimensional. The above transformation rules for the vector field strength and period matrix
imply

\begin{equation}
\Im{\rm m} \mathcal{N}^{\prime} =(C\mathcal{N}^{*}+D)^{-1\, T}   
\Im{\rm m} \mathcal{N} (C\mathcal{N}+D)^{-1}\, ,
\hspace{1cm}
F^{\prime\Lambda\, +} = (C\mathcal{N}^{*}+D)_{\Lambda\Sigma}F^{\Sigma\, +}\, ,
\end{equation}

\noindent
so the combination $\Im {\rm m}\mathcal{N}_{\Lambda\Sigma} F^{\Lambda\,+}{}_{\mu}{}^{\rho}F^{\Lambda\, -}{}_{\nu\rho}$ that appears in the
energy-momentum tensor is automatically invariant.

So far, the situation is the following: we have noticed that the set of Maxwell and Bianchi identities admit a global group of linear symmetries given by $\mathrm{GL}\left(2n_{v}+2,\mathbb{R}\right)$. In order to define an action consistent with (\ref{eq:defFmagprime}) we have to impose three conditions

\begin{enumerate}

\item The transformation group mas be reduced from $\mathrm{GL}\left(2n_{v}+2,\mathbb{R}\right)$ to $\mathrm{Sp}\left(2n_{v}+2,\mathbb{R}\right)$

\item The $\mathrm{Sp}\left(2n_{v}+2,\mathbb{R}\right)$ rotation must be performed together with a transformation on the scalars, given by a diffeomorphism $\xi\in {\rm Diff}\left(\mathcal{M}_{{\rm scalar}}\right)$.

\item The couplings of scalars and vector fields $\mathcal{N}_{\Lambda\Sigma}$ must transform as indicated in (\ref{eq:Ntransformation}).

\end{enumerate}
So, as long as our \emph{kinetic matrix} $\mathcal{N}_{\Lambda\Sigma}$ obeys (\ref{eq:Ntransformation}) we can consistently define \emph{symplectic/duality rotations} on the theory in such a way that they act as symmetries of the Maxwell and the Bianchi identities. Notice that this does not mean at all that the duality rotations are symmetries of the action: duality rotations are not even a symmetry of

\begin{equation}
\label{eq:vectoraction}
I_{{\rm Maxwell}}
=
\int d^{4}x \sqrt{|g|}
\left\{
+2 \Im{\rm m} \mathcal{N}_{\Lambda\Sigma}
F^{\Lambda}{}_{\mu\nu}F^{\Sigma\, \mu\nu}
-2 \Re{\rm e} \mathcal{N}_{\Lambda\Sigma}
F^{\Lambda}{}_{\mu\nu}\star F^{\Sigma\, \mu\nu}
\right\}\, ,   
\end{equation}

\noindent
\emph{i.e.}, the part of the Lagrangian corresponding to the vector fields. But on top of that, we are considering arbitrary scalar diffeomorphisms, which will not preserve the non-linear $\sigma$-model nor the equations of motion for the scalars. Therefore, if we require at least the duality rotations to be symmetries of the equations of motion, we have to consider not arbitrary diffeomorphisms but only isometries of the scalar metric $\mathcal{G}_{ij}(\phi)$, which are exact symmetries of 

\begin{equation}
\label{eq:scalaraction}
I_{{\rm Scalars}}
=
\int d^{4}x \sqrt{|g|}
\left\{
\mathcal{G}_{ij}(\phi)\partial_{\mu}\phi^{i}\partial^{\mu}\phi^{j}\right\}\, .   
\end{equation}

\noindent
Therefore, the homomorphism $\mathfrak{i}$ must be again reduced to 

\begin{equation}
\label{eq:homi2}
\mathfrak{i}:~{\rm Isometries}\left(\mathcal{M}_{{\rm scalar}},\mathcal{G}_{ij}\right) \rightarrow \mathrm{Sp}\left(2n_{v}+2,\mathbb{R}\right)\, .
\end{equation}

\noindent
Thus, the duality/symplectic transformations, \emph{i.e.} global symmetries of the equations of motion, are the isometries of the scalar manifold. considered as a Riemannian manifold equipped with the metric $\mathcal{G}_{ij}$, which act on the scalars as usual diffeomorphisms and on the vector fields linearly through the homomorphism (\ref{eq:homi2}), as long as the matrix $\mathcal{N}_{\Lambda|sigma}$ transforms as prescribed by (\ref{eq:Ntransformation}).

It can checked that the strict symmetries of the Lagrangian are the isometries of the scalar manifold with a block diagonal embedding on the symplectic group $\mathrm{Sp}\left(2n_{v}+2,\mathbb{R}\right)$, and the symmetries of the Lagrangian up to a total derivative are those whose embedding obeys $C=0$.

%%%%%%%%%%%%%%%%%%%%%%%%%%%%%%%%%%%%%%%%%%%%%%%%%%%%%%%%%%%%%%%%%%%%%%
%%%%%%%%%%%%%%%%%%%%%%%%%%%%%%%%%%%%%%%%%%%%%%%%%%%%%%%%%%%%%%%%%%%%%%
%%%%%%%%%%%%%%%%%%%%%%%%%%%%%%%%%%%%%%%%%%%%%%%%%%%%%%%%%%%%%%%%%%%%%%
%%%%%%%%%%%%%%%%%%%%%%%%%%%%%%%%%%%%%%%%%%%%%%%%%%%%%%%%%%%%%%%%%%%%%%

\section{\texorpdfstring{$\mathcal{N} = 2,~d = 4$}{ N = 2,~d = 4} ungauged Supergravity}
\label{sec:N2sugra}

%%%%%%%%%%%%%%%%%%%%%%%%%%%%%%%%%%%%%%%%%%%%%%%%%%%%%%%%%%%%%%%%%%%%%%
%%%%%%%%%%%%%%%%%%%%%%%%%%%%%%%%%%%%%%%%%%%%%%%%%%%%%%%%%%%%%%%%%%%%%%
%%%%%%%%%%%%%%%%%%%%%%%%%%%%%%%%%%%%%%%%%%%%%%%%%%%%%%%%%%%%%%%%%%%%%%
%%%%%%%%%%%%%%%%%%%%%%%%%%%%%%%%%%%%%%%%%%%%%%%%%%%%%%%%%%%%%%%%%%%%%%

$\mathcal{N}=2$ four-dimensional Supergravity makes reference generically to \emph{any four-dimensional theory} of gravity invariant under two supersymmetries. Here we will consider such theory up to two derivatives in the Lagrangian and in the absence of gauings, and in consequence, we call it ungauged four-dimensional classical Supergravity. The matter content of the theory is the following

\begin{enumerate}

\item \emph{Gravitational multiplet:} $\left(e^{a}_{\, \mu}, \bar{\psi}^{I},\psi_{I},A^0\right)$, where $e^{a}_{\, \mu}$ is the Vielbein (together with the spin connection one-form $\omega^{ab}$), $\psi_{I}$ is an $SU(2)$ doublet of gravitino one-forms, and $A^0$ is the graviphoton one-form. 

\item \emph{$n_{v}$ vector supermultiples:} $\left(A^{i}, \bar{\lambda}^{i}_{I}, \lambda^{i\, I}, z^{i}\right)$, where $A^{i}$ $i=1,\cdots,n_{v}$ is a one-form, $\lambda^{i}_{I}$ is a zero-form spinor, and $z^{i}$ is a complex scalar. The scalars $z^{i}$ parametrize the $n_{v}$-dimensional base of a Special K\"ahler bundle $\mathcal{S}\mathcal{V}$.

\item \emph{$n_{h}$ hypermultiplets:} $\left(\chi_{\alpha},\chi^{\alpha},q^{u}\right)$, where $\chi_{\alpha}$ is zero-form spinor $\left(\alpha=1,\cdots,2n_{h}\right)$, and four real scalars $\left(u=1,\cdots,4n_{h}\right)$ which parametrize a $4n_{h}$-dimensional Quaternionic manifold $\mathcal{H}\mathcal{M}$ .

\end{enumerate}

\noindent
It is convenient tu define a new index $\Lambda = (0,i)$, which allows to write all the vector fields of the theory as $\left\{ A^{\Lambda}_{\mu}\, , \Lambda = 0,\cdots , n_{v}\right\}$. Since we are interested in bosonic solutions, we will set all the fermions to zero, which is always a consisten truncation. The general bosonic Lagrangian is given by

\begin{eqnarray}
\label{eq:generalactionN2}
&S&  =  \int d^{4}x \sqrt{|g|} \left( R +\mathcal{G}_{i\bar{j}}\left(z,\bar{z}\right)\partial_{\mu}z^{i}\partial^{\mu}\bar{z}^{\bar{j}} + h_{uv}\left(q\right)\partial_{\mu}q^{u}\partial^{\mu}q^v \right.\nonumber\\
&+&\left. 2 \Im{\rm m} \mathcal{N}_{\Lambda\Sigma}\left(z,\bar{z}\right) F^{\Lambda}{}_{\mu\nu}F^{\Sigma\, \mu\nu} -2 \Re{\rm e} \mathcal{N}_{\Lambda\Sigma}\left(z,\bar{z}\right) F^{\Lambda}{}_{\mu\nu}\star F^{\Sigma\, \mu\nu}\right) \nonumber \, ,  
\end{eqnarray}

\noindent
Observe that the canonical normalization of the vector fields kinetic
terms implies that $\Im{\rm m}\mathcal{N}_{\Lambda\Sigma}$ is negative
definite, as is guaranteed by special geometry \cite{Craps:1997gp}. The equations of motion for the hyper-scalars corresponding to (\ref{eq:generalactionN2}) are given by

\begin{equation}
\label{eq:EOMMierdini}
\mathfrak{D}_{\mu}\partial^{\mu}q^{u}=
\nabla_{\mu}\partial^{\mu}q^{u} 
+\Gamma_{vw}{}^{u}\partial^{\mu}q^{v}\partial_{\mu}q^{w}=0\, ,  
\end{equation}

\noindent
where $\Gamma_{vw}{}^{u}$ are the Christoffel symbols of the 2$^{nd}$ kind for the metric $h_{uv}$. Therefore, it is always consistent to truncate the hyper-scalars to a constant value $q^{u}=q^{u}_{0}$, and we will do so in the sequel, since black hole solutions with non-trivial hyper-scalars are believed to be singular, since the would develop \emph{scalar hair}. Let's state for completeness that the scalars $q^{u}$ parametrize a Quaternionic manifold \cite{Besse}, \emph{i.e.} a Riemannian manifold of special holonomy, which will not be discussed here.

\noindent
As a consequence of supersymmetry, the metric $\mathcal{G}_{i\bar{j}}\left(z,\bar{z}\right)$ of the non-linear $\sigma$ model and \emph{period matrix} $\mathcal{N}_{\Lambda\Sigma}\left(z,\bar{z}\right)$ are constrained in a very precise way. We have the following structure 

\begin{enumerate}

\item The scalars $z^ i$ parametrize a Special K\"ahler manifold\footnote{For more details about Special K\"ahler geometry, see chapter \ref{chapter:mathpreliminaries} and references therein.}: a holomorphic non-trivial flat tensor bundle $\mathcal{S}\mathcal{V}= \mathcal{S}\mathcal{M}\otimes \mathcal{L}$ with structural group $Sp\left(2n_v+2,\mathbb{R}\right)\otimes U(1)$.

\item All the couplings of the theory (in the absence of hyper-multiplets) can be constructed in terms of the holomorphic section $\Omega\in \Gamma\left(\mathcal{S}\mathcal{M}\right)$ or the covariantly holomorphic symplectic section $\mathcal{V}$. 

\end{enumerate}

\noindent
Since we are interested in bosonic configurations, the only couplings that we need to consider are those of vector fields and scalars, given by the period matrix $\mathcal{N}_{\Lambda\Sigma}\left(z,\bar{z}\right)$. It can be shown that, in terms of the covariantly holomorphic symplectic section $\mathcal{V}$, $\mathcal{N}_{\Lambda\Sigma}\left(z,\bar{z}\right)$ is given by \cite{Castellani:1990zd} 

\begin{equation}
\label{eq:SGDefN}
\mathcal{M}_{\Lambda}  = \mathcal{N}_{\Lambda\Sigma} \mathcal{L}^{\Sigma}\, ,
\hspace{1cm}
h_{\Lambda\, i}  = \mathcal{N}^{*}{}_{\Lambda\Sigma} f^{\Sigma}{}_{i} \, ,
\end{equation}

\noindent
where $\mathcal{M}_{\Lambda}, \mathcal{L}^{\Sigma}, h_{\Lambda\, i}$ and $f^{\Sigma}{}_{i}$ have been defined in (\ref{eq:SGDefFund}) and (\ref{eq:SGDefU}).

Notice that Eq. (\ref{eq:SGDefN}) implies that $\mathcal{N}_{\Lambda\Sigma}\left(z,\bar{z}\right)$ transforms under diffeomorphisms of the base space as required by (\ref{eq:Ntransformation}). Therefore, we can apply the formalism of section \ref{sec:duality}\footnote{Obviously the action (\ref{eq:generalactionN2}) is a particular case of (\ref{eq:generalaction}).} and conclude that the equations of motion of ungauged $\mathcal{N}=2$ Supergravity enjoy duality invariance. The same conclusion holds in the presence of hyper-multiplets.

To summarize, four-dimensional ungauged $\mathcal{N}=2$ Supergravity in the absence of hyper-multiplets is completely specified once the Special K\"ahler bundle $\mathcal{S}\mathcal{M}$ describing the self-interactions of the vector multiplets is given. As explained in chapter \ref{chapter:mathpreliminaries}, specifying such bundle is equivalent to specify the holomorphic symplectic section, which is equivalente, when it exists, to specify the second-order homogeneous prepotential $F(\mathcal{X})$. Therefore, four-dimensional ungauged $\mathcal{N}=2$ Supergravity in the absence of hyper-multiplets can be completely specified in terms of just a function, the prepotential.

%%%%%%%%%%%%%%%%%%%%%%%%%%%%%%%%%%%%%%%%%%%%%%%%%%%%%%%%%%%%%%%%%%%%%%
%%%%%%%%%%%%%%%%%%%%%%%%%%%%%%%%%%%%%%%%%%%%%%%%%%%%%%%%%%%%%%%%%%%%%%

\subsection{Type-IIA String Theory on a Calabi Yau manifold}
\label{sec:IIACY}

%%%%%%%%%%%%%%%%%%%%%%%%%%%%%%%%%%%%%%%%%%%%%%%%%%%%%%%%%%%%%%%%%%%%%%
%%%%%%%%%%%%%%%%%%%%%%%%%%%%%%%%%%%%%%%%%%%%%%%%%%%%%%%%%%%%%%%%%%%%%%

Let's consider a particular example of ungauged $\mathcal{N}=2$ Supergravity embeddable in String Theory. In particular, we are going to consider the class of Supergravities that are obtained by compactifying Type-IIA String Theory on a Calabi Yau manifold in the absence of fluxes. We will come back to this example in chapter \ref{chapter:quantumbhs}, obtaining some of its black hole solutions. As in the previous section, we will omit the hyper-scalar sector.

Type-IIA String Theory on a Calabi-Yau threefold yields four-dimensional, ungauged, $\mathcal{N}=2$ Supergravity, given by \cite{Candelas:1990qd,Candelas:1990rm,Hosono:1993qy,Hosono:1994av}

\begin{equation}
F(\mathcal{X}) =  -\frac{1}{3!}\kappa^{0}_{ijk} \frac{\mathcal{X}^i \mathcal{X}^j \mathcal{X}^k}{\mathcal{X}^0} +i\frac{\chi\zeta(3)}{2(2\pi)^3}+\frac{i(\mathcal{X}^0)^2}{(2\pi)^3}\sum_{\{d_{i}\}} n_{\{d_{i}\}} Li_{3}\left(e^{2\pi i d_{i} \frac{\mathcal{X}^{i}}{\mathcal{X}^0}}\right)\nonumber \ \, ,
\end{equation}

\noindent
where $Li_3(x)=\sum^{\infty}_{k=1}\frac{x^k}{k^3}$ is the third polylogarithmic function, $\chi$ is the Euler characteristic of the Calabi-Yau, $\zeta(3)$ is the Riemann zeta function of 3, $n_{\{d_{i}\}}$ is the number of genus $0$ instantons\footnote{That is, the number of distinct holomorphic mappings of the genus $0$ world-sheet onto holomorphic two-cycles with degrees $\{d_{i}\}$} and $\kappa^0_{ijk}$ are the classical intersection numbers. Using special coordinates $z^{i}=\frac{\mathcal{X}^{i}}{\mathcal{X}^0}$ we can write the prepotential as follows (gauge $\mathcal{X}^{0}=1$)

\begin{equation}
\label{eq:IIACYprep}
\mathcal{F} =  -\frac{1}{3!}\kappa^{0}_{ijk} z^i z^j z^k +i\frac{\chi\zeta(3)}{ 2(2\pi)^3}+\frac{i}{(2\pi)^3}\sum_{\{d_{i}\}} n_{\{d_{i}\}} Li_{3}\left(e^{2\pi i d_{i} z^{i}}\right) \nonumber\ \, ,
\end{equation}

\noindent
The theory defined by (\ref{eq:IIACYprep}) is extremely involved due to the infinite sum of polylogarithms. We can simplify it by considering the \emph{large-volume} compactification limit $\Im{\rm m}z^{i}\rightarrow\infty$, where the prepotential is given by

\begin{equation}
\mathcal{F}_{0} =  -\frac{1}{3!}\kappa^{0}_{ijk} z^i z^j z^k  \nonumber\ \, .
\end{equation}

\noindent
Let's consider now the compactification on a specific Calabi-Yau threefold, the Quintic manifold. The effective theory of Type-IIA String Theory compactified on the Quintic C.Y. three-fold, in the large-volume compactification limit, is given by\footnote{Up to an unimportant constat for our purposes,} \cite{Candelas:1990qd,Candelas:1990rm}

\begin{equation}
\label{eq:t3}
\mathcal{F}_{0} =  -t^3  \nonumber\ \, .
\end{equation}

\noindent
Let's construct the bosonic Lagrangian. First, we need the geometric data of the Special K\"ahler manifold relevant to construct the Lagrangian, that is, the scalar metric and the covariantly holomorphic symplectic section. Using the formulae of chapter (\ref{chapter:mathpreliminaries}) we obtain

\begin{equation}
\label{eq:t3data}
\mathcal{G}_{t\bar{t}}=\frac{-3}{(t-\bar{t})^2}\, ,\qquad \mathcal{V}^{T}=\left(1,t,t^3,-3t^2\right)\, .  \nonumber
\end{equation}

\noindent
We can identify the Special K\"ahler manifold corresponding to (\ref{eq:t3}), which is an homogeneous, symmetric space:

\begin{equation}
\mathcal{M}=\frac{SU(1,1)}{U(1)}
\end{equation}

\noindent
Such geometry would be completely modified if we were to introduce the constant and the polilogarithmic corrections on the prepotential: the scalar manifold would not be an homogeneous space anymore!

With (\ref{eq:t3data}) we can now compute $\mathcal{N}_{\Lambda\Omega}$, using (\ref{eq:SGDefN}). The result is

\begin{equation}
{\rm Re}\,\mathcal N_{IJ} \,=\, \left(\begin{array}{cc}
-2\Re^3  &   3 \Re^2\\ [2mm]
3 \Re^2  &  -6 \Re
\end{array}\right)\,,
\end{equation}

\begin{equation}
{\rm Im}\,\mathcal N_{IJ} \,=\, \left(\begin{array}{cc}
-(\Im^3 + 3 \Re^2 \Im)  & 3\Re \Im  \\ [2mm]
 3\Re \Im  &  -3\Im
\end{array}\right)\, ,
\end{equation}

\noindent
where $\Re = {\rm Re}(t)$ and $\Im = {\rm Im}(t)$. Therefore, the bosonic Lagrangian, in the absence of hyper-multiplets, of Type-IIA String Theory compactified on the mirror Quintic manifold, is finally given by

\begin{eqnarray}
\label{eq:generalaction}
&S&  =  \int d^{4}x \sqrt{|g|} \left( R -\frac{3\partial_{\mu}t\partial^{\mu}\bar{t}}{(t-\bar{t})^2}+2 \Im{\rm m} \mathcal{N}_{\Lambda\Sigma}\left(z,\bar{z}\right) F^{\Lambda}{}_{\mu\nu}F^{\Sigma\, \mu\nu} - 2 \Re{\rm e} \mathcal{N}_{\Lambda\Sigma}\left(z,\bar{z}\right) F^{\Lambda}{}_{\mu\nu}\star F^{\Sigma\, \mu\nu}\right) \nonumber \, .  
\end{eqnarray}

%%%%%%%%%%%%%%%%%%%%%%%%%%%%%%%%%%%%%%%%%%%%%%%%%%%%%%%%%%%%%%%%%%%%%%
%%%%%%%%%%%%%%%%%%%%%%%%%%%%%%%%%%%%%%%%%%%%%%%%%%%%%%%%%%%%%%%%%%%%%%
%%%%%%%%%%%%%%%%%%%%%%%%%%%%%%%%%%%%%%%%%%%%%%%%%%%%%%%%%%%%%%%%%%%%%%
%%%%%%%%%%%%%%%%%%%%%%%%%%%%%%%%%%%%%%%%%%%%%%%%%%%%%%%%%%%%%%%%%%%%%%

\section{\texorpdfstring{$\mathcal{N} > 2,~d = 4$}{ N > 2,~d = 4} ungauged Supergravity}
\label{sec:Nesugra}

%%%%%%%%%%%%%%%%%%%%%%%%%%%%%%%%%%%%%%%%%%%%%%%%%%%%%%%%%%%%%%%%%%%%%%
%%%%%%%%%%%%%%%%%%%%%%%%%%%%%%%%%%%%%%%%%%%%%%%%%%%%%%%%%%%%%%%%%%%%%%
%%%%%%%%%%%%%%%%%%%%%%%%%%%%%%%%%%%%%%%%%%%%%%%%%%%%%%%%%%%%%%%%%%%%%%
%%%%%%%%%%%%%%%%%%%%%%%%%%%%%%%%%%%%%%%%%%%%%%%%%%%%%%%%%%%%%%%%%%%%%%

The structure of four-dimensional ungauged $\mathcal{N} > 2$ Supergravity is very similar to the structure of four-dimensional ungauged $\mathcal{N} = 2$ Supergravity coupled to vector multiplets. A similar \emph{symplectic formulation}, that is, based on the construction of a vector bundle with a symplectic structure group over the scalar manifold can be realized, and all the couplings of the theory written in terms of a section of such bundle. The basic reference for this section is \cite{Andrianopoli:1996ve}, where a much more detailed exposition can be found. We will use the notation and conventions of \cite{Bellorin:2005zc,Meessen:2010fh}. 

If we restrict ourselves to theories with maximum spin two, we can conclude that the amount of supersymmetry is constrained to be less or equal than eight $\mathcal{N}\leq 8$\footnote{That is, thirty two real charges, if we consider Majorana spinors.}. In addition, we will consider only terms up to two derivatives on the Lagrangian. As in the $\mathcal{N}=2$ case, all the $\mathcal{N}>2$ Supergravities contain in the lagrangian a non-linear $\sigma$ model of the form

\begin{equation}
\label{eq:scalaractionN}
I_{{\rm Scalars}}
=
\int d^{4}x \sqrt{|g|}
\left\{
\mathcal{G}_{ij}(\phi)\partial_{\mu}\phi^{i}\partial^{\mu}\phi^{j}\right\}\, .   
\end{equation}

\noindent
As we saw, for $\mathcal{N}=2$ ungauged Supergravity coupled to vector multiplets, the scalars parametrize the base space of a Special K\"ahler bundle, whose defining section $\mathcal{V}$ specifies the $\mathcal{N}=2$ model, since the Lagrangian, couplings and scalar metric, can be completely specified in terms of $\mathcal{V}$. Here the situation is similar. However the \emph{extra} amount supersymmetry imposes stronger constraints on the theory, and now the Riemannian scalar manifold $\left(\mathcal{M}_{{\rm Scalar}},\mathcal{G}_{ij}\right)$ has to be an Irreducible Riemannian Globally Symmetric space\footnote{See section \ref{sec:homspaces} for more details.}\footnote{Notice that $\mathcal{G}_{ij}$ is minus the metric of the Supergravity $\sigma$-model.} \cite{Ferrara:2008de}. Therefore, all the $\sigma$-models for $\mathcal{N}>2$ Supergravity are of the form

\begin{equation}
\mathcal{M}_{scalar} =\frac{G}{H}
\end{equation}

\noindent
where $G$ is the group of isometries of the scalar manifold $\left(\mathcal{M}_{{\rm Scalar}},\mathcal{G}_{ij}\right)$ and $H$ is its isotropy group. In particular, for $\mathcal{N}>2$ Supergravity $G$ is the maximally non-compact real form of a simple Lie group, which depends on the particular $\mathcal{N}>2$ Supergravity. The different groups appearing in Supergravity $\mathcal{N}>2$ are summarized in table \ref{table:sugragroups}. The isotropy group $H$ is in turn of the form

\begin{equation}
H=H_{{\rm A}}\times H_{{\rm M}}\, ,
\end{equation}

\noindent
where $H_{{\rm A}}$ corresponds to the automorphisms group of the corresponding supersymmetry algebra, and $H_{{\rm M}}$ is related to the matter vector multiplets. When matter cannot be incorporated to the theory, namely for $\mathcal{N}>4$, we have $H_{{\rm M}}={\rm I}$. The geometric formulation of $\mathcal{N}>2$ can be given in terms of an specific fibre bundle over $\frac{G}{H}$, namely, a trivial flat symplectic bundle of the form \cite{Andrianopoli:1996ve,Aschieri:2008ns}

\begin{equation}
G\times_{H}\mathbb{R}^{2n}\rightarrow \frac{G}{H}\, ,
\end{equation}

\noindent
which justifies its construction in section \ref{sec:homspaces}. The couplings of the theory can be obtained now from the symplectic section $\mathcal{V}_{IJ}$ defined in \ref{eq:symsec}. 

All the four-dimensional ungauged $\mathcal{N} > 2$ Supergravities can be described in a unified way. In particular, all the Supergravity $\mathcal{N}>2$ matter contents can be written in the same generic form; we only need to take into account the range of values taken by the $\mathrm{U}(\mathcal{N})$ R-symmetry indices, denoted by uppercase Latin letters $I$ \textit{etc.} taking on values $1,\cdots ,\mathcal{N}$, in each particular case. Only fields and terms that should be considered are those whose number of antisymmetric $\mathrm{SU}(\mathcal{N})$ indices is correct, i.e.~objects with more than $\mathcal{N}$ antisymmetric indices are zero and terms with Levi-Civit\`a symbols $\epsilon^{I_{1}\cdots I_{M}}$ should only be considered when $M$ equals the $\mathcal{N}$ of the supergravity theory under consideration. There are also constraints on the generic fields for specific values of $\mathcal{N}$ that we are going to review.

The generic supergravity multiplet in four dimensions is

\begin{equation}
\label{Ngrav}
\left\{ e^{a}{}_{\mu},\psi_{I\, \mu},A^{IJ}{}_{\mu},\chi_{IJK},
\chi^{IJKLM},\,P_{IJKL\, \mu}\right\}\, ,\,\,\,\,
I,J,\dots=1,\cdots, N\, ,
\end{equation}

\noindent
and the generic vector multiplets (labeled by $i=1,\cdots,n$) are

\begin{equation}
\left\{A_{i\, \mu},\lambda_{iI},\lambda_{i}{}^{IJK},P_{iIJ\, \mu}\right\}\, .
\end{equation}

\noindent
The spinor fields $\psi_{I\, \mu},\chi_{IJK},\chi^{IJKLM},
\lambda_{iI},\lambda_{i}{}^{IJK}$ have positive chirality with the given
positions of the $\mathrm{SU}(\mathcal{N})$ indices. 

The scalars of these theories are encoded into the $2\bar{n}$-dimensional
($\bar{n}\equiv n+\frac{\mathcal{N}(\mathcal{N}-1)}{2}$) symplectic sections ($\Lambda=1,\dots
\bar{n}$) $\mathcal{V}_{IJ}$ and $\mathcal{V}_{i}$ (see section \ref{sec:homspaces}). They appear in the bosonic sector of the theory via the pullbacks of the Vielbeine $P_{IJKL\mu}$ (Supergravity
multiplet) and $P_{iIJ\, \mu}$ (matter multiplets)\footnote{The Vielbeine $P_{ij\, \mu}$ either vanish identically or depend on $P_{IJKL\mu}$ and $P_{iIJ\, \mu}$, depending on the specific value of $\mathcal{N}$. Thus, they are not needed as independent variables to construct the theories.}.  There are three instances of theories for which the scalar Vielbeine are constrained:
first, when $\mathcal{N}=4$ the matter scalar Vielbeine are constrained by the
$\mathrm{SU}(4)$ complex self-duality relation\footnote{ In order to highlight the fact that an equation holds for a specific $\mathcal{N}$ only, we write a numerical variation of the token ``$\mathcal{N}=4::$'' to the left of the equation.  }

\begin{equation}
\mathcal{N}\ =\ 4 ::\hspace{.6cm}
P^{*\, i\, IJ} \ =\  \tfrac{1}{2}\varepsilon^{IJKL}\ P_{i\, KL}\, .
\end{equation}

Secondly, in $\mathcal{N}=6$ the scalars in the supergravity multiplet are represented by one Vielbein $P_{IJ}$ and one Vielbein $P_{IJKL}$ related by the $\mathrm{SU}(6)$ duality relation

\begin{equation}
\mathcal{N}\ =\ 6 ::\hspace{.6cm}
P^{*\, IJ} \ =\ \tfrac{1}{4!}\varepsilon^{IJK_{1}\cdots K_{4}} \ P_{K_{1}\cdots K_{4}}\, ,
\end{equation}

\noindent
and lastly the $\mathcal{N}=8$ case, in which the Vielbeine is constrained by the $\mathrm{SU}(8)$ complex self-duality relation

\begin{equation}
\mathcal{N}\ =\ 8 ::\hspace{.6cm}
P^{*\, I_{1}\cdots I_{4}} =\tfrac{1}{4!}\varepsilon^{I_{1}\cdots
  I_{4}J_{1}\cdots J_{4}} \ P_{J_{1}\cdots J_{4}}\, .
\end{equation}

\noindent
These constraints must be taken into account in the action.

The graviphotons $A^{IJ}{}_{\mu}$ do not appear directly in the theory, rather they only appear through the ``dressed'' vectors, which are defined by

\begin{equation} 
A^{\Lambda}{}_{\mu}\equiv 
{\textstyle\frac{1}{2}}f^{\Lambda}{}_{IJ}A^{IJ}{}_{\mu}
+f^{\Lambda}{}_{i} A^{i}{}_{\mu}\, .
\end{equation}

\noindent
For $\mathcal{N}>4$ the larger amount of supersymmetry implies that the theory is unique: no matter can be added and all the fields belong to the gravitational multiplet. Therefore, the scalar manifold is also completely fixed, as can be seen in table \ref{table:sugragroups}.

As in the $\mathcal{N}=2$ Supergravity case, we are only interested in the bosonic part of the Lagrangian, since we are going to consider exclusively bosonic solutions. The bosonic Lagrangian is again of the form

\begin{equation}
\label{eq:generalactionN}
  \begin{array}{rcl}
S & = & {\displaystyle\int} d^{4}x\sqrt{|g|}
\left[
R
+2\Im{\rm m}\mathcal{N}_{\Lambda\Sigma} 
F^{\Lambda\, \mu\nu}F^{\Sigma}{}_{\mu\nu}
-2\Re{\rm e}\mathcal{N}_{\Lambda\Sigma} 
F^{\Lambda\, \mu\nu}\star F^{\Sigma}{}_{\mu\nu}
\right.
\\
& & \\
& &
\hspace{2cm}
\left.
+\frac{2}{4!}\alpha_{1}P^{*\, IJKL}{}_{\mu}P_{IJKL}{}^{\mu}
+\alpha_{2}P^{*\, iIJ}{}_{\mu}P_{iIJ}{}^{\mu}
\right]\, ,
\end{array}
\end{equation}

\noindent
where $\mathcal{N}_{\Lambda\Sigma}$ is the generalization of the $\mathcal{N}=2$ period
matrix, defined in Eq.~(\ref{eq:periodmatrix}), and where the parameters $\alpha_{1},\alpha_{2}$ are equal to $1$ in all cases except for $\mathcal{N}=4,6$ and $8$ as one needs to take into account the above constraints on the Vielbeine: $\alpha_{2}=1/2$ for $\mathcal{N}=4$, $\alpha_{1}+\alpha_{2}=1$ for $\mathcal{N}=6$ (the simplest choice being $\alpha_{2}=0$) and $\alpha_{1}=1/2$ for $\mathcal{N}=8$. The action is
good enough to compute the Einstein and Maxwell equations, but not the scalars' equations of motion in the cases in which the scalar Vielbeine are constrained: these constraints have to be properly dealt with and the resulting equations of motion are given below. The period matrix $\mathcal{N}_{\Lambda\Sigma}$ is defined by\footnote{See section \ref{sec:homspaces} for more details.}

\begin{equation}
\label{eq:periodmatrix}
\mathcal{N}=hf^{-1}=\mathcal{N}^{T}\, ,   
\end{equation}

\noindent
Notice that, as in the $\mathcal{N}=2$ case, Eq. (\ref{eq:periodmatrix}) implies that $\mathcal{N}_{\Lambda\Sigma}$ transforms under diffeomorphisms of the base space as required by (\ref{eq:Ntransformation}). Therefore, we can apply the formalism of section \ref{sec:duality}\footnote{Obviously the action (\ref{eq:generalactionN}) is a particular case of (\ref{eq:generalaction}).} and conclude that the equations of motion of ungauged $\mathcal{N}>2$ Supergravity enjoy duality invariance.

As corresponds to the general formalism explained in section \ref{sec:duality}, the isometry group, for each $\mathcal{N}=3,\dots,8$ must be embedded in the corresponding symplectic group, something which is always possible. As explained in section \ref{sec:homspaces}, the isometry group of I.G.R.S spaces is $G$ itself. The scalar manifold for all $\mathcal{N}>2$ Supergravities, together its symplectic representation, is detailed in table \ref{table:sugragroups}.

\begin{table}[t]
\label{table:sugragroups}
\begin{center}
\begin{tabular}{|c||c|c|}
\hline
$\mathcal{N}$& $
\begin{array}{c}
\\
$G/H$ \\
~
\end{array}
$ & $
\begin{array}{c}
\\
  \mathbf{ R}
\\
~
\end{array}

$ \\ \hline\hline
$
\begin{array}{c}
\\
\mathcal{N}=3 \\
~
\end{array}
$ & $\frac{SU(3,n_{v})}{SU(3)\times SU(n_{v})}$ & $ \mathbf{(3+n_{v})_{c}}$     \\ \hline
$
\begin{array}{c}
\\
\mathcal{N}=4 \\
~
\end{array}
$ & $\frac{SL(2, \mathbb{R})}{U(1)}\times \frac{SO(6,n_{v})}{SO(6)\times SO(n_{v})}$ & $\mathbf{(2, 6+n_{v})}$   \\ \hline
$
\begin{array}{c}
\\
\mathcal{N}=5 \\
~
\end{array}
$ & $\frac{SU(1,5)}{U(5)}$ & $ \mathbf{20}$   \\ \hline
$
\begin{array}{c}
\\
\mathcal{N}=6 \\
~
\end{array}
$ & $\frac{SO^{\ast }(12)}{U(6)}$ & $\mathbf{ 32}$
 \\ \hline
$
\begin{array}{c}
\\
\mathcal{N}=8 \\
~
\end{array}
$ & $\frac{E_{7\left( 7\right)}}{SU(8)/\mathbb{Z}_{2}}$ & $\mathbf{ 56}$   \\ \hline
\end{tabular}
\end{center}
\caption{ $N\geqslant 3$ supergravity sequence of groups $G$ of  the corresponding ${G\over H}$ symmetric spaces, and their symplectic representations  $\mathbf{R}$}
\end{table}

\cleardoublepage

%%%%%%%%%%%%%%%%%%%%%%%%%%%%%%%%%%%%%%%%%%%%%%%%%%%%%%%%%%%%%%%%%%%%%%%
%%% CHAPTER 4: SUPERGRAVITY BLACK HOLES
%%%%%%%%%%%%%%%%%%%%%%%%%%%%%%%%%%%%%%%%%%%%%%%%%%%%%%%%%%%%%%%%%%%%%%%

\renewcommand{\leftmark}{\MakeUppercase{Chapter \thechapter. Supergravity Black Holes}}
\chapter{Supergravity black holes}
\label{chapter:sugrabhs}

%%%%%%%%%%%%%%%%%%%%%%%%%%%%%%%%%%%%%%%%%%%%%%%%%%%%%%%%%%%%%%%%%%%%%%
%%%%%%%%%%%%%%%%%%%%%%%%%%%%%%%%%%%%%%%%%%%%%%%%%%%%%%%%%%%%%%%%%%%%%%
%%%%%%%%%%%%%%%%%%%%%%%%%%%%%%%%%%%%%%%%%%%%%%%%%%%%%%%%%%%%%%%%%%%%%%
%%%%%%%%%%%%%%%%%%%%%%%%%%%%%%%%%%%%%%%%%%%%%%%%%%%%%%%%%%%%%%%%%%%%%%

In this chapter we are going to obtain the form of the most general black hole solution of four-dimensional, ungauged Supergravity\footnote{In fact, we are going to obtain the form of the most general static, spherically symmetric solution of the action (\ref{eq:generalaction4}), which basically covers any theory of gravity coupled to scalars and vector fields, up to two derivatives.}, using the so-called \emph{conform-static} coordinates, since they provide a \emph{universal} way, that is, model independent, to identify the extremal limits and the horizon of the black hole. As a result, these coordinates allow a general study of the attractor behavior of several physical quantities on the event horizon of the black hole, yielding, as a plus, simplified equations of motion.

We will then use the (previously obtained) general form of a Supergravity black hole to identify its \emph{hidden} conformal symmetries, extending them to a full Virasoro algebra, which provides the link to the dual conformal description of the microscopic degrees of freedom of the entropy.

%%%%%%%%%%%%%%%%%%%%%%%%%%%%%%%%%%%%%%%%%%%%%%%%%%%%%%%%%%%%%%%%%%%%%%%%%%%%%%%%%%%%%%%%%%%%%%%%%%%%%%%%%%%%%%%
%%%%%%%%%%%%%%%%%%%%%%%%%%%%%%%%%%%%%%%%%%%%%%%%%%%%%%%%%%%%%%%%%%%%%%%%%%%%%%%%%%%%%%%%%%%%%%%%%%%%%%%%%%%%%%%
%%%%%%%%%%%%%%%%%%%%%%%%%%%%%%%%%%%%%%%%%%%%%%%%%%%%%%%%%%%%%%%%%%%%%%%%%%%%%%%%%%%%%%%%%%%%%%%%%%%%%%%%%%%%%%%
%%%%%%%%%%%%%%%%%%%%%%%%%%%%%%%%%%%%%%%%%%%%%%%%%%%%%%%%%%%%%%%%%%%%%%%%%%%%%%%%%%%%%%%%%%%%%%%%%%%%%%%%%%%%%%%

\section{The general form a Supergravity black hole}
\label{sec:sugrablackhole}

%%%%%%%%%%%%%%%%%%%%%%%%%%%%%%%%%%%%%%%%%%%%%%%%%%%%%%%%%%%%%%%%%%%%%%%%%%%%%%%%%%%%%%%%%%%%%%%%%%%%%%%%%%%%%%%
%%%%%%%%%%%%%%%%%%%%%%%%%%%%%%%%%%%%%%%%%%%%%%%%%%%%%%%%%%%%%%%%%%%%%%%%%%%%%%%%%%%%%%%%%%%%%%%%%%%%%%%%%%%%%%%
%%%%%%%%%%%%%%%%%%%%%%%%%%%%%%%%%%%%%%%%%%%%%%%%%%%%%%%%%%%%%%%%%%%%%%%%%%%%%%%%%%%%%%%%%%%%%%%%%%%%%%%%%%%%%%%
%%%%%%%%%%%%%%%%%%%%%%%%%%%%%%%%%%%%%%%%%%%%%%%%%%%%%%%%%%%%%%%%%%%%%%%%%%%%%%%%%%%%%%%%%%%%%%%%%%%%%%%%%%%%%%%

We are going to consider black-hole solutions of four-dimensional theories of the
general form
\begin{equation}
\label{eq:generalaction4}
I
=
\int d^{4}x \sqrt{|g|}
\left\{
R +\mathcal{G}_{ij}(\phi)\partial_{\mu}\phi^{i}\partial^{\mu}\phi^{j}
+2 \Im{\rm m} \mathcal{N}_{\Lambda\Sigma}
F^{\Lambda}{}_{\mu\nu}F^{\Sigma\, \mu\nu}
-2 \Re{\rm e} \mathcal{N}_{\Lambda\Sigma}
F^{\Lambda}{}_{\mu\nu}\star F^{\Sigma\, \mu\nu}
\right\}\, ,   
\end{equation}
which, as explained in chapter \ref{chapter:sugra}, includes the bosonic sectors of all four-dimensional ungauged
supergravities for appropriate $\sigma$-model metrics $\mathcal{G}_{ij}(\phi)$
and (complex) kinetic matrix $\mathcal{N}_{\Lambda\Sigma}(\phi)$, with
negative-definite imaginary part. The indices $i,j,\dots~= 1,\dots,n_{v}$ run over the scalar
fields and the indices $\Lambda, \Sigma,\dots~= 0,\dots,n_{v}$ over the 1-form fields. Their
numbers are related only for $N\geq 2$ supergravity theories.
Since we want to obtain static solutions, we consider the metric

\begin{equation}
\label{eq:generalbhmetric}
ds^{2} 
 = 
e^{2U} dt^{2} - e^{-2U} \gamma_{\underline{m}\underline{n}}
dx^{\underline{m}}dx^{\underline{n}}\, , 
\end{equation}

\noindent
where $\gamma_{\underline{m}\underline{n}}$ is a 3-dimensional
(\textit{transverse}) Riemannian metric, to be specified later. Using
Eq.~(\ref{eq:generalbhmetric}) and the assumption of staticity for all the
fields, we perform a dimensional reduction over time in the equations of
motion that follow from the above general action. We obtain a set of reduced
equations of motion that can be written in the form\footnote{See
  Ref.~\cite{Ferrara:1997tw} for more details.}
\begin{eqnarray}
\label{eq:Eq3dim1}
\nabla_{\underline{m}}
\left(\mathcal{G}_{AB} \partial^{\underline{m}}\tilde{\phi}^{B}\right)
-\tfrac{1}{2}\partial_{A} \mathcal{G}_{BC}
\partial_{\underline{m}}\tilde{\phi}^{B}\partial^{\underline{m}}\tilde{\phi}^{C} 
& = & 
0\, ,
\\
& & \nonumber \\
\label{eq:Eq3dim2}
R_{\underline{m}\underline{n}}
+\mathcal{G}_{AB}\partial_{\underline{m}}\tilde{\phi}^{A}
\partial_{\underline{n}}\tilde{\phi}^{B} 
& = & 
0\, ,
\\
& & \nonumber \\
\label{eq:Eq3dim3}
\partial_{[\underline{m}}\psi^{\Lambda}\partial_{\underline{n}]}\chi_{\Lambda} 
& = & 
0\, ,
\end{eqnarray}
where all the tensor quantities refer to the 3-dimensional metric
$\gamma_{\underline{m}\underline{n}}$ and where we have defined the metric
$\mathcal{G}_{AB}$
\begin{equation}
\mathcal{G}_{AB}
\equiv
\left(
  \begin{array}{ccc}
   2 &  &  \\
  & \mathcal{G}_{ij} &  \\
  &   & 4 e^{-2U}\mathcal{M}_{MN} 
  \end{array}
\right)\, ,
\end{equation}

\begin{equation}
(\mathcal{M}_{MN})
\equiv
\left( 
\begin{array}{lr}
    (\mathfrak{I}+\mathfrak{R}\mathfrak{I}^{-1}\mathfrak{R})_{\Lambda\Sigma} &
    -(\mathfrak{R}\mathfrak{I}^{-1})_{\Lambda}{}^{\Sigma} \\
    & \\
    -(\mathfrak{I}^{-1}\mathfrak{R})^{\Lambda}{}_{\Sigma} &
    (\mathfrak{I}^{-1})^{\Lambda\Sigma} \\   
  \end{array}
\right)
\, ,
\hspace{.3cm}
\mathfrak{R}_{\Lambda\Sigma} \equiv \Re{\rm e}\mathcal{N}_{\Lambda\Sigma}\, ,
\hspace{.3cm}
\mathfrak{I}_{\Lambda\Sigma} \equiv \Im{\rm m}\mathcal{N}_{\Lambda\Sigma}\, ,
\end{equation}

\noindent
in the \emph{extended} manifold of coordinates
$\tilde{\phi}^{A}=\left(U,\phi^{i},\psi^{\Lambda},\chi_{\Lambda}\right)$.

Eqs.~(\ref{eq:Eq3dim1}) and (\ref{eq:Eq3dim2}) can be obtained from the
three-dimensional effective action

\begin{equation}
\label{eq:Eq3dim3action3dim}
I=\int\! d^{3}x \sqrt{|\gamma|}
\left\{ R 
+\mathcal{G}_{AB}\partial_{\underline{m}}\tilde{\phi}^{A}
\partial^{\underline{m}}\tilde{\phi}^{B}\right\}\, ,
\end{equation}

\noindent
to which we have to add as a constraint the Eq.~(\ref{eq:Eq3dim3}).

In order to further dimensionally reduce the theory to a mechanical, one-dimensional, problem, we introduce the following transverse metric 
\begin{equation}
\label{eq:gammak}
\gamma_{\underline{m}\underline{n}}
dx^{\underline{m}}dx^{\underline{n}}
 = 
\frac{d\tau^{2}}{W_{\kappa}^{4}} 
+
\frac{d\Omega^{2}_{\kappa}}{W^{2}_{\kappa}}\, ,
\end{equation}
where $W_{\kappa}$ is an arbitrary function of $\tau$ and $d\Omega^{2}_{\kappa}$ is the metric of
the 2-dimensional symmetric space of curvature $\kappa$ and unit radius:
\begin{eqnarray}
d\Omega^{2}_{(1)} & \equiv & d\theta^{2}+\sin^{2}{\! \theta}\, d\phi^{2}\, ,
\\
& & \nonumber \\
\label{eq:domega1}
d\Omega^{2}_{(-1)} & \equiv & d\theta^{2}+\sinh^{2}{\! \theta}\, d\phi^{2}\, ,
\\
& & \nonumber \\
\label{eq:domega0}
d\Omega^{2}_{(0)} & \equiv & d\theta^{2}+d\phi^{2}\, .
\end{eqnarray}
Notice that for $k=1$, Eq. (\ref{eq:generalbhmetric}), assuming Eq. (\ref{eq:gammak}), is the most general spherically symmetric, static metric of a four-dimensional space-time. In the three cases $k=1,0,-1$ the equation for $W_{\kappa}(\tau)$ can be integrated and the
result is 
\begin{eqnarray}
\label{eq:sinh}
W_{1} 
& = & 
\frac{\sinh{r_{0} \tau}}{r_{0}}\, ,\\
& & \nonumber \\
\label{eq:cosh}
W_{-1}
& = &
\frac{\cosh{r_{0} \tau}}{r_{0}}\, , \\
& & \nonumber \\
\label{eq:exp}
W^{\pm}_{0}
& = & 
a e^{\mp r_{0}\tau}\, .
\end{eqnarray}
$a$ is an arbitrary real constant with dimensions of inverse length and $r_0$ is an integration constant whose interpretation depends on $k$. We are interested in the case $k=1$, corresponds to asymptotically flat, spherically symmetric, static black holes. The case $k=0$ has been recently studied in Ref.~ \cite{Bueno:2012sd} and corresponds to a rich spectrum of Lifshitz-like solutions with hyper-scaling violation. The case $k=-1$ has been studied in \cite{delaCruz-Dombriz:2013dha} and corresponds to topological solutions with a particular singular behavior. 

Remarkably enough, in the three cases (\ref{eq:sinh}), (\ref{eq:cosh}) and (\ref{eq:exp}) we are left with the same equations for the one-dimensional fields, which are given by
\begin{eqnarray}
\label{eq:Eq1tau}
\frac{d}{d\tau} \left(\mathcal{G}_{AB}
  \frac{d\tilde{\phi}^{B}}{d\tau}\right)
-\tfrac{1}{2}\partial_{A}\mathcal{G}_{BC}\frac{d\tilde{\phi}^{B}}{d\tau}
\frac{d\tilde{\phi}^{C}}{d\tau}
& = & 
0\, ,
\\
& & \nonumber \\
\label{eq:Eq2tau}
\mathcal{G}_{BC}\frac{d\tilde{\phi}^{B}}{d\tau}
\frac{d\tilde{\phi}^{C}}{d\tau}
& = & 
2 r^{2}_{0}\, .
\end{eqnarray}
The electrostatic and magnetostatic potentials $\psi^{\Lambda},\chi_{\Lambda}$
only appear through their $\tau$-derivatives. The associated conserved
quantities are the magnetic and electric charges
$p^{\Lambda},q_{\Lambda}$ and can be used to eliminate completely the
potentials. The remaining equations of motion can be put in the convenient
form 
\begin{eqnarray}
\label{eq:e1}
U^{\prime\prime}
+e^{2U}V_{\rm bh}
& = & 0\, ,\\ 
& & \nonumber \\
\label{eq:Vbh-r0-real}
(U^{\prime})^{2} 
+\tfrac{1}{2}\mathcal{G}_{ij}\phi^{i\, \prime}  \phi^{j\, \prime}  
+e^{2U} V_{\rm bh}
& = & r_{0}^{2}\, ,\\
& & \nonumber \\
\label{eq:e3}
(\mathcal{G}_{ij}\phi^{j\, \prime})^{\prime}
-\tfrac{1}{2} \partial_{i}\mathcal{G}_{jk}\phi^{j\, \prime}\phi^{k\, \prime}
+e^{2U}\partial_{i}V_{\rm bh}
& = & 0\, ,
\end{eqnarray}
in which the primes indicate differentiation with respect to $\tau$ and the
so-called \textit{black-hole potential} $V_{\rm bh}$ is given by
\begin{equation}
-V_{\rm bh}(\phi,\mathcal{Q})
\equiv
-\tfrac{1}{2}\mathcal{Q}^{M}\mathcal{Q}^{N} \mathcal{M}_{MN}\, ,
\hspace{1cm}
(\mathcal{Q}^{M})
\equiv
\left(
  \begin{array}{c}
   p^{\Lambda} \\ q_{\Lambda} \\ 
  \end{array}
\right)\, .
\end{equation}
Eqs.~(\ref{eq:e1}) and (\ref{eq:e3}) can be derived from the effective action
\begin{equation}
\label{eq:effectiveaction}
I_{\rm eff}[U,\phi^{i}] = \int d\tau \left\{ 
(U^{\prime})^{2}  
+\tfrac{1}{2}\mathcal{G}_{ij}\phi^{i\, \prime}  \phi^{j\, \prime}  
-e^{2U} V_{\rm bh}
  \right\}\, .  
\end{equation}
Eq.~(\ref{eq:Vbh-r0-real}) is nothing but the conservation of the Hamiltonian
(due to absence of explicit $\tau$-dependence of the Lagrangian) with a
particular value of the integration constant $r_{0}^{2}$.

A large number of solutions of this system, for different theories of
$\mathcal{N}=2,d=4$ supergravity coupled to vector supermultiplets, have been found
(see \textit{e.g.}~Refs.~\cite{Galli:2011fq,Mohaupt:2011aa,
Meessen:2011aa,Bueno:2012jc,Galli:2012pt}), focusing always on the case $k=1$. With this choice of transverse metric, they describe
single, charged, static, spherically-symmetric, asymptotically-flat,
non-extremal black holes. However, since the equations of motion are exactly the same in the three cases $k=1,0,-1$, these solutions are still solutions if we set $\kappa=0,-1$ in the transverse
metric, and therefore they can be used to describe Lifshitz-like and topological solutions. Hence, for each solution of the effective system of equations we can build three different metrics, representing three different, non-equivalent space-times, that solve the equations of motion of the original theory (this is the essence of the \emph{Bh-hvLif-T triality} \cite{Bueno:2012vx}).   

To summarize, and focusing only in the $k=1$ case, the metrics of all spherically symmetric, static, black-hole solutions of the
action~(\ref{eq:generalaction4}) have the general form

\begin{equation}
\label{eq:generalbhmetric4}
\begin{array}{rcl}
ds^{2} 
& = & 
e^{2U} dt^{2} - e^{-2U} \gamma_{\underline{m}\underline{n}}
dx^{\underline{m}}dx^{\underline{n}}\, ,  \\
& & \\
\gamma_{\underline{m}\underline{n}}
dx^{\underline{m}}dx^{\underline{n}}
& = & 
{\displaystyle \left(\frac{r_{0}}{\sinh{r_{0}\tau}}\right)^{2}}
\left[
{\displaystyle \left(\frac{r_{0}}{\sinh{r_{0}\tau}}\right)^{2}} d\tau^{2} 
+
d\Omega^{2}_{(2)}
\right]\, ,\\
\end{array}
\end{equation}

\noindent
where $r_{0}$ is the non-extremality parameter and $U(\tau)$ is a function of
the radial coordinate $\tau$ that characterizes each particular solution. In
these coordinates the exterior of the event horizon is covered by $\tau\in\left(-\infty,0\right)$, the event horizon being located at $\tau\rightarrow -\infty$
and the spatial infinity at $\tau \rightarrow 0^{-}$. The interior of the Cauchy
horizon (if any) is covered by $\tau\in\left(\tau_{S},\infty\right)$, the
inner horizon being located at $\tau\rightarrow +\infty$ while the singularity
is located at some finite, positive, value $\tau_S$ of the radial coordinate $\tau$ 
\cite{Galli:2011fq}.

The task of obtaining black hole solutions to the action (\ref{eq:generalaction4}) is therefore reduced to find the solution $(U(\tau),\phi^{i}(\tau))$ of the corresponding effective, one-dimensional, system of ordinary differential equations. All the four-dimensional fields, solving the original, four-dimensional, equations of motion, can be constructed from $(U(\tau),\phi^{i}(\tau))$\footnote{For $\mathcal{N}=2$ Supergravity, we will use a complex notation $(U(\tau),z^{i}(\tau))$, where the $z^{i}$ are complex scalars.}.

Using (\ref{eq:generalbhmetric4}), we can compute the area of a 2-sphere at fixed radial coordinate
$\tau=\tau_{0}$, which is given by
\begin{equation}
\label{eq:areatau}
A(\tau_{0}) = 4\pi f^2(\tau_{0}) e^{-2 U(\tau_{0})}\, ,
\end{equation}
where
\begin{equation}
\label{eq:ftau}
f(\tau)\equiv \frac{r_{0}}{\sinh{r_{0}\tau}}\, .  
\end{equation}
Therefore, the areas of the event and Cauchy horizons, $A_{+}$ and $A_{-}$
respectively, read
\begin{equation}
\label{eq:areahorizons}
A_{\pm}= \lim_{\tau_{0}\rightarrow \mp \infty} A(\tau_{0}) \, .
\end{equation}

\noindent
We will use Eq. (\ref{eq:areahorizons}) later in order to correctly interpret the near-horizon limits of the massless Klein-Gordon equation.

%%%%%%%%%%%%%%%%%%%%%%%%%%%%%%%%%%%%%%%%%%%%%%%%%%%%%%%%%%%%%%%%%%%%%%%%%%%%%%%%%%%%%%%%%%%%%%%%%%%%%%%%%%%%%%%
%%%%%%%%%%%%%%%%%%%%%%%%%%%%%%%%%%%%%%%%%%%%%%%%%%%%%%%%%%%%%%%%%%%%%%%%%%%%%%%%%%%%%%%%%%%%%%%%%%%%%%%%%%%%%%%
%%%%%%%%%%%%%%%%%%%%%%%%%%%%%%%%%%%%%%%%%%%%%%%%%%%%%%%%%%%%%%%%%%%%%%%%%%%%%%%%%%%%%%%%%%%%%%%%%%%%%%%%%%%%%%%
%%%%%%%%%%%%%%%%%%%%%%%%%%%%%%%%%%%%%%%%%%%%%%%%%%%%%%%%%%%%%%%%%%%%%%%%%%%%%%%%%%%%%%%%%%%%%%%%%%%%%%%%%%%%%%%

\section{Hidden symmetry and the microscopic description of the entropy}

%%%%%%%%%%%%%%%%%%%%%%%%%%%%%%%%%%%%%%%%%%%%%%%%%%%%%%%%%%%%%%%%%%%%%%%%%%%%%%%%%%%%%%%%%%%%%%%%%%%%%%%%%%%%%%%
%%%%%%%%%%%%%%%%%%%%%%%%%%%%%%%%%%%%%%%%%%%%%%%%%%%%%%%%%%%%%%%%%%%%%%%%%%%%%%%%%%%%%%%%%%%%%%%%%%%%%%%%%%%%%%%
%%%%%%%%%%%%%%%%%%%%%%%%%%%%%%%%%%%%%%%%%%%%%%%%%%%%%%%%%%%%%%%%%%%%%%%%%%%%%%%%%%%%%%%%%%%%%%%%%%%%%%%%%%%%%%%
%%%%%%%%%%%%%%%%%%%%%%%%%%%%%%%%%%%%%%%%%%%%%%%%%%%%%%%%%%%%%%%%%%%%%%%%%%%%%%%%%%%%%%%%%%%%%%%%%%%%%%%%%%%%%%%

In Ref.~\cite{Bertini:2011ga} it was shown that the massless Klein-Gordon equation
in the background of the four-dimensional Schwarzschild black hole exhibits a $SL(2,\mathbb{R})$
invariance in the near-horizon limit which extends to spatial infinity at
sufficiently low frequencies.  Here we will generalize these results to every
charged, static, spherically symmetric black-hole solution of ~(\ref{eq:generalaction4}), whose general black hole solution is of the form Eq.~(\ref{eq:generalbhmetric4}).

In the space-time background given by the metric (\ref{eq:generalbhmetric4}),
the massless Klein-Gordon equation
\begin{equation}
\label{eq:KG}
\frac{1}{\sqrt{|g|}}\partial_{\mu}\left(\sqrt{|g|}g^{\mu\nu}\partial_{\nu} \Phi\right)=0\, ,
\end{equation}
can be written in  the form
\begin{equation}
\label{eq:KG4d}
e^{-2 U}\partial^2_{t}\Phi-e^{2U}f^{-4}\partial^2_{\tau}\Phi 
-e^{2U} f^{-2}\Delta_{S^2}\Phi=0\, ,
\end{equation}
where $f(\tau)$ has been defined in Eq.~(\ref{eq:ftau}) and 
\begin{equation}
\label{eq:laplace}
\Delta_{S^2}\Phi
=
\frac{1}{\sin\theta}\partial_{\theta}\left(\sin\theta\partial_{\theta}\Phi\right)
+\frac{1}{\sin^2\theta}\partial^2_{\phi}\Phi\, ,
\end{equation}
is the Laplacian on the round 2-sphere of unit radius.  Using the separation
ansatz
\begin{equation}
\label{eq:separationvariables}
\Phi= e^{-i\omega t} R (\tau) Y^{l}_{m} (\theta,\phi)\, ,
\end{equation}
and
\begin{equation}
\label{eq:delta}
\Delta_{S^2}Y^{l}_{m} (\theta,\phi)=-l(l+1)Y^{l}_{m} (\theta,\phi) \, ,
\end{equation}
we find 
\begin{equation}
\omega^{2} e^{-4 U} f^2 R(\tau)+f^{-2}\partial^2_{\tau}R(\tau) = 
l (l+1)R(\tau)\, ,
\end{equation}
so we can write Eq. (\ref{eq:KG4d}) as
\begin{equation}
\label{eq:KG4dseparated1}
\mathcal{K}_{4}\Phi=l(l+1)\Phi\, ,
\end{equation}
where $\mathcal{K}_{4}$ is the second-order differential operator
\begin{equation}
\label{eq:KG4dH}
\mathcal{K}_{4} \equiv - e^{-4 U} f^{2} \partial^{2}_{t}+f^{-2}\partial^{2}_{\tau}\, .
\end{equation}

In order to exhibit the hidden conformal structure of the given space-time, we
want to find a representation of $SL(2,\mathbb{R})$ in terms of first-order
differential operators (vector fields) in the $t-\tau$ submanifold, such as
the $SL(2,\mathbb{R})$ quadratic Casimir, constructed from those vector fields
is equal to the second-order differential operator $\mathcal{K}_{4}$. Thus, we
want to find three real vector fields
\begin{equation}
\label{eq:Hs}
L_{m} =  a_{mt}\partial_{t}+a_{m\tau}\partial_{\tau}\, ,
\hspace{1cm}
m=0,\pm 1\, , 
\end{equation}
for some functions $a_{mt}(t,\tau),a_{m\tau}(t,\tau)$, whose Lie brackets
satisfy $\mathfrak{sl}(2)$ Lie algebra
\begin{eqnarray}
\label{eq:Hsconmutation}
[L_{m},L_{n}]=(m-n)  L_{m+n}\, ,
\hspace{1cm}
m=0,\pm 1\, ,
\end{eqnarray}
and such that 
\begin{eqnarray}
\label{eq:CasimirK}
\mathcal{H}^{2}
\equiv
L^{2}_{0}-\tfrac{1}{2}\left(L_{1} L_{-1}+L_{-1} L_{1}\right) 
= 
\mathcal{K}_{4}\, .
\end{eqnarray}

In order to simplify the problem, following \cite{Bertini:2011ga}, we have to
make some additional assumptions on the functions
$a_{It}(t,\tau),a_{I\tau}(t,\tau)$, Thus, we make the following ansatz
\begin{eqnarray}
\label{eq:Hs2}
L_{1} 
& = & 
l(t)\left[-m(\tau)\partial_{t}+n(\tau)\partial_{\tau}\right]\, ,
\\
& & \nonumber \\
L_{0} 
& = & 
-\frac{c}{r_{0}}\partial_{t}\, ,
\\
& & \nonumber \\
L_{-1} 
& = & 
-l^{-1}(t)\left[ m(\tau)\partial_{t}+n(\tau)\partial_{\tau}\right]\, ,
\end{eqnarray}
where $m$ and $n$ are functions of $\tau$, $l$ is a function of $t$ and $c$ is
a real constant.

Plugging this ansatz into Eq.~(\ref{eq:Hsconmutation}) we obtain two
differential equations
\begin{eqnarray}
\label{eq:diff1}
m^{2}\partial_{t} \log l + n\partial_{\tau} m 
& = & 
\frac{c}{r_{0}}\, ,
\\
& & \nonumber \\
\frac{c}{r_{0}}\partial_{t}\log l = 1\, ,
\end{eqnarray}
and plugging it into Eq.~(\ref{eq:CasimirK}) we obtain three equations
\begin{eqnarray}
\label{eq:diff2}
m 
& = & 
h\partial_{\tau} n\, ,
\\
& & \nonumber \\
m^{2} 
& = &
e^{-4 U} f^{2}+\left(c/r_{0}\right)^{2}\, ,
\\
& & \nonumber \\
n^{2}
& =& f^{-2}\, .
\end{eqnarray}

These equations cannot be solved for arbitrary $U(\tau)$: we can find $l,m,n$
as functions of $f(\tau)$ and the constant $c$
\begin{equation}
\label{eq:sol1}
l(t) = c_{0} e^{r_{0}t/c}\, ,
\hspace{1cm}
 n^{2}(\tau)=f^{-2}\, ,
\hspace{1cm}
m(\tau)=h \cosh{(r_{0}\tau)}\, ,
\end{equation}
for some real constant $c_{0}$, leaving the following equation for the
constant $c$ to be solved:
\begin{eqnarray}
\label{eq:constraint2}
c^{2} = \left(e^{-2U} f^{2}\right)^{2}\, .
\end{eqnarray}

This equation has only one exact solution, given by $e^{U}\sim f$, which does not correspond to any
asymptotically flat black hole. We have to content ourselves with a range of
values of the coordinate $\tau$ in which the above equation can be solved
approximately. The two ranges that we have identified correspond to the two
near-horizon regions (event and Cauchy horizons $\tau\rightarrow -\infty$ or
$\tau\rightarrow -\infty$, respectively) in which 
\begin{equation}
\label{eq:constraint3}
\left(e^{-2U} f^{2}\right)^{2}
\stackrel{\tau \rightarrow \mp \infty}{\sim}
\left(\frac{A_{\pm}}{4\pi}\right)^{2}+\mathcal{O}(e^{\pm r_{0}\tau})
=
c^{2}+\mathcal{O}(e^{\pm r_{0}\tau})\, ,
\end{equation}
according to Eq.~(\ref{eq:areahorizons}).

We conclude that in the geometry of any four-dimensional, charged, static,
black-hole solution of a theory of the form~(\ref{eq:generalaction4}),
there are two triplets of vector fields $L^{+}_{m}$ and $L^{-}_{m}$, $m=0,\pm
1$ given by
\begin{eqnarray}
\label{eq:Hs+}
L^{\pm}_{1}  
& = &
-\frac{e^{r_{0}\pi t /S_{\pm}}}{r_{0}}\left(\frac{S_{\pm}}{\pi} \cosh{(r_{0} \tau)}
\partial_{t}+\sinh{(r_{0}\tau)}\partial_{\tau}\right)\\
& & \nonumber \\
L^{\pm}_{0} 
& = &
 -\frac{S_{\pm}}{r_{0}\pi}\partial_{t}\, ,
\\
& & \nonumber \\
L^{\pm}_{-1} 
& = &
-\frac{e^{-r_{0}\pi t /S_{\pm}}}{r_{0}}
\left(\frac{S_{\pm}}{\pi} \cosh{(r_{0}\tau)}\partial_{t} 
-\sinh{(r_{0}\tau)}\partial_{\tau}\right)\, ,
\end{eqnarray}
where $S_{\pm}=\frac{A_{\pm}}{4}$, which generate two $\mathfrak{sl}(2)$
algebras whose quadratic Casimirs
\begin{equation}
\mathcal{H}^{\pm\, 2}
\equiv
(L^{\pm}_{0})^{2}
-\tfrac{1}{2}\left(L^{\pm}_{1} L^{\pm}_{-1}+L^{\pm}_{-1} L^{\pm}_{1}\right)\, ,   
\end{equation}
approximate the massless Klein-Gordon equation in the two near-horizon
regions\footnote{Observe that we only approximate some terms (i.e.~we keep some
  sub-dominating terms): 
\begin{equation}
e^{-4U}f^{2} = f^{-2} (e^{-2U}f^{2})^{2} 
\sim     
f^{-2}
\left[\left(\frac{A_{\pm}}{4\pi}\right)^{2}+\mathcal{O}(e^{\pm
    r_{0}\tau})\right]
\sim f^{-2}
\left(\frac{A_{\pm}}{4\pi}\right)^{2}+\mathcal{O}(e^{\pm
    r_{0}\tau})\, ,
\end{equation}
which is correct to that order. On the other hand, we do not need to restrict
ourselves to any particular range of frequencies.  
}:
\begin{equation}
\mathcal{K}_{4}\Phi 
= \left\{- e^{-4 U} f^{2} \partial^{2}_{t}+f^{-2}\partial^{2}_{\tau} \right\}
\Phi\,\, \stackrel{\tau\rightarrow \mp \infty}{\longrightarrow}\,\,
f^{-2}\left\{-\left(S_{\pm}/\pi \right)^{2}\partial^{2}_{t}+\partial^{2}_{\tau} \right\}
\Phi
=
\mathcal{H}^{\pm\, 2} \Phi\, . 
\end{equation}

We can see from Eq.~(\ref{eq:Hs+}) that the extremal limit $r_{0}\rightarrow
0$ is singular. The reason is that the operations of taking the near-horizon
limit and of taking the extremal limit $r_{0}\rightarrow 0$ do not commute.

The $\mathfrak{sl}(2)$ algebra that we have just found can be immediately
extended to a complete Witt algebra (or a Virasoro algebra with vanishing
central charge) with the commutation relations (\ref{eq:Hsconmutation}) for
all $m\in \mathbb{Z}$. The generators of the Witt algebra are given by
\begin{eqnarray}
\label{eq:Lm}
L^{\pm}_{m}  
& = &
-\frac{e^{m r_{0}\pi t /S_{\pm}}}{r_{0}}\left(\frac{S_{\pm}}{\pi} \cosh{(m r_{0} \tau)}
\partial_{t}+\sinh{(m r_{0}\tau)}\partial_{\tau}\right) \, .
\end{eqnarray}

To summarize, we have constructed two Witt algebras which have a well-defined
action in the space of solutions to the wave equation in the background of the
exterior and interior near-horizon limits of a generic, charged, static black
hole solution of (\ref{eq:generalaction4}). The two $\mathfrak{sl}(2)$ subalgebras are symmetries of these wave
equations, since the wave operators can be seen as their Casimirs, but they
are not symmetries of the background metrics which, being essentially the
products of Rindler spacetime (locally Minkowski) and spheres, have abelian
(in the time-radial part) isometry algebras.

This result generalizes those obtained in
Refs.~\cite{Bertini:2011ga,Wang:2010qv,Lowe:2011aa,Camblong:2004ye}, and present an opportunity to put to test some conjectures and common lore of this field. To
start with, is there a CFT associated to the Witt algebras and can one compute
the central charge of the Virasoro algebra? A most naive computation does not
seem to give meaningful results. This, of course, does not preclude the
possibility that a more rigorous calculation, preceded of careful definitions
of the boundary conditions of the fields at the relevant boundaries (which
have to be identified first) may give a meaningful answer.

Meanwhile, it is amusing to speculate on the possible consequences of the
existence of such a CFT withe left and right sectors whose entropies $S_{\rm
  R},S_{\rm L}$ and temperatures would be related to the temperatures and
entropies of the outer and inner horizons ($T_{+},T_{-}$ and $S_{+},S_{-}$,
respectively) by

\begin{eqnarray}
S_{\pm} & = & S_{\rm R} \pm S_{\rm L}\, , \\
& & \nonumber \\
\frac{1}{T_{\pm}} & = & 
\tfrac{1}{2}
\left(
\frac{1}{T_{\rm R}} \pm \frac{1}{T_{\rm L}}
\right)\, , 
\end{eqnarray}

\noindent
and obeying the fundamental relation 

\begin{equation}
S_{+} = \frac{\pi^{2}}{12}(c_{\rm R}T_{R} +c_{\rm L}T_{\rm L})\, ,  
\end{equation}

\noindent
where $c_{\rm L,R}$ are the central charges of the left and right sectors,
which will be assumed to be equal $c_{\rm R}= c_{\rm L} = c$.

The temperatures and entropies of the outer and inner horizons are related to
the non-extremality parameter $r_{0}$ by

\begin{equation}
2S_{\pm}T_{\pm} = r_{0}\, ,  
\end{equation}

\noindent
which implies for the temperatures of the left and right sectors

\begin{equation}
4S_{\rm L,R}T_{\rm L,R} = r_{0}\, .  
\end{equation}

\noindent
In the extremal limit 

\begin{equation}
S_{\rm L} \rightarrow 0\, ,
\hspace{.5cm}
T_{\rm R} \rightarrow 0\, ,
\hspace{.5cm}
T_{\pm} \rightarrow 0\, ,
\hspace{.5cm}
S_{\pm} \rightarrow S_{\rm R}\, ,
\end{equation}

\noindent
and both $S_{\rm R}$ and $T_{\rm L}$ remain finite and are convenient
quantities to work with. In particular, we can express the central charge that
the CFT should have in order to reproduce the Bekenstein-Hawking entropy 
consistently with this picture, in terms of these two parameters:

\begin{equation}
c = \frac{12}{\pi^{2}} \frac{S_{\rm R}}{T_{\rm L}}\, .  
\end{equation}
%\cleardoublepage

%%%%%%%%%%%%%%%%%%%%%%%%%%%%%%%%%%%%%%%%%%%%%%%%%%%%%%%%%%%%%%%%%%%%%%%
%%% CHAPTER 5: SUPERSYMMETRIC SOLUTIONS
%%%%%%%%%%%%%%%%%%%%%%%%%%%%%%%%%%%%%%%%%%%%%%%%%%%%%%%%%%%%%%%%%%%%%%%

\renewcommand{\leftmark}{\MakeUppercase{Chapter \thechapter. Supersymmetric Solutions}}
\chapter{All the supersymmetric black holes of extended Supergravity}
\label{chapter:susysolutions}

In this chapter we are going to explicitly construct the most general (single and multi-center) supersymmetric black hole metric of $\mathcal{N}>2$  ungauged Supergravity in four dimensions, using the algorithm provided to that effect in \cite{Meessen:2010fh}\footnote{Generalizing the results of the seminal work \cite{Gauntlett:2002nw} by Gauntlett \emph{et al}. }, where the exhaustive classification of all the time-like supersymmetric solutions of any extended four-dimensional ungauged Supergravity was performed. Although we are going first to specialize to the case of $\mathcal{N}=8$ Supergravity, thanks to the properties of the \emph{groups of Type} $E_7$ we will see that our results also apply to all $\mathcal{N}>2$ Supergravities and also to specifc instances of $\mathcal{N}=2$, namely those with symmetric scalar manifolds. For the remaining $\mathcal{N}=2$ Supergravity cases, since the theory (in the absence of hypermultiplets) is specified upon the choice of a Special K\"ahler manifold, we can only characterize the form of the supersymmetric solution, whose details depend on the particular $\mathcal{N}=2$ model. Previous results on black holes and attractors in $\mathcal{N}=8$ Supergravity can be found in \cite{Khuri:1995xk,Arcioni:1998mn,Bertolini:1998mt,Bertolini:1999je,Bertolini:1999uz,Bertolini:2000ei,Ferrara:2006em,Ferrara:2009bw,Ceresole:2009jc,Ortin:2011vm,Cacciatori:2010ws,Bossard:2012ge,Cacciatori:2012wu}.

%%%%%%%%%%%%%%%%%%%%%%%%%%%%%%%%%%%%%%%%%%%%%%%%%%%%%%%%%%%%%%%%%%%%%%
%%%%%%%%%%%%%%%%%%%%%%%%%%%%%%%%%%%%%%%%%%%%%%%%%%%%%%%%%%%%%%%%%%%%%%
%%%%%%%%%%%%%%%%%%%%%%%%%%%%%%%%%%%%%%%%%%%%%%%%%%%%%%%%%%%%%%%%%%%%%% 
%%%%%%%%%%%%%%%%%%%%%%%%%%%%%%%%%%%%%%%%%%%%%%%%%%%%%%%%%%%%%%%%%%%%%%

\section{The mathematical formalism}
\label{sec:mathformalism}

%%%%%%%%%%%%%%%%%%%%%%%%%%%%%%%%%%%%%%%%%%%%%%%%%%%%%%%%%%%%%%%%%%%%%%
%%%%%%%%%%%%%%%%%%%%%%%%%%%%%%%%%%%%%%%%%%%%%%%%%%%%%%%%%%%%%%%%%%%%%%
%%%%%%%%%%%%%%%%%%%%%%%%%%%%%%%%%%%%%%%%%%%%%%%%%%%%%%%%%%%%%%%%%%%%%% 
%%%%%%%%%%%%%%%%%%%%%%%%%%%%%%%%%%%%%%%%%%%%%%%%%%%%%%%%%%%%%%%%%%%%%%

According to the results of \cite{Meessen:2010fh}, in order to construct a
timelike black-hole-type supersymmetric solution of $\mathcal{N}=8$
supergravity we may proceed as follows\footnote{We have included in this
  recipe, to simplify it, the vanishing of the ``hyperscalars''.}:

\begin{enumerate}
\item Choose an $x$-dependent rank-2, $8 \times 8$ complex antisymmetric
  $M_{IJ}$, These matrices must satisfy a number of constraints that are
  difficult to solve. This implies that, in practice, we cannot construct the
  most general matrices that satisfy them. Nevertheless, with those matrices
  we can proceed to the next step.

\item The scalars are encoded into the $56$-dimensional symplectic vector

\begin{equation}
(\mathcal{V}^{M}{}_{IJ})
=
\left( 
  \begin{array}{c}
f^{ij}{}_{IJ} \\ h_{ij\, IJ} \\
  \end{array}
\right)\, ,
\end{equation}

\noindent
antisymmetric in the \textit{local} $SU(8)$ indices $I,J=1,\cdots 8$. It
transforms in the fundamental (\textbf{56}) of $E_{7(7)}$ ($ij$ indices) and
as antisymmetric $U(8)$ tensor ($IJ$ indices), It satisfies\footnote{The
  symplectic product of two vectors $\langle \mathcal{A}\mid\mathcal{B}\rangle$
is defined by 
\begin{equation}
\langle \mathcal{A}\mid\mathcal{B}\rangle 
\equiv 
\mathcal{A}_{M}\mathcal{B}^{M}
\equiv
\mathcal{A}^{N}\mathcal{B}^{M}\Omega_{MN}\, ,
\end{equation}
where
\begin{equation}
(\Omega_{MN})
\equiv
\left(
  \begin{array}{cc}
 0 & \mathbbm{1}_{28\times 28} \\
-\mathbbm{1}_{28\times 28}   & 0 \\
  \end{array}
\right)\, ,
\end{equation}
is the skew metric of Sp$(56,\mathbb{R})$ that we use to lower (as above) or
raise symplectic indices.  
}

\begin{equation}
\langle \mathcal{V}_{IJ}\mid\bar{\mathcal{V}}^{KL}\rangle 
=
\tfrac{1}{2}\bar{f}^{ij\, KL}h_{ij\, IJ} 
- 
\tfrac{1}{2} \bar{h}_{ij}{}^{KL}f^{ij}{}_{IJ}
=    
-2i\delta^{KL}{}_{IJ}\, ,
\hspace{1cm}
\langle \mathcal{V}_{IJ}\mid\mathcal{V}_{KL}\rangle  
= 
0\, ,
\end{equation}

Using the matrix $M_{IJ}$ chosen in the previous step, we define the real
symplectic vectors $\mathcal{R}^{M}$ and $\mathcal{I}^{M}$

\begin{equation}
\mathcal{R}^{M}+i\mathcal{I}^{M} 
\equiv  
\mathcal{V}^{M}{}_{IJ}\frac{M^{IJ}}{|M|^{2}}\, ,
\hspace{1cm}
|M|^{2} = M_{IJ}M^{IJ}\, .
\end{equation}

\noindent
These two are, by definition, $U(8)$ singlets (no $U(8)$ gauge-fixing
necessary) and only transform in the fundamental of $E_{7(7)}$.

\item The components of $\mathcal{I}$ are $56$ real functions
  $\mathcal{H}^{M}$ harmonic in the Euclidean $\mathbb{R}^{3}$ transverse
  space.

\item $\mathcal{R}$ is to be be found from $\mathcal{I}$ exploiting the
  redundancy in the description of the scalars by the sections
  $\mathcal{V}^{M}{}_{IJ}$\footnote{$\mathcal{V}^{M}{}_{IJ}$ uses $56^{2}$
    complex components to describe just $70$ physical scalars. The constraints
    that it satisfies imply a large number of relations between the
    components. The same is true for the components projected with
    $M_{IJ}$. This step is equivalent to the resolution of the
    \textit{stabilization equations} in $\mathcal{N}=2$ theories.}. Even with
  the knowledge of $M_{IJ}$ this is a very difficult step.

\item The metric is

\begin{equation}
\label{eq:themetric}
ds^{2} \; =\; e^{2U} (dt+ \omega)^{2} -e^{-2U}d\vec{x}^{\, 2}\, ,
\end{equation}

\noindent
where
  
\begin{eqnarray}
e^{-2U} 
& = &    
|M|^{-2}
=
\langle\,\mathcal{R}\mid \mathcal{I}\, \rangle
= 
\tfrac{1}{2}\mathcal{I}^{ij}\mathcal{R}_{ij}
-
\tfrac{1}{2}\mathcal{I}_{ij}\mathcal{R}^{ij}\, ,
\\
& & \nonumber \\
(d \omega)_{mn} 
& = &  
2\epsilon_{mnp}
\langle\,\mathcal{I}\mid \partial_{p}\mathcal{I}\, \rangle\, ,
\label{eq:omegaequation}
\end{eqnarray}

\noindent
and can be constructed automatically provided one has been given the harmonic
functions corresponding to $\mathcal{I}$ and $\mathcal{R}(\mathcal{I})$, quite
independently of the construction of these objects from $M_{IJ}$ and
$\mathcal{V}^{M}{}_{IJ}$. The same is true for the vector field strengths.

\item The vector field strengths are given by 

\begin{equation}
\mathcal{F}
=  
-{\textstyle\frac{1}{2}} d (\mathcal{R}\hat{V})   
-{\textstyle\frac{1}{2}}\star(\hat{V}\wedge d\mathcal{I}) \, ,  
\hspace{1.5cm}
\hat{V} = \sqrt{2} e^{2U}(dt+ \omega)\, . 
\end{equation}

\item The Vielbeins describing the scalars in the coset $E_{7(7)}/SU(8)$
  $P_{IJKL, \mu}$ are split into two complementary sets:

  \begin{equation}
    P_{IJKL}\, 
    \mathcal{J}^{I}{}_{[M}\mathcal{J}^{J}{}_{N} \mathcal{J}{}^{K}{}_{P} 
    \tilde{\mathcal{J}}{}^{L}{}_{Q]}\, , \,\,\,\,\,
    \mathrm{and}\,\,\,\,\,\, 
    P_{IJKL}\, \mathcal{J}^{I}{}_{[M} \tilde{\mathcal{J}}{}^{J}{}_{N}
    \tilde{\mathcal{J}}{}^{K}{}_{P} \tilde{\mathcal{J}}{}^{L}{}_{Q]}\, ,
  \end{equation}

\noindent
where we have defined the projectors 

\begin{equation}
\mathcal{J}^{I}{}_{J} \equiv \frac{2 M^{IK} M_{JK}}{|M|^{2}}\, ,
\hspace{1cm}
\mathcal{J}^{I}{}_{J}=\delta^{I}{}_{J}-\tilde{\mathcal{J}}^{I}{}_{J}\, .    
\end{equation}

  All those in the second set have been assumed to vanish from the start,
  since they would lead to a non-trivial metric in the transverse
  3-dimensional space, while those in the first set can in principle be found
  from $\mathcal{R}$ and $\mathcal{I}$, using the definitions of these vectors
  and of the Vielbein and the explicit form of the chosen $M_{IJ}$, setting
  $\mathcal{I}^{M}=H^{M}(x)$ and confronting the third step: the resolution of
  the stabilization equations.

\end{enumerate}

%%%%%%%%%%%%%%%%%%%%%%%%%%%%%%%%%%%%%%%%%%%%%%%%%%%%%%%%%%%%%%%%%%%%%%
%%%%%%%%%%%%%%%%%%%%%%%%%%%%%%%%%%%%%%%%%%%%%%%%%%%%%%%%%%%%%%%%%%%%%%
%%%%%%%%%%%%%%%%%%%%%%%%%%%%%%%%%%%%%%%%%%%%%%%%%%%%%%%%%%%%%%%%%%%%%% 
%%%%%%%%%%%%%%%%%%%%%%%%%%%%%%%%%%%%%%%%%%%%%%%%%%%%%%%%%%%%%%%%%%%%%%

\section{The supersymmetric black hole solution of \texorpdfstring{$\mathcal{N}=8$}{mathcal{N}=8} Supergravity}

%%%%%%%%%%%%%%%%%%%%%%%%%%%%%%%%%%%%%%%%%%%%%%%%%%%%%%%%%%%%%%%%%%%%%%
%%%%%%%%%%%%%%%%%%%%%%%%%%%%%%%%%%%%%%%%%%%%%%%%%%%%%%%%%%%%%%%%%%%%%%
%%%%%%%%%%%%%%%%%%%%%%%%%%%%%%%%%%%%%%%%%%%%%%%%%%%%%%%%%%%%%%%%%%%%%% 
%%%%%%%%%%%%%%%%%%%%%%%%%%%%%%%%%%%%%%%%%%%%%%%%%%%%%%%%%%%%%%%%%%%%%%

For the last 20 years, black holes have been intensively studied in string
theory and supergravity with never-decreasing interest. A large part of effort
has been focused on two subjects: the construction of the most general
black-hole solutions of these theories and the understanding and computation
of different physical properties, specially the entropy, of the black-hole
solutions, following the seminal result of Strominger and Vafa
\cite{Strominger:1996sh}.

The attractor mechanism \cite{Ferrara:1997tw,Ferrara:1995ih} has provided a
bridge between these two subjects, allowing the computation of the entropy and
other black-hole properties on the black-hole horizon without the knowledge of
the complete black-hole solutions, at least in the extremal cases. In theories
with a very high degree of (super-) symmetry, though, it is not necessary to
use this mechanism and the entropy of the extremal black holes can be
determined requiring duality-invariance, correct dimensionality and
moduli-independence (which is a consequence of the attractor mechanism
\cite{Ferrara:1997tw}). In particular, the entropy of the extremal black holes
of $\mathcal{N}=8$ supergravity \cite{Cremmer:1979up,deWit:1982ig} was found
in \cite{Kallosh:1996uy} to be given by the unique quartic invariant of the
$E_{7(7)}$ duality group. If we use the real basis 

\begin{equation}
\mathcal{Q} \equiv   
\left(
  \begin{array}{c}
  p^{ij} \\ q_{ij} \\  
  \end{array}
\right)\,  , 
\end{equation}

\noindent
for the charges, where the indices $i,j=1,\cdots,8$ transform homogeneously
under the $SL(8,\mathbb{R})\subset E_{7(7)}$ and each pair of indices is
antisymmetrized (so there are $28$ electric plus $28$ magnetic independent
charges), the quartic invariant is known as the Cartan invariant
$J_{4}(\mathcal{Q})$ \cite{CARTAN}

\begin{equation}
 \label{Cartan} 
J_{4}(\mathcal{Q})
= 
p^{ij}q_{jk}p^{kl}q_{li}
-\tfrac{1}{4} (p^{ij}q_{ij})^{2} 
+\tfrac{1}{96}\, \varepsilon_{ijklmnpq}p^{ij}p^{kl}p^{mn}p^{pq} 
+\tfrac{1}{96}\, \varepsilon^{ijklmnpq}q_{ij}q_{kl}q_{mn}q_{pq}\, . 
\end{equation}

In the complex basis, the quartic invariant is known as the Julia-Cremmer
invariant $\diamondsuit (\mathcal{Q})$ \cite{Cremmer:1979up}. They are
equal up to a sign \cite{Balasubramanian:1997az,Gunaydin:2000xr} and we will
not be concerned with its explicit form.

Although it has not been proven directly\footnote{To the best of our
  knowledge, not even within the FGK formalism of \cite{Ferrara:1997tw}.}, the
entropy formula for the extremal black holes of $\mathcal{N}=8$
supergravity

\begin{equation}
\label{eq:entropyformula}
S = \pi \sqrt{|J_{4}(\mathcal{Q})|}\, ,   
\end{equation}

\noindent
has passed all checks and, in particular, it has been shown to reproduce the
entropies of black holes of supergravity theories with $\mathcal{N}<8$
(specially $\mathcal{N}=2$) obtained by truncation of $\mathcal{N}=8$. For
supersymmetric black holes $J_{4}(\mathcal{Q})>0$ and one does not need to
take the absolute value.

One of the main obstructions for proving this formula is our lack of knowledge
of the general extremal black-hole solutions of $\mathcal{N}=8$ supergravity
as opposite to our complete knowledge of those of the $\mathcal{N}=2$ theories
\cite{Behrndt:1997ny,LopesCardoso:2000qm,Denef:2000nb,Bates:2003vx,Meessen:2006tu}. This,
and the standard lore that all the $1/8$ supersymmetric (the ones with a
potentially regular horizon) black-hole solutions of $\mathcal{N}=8$ are
supersymmetric black-hole solutions of some of the $\mathcal{N}=2$ truncations
of that theory (which seems to have been disproven by the explicit examples of
\cite{Bena:2011pi,Bena:2012ub}) explains why most of the literature on
$\mathcal{N}=8$ black holes deals with such truncations.

The supersymmetric black-hole solutions of $\mathcal{N}=2$ supergravity were
re-discovered in \cite{Meessen:2006tu} among the time-like supersymmetric
solutions of the theory, which were found by exploiting the integrability
conditions of the Killing spinor equations following Tod \cite{Tod:1983pm}
along the lines of \cite{Gauntlett:2002nw}. The same procedure was followed in
\cite{Meessen:2010fh} for all $\mathcal{N}\geq 2, d=4$ ungauged
supergravities, using the (almost) $\mathcal{N}$-independent formalism of
\cite{Andrianopoli:1996ve}, but the result, which we are going to explain in
the next section, looked too complicated to be used in the explicit
construction of the solutions, in spite to its similarity to the result found
in the $\mathcal{N}=2$ case.

We have recently realized, though, that the results found in
\cite{Meessen:2010fh} do permit the explicit construction of the metric of the
most general single and multi-black-hole solutions of ungauged $\mathcal{N}=8$
supergravity. The complications are restricted to the explicit construction of
the scalar fields. Thus, we are going to show how to construct the metrics of
the most general black holes ungauged $\mathcal{N}=8$ supergravity, but we
will not be able to provide a simple algorithm to find the scalar fields
corresponding to those solution. Nevertheless, the consistency of the formalism
ensures their existence and there is much that can be learned from the
metrics.

In the next section we are going to give the general form of the supersymmetric metric solution of $\mathcal{N}=8$ ungauged Supergravity, after which we will discuss the black-hole case, showing how the entropy formula (\ref{eq:entropyformula}) arises for supersymmetric black holes and which of $E_{7(7)}$ invariants studied in \cite{Andrianopoli:2011gy} actually arise in the two-center case.

%%%%%%%%%%%%%%%%%%%%%%%%%%%%%%%%%%%%%%%%%%%%%%%%%%%%%%%%%%%%%%%%%%%%%%
%%%%%%%%%%%%%%%%%%%%%%%%%%%%%%%%%%%%%%%%%%%%%%%%%%%%%%%%%%%%%%%%%%%%%%
%%%%%%%%%%%%%%%%%%%%%%%%%%%%%%%%%%%%%%%%%%%%%%%%%%%%%%%%%%%%%%%%%%%%%% 
%%%%%%%%%%%%%%%%%%%%%%%%%%%%%%%%%%%%%%%%%%%%%%%%%%%%%%%%%%%%%%%%%%%%%%

\subsection{The metrics of the supersymmetric black-hole solutions of
  \texorpdfstring{$\mathcal{N}=8$}{mathcal{N}=8} supergravity}
  \label{sec:N8stabilization}

%%%%%%%%%%%%%%%%%%%%%%%%%%%%%%%%%%%%%%%%%%%%%%%%%%%%%%%%%%%%%%%%%%%%%%
%%%%%%%%%%%%%%%%%%%%%%%%%%%%%%%%%%%%%%%%%%%%%%%%%%%%%%%%%%%%%%%%%%%%%%
%%%%%%%%%%%%%%%%%%%%%%%%%%%%%%%%%%%%%%%%%%%%%%%%%%%%%%%%%%%%%%%%%%%%%% 
%%%%%%%%%%%%%%%%%%%%%%%%%%%%%%%%%%%%%%%%%%%%%%%%%%%%%%%%%%%%%%%%%%%%%%

If we want to construct the most general black-hole solutions of
$\mathcal{N}=8$ supergravity, the recipe demands a parametrization of the
space of all the matrices $M_{IJ}(x)$ that satisfy all the technical
requirements, which is very difficult to find.

We have realized, however, that this is a problem that we only need to solve
explicitly if we want to construct explicitly the scalar fields.  If we are
only interested in constructing the metric (and perhaps the vector fields) all
we really need is to assume that the problem has been solved and the resulting
$M_{IJ}(x)$ has been used to define $\mathcal{R}$ and $\mathcal{I}$.

One may naively think that both the explicit form of $M_{IJ}(x)$ and the
explicit expression of the components $\mathcal{V}^{M}{}_{IJ}$ are needed to
set up the stabilization equations and to solve them, finding $\mathcal{R}$ as
a function of $\mathcal{I}$.  Fortunately, this problem can be reformulated as
follows: with a real vector in the $\mathbf{56}$ of $E_{7(7)}$, $\mathcal{I}$,
we want to construct another one in the same representation which is a
non-trivial function of the former, $\mathcal{R}(\mathcal{I})$. For a single
$\mathcal{I}$, there is a unique way of constructing a $\mathbf{56}$ from
another $\mathbf{56}$, provided by the Jordan triple product\footnote{The
  Jordan triple product of three different $\mathbf{56}$s is defined only up
  to terms proportional to the symplectic products of two of the three
  $\mathbf{56}$s. The ambiguity disappears when we consider them to be equal,
  since the symplectic products will automatically vanish.}. Thus,
$\mathcal{R}^{M}(\mathcal{I})$ must be given by

\begin{equation}
  \mathcal{R}^{M}(\mathcal{I}) \sim (\mathcal{I},\mathcal{I},\mathcal{I})^{M}\, ,  
\end{equation}

\noindent
where

\begin{equation}
  \begin{array}{rcl}
    (\mathcal{I},\mathcal{I},\mathcal{I})^{ij}
    & = &  
    \tfrac{1}{2}
    \mathcal{I}^{ik}\mathcal{I}_{kl}\mathcal{I}^{lj}
    +\tfrac{1}{8}
    \mathcal{I}^{ij}\mathcal{I}_{kl}\mathcal{I}^{kl}
    -\tfrac{1}{96}\varepsilon^{ijklmnpq}\mathcal{I}_{kl}\mathcal{I}_{mn}\mathcal{I}_{pq}\, ,
    \\    
    & & \\
    (\mathcal{I},\mathcal{I},\mathcal{I})_{ij}
    & = &  
    -\tfrac{1}{2}
    \mathcal{I}_{ik}\mathcal{I}^{kl}\mathcal{I}_{lj}
    -\tfrac{1}{8}
    \mathcal{I}_{ij}\mathcal{I}_{kl}\mathcal{I}^{kl}
    +\tfrac{1}{96}\varepsilon_{ijklmnpq}\mathcal{I}^{kl}\mathcal{I}^{mn}\mathcal{I}^{pq}\, .
    \\    
  \end{array}
\end{equation}

To determine the proportionality factor we must first take into account that
we expect the $\mathcal{R}^{M}(\mathcal{I})$ to be homogenous of first order
in $\mathcal{I}$, which requires that we divide
$(\mathcal{I},\mathcal{I},\mathcal{I})$ by an $E_{7(7)}$ invariant (to
preserve the symmetry properties) homogenous of second degree in
$\mathcal{I}$, which can only be $\sqrt{J_{4}(\mathcal{I})}$.

We, thus, conclude that, up to a normalization constant $\beta$ to be
determined later, the solution to the stabilization equations of
$\mathcal{N}=8$ supergravity defined in the previous section is

\begin{equation}
\label{eq:R}
  \mathcal{R}^{M}(\mathcal{I}) = 
  \beta \frac{(\mathcal{I},\mathcal{I},\mathcal{I})^{M}}{\sqrt{J_{4}(\mathcal{I})}}\, ,  
\end{equation}

\noindent
which is our main result and allows the complete construction of the metrics
of all the supersymmetric black holes of the theory. 

Actually, since, as we are going to show in the next section, $\beta=2$,
$\mathcal{R}^{M}$ coincides exactly with the \textit{Freudenthal
  dual}\footnote{We thank M.~Duff and L.~Borsten for pointing out this fact to
  us.} of $\mathcal{I}^{M}$, which we can denote by $\tilde{I}^{M}$ defined in
\cite{Borsten:2009zy}. The Freudenthal dual $\tilde{Q}$ enjoys several
remarkable properties. Firstly,

\begin{equation}
\label{eq:firstproperty}
\langle\, \tilde{\mathcal{Q}} \mid \mathcal{Q}\, \rangle  = 2
J_{4}(\mathcal{Q})\, ,   
\end{equation}

\noindent
which follows from the property of the Jordan triple product

\begin{equation}
  \langle\, (\mathcal{Q},\mathcal{Q},\mathcal{Q})\mid \mathcal{Q}\, \rangle  
  =
  J_{4}(\mathcal{Q})\, .
\end{equation}

Secondly, 

\begin{equation}
\label{eq:secondproperty}
\tilde{\tilde{\cal Q}} = -\mathcal{Q}\, ,  
\end{equation}

\noindent
which eliminates a possible solution to the stabilization equations (namely
$\mathcal{R}^{M}=\tilde{\tilde{\cal I}}^{M}$) because $e^{-2U} = \langle\
\mathcal{R} \mid\mathcal{I} \rangle$ would vanish identically.

Thirdly,

\begin{equation}
\label{eq:thirdproperty}
J_{4}(\tilde{\cal Q}) =  J_{4}(\mathcal{Q})\, . 
\end{equation}

Finally, in \cite{Ferrara:2011gv} (where the definition of Freudenthal dual
was generalized to all $\mathcal{N}\geq 2$ theories) is has been shown to be a
symmetry of the space of critical points of the black-hole potential
introduced in \cite{Ferrara:1997tw}.

Thus, following the recipe, and choosing some harmonic functions $H^{M}(x)$,
the metric function $e^{-2U}$ is always given by 

\begin{equation}
\label{eq:metricfunctionN8}
e^{-2U} =  \beta \sqrt{J_{4}(H)}\, ,  
\end{equation}

\noindent
and the 1-form $\omega$ is always given by the solution to

\begin{equation}
\label{eq:omega}
(d \omega)_{mn} 
 =   
\varepsilon_{mnp}
\left(\mathcal{I}_{ij}\partial_{p}\mathcal{I}^{ij}
-
\mathcal{I}^{ij}\partial_{p}\mathcal{I}_{ij}\right)\, .
\end{equation}

\noindent
the vector field strengths follow from the general formula and the scalars, as
mentioned before, cannot be easily recover, even if we now introduce an
$M_{IJ}$ with all the required properties. This is an evident shortcoming of
this procedure, but we believe it is compensated by the possibility of
studying explicitly the general black-hole metric.

Observe that, as expected, $\mathcal{R}_{M}$ can be obtained from the metric
function as

\begin{equation}
2\mathcal{R}_{M}(\mathcal{I}) = \frac{\partial\, e^{-2U}}{\partial
  \mathcal{I}^{M}}\, .  
\end{equation}

Furthermore, observe that the expression that we have given for the metric
function reduces to those found in \cite{Ferrara:2006yb} for all the magic
$\mathcal{N}=2$ truncations of $\mathcal{N}=8$ supergravity and another simple
truncation also reduces it to that of the well-known $STU$ model. The solution
to the stabilization equations of the 4-dimensional supergravities with
duality groups of Type $E7$
\cite{Ferrara:2011dz,Ferrara:2011aa,Ferrara:2012qp} is also given by an
analogous expression.

In the next sections we analyze what these formulae mean for 1- and
2-center solutions.

%%%%%%%%%%%%%%%%%%%%%%%%%%%%%%%%%%%%%%%%%%%%%%%%%%%%%%%%%%%%%%%%%%%%%%
%%%%%%%%%%%%%%%%%%%%%%%%%%%%%%%%%%%%%%%%%%%%%%%%%%%%%%%%%%%%%%%%%%%%%%
%%%%%%%%%%%%%%%%%%%%%%%%%%%%%%%%%%%%%%%%%%%%%%%%%%%%%%%%%%%%%%%%%%%%%% 
%%%%%%%%%%%%%%%%%%%%%%%%%%%%%%%%%%%%%%%%%%%%%%%%%%%%%%%%%%%%%%%%%%%%%%

\subsection{Single supersymmetric black-hole solutions}

%%%%%%%%%%%%%%%%%%%%%%%%%%%%%%%%%%%%%%%%%%%%%%%%%%%%%%%%%%%%%%%%%%%%%%
%%%%%%%%%%%%%%%%%%%%%%%%%%%%%%%%%%%%%%%%%%%%%%%%%%%%%%%%%%%%%%%%%%%%%%
%%%%%%%%%%%%%%%%%%%%%%%%%%%%%%%%%%%%%%%%%%%%%%%%%%%%%%%%%%%%%%%%%%%%%% 
%%%%%%%%%%%%%%%%%%%%%%%%%%%%%%%%%%%%%%%%%%%%%%%%%%%%%%%%%%%%%%%%%%%%%%

To study more closely these black-hole metrics it is convenient to introduce
the so-called $\mathbb{K}$-tensor \cite{Marrani:2010de,Andrianopoli:2011gy},
which is associated to the completely symmetric linearization of the Cartan
invariant performed in \cite{Faulkner} (see \cite{Kallosh:2012yy} for more
details):

\begin{equation}
\label{eq:J4prime}
  \begin{array}{rcl}
J^{\prime}_{4}(\mathcal{Q}_{1},\mathcal{Q}_{2},\mathcal{Q}_{3},\mathcal{Q}_{4})
& \equiv &
\tfrac{1}{6}
\mathrm{Tr}_{SL(8,\mathbb{R})}
\left\{
 p_{1} \cdot  q_{2} \cdot p_{3} \cdot q_{4} 
+p_{1} \cdot q_{3} \cdot p_{4} \cdot q_{2}
+p_{1} \cdot q_{4} \cdot p_{2} \cdot q_{3}
+(p\leftrightarrow q)
\right\}
\\
& & \\
& &   
-\tfrac{1}{12}
\left\{
[\mathcal{Q}_{1}\mid \mathcal{Q}_{2} ] [\mathcal{Q}_{3} \mid \mathcal{Q}_{4} ]
+
[\mathcal{Q}_{1}\mid \mathcal{Q}_{3} ] [\mathcal{Q}_{2} \mid \mathcal{Q}_{4} ]
+
[\mathcal{Q}_{1}\mid \mathcal{Q}_{4} ] [\mathcal{Q}_{2} \mid \mathcal{Q}_{3} ]
\right\}
\\
& & \\
& & 
+\tfrac{1}{4}
\left[
\mathrm{Pf}_{SL(8,\mathbb{R})}||p_{1} p_{2} p_{3} p_{4}||
+
(p\leftrightarrow q)
\right]\, ,
\end{array}
\end{equation}

\noindent
where $\mathrm{Tr}_{SL(8,\mathbb{R})}$ stands for the trace of the products of
$p$ and $q$ matrices (always one upper and one lower index), we have defined,
for convenience, the symmetric product

\begin{equation}
[\mathcal{Q}_{1}\mid \mathcal{Q}_{2} ]
\equiv  
-\tfrac{1}{2}\mathrm{Tr}_{SL(8,\mathbb{R})}[p_{1}\cdot q_{2} +
(p\leftrightarrow q) ]\, ,  
\end{equation}

\noindent
and

\begin{equation}
\label{Pf}
  \begin{array}{rcl}
\mathrm{Pf}||p_{1} p_{2} p_{3} p_{4}||
& \equiv & 
\tfrac{1}{4!}
\varepsilon_{ijklmnop}
p_{1}^{ij}p_{2}^{kl}p_{3}^{mn}p_{4}^{op}\, ,
\\
& & \\
\mathrm{Pf}||q_{1} q_{2} q_{3} q_{4}||
& \equiv & 
\tfrac{1}{4!}
\varepsilon^{ijklmnop}
q_{1\, ij}q_{2\, kl}q_{3\, mn}q_{4\, op}\, .
\end{array}
\end{equation}

The $\mathcal{K}$-tensor can be defined by its contraction with four different
fundamentals:
\begin{equation}
\label{us}
 \mathbb{K}_{MNPQ} \mathcal{Q}_{1}{}^{M} \mathcal{Q}_{2}{}^{N}
 \mathcal{Q}_{3}{}^{P} \mathcal{Q}_{4}{}^{Q}
\equiv
J^{\prime}_{4}(\mathcal{Q}_{1},\mathcal{Q}_{2},\mathcal{Q}_{3},\mathcal{Q}_{4})
\, ,
\end{equation}

\noindent
and, since $J^{\prime}_{4}$ is completely symmetric in the four
$\mathbf{56}$s, the $\mathbb{K}$-tensor is also completely symmetric in the
four symplectic indices 

\begin{equation}
  \mathbb{K}_{MNPQ}= \mathbb{K}_{(MNPQ)}\, .
\end{equation}

By construction

\begin{equation}
J^{\prime}_{4}(\mathcal{Q},\mathcal{Q},\mathcal{Q},\mathcal{Q})
=
J_{4}(\mathcal{Q})
=  
\mathbb{K}_{MNPQ} \mathcal{Q}^{M} \mathcal{Q}^{N}
 \mathcal{Q}^{P} \mathcal{Q}^{Q}\, ,
\end{equation}

\noindent
and the Jordan triple product can be also written in terms of this tensor as

\begin{equation}
(\mathcal{Q},\mathcal{Q},\mathcal{Q})^{M} =   
\mathbb{K}^{M}{}_{NPQ} \mathcal{Q}^{N}
 \mathcal{Q}^{P} \mathcal{Q}^{Q}\, ,
\end{equation}

\noindent
so we can write the symplectic vector $\mathcal{R}$ (\ref{eq:R}) and the
metric function $e^{-2U}$ (\ref{eq:metricfunctionN8}) in the more useful form

\begin{eqnarray}
 \mathcal{R}_{M} 
& = &
{\displaystyle\beta\frac{\mathbb{K}_{MNPQ}  H^{N} H^{P} H^{Q}}{\sqrt{J_{4}(H)}}}\, ,
\\
& & \nonumber \\
e^{-2U} 
& = &
\beta \sqrt{\mathbb{K}_{MNPQ} H^{M} H^{N} H^{P} H^{Q}}\, .
\end{eqnarray}

Single, extremal, static ($\omega=0$) black-hole solutions are associated to
harmonic functions of the form

\begin{equation}
H^{M}= A^{M} + \frac{\mathcal{Q}^{M}/\sqrt{2}}{r}\, ,  
\hspace{1cm}
r \equiv |\vec{x}|\, ,
\end{equation}

\noindent
where the $A^{M}$ are constants to be determined in terms of the
physical constants of the solution. This is done by requiring asymptotic
flatness and absence of sources of NUT charge and using the relation between
these constants and the asymptotic values of the scalars (which we do not know
explicitly). This means that we will not be able to find the general form of
these constants. Nevertheless, let us see how far we can go.

Asymptotic flatness implies

\begin{equation}
|M_{\infty}|^{-2} 
=
\langle\, \mathcal{R}_{\infty} \mid \mathcal{I}_{\infty} \,\rangle
=
e^{-2U_{\infty}}
=
1\, ,
\end{equation}

\noindent
and requires the normalization

\begin{equation}
\label{eq:KA4=1}
\mathbb{K}_{MNPQ} A^{M} A^{N} A^{P} A^{Q}=
\beta^{-2}\, .
\end{equation}
 
The absence of sources of NUT charge follows from setting $\omega=0$ in
Eq.~(\ref{eq:omegaequation}):

\begin{equation}
\label{eq:AQ=0}
0
=
\langle\, A \mid \mathcal{Q} \, \rangle
= 
\Im{\rm m}\, \left(\mathcal{Z}_{\infty\, IJ}M^{IJ}_{\infty} \right)\, ,  
\end{equation}

\noindent
where we have used the definition of $\mathcal{I}$, we have also used
asymptotic flatness and the definition of the central charge matrix of
$\mathcal{N}=8$ supergravity

\begin{equation}
\mathcal{Z}_{IJ}\equiv \langle\, \mathcal{V}_{IJ} \mid \mathcal{Q} \,
\rangle\, .  
\end{equation}

The projection

\begin{equation}
\mathcal{Z} \equiv \tfrac{1}{\sqrt{2}}\mathcal{Z}_{IJ}\frac{M^{IJ}}{|M|^{2}}\, ,
\end{equation}

\noindent
plays the r\^ole of central charge for the solutions associated to $M^{IJ}$,
which projects in the $U(8)$ directions in which supersymmetry is preserved.
As shown in \cite{Ortin:2011vm}, it drives the flow of the metric function
(but not that of the $\mathcal{N}=8$ scalars).  The condition of vanishing NUT
charge can be written in the form

\begin{equation}
N = \Im{\rm m}\, \mathcal{Z}_{\infty}=0\, ,   
\end{equation}

\noindent
as in an $N=2$ theory with central charge $\mathcal{Z}$. As we are going to
see the mass of the black hole is given by the real part of
$\mathcal{Z}_{\infty}$ which coincides with the absolute value (because the
imaginary part vanishes)\footnote{Entirely analogous expressions have been
  given in \cite{Ferrara:2006yb} for the masses of the black holes of the
  magic $\mathcal{N}=2$ truncations of $\mathcal{N}=8$ supergravity.}:

\begin{equation}
M = 
| \mathcal{Z}_{\infty}|
=
\Re{\rm e}\, \mathcal{Z}_{\infty} 
=   
\tfrac{1}{\sqrt{2}}\langle\, \mathcal{R}_{\infty} \mid \mathcal{Q} \, \rangle
= 
\tfrac{1}{\sqrt{2}}\beta^{2}
\mathbb{K}_{MNPQ} A^{M} A^{N} A^{P}
\mathcal{Q}^{Q}\, .
\end{equation}

Taking these conditions and relations into account\footnote{We will have to
  impose additional conditions, like the positivity of the mass, to ensure the
  regularity of the metric.}, we find that the metric function has the form

\begin{equation}
e^{-2U}
=
\sqrt{
1+\frac{4M}{r} 
+
\frac{3\beta^{2}
\mathbb{K}_{MNPQ}A^{M}A^{N}\mathcal{Q}^{P}\mathcal{Q}^{Q}}{r^{2}}
+
\frac{\sqrt{2}\beta^{2} 
\mathbb{K}_{MNPQ}A^{M}\mathcal{Q}^{N}\mathcal{Q}^{P}\mathcal{Q}^{Q}}{r^{3}}
+\frac{\beta^{2} J_{4}(\mathcal{Q})/4}{r^{4}} 
}\, .  
\end{equation}

The asymptotic behavior confirms the identification of the mass parameter,
which, as all the other coefficients of the $1/r^{n}$ terms in the square root
(in particular $J_{4}(\mathcal{Q})$), has to be positive for the metric to be
regular. In the near-horizon limit $r\rightarrow 0$, the last term dominates
the metric function and we recover the well-known entropy formula
(\ref{eq:entropyformula}) setting $\beta=2$. The coefficients of $1/r^{2}$ and
$1/r^{3}$ do not have a simple expression in terms of the physical parameters.

%%%%%%%%%%%%%%%%%%%%%%%%%%%%%%%%%%%%%%%%%%%%%%%%%%%%%%%%%%%%%%%%%%%%%%
%%%%%%%%%%%%%%%%%%%%%%%%%%%%%%%%%%%%%%%%%%%%%%%%%%%%%%%%%%%%%%%%%%%%%%
%%%%%%%%%%%%%%%%%%%%%%%%%%%%%%%%%%%%%%%%%%%%%%%%%%%%%%%%%%%%%%%%%%%%%% 
%%%%%%%%%%%%%%%%%%%%%%%%%%%%%%%%%%%%%%%%%%%%%%%%%%%%%%%%%%%%%%%%%%%%%%

\subsection{Supersymmetric 2-center solutions}

%%%%%%%%%%%%%%%%%%%%%%%%%%%%%%%%%%%%%%%%%%%%%%%%%%%%%%%%%%%%%%%%%%%%%%
%%%%%%%%%%%%%%%%%%%%%%%%%%%%%%%%%%%%%%%%%%%%%%%%%%%%%%%%%%%%%%%%%%%%%%
%%%%%%%%%%%%%%%%%%%%%%%%%%%%%%%%%%%%%%%%%%%%%%%%%%%%%%%%%%%%%%%%%%%%%% 
%%%%%%%%%%%%%%%%%%%%%%%%%%%%%%%%%%%%%%%%%%%%%%%%%%%%%%%%%%%%%%%%%%%%%%

Multicenter solutions  can be constructed by choosing harmonic functions with
several poles, as in $\mathcal{N}=2$ theories \cite{Denef:2000nb,Bates:2003vx},  

\begin{equation}
H^{M}= A^{M} +\sum_{a}
\frac{\mathcal{Q}^{M}_{a}/\sqrt{2}}{|\vec{x}-\vec{x_{a}}|}\, ,  
\end{equation}

\noindent
and tuning the parameters $A^{M},\mathcal{Q}^{M}_{a},\vec{x}_{a}$,so the
integrability conditions of the equation for $\omega$ (\ref{eq:omegaequation})

\begin{equation}
  \langle\,  A \mid \mathcal{Q}_{a}\, \rangle
  +\sum_{b} \frac{\langle\,  \mathcal{Q}_{b} \mid \mathcal{Q}_{a}\,
    \rangle/\sqrt{2}}{|\vec{x}_{a}-\vec{x}_{b}|}  =0\, .
\end{equation}

\noindent
Summing the above equations over $a$ and taking into account the antisymmetry
of the symplectic product, we find that the constants $A$, apart from
satisfying (\ref{eq:KA4=1}), also satisfy the condition (\ref{eq:AQ=0}) where
$\mathcal{Q}=\sum_{a}\mathcal{Q}_{a}$. 

When these equations are satisfied, $\omega$ exists and describes the total
angular momentum of the multi-black-hole system, just as in the
$\mathcal{N}=2$ cases, since the equations are identical.

The (square of) the metric function will contain many terms, up to order
$|\vec{x}-\vec{x}_{a}|^{-4}$.  The term of order $|\vec{x}-\vec{x}_{a}|^{-1}$
has the coefficient

\begin{equation}
M_{a}
\equiv
2\sqrt{2}
\mathbb{K}_{MNPQ} A^{M} A^{N} A^{P}
\mathcal{Q}_{a}^{Q}\, ,
\end{equation}

\noindent
which corresponds to the mass that the $a^{\rm th}$ center if it was
isolated. The mass of the solution is the sum of these parameters
$M=\sum_{a}M_{a}$. 
 
The coefficient of $|\vec{x}-\vec{x}_{a}|^{-n}|\vec{x}-\vec{x}_{b}|^{-m}$ with
$m+n=4$ is one the five quartic invariants listed in
\cite{Andrianopoli:2011gy} for 2-center solutions
\begin{equation}
  \begin{array}{rcl}
I_{+2}
& = & 
 \mathbb{K}_{MNPQ} \mathcal{Q}_{a}{}^{M} \mathcal{Q}_{a}{}^{N}
 \mathcal{Q}_{a}{}^{P} \mathcal{Q}_{a}{}^{Q}
=
J^{\prime}_{4}(\mathcal{Q}_{a},\mathcal{Q}_{a},\mathcal{Q}_{a},\mathcal{Q}_{a})
=
J_{4}(\mathcal{Q}_{a})\, ,
\\
& & \\
I_{+1}
& = & 
 \mathbb{K}_{MNPQ} \mathcal{Q}_{a}{}^{M} \mathcal{Q}_{a}{}^{N}
 \mathcal{Q}_{a}{}^{P} \mathcal{Q}_{b}{}^{Q}
=
J^{\prime}_{4}(\mathcal{Q}_{a},\mathcal{Q}_{a},\mathcal{Q}_{a},\mathcal{Q}_{b})\, ,
\\
& & \\
I_{0}
& = & 
 \mathbb{K}_{MNPQ} \mathcal{Q}_{a}{}^{M} \mathcal{Q}_{a}{}^{N}
 \mathcal{Q}_{b}{}^{P} \mathcal{Q}_{b}{}^{Q}
=
J^{\prime}_{4}(\mathcal{Q}_{a},\mathcal{Q}_{a},\mathcal{Q}_{b},\mathcal{Q}_{b})\, ,
\\
& & \\
I_{-1}
& = & 
\mathbb{K}_{MNPQ} \mathcal{Q}_{a}{}^{M} \mathcal{Q}_{b}{}^{N}
 \mathcal{Q}_{b}{}^{P} \mathcal{Q}_{b}{}^{Q}
=
J^{\prime}_{4}(\mathcal{Q}_{a},\mathcal{Q}_{b},\mathcal{Q}_{b},\mathcal{Q}_{b})\, ,
\\
& & \\
I_{-2}
& = & 
\mathbb{K}_{MNPQ} \mathcal{Q}_{b}{}^{M} \mathcal{Q}_{b}{}^{N}
\mathcal{Q}_{b}{}^{P} \mathcal{Q}_{b}{}^{Q}
=
J^{\prime}_{4}(\mathcal{Q}_{b},\mathcal{Q}_{b},\mathcal{Q}_{b},\mathcal{Q}_{b})\, ,
=
J_{4}(\mathcal{Q}_{b})\, .
\end{array}
\end{equation}

\noindent
The $I_{+2}$ $I_{-2}$ give the contributions of each center to the entropy.

With more than two centers, other combinations will appear based on the
quartic invariant. The sextic invariant found in \cite{Andrianopoli:2011gy}
does not seem to occur in these solutions.

%%%%%%%%%%%%%%%%%%%%%%%%%%%%%%%%%%%%%%%%%%%%%%%%%%%%%%%%%%%%%%%%%%%%%%
%%%%%%%%%%%%%%%%%%%%%%%%%%%%%%%%%%%%%%%%%%%%%%%%%%%%%%%%%%%%%%%%%%%%%%
%%%%%%%%%%%%%%%%%%%%%%%%%%%%%%%%%%%%%%%%%%%%%%%%%%%%%%%%%%%%%%%%%%%%%% 
%%%%%%%%%%%%%%%%%%%%%%%%%%%%%%%%%%%%%%%%%%%%%%%%%%%%%%%%%%%%%%%%%%%%%%

\section{Supersymmetric black holes and groups of Type $E_{7}$}

%%%%%%%%%%%%%%%%%%%%%%%%%%%%%%%%%%%%%%%%%%%%%%%%%%%%%%%%%%%%%%%%%%%%%%
%%%%%%%%%%%%%%%%%%%%%%%%%%%%%%%%%%%%%%%%%%%%%%%%%%%%%%%%%%%%%%%%%%%%%%
%%%%%%%%%%%%%%%%%%%%%%%%%%%%%%%%%%%%%%%%%%%%%%%%%%%%%%%%%%%%%%%%%%%%%% 
%%%%%%%%%%%%%%%%%%%%%%%%%%%%%%%%%%%%%%%%%%%%%%%%%%%%%%%%%%%%%%%%%%%%%%

The concept of group of Type $E_{7}$ axiomatizes the key properties of the fundamental (symplectic) representation the exceptional group $E_{7}$.  

$E_7$ is one of the exceptional simple groups on the classification of all the simple compact groups (or, analogously, complex simple Lie algebras) made by \'Elie Cartan \cite{CARTAN}. The maximally non-compact real form $G=E_{7(7)}$ of $E_{7}$ is precisely the group that appears in the coset scalar manifold $G/H$ of $\mathcal{N}=8$ Supergravity.

%A real Lie algebra $\mathfrak{g}_{\mathbb{R}}$ is the complex form of a complex Lie algebra if $\mathfrak{g}\cong \mathfrak{g}_{\mathbb{R}}\otimes \mathbb{C}$. The corresponding group is obtained by usual exponentiation. The maximally non-compact real form $G=E_{7(7)}$ is precisely the group that appears in the coset scalar manifold $G/H$ of $\mathcal{N}=8$ Supergravity. 

%It is the maximally non-compact real form since, intuitively, it corresponds to a Lie group that is as far as possible from being compact. More precisely, the $(7)$ in $E_{7(7)}$ is the difference of the number of non-compact generators of $E_{7(7)}$ and the number of compact ones. 

The first axiomatic characterization of groups \textit{\textquotedblleft of
Type }$E_{7}$\textit{"} through a module (irreducible representation) was
given in 1967 by Brown \cite{brown} (here we will follow \cite{Ferrara:2012qp}, see also \cite{Ferrara:2011aa}).

A group $G$ of Type $E_{7}$ is a Lie group endowed with a representation $%
\mathbf{R}$ such that:

\begin{enumerate}

\item $\mathbf{R}$ is \textit{symplectic}, \textit{i.e.} (the subscripts
\textquotedblleft $s$\textquotedblright\ and \textquotedblleft $a$%
\textquotedblright\ stand for symmetric and skew-symmetric throughout):
\begin{equation}
\exists !\Omega_{\left[ MN\right] }\equiv \mathbf{1\in R\times }_{a}%
\mathbf{R;}  \label{sympl-metric}
\end{equation}%
$\Omega_{\left[ MN\right] }$ defines a non-degenerate skew-symmetric
bilinear form, that is, a \emph{symplectic product}$\left\langle \cdot,\cdot\right\rangle$: given two different vectors $\mathcal{Q}_{1},\mathcal{Q}_{2}\in\mathbf{R}$, $\left\langle \cdot
\cdot\right\rangle$ is defined as

\begin{equation}
\left\langle \mathcal{Q}_{1},\mathcal{Q}_{2}\right\rangle \equiv \mathcal{Q}
_{1}^{M}\mathcal{Q}_{2}^{N}\Omega_{MN}=-\left\langle \mathcal{Q}_{2},
\mathcal{Q}_{1}\right\rangle .
\end{equation}

\item $\mathbf{R}$ admits a unique rank-$4$ completely symmetric primitive $%
G $-invariant structure, the so-called $K$-tensor

\begin{equation}
\exists !\mathbb{K}_{\left( MNPQ\right) }\equiv \mathbf{1\in }\left[ \mathbf{%
R\times R\times R\times R}\right] _{s}\mathbf{;}
\end{equation}%
thus, by contracting the $K$-tensor with the same vector $\mathcal{Q}\in\mathbf{R}$, we obtain a rank-four homogeneous $G$-invariant polynomial (here $\varsigma $ is a normalization constant):
\begin{equation}
\mathbf{q}\left( \mathcal{Q}\right) \equiv \varsigma \mathbb{K}_{MNPO}
\mathcal{Q}^{M}\mathcal{Q}^{N}\mathcal{Q}^{P}\mathcal{Q}^{O},  
\end{equation}

\noindent
which corresponds to the evaluation of the rank-four symmetric invariant $%
\mathbf{q}$-structure induced by the $K$-tensor on four identical modules $\mathbf{R}$.

\item If a trilinear map $t\mathbf{:R\times R\times R}\rightarrow \mathbf{R}$
is defined such that

\begin{equation}
\left\langle t\left( \mathcal{Q}_{1},\mathcal{Q}_{2},\mathcal{Q}_{3}\right) ,
\mathcal{Q}_{4}\right\rangle =\mathbf{q}\left( \mathcal{Q}_{1},\mathcal{Q}
_{2},\mathcal{Q}_{3},\mathcal{Q}_{4}\right) ,
\end{equation}%

\noindent
then it holds that

\begin{equation}
\left\langle t\left( \mathcal{Q}_{1},\mathcal{Q}_{1},\mathcal{Q}_{2}\right)
,t\left( \mathcal{Q}_{2},\mathcal{Q}_{2},\mathcal{Q}_{2}\right)
\right\rangle =-2\left\langle \mathcal{Q}_{1},\mathcal{Q}_{2}\right\rangle
\mathbf{q}\left( \mathcal{Q}_{1},\mathcal{Q}_{2},\mathcal{Q}_{2},\mathcal{Q}
_{2}\right) .  \label{FTS3}
\end{equation}%

\end{enumerate}

\noindent
Therefore, for any Supergravity whose scalar manifold is a coset space $G/H$ with $G$ of the $E_{7}$ type, we can apply the same discussion that we made in section \ref{sec:N8stabilization} for the $\mathcal{N}=8$ case and conclude that \emph{a solution} to the stabilization equations is given by

\begin{equation}
  \mathcal{R}^{M}(\mathcal{I}) = \beta(\mathcal{I},\mathcal{I},\mathcal{I})^{M}\, ,  
\end{equation}

\noindent
where now $t(\cdot,\cdot,\cdot)=(\cdot,\cdot,\cdot)$ is the trilinear product of the corresponding group $G$ of Type $E_{7}$. Once we have $t(\cdot,\cdot,\cdot)$ for a particular $G$ corresponding to a Supergravity, the supersymmetric black hole metric can be constructed following exactly the same steps as in the $\mathcal{N}=8$ case.

Remarkably enough, all $2\leq \mathcal{N}\leq 8$-extended four-dimensional Supergravities with symmetric scalar manifolds $\frac{G}{H}$ have $G$ of Type $E_{7}$ \cite{Borsten:2009zy,Ferrara:2012qp}. The Supergravity duality groups $G$ can be then classified into three different classes \cite{1998math.....11035G,1998math.....11056S,Ferrara:2012qp}, simple non-degenerate, simple degenerate and semi-simple non-degenerate, all of them belonging to the class of groups of Type $E_{7}$.

\paragraph{Simple non-degenerate}

A group of Type $E_{7}$ is non-degenerate if the quartic form $q(\cdot,\cdot,\cdot,\cdot)$ is absolutely irreducible (irreducible over a separable closure of the base field). The simple non-degenerate four-dimensional Supergravity groups $G$ of Type $E_{7}$ are given by \cite{Ferrara:2012qp}

\begin{enumerate}

\item $G=E_{7(7)}\rightarrow \mathcal{N}=8 ~~{\rm Supergravity.}$

\item $G=SO^{\ast}(12)\rightarrow \mathcal{N}=6 ~~{\rm Supergravity.}$

\item $G=SU(1,5)\rightarrow \mathcal{N}=5 ~~{\rm Supergravity.}$

\end{enumerate}

\paragraph{Simple degenerate}

The simple degenerate four-dimensional Supergravity groups $G$ of Type $E_{7}$ are given by \cite{Ferrara:2012qp}

\begin{enumerate}

\item $G=U(3,n_v)\rightarrow \mathcal{N}=3 ~~{\rm Supergravity~coupled~to}~ n_{v}~ {\rm vector~ multiplets.}$

\item $G=U(1,n_v)\rightarrow \mathcal{N}=2 ~~{\rm Supergravity ~\emph{minimally} ~coupled ~to}~ n_{v}~{\rm~vector~multiplets.}$

\end{enumerate}

For either $G=U(3,n_v)$ or $G=U(1,n_v)$, as well as for generic reducible groups of Type $E_{7}$, one can see that the quartic form is reducible \cite{1998math.....11056S,Ferrara:2012qp} as follows

\begin{equation}
\mathbb{K}_{MNPQ}=\alpha \mathbb{S}_{M(N}\mathbb{S}_{PQ)}\, ,
\end{equation}

\noindent
where $\alpha\in \mathbb{R}$ is a constant and $\mathbb{S}$ is the rank-two symmetric symplectic tensor defined by

\begin{equation}
\mathbf{Q}_{1}^{i}\overline{\mathbf{Q}}_{2}^{\overline{j}}\eta _{i\overline{j%
}}=\mathbb{S}_{MN}\mathcal{Q}_{1}^{M}\mathcal{Q}_{2}^{N}+i\mathbb{C}_{MN}%
\mathcal{Q}_{1}^{M}\mathcal{Q}_{2}^{N},
\end{equation}%

\noindent
where $\eta _{i\overline{j}}$ is the invariant metric of the fundamental
irrep. $\mathbf{r}+\mathbf{s}$ of $U\left( r,s\right) $, and $\mathbf{Q}%
_{x}^{i}$ and $\mathbf{Q}_{x}^{i}$ are the charge vectors in the \textit{%
complex} (manifestly $U\left( r,s\right) $-covariant) symplectic frame. It follows that

\begin{equation}
q(\mathcal{Q},\mathcal{Q},\mathcal{Q},\mathcal{Q})=\gamma \left(\mathbb{S}_{MN}\mathcal{Q}^{M}\mathcal{Q}^{N}\right)^2\,,
\end{equation}%

\noindent
for an appropriate number $\gamma$. This means that for these Supergravities the metric factor $e^{-2U}$ can be written simply as

\begin{equation}
e^{-2U}\sim\left|\mathbb{S}_{MN}H^{M}H^{N}\right|\,,
\end{equation}%

\noindent
which indeed can be explicitly checked by solving the stabilization equations by algebraic methods.

\paragraph{Semi-simple non-degenerate}

The semi-simple non-degenerate four-dimensional Supergravity groups $G$ of Type $E_{7}$ are given by \cite{Ferrara:2012qp}

\begin{enumerate}

\item $G=SL(2,\mathbb{R})\times SO(6,n_{v})\rightarrow \mathcal{N}=4~~{\rm Supergravity ~coupled ~to} ~n_{v} ~{\rm ~vector~multiplets.}$

\item $G=SL(2,\mathbb{R})\times SO(2,n_{v})\rightarrow \mathcal{N}=2~~{\rm Supergravity ~coupled ~to}~ n_{v}+1~ {\rm vector~multiplets.}$

\end{enumerate}
%%%%%%%%%%%%%%%%%%%%%%%%%%%%%%%%%%%%%%%%%%%%%%%%%%%%%%%%%%%%%%%%%%%%%%
%%%%%%%%%%%%%%%%%%%%%%%%%%%%%%%%%%%%%%%%%%%%%%%%%%%%%%%%%%%%%%%%%%%%%%
%%%%%%%%%%%%%%%%%%%%%%%%%%%%%%%%%%%%%%%%%%%%%%%%%%%%%%%%%%%%%%%%%%%%%% 
%%%%%%%%%%%%%%%%%%%%%%%%%%%%%%%%%%%%%%%%%%%%%%%%%%%%%%%%%%%%%%%%%%%%%%

\section{$\mathcal{N}=2$ Supergravity supersymmetric black holes}

%%%%%%%%%%%%%%%%%%%%%%%%%%%%%%%%%%%%%%%%%%%%%%%%%%%%%%%%%%%%%%%%%%%%%%
%%%%%%%%%%%%%%%%%%%%%%%%%%%%%%%%%%%%%%%%%%%%%%%%%%%%%%%%%%%%%%%%%%%%%%
%%%%%%%%%%%%%%%%%%%%%%%%%%%%%%%%%%%%%%%%%%%%%%%%%%%%%%%%%%%%%%%%%%%%%% 
%%%%%%%%%%%%%%%%%%%%%%%%%%%%%%%%%%%%%%%%%%%%%%%%%%%%%%%%%%%%%%%%%%%%%%

The case of $\mathcal{N}=2$ Supergravity cannot be solved in general since the scalar manifold is not specified unambiguously. As explained in chapter \ref{chapter:sugra}, supersymmetry only constraint the manifold spanned by the scalar fields on the vector multiplets to be of Special K\"ahler type, a condition which does not fix the manifold unequivocally. What can been done, however, is to characterize and classify the most general supersymmetric solutions of the theory, a knowledge that will be of outermost importance when trying to propose a formalism to deal with general (not only supersymmetric) black hole solutions . 

The supersymmetric solutions of Supergravity, and in particular of $\mathcal{N}=2$, can be classified in two classes, the time-like and null, regarding the casual character of the Killing vector constructed from the Killing spinors and the Clifford-algebra $\gamma$-matrices \cite{Tod:1983pm,Gauntlett:2002nw,Gauntlett:2003fk,Bellorin:2006yr,Caldarelli:2003pb,Gutowski:2003rg,Cariglia:2004qi,
Gauntlett:2002fz,Klemm:2010mc,Deger:2010rb}. The black hole solutions belong to the time-like class, so that's the particular class that we are going to review here, assuming from the beginning that the hyper-scalars have been set to a constant value, since otherwise we don't expect regular solutions. Notice that this is a particular case of the exposition made in section \ref{sec:mathformalism}, but we include it here since it is of capital importance for chapter \ref{chapter:HFGK}. 

The black hole supersymmetric metric is given by

\begin{equation}
ds^{2} = e^{2U}(dt+\omega)^{2} 
-e^{-2U}\delta_{mn}dx^{m}dx^{n}\, .
\end{equation}

\noindent
Notice that the only difference respect to the general Supergravity black hole metric (\ref{eq:generalbhmetric}) (assuming staticity $\omega=0$) is that now the transverse metric is flat, as corresponds in (\ref{eq:generalbhmetric}) to the extremal limit $r_{0}\rightarrow 0$. That is, supersymmetry always implies extremalily\footnote{Of course, this conclusion relies on the assumption that $r_{0}$ can be interpreted as the extremality parameter for all the Supergravity black holes.}, but not the other way around \cite{Khuri:1995xq,Ortin:1997yn}.

We take now the covariantly holomorphic section $\mathcal{V}$ of Special Kahler geometry and a function $X\left(z,\bar{z}\right)$, with K\"ahler weight such that $\mathcal{V}/X$ is K\"ahler neutral, and define

\begin{equation}
\mathcal{R}+i\mathcal{I}=\mathcal{V}/X\, .
\end{equation}

\noindent
Then, it can be shown that a for a supersymmetric black hole we have

\begin{eqnarray}
e^{-2U} & = & \langle \mathcal{R} \mid \mathcal{I} \rangle\, ,  \\
& & \nonumber \\
(d\omega)_{xy} & = & 2 \epsilon_{xyz}
\langle\,\mathcal{I}\mid \partial^{z}\mathcal{I}\, \rangle\, .
\end{eqnarray}

\begin{equation}
\mathcal{I}^M=a^M-\frac{\mathcal{Q}^M}{\sqrt{2}}\tau\, .  
\end{equation}

\noindent
The vector field strengths are given by

\begin{equation}
F = -{\textstyle\frac{1}{\sqrt{2}}} \{d[e^{2U}\mathcal{R}(dt+\omega) ] 
-{}^{\star}[ e^{-2U} d\mathcal{I}\wedge(dt+\omega)] \}\, ,
\end{equation}

\noindent
and the scalar fields $z^{i}$ can be computed by taking the quotients

\begin{equation}
z^{i}=(\mathcal{V}/X)^{i}/(\mathcal{V}/X)^{0}\, .
\end{equation}

\noindent
Given $\mathcal{I}$, $\mathcal{R}\equiv \Re{\rm e}(\mathcal{V}/X)$ can in principle be found by solving the generalized stabilization, which depend on the specific model under consideration. Solving the stabilization equations completely determines the solution, since all the physical fields can be constructed in terms of $\mathcal{I}^M=a^M-\frac{\mathcal{Q}^M}{\sqrt{2}}\tau$, as it can be checked from the previous formulae.
%\cleardoublepage

%%%%%%%%%%%%%%%%%%%%%%%%%%%%%%%%%%%%%%%%%%%%%%%%%%%%%%%%%%%%%%%%%%%%%%%
%%% CHAPTER 6: HFGK FORMALISM
%%%%%%%%%%%%%%%%%%%%%%%%%%%%%%%%%%%%%%%%%%%%%%%%%%%%%%%%%%%%%%%%%%%%%%%

\renewcommand{\leftmark}{\MakeUppercase{Chapter \thechapter. H.F.G.K. formalism}}
\chapter{The H-F.G.K. formalism}
\label{chapter:HFGK}

In this chapter we are going to introduce a new formalism, the so-called H-F.G.K. formalism, based on a dimensional reduction of the original four-dimensional action and the use of a new set of duality-covariant variables inspired by supersymmetry, which eases the explicit construction of non-supersymmetric black hole solutions of $\mathcal{N}=2$ four-dimensional ungauged Supergravity.

%%%%%%%%%%%%%%%%%%%%%%%%%%%%%%%%%%%%%%%%%%%%%%%%%%%%%%%%%%%%%%%%%%%%%%
%%%%%%%%%%%%%%%%%%%%%%%%%%%%%%%%%%%%%%%%%%%%%%%%%%%%%%%%%%%%%%%%%%%%%%
%%%%%%%%%%%%%%%%%%%%%%%%%%%%%%%%%%%%%%%%%%%%%%%%%%%%%%%%%%%%%%%%%%%%%%
%%%%%%%%%%%%%%%%%%%%%%%%%%%%%%%%%%%%%%%%%%%%%%%%%%%%%%%%%%%%%%%%%%%%%%

\section{H-FGK for \texorpdfstring{$\mathcal{N}=2$}{N=2}, \texorpdfstring{$d=4$}{d=4} supergravity}
\label{sec:HFGK}

%%%%%%%%%%%%%%%%%%%%%%%%%%%%%%%%%%%%%%%%%%%%%%%%%%%%%%%%%%%%%%%%%%%%%%
%%%%%%%%%%%%%%%%%%%%%%%%%%%%%%%%%%%%%%%%%%%%%%%%%%%%%%%%%%%%%%%%%%%%%%
%%%%%%%%%%%%%%%%%%%%%%%%%%%%%%%%%%%%%%%%%%%%%%%%%%%%%%%%%%%%%%%%%%%%%%
%%%%%%%%%%%%%%%%%%%%%%%%%%%%%%%%%%%%%%%%%%%%%%%%%%%%%%%%%%%%%%%%%%%%%%

In \cite{Mohaupt:2011aa,Meessen:2011aa} it was shown that the search of
single, static, spherically-symmetric black-hole solutions of an
$\mathcal{N}=2$, $d=4$ supergravity coupled to $n_{v}$ vector multiplets (and,
correspondingly, including $n_{v}$ complex scalars $z^{i}$ and $n_{v}+1$ Abelian
vector fields $A^{\Lambda}{}_{\mu}$) with electric ($q_{\Lambda}$) and
magnetic ($p^{\Lambda}$) charges described by the $2(n_{v}+1)$-dimensional
symplectic vector $(\mathcal{Q}^{M}) \equiv (p^{\Lambda}, q_{\Lambda})^{T}$, is remarkably simplified by going to a new set of $2(n_{v}+1)$ variables $H^{M}$ which form a linear, symplectic, representation of the $U$-duality group and that become harmonic functions on euclidean $\mathbb{R}^3$ for supersymmetric black hole solutions.

We proceed now to describe the change of variables, from those defining a black-hole solution for given electric and magnetic charges $(\mathcal{Q}^{M}) = (p^{\Lambda},q_{\Lambda})^{\mathrm{T}}$,
namely the metric function $U$ and the complex scalars $z^{i}$\footnote{See section \ref{sec:sugrablackhole} for more details.}, to the
variables $(H^{M}) = (H^{\Lambda},H_{\Lambda})^{\mathrm{T}}$ that have the same
transformation properties as the charges. There is an evident mismatch between
these two sets of variables, because $U$ is real. For consistency we will
introduce a complex variable $X$ of the form\footnote{ In this section we will
  be following the conventions of Ref.~\cite{Meessen:2006tu}, where the
  function $X$ appears as a scalar bilinear built out of the Killing spinors.  }
\begin{equation}
X = \tfrac{1}{\sqrt{2}}e^{U+i\alpha}\, ,
\end{equation}
\noindent
although the phase $\alpha$ does not occur in the original FGK formalism.  The
change of variables will then be well defined, and the absence of $\alpha$ will
lead to a constraint on the new set of variables: this constraint is related to
the absence of NUT charge, a possibility which in $d=4$ is allowed for by
spherical symmetry.

The theory is specified by the prepotential\footnote{We only use the
  prepotential here to determine quickly the homogeneity properties of the
  objects we are going to deal with. These properties are, however, valid for
  any $\mathcal{N}=2$ theory in any symplectic frame, whether or not a prepotential
  exists.}  $\mathcal{F}$, a homogeneous function of second degree in the
complex coordinates $\mathcal{X}^{\Lambda}$. Consequently, defining
\begin{equation}
\mathcal{F}_{\Lambda} 
\equiv 
\frac{\partial\mathcal{F}}{\partial\mathcal{X}^{\Lambda}} \hspace{.5cm}\mbox{and}\hspace{.5cm}
\mathcal{F}_{\Lambda\Sigma} 
\equiv 
\frac{\partial^{2}\mathcal{F}}{\partial\mathcal{X}^{\Lambda}\partial\mathcal{X}^{\Sigma}}\,,
\hspace{.5cm}\mbox{we have:}\hspace{.5cm} 
\mathcal{F}_{\Lambda} = \mathcal{F}_{\Lambda\Sigma}\mathcal{X}^{\Sigma}\, .  
\end{equation}
\noindent
Since the matrix $\mathcal{F}_{\Lambda\Sigma}$ is homogenous of degree zero and
$X$ has the same K\"ahler weight as the covariantly holomorphic section
\begin{equation}
\left(\mathcal{V}^{M}\right) = 
\left(
  \begin{array}{c}
   \mathcal{L}^{\Lambda} \\
   \mathcal{M}_{\Lambda} \\
  \end{array}
\right)  
=
e^{\mathcal{K}/2}
\left(
  \begin{array}{c}
   \mathcal{X}^{\Lambda} \\
   \mathcal{F}_{\Lambda} \\
  \end{array}
\right),  
\end{equation}
\noindent
where $\mathcal{K}$ is the K\"ahler potential, we also find
\begin{equation}
\label{eq:MFL}
\frac{\mathcal{M}_{\Lambda}}{X} = \mathcal{F}_{\Lambda\Sigma}\, \frac{\mathcal{L}^{\Sigma}}{X}\, .
\end{equation}
\noindent
Defining the K\"ahler-neutral, real, symplectic vectors
$\mathcal{R}^{M}$ and $\mathcal{I}^{M}$ by
\begin{equation}
\label{eq:RandIdef}
\mathcal{R}^{M}  = \Re\mathfrak{e}\, \mathcal{V}^{M}/X \, ,
\hspace{1cm}
\mathcal{I}^{M} = \Im\mathfrak{m}\,  \mathcal{V}^{M}/X\, , 
\end{equation}
\noindent
and using the symplectic metric 
\begin{equation}
\left(\Omega_{MN} \right) 
\equiv 
\left(
  \begin{array}{cc}
0 
& 
\mathbb{I}
\\
-\mathbb{I}
& 
0
\end{array}
\right)
\end{equation}
\noindent
as well as its inverse $\Omega^{MN}$ to lower and raise the symplectic indices
according to the convention
\begin{equation}
\mathcal{R}_{M} = \Omega_{MN}\mathcal{R}^{N}\, ,
\hspace{1cm}  
\mathcal{R}^{M} = \mathcal{R}_{N}\Omega^{NM}\, ,
\end{equation}
one can rewrite the complex relation (\ref{eq:MFL}) in the real form
\begin{equation}
\label{eq:RMI}
\mathcal{R}_{M} = -\mathcal{M}_{MN}(\mathcal{F})\mathcal{I}^{N}\, .
\end{equation}
\noindent
The symmetric symplectic matrix
\begin{equation}
\mathcal{M}(\mathcal{A}) 
\equiv 
\left(
  \begin{array}{cc}
\Im\mathfrak{m}\, \mathcal{A}_{\Lambda\Sigma} +
\Re\mathfrak{e}\, \mathcal{A}_{\Lambda\Omega}\, 
\Im\mathfrak{m}\, \mathcal{A}^{-1|\, \Omega\Gamma}\,
\Re\mathfrak{e}\, \mathcal{A}_{\Gamma\Sigma} 
& 
\hspace{.5cm}
-\Re\mathfrak{e}\, \mathcal{A}_{\Lambda\Omega}\,
\Im\mathfrak{m}\, \mathcal{A}^{-1 |\, \Omega\Sigma}
\\
\\
-
\Im\mathfrak{m}\, \mathcal{A}^{-1 |\, \Lambda\Omega}\,
\Re\mathfrak{e}\, \mathcal{A}_{\Omega\Sigma}
&
\Im\mathfrak{m}\, \mathcal{A}^{-1|\, \Lambda\Sigma}
\end{array}
\right),
\end{equation}
\noindent
can be associated with any symmetric complex matrix
$\mathcal{A}_{\Lambda\Sigma}$ with a non-degenerate imaginary part (such as
$\mathcal{F}_{\Lambda\Sigma}$ and the period matrix
$\mathcal{N}_{\Lambda\Sigma}$). The inverse of $\mathcal{M}_{MN}$, denoted by
$\mathcal{M}^{MN}$, is the result of raising the indices with the inverse
symplectic metric.

It is also immediate to prove the relation 
\begin{equation}
\label{eq:dRMdI}
d\mathcal{R}_{M} = -\mathcal{M}_{MN}(\mathcal{F})\, d\mathcal{I}^{N}\, .
\end{equation}
From this equality, its inverse and the symmetry properties of
$\mathcal{M}_{MN}$ we can derive the following relation between partial
derivatives (see {\it e.g.\/} \cite{Bellorin:2006xr}):
\begin{equation}
\label{eq:partials}
\frac{\partial \mathcal{I}^{M}}{\partial\mathcal{R}_{N}}  
=
\frac{\partial \mathcal{I}^{N}}{\partial\mathcal{R}_{M}}  
=
-\frac{\partial \mathcal{R}^{M}}{\partial\mathcal{I}_{N}}  
=
-\frac{\partial \mathcal{R}^{N}}{\partial\mathcal{I}_{M}}= -\mathcal{M}^{MN}(\mathcal{F}) \, .
\end{equation}

Since we want the new variables to become harmonic functions on euclidean $\mathbb{R}^3$, we introduce two dual sets of variables $H^{M}$ and $\tilde{H}_{M}$ and replace the original $n+1$ fields $X,z^{i}$ by the $2n+2$ real variables $H^{M}(\tau)$

\begin{equation}
\mathcal{I}^{M}(X,Z,X^{*},Z^{*}) = H^{M}\, .  
\end{equation}

\noindent
The dual variables $\tilde{H}^{M}$ can be identified with $\mathcal{R}^{M}$,
which we can express as functions of the $H^{M}$ through
Eq.~(\ref{eq:RMI}). This gives $\mathcal{V}^{M}/X$ as a function of the
$H^{M}$. The physical fields can then be recovered by
\begin{equation}
z^{i}\ =\ \frac{\mathcal{V}^{i}/X}{\mathcal{V}^{0}/X} \hspace{.7cm}\mbox{and}\hspace{.7cm}
e^{-2U}\ =\ \frac{1}{2|X|^{2}} \ =\ \mathcal{R}_{M}\mathcal{I}^{M}\, .  
\end{equation}
\noindent
The phase of $X$, $\alpha$, can be found by solving the differential equation
({\it cf.\/}~Eqs.~(3.8), (3.28) in Ref.~\cite{Galli:2010mg})
\begin{equation}
\dot{\alpha} =2|X|^{2} \dot{H}^{M}H_{M} -\mathcal{Q}_{\star}\, ,
\hspace{.5cm}\mbox{where}\hspace{.5cm}
\mathcal{Q}_{\star} = \tfrac{1}{2i}\dot{z}^{i}\partial_{i}\mathcal{K} +\mathrm{c.c.} 
\end{equation}
\noindent
is the pullback of the K\"ahler connection 1-form
\begin{equation}
\mathcal{Q}_{\star} = \tfrac{1}{2i}\dot{z}^{i}\partial_{i}\mathcal{K} +\mathrm{c.c.}  
\end{equation}

Having detailed the change of variables, we want to rewrite the FGK action for
static, spherically symmetric solutions of $\mathcal{N}=2$, $d=4$ Supergravity
\cite{Ferrara:1997tw}, {\it i.e.\/}
\begin{equation}
\label{eq:effectiveaction}
I_{\text{FGK}}[U,z^{i}] = \int d\tau \left\{ 
(\dot{U})^{2}  
+\mathcal{G}_{ij^{*}}\dot{z}^{i}  \dot{z}^{*\, j^{*}}  
-\tfrac{1}{2}e^{2U}
\mathcal{M}_{MN}(\mathcal{N})\mathcal{Q}^{M}\mathcal{Q}^{N} 
+r_{0}^{2}
\right\} ,  
\end{equation}
\noindent
in terms of the variables $H^{M}$. We start by defining the function $\mathsf{W}(H)$, which can be identified with the \emph{Hesse potential}

\begin{equation}
\label{eq:Wdef}
\mathsf{W}(H) \equiv \tilde{H}_{M}(H)H^{M} = e^{-2U} = \frac{1}{2|X|^{2}}\, ,
\end{equation}

\noindent
which is homogenous of second degree in the $H^{M}$. Using the properties
(\ref{eq:partials}) one can show that
\begin{eqnarray}
\partial_{M}\mathsf{W}
& \equiv & 
\frac{\partial\mathsf{W}}{\partial H^{M}}
= 
2\tilde{H}_{M}\, , 
\\
& & \nonumber \\
\partial^{M}\mathsf{W}
& \equiv & 
\frac{\partial\mathsf{W}}{\partial \tilde{H}_{M}}
= 
2H^{M}\, , 
\\
& & \nonumber \\
\partial_{M}\partial_{N}\mathsf{W} 
& = & 
-2\mathcal{M}_{MN}(\mathcal{F})\, ,
\\
& & \nonumber \\
\mathsf{W}\,\partial_{M}\partial_{N}\log \mathsf{W} 
& = & 
2\mathcal{M}_{MN}(\mathcal{N})+4\mathsf{W}^{-1}H_{M}H_{N}\, ,
\end{eqnarray}
\noindent
where the last property is based on the following relation\footnote{ This
  relation can be derived from the identities in Ref.~\cite{Ceresole:1995ca}.
}
\begin{equation}
-\mathcal{M}_{MN}(\mathcal{N})  
=
\mathcal{M}_{MN}(\mathcal{F})  
+4\mathcal{V}_{(M}\mathcal{V}^{*}_{N)}\, .
\end{equation}

Using the special geometry identity $\mathcal{G}_{ij^{*}} =
-i\mathcal{D}_{i}\mathcal{V}_{M}\mathcal{D}_{j^{*}}\mathcal{V}^{*\, M}$, we
can rewrite the effective action in the form
\begin{equation}
\label{eq:effectiveaction2}
-I_{\rm eff}[H] 
= 
\int d\tau 
\left\{ 
\tfrac{1}{2}\partial_{M}\partial_{N}\log\mathsf{W} 
\left(\dot{H}^{M}\dot{H}^{N}+\tfrac{1}{2}\mathcal{Q}^{M}\mathcal{Q}^{N} \right) 
-\Lambda
-r_{0}^{2}
\right\} ,  
\end{equation}
\noindent
where we have defined
\begin{equation}
\Lambda \equiv \left(\frac{\dot{H}^{M}H_{M}}{ \mathsf{W}}\right)^{2} 
+\left(\frac{\mathcal{Q}^{M}H_{M}}{ \mathsf{W}}\right)^{2}\, .
\end{equation}
\noindent
The $\tau$-independence of the Lagrangian implies the conservation of the
Hamiltonian $\mathcal{H}$
\begin{equation}
\label{eq:hamiltonianconstraint2}
\mathcal{H}\equiv 
-\tfrac{1}{2}\partial_{M}\partial_{N}\log\mathsf{W} 
\left(\dot{H}^{M}\dot{H}^{N}-\tfrac{1}{2}\mathcal{Q}^{M}\mathcal{Q}^{N} \right) 
+\left(\frac{\dot{H}^{M}H_{M}}{ \mathsf{W}}\right)^{2} 
-\left(\frac{\mathcal{Q}^{M}H_{M}}{ \mathsf{W}}\right)^{2}
-r_{0}^{2}=0\, .  
\end{equation}

\noindent
The equations of motion can be written in the form
\begin{equation}
\label{eq:equationsofmotion2}
\tfrac{1}{2}\partial_{P}\partial_{M}\partial_{N}\log \mathsf{W} 
\left(\dot{H}^{M}\dot{H}^{N} -\tfrac{1}{2}\mathcal{Q}^{M}\mathcal{Q}^{N}
\right)  
+\partial_{P}\partial_{M}\log \mathsf{W}\, \ddot{H}^{M}
-\frac{d}{d\tau}\left(\frac{\partial \Lambda}{\partial \dot{H}^{P}}\right)
+\frac{\partial \Lambda}{\partial H^{P}}=0\, .
\end{equation}
\noindent
Contracting them with $H^{P}$ and using the homogeneity properties of the
different terms as well as the Hamiltonian constraint above, we find the
equation ({\it cf.\/}~Eq.~(3.31) of Ref.~\cite{Galli:2010mg} for the stationary
extremal case)
\begin{equation}
\label{eq:Urewriten}
\tfrac{1}{2}\partial_{M}\log \mathsf{W} \left(\ddot{H}^{M}
  -r_{0}^{2}H^{M}\right) +\left(\frac{\dot{H}^{M}H_{M}}{ \mathsf{W}}\right)^{2}  = 0\, ,
\end{equation}
\noindent
which corresponds to the equation of motion of the variable $U$ in the standard
formulation.

Note that in the extremal case ($r_{0}=0$) and in the absence of the NUT charge
 
\begin{equation}
\label{eq:noNUTconstraint}
\dot{H}^{M}H_{M}=0\, ,
\end{equation}

\noindent
the equations of motion are solved by harmonic functions
$\dot{H}^{M} = \mathcal{Q}^{M}$ \cite{Bellorin:2006xr}.

%%%%%%%%%%%%%%%%%%%%%%%%%%%%%%%%%%%%%%%%%%%%%%%%%%%%%%%%%%%%%%%%%%%%%%
%%%%%%%%%%%%%%%%%%%%%%%%%%%%%%%%%%%%%%%%%%%%%%%%%%%%%%%%%%%%%%%%%%%%%%
%%%%%%%%%%%%%%%%%%%%%%%%%%%%%%%%%%%%%%%%%%%%%%%%%%%%%%%%%%%%%%%%%%%%%%
%%%%%%%%%%%%%%%%%%%%%%%%%%%%%%%%%%%%%%%%%%%%%%%%%%%%%%%%%%%%%%%%%%%%%%

\subsection{Extremal black holes}
\label{sec-d4extremal}

%%%%%%%%%%%%%%%%%%%%%%%%%%%%%%%%%%%%%%%%%%%%%%%%%%%%%%%%%%%%%%%%%%%%%%
%%%%%%%%%%%%%%%%%%%%%%%%%%%%%%%%%%%%%%%%%%%%%%%%%%%%%%%%%%%%%%%%%%%%%%
%%%%%%%%%%%%%%%%%%%%%%%%%%%%%%%%%%%%%%%%%%%%%%%%%%%%%%%%%%%%%%%%%%%%%%
%%%%%%%%%%%%%%%%%%%%%%%%%%%%%%%%%%%%%%%%%%%%%%%%%%%%%%%%%%%%%%%%%%%%%%

Extremal supersymmetric black holes are expected to be described by
$H^{M}(\tau)$ which are harmonic in Euclidean $\mathbb{R}^{3}$, i.e.~linear in
$\tau$\footnote{\label{foot:laotra} As mentioned above, this anstaz arises
  quite naturally when one imposes the constraint
  Eq.~(\ref{eq:noNUTconstraint}), but it may not be the most general one. The
  known extremal solutions (usually non-supersymmetric) that do not conform to
  this ansatz do not satisfy that constrained, either
  \cite{Gimon:2009gk,Galli:2010mg}. On the other hand, the representation of a
  solution in terms of the $H^{M}$may not be unique and the harmonicity or the
  fact that the constraint Eq.~(\ref{eq:noNUTconstraint}) is satisfied or not,
  may not always be a characteristic feature of a solution. In
  what follows we are going to explore the (large) sector of the space of
  black-hole solutions which can be described by harmonc $H^{M}$s and,
  therefore, satisfy the constraint Eq.~(\ref{eq:noNUTconstraint}). Analogous
  remarks apply to the non-extremal hyperbolic ansatz to be studied later.}:

\begin{equation}
\label{eq:harmonicfunctions}
H^{M}= A^{M} -\tfrac{1}{\sqrt{2}}B^{M}\tau\, ,  
\end{equation}

\noindent
where $A^{M}$ and $B^{M}$ are integration constants to be determined as
functions of the independent physical constants (namely, the charges
$\mathcal{Q}^{M}$ and the values of the scalars at spatial infinity
$z^{i}_{\infty}$) by using the equations of motion
(\ref{eq:equationsofmotion2})-(\ref{eq:noNUTconstraint}) and the asymptotic
conditions.  

The equations of motion for the above ansatz can be written in a simple and
suggestive form

\begin{eqnarray}
\partial_{P}\left[V_{\rm bh}(H,\mathcal{Q})- V_{\rm bh}(H,B)\right]
& = & 
0\, ,
\label{eq:dV=dV}
\\
& & \nonumber \\
V_{\rm bh}(H,\mathcal{Q})- V_{\rm bh}(H,B)
& = & 
0\, ,
\label{eq:V=V}
\\
& & \nonumber \\
A^{M}B_{M}
& = & 
0\, .  
\label{eq:AB=0}
\end{eqnarray}

\noindent
Observe that the first two equations are automatically solved for
$B^{M}=\mathcal{Q}^{M}$, which corresponds to the supersymmetric
case. The third equation (enforcing absence of NUT charge)
takes the form $A^{M}\mathcal{Q}_{M}$ and still has to be solved,
which can be done generically \cite{Denef:2000nb,Bates:2003vx} as we
are going to show.

Furthermore, observe that the Hamiltonian constraint (\ref{eq:V=V}) is
equivalent to the requirement that the black-hole potential, \textit{evaluated
  on the solutions} has the same form in terms of the true or the \textit{fake
  central charge}\footnote{\label{foot:esa} It is worth stressing that, even though
  the first equation is the derivative of the second with respect to
  $H^{P}$, solving the second for some functions $H^{M}$ does not
  imply having solved the first. Only if we find a $B^{M}$ such that
  the second equation is satisfied identically for any $H^{M}$ will
  the first equation be satisfied as well. The number of $B^{M}$s with
  this property and their value depend on the particular theory under
  consideration, but their existence is a quite general
  phenomenon.}

\begin{equation}
\label{eq:fakecentralchargedef}
\tilde{\mathcal{Z}}(\phi,B) \equiv \langle\, \mathcal{V}\mid B\, \rangle\, ,  
\end{equation}

\noindent
that is

\begin{equation}
\label{eq:VandfakeZ}
-V_{\rm bh}(\phi,\mathcal{Q})  
=
|\tilde{\cal Z}|^{2} 
+\mathcal{G}^{ij^{*}}\mathcal{D}_{i}\tilde{\cal Z}\mathcal{D}_{j^{*}}\tilde{\cal Z}^{*}\, .
\end{equation}

The asymptotic conditions take the form 

\begin{eqnarray}
\label{eq:asympflat}
\mathsf{W}(A) 
& = &   
1\, ,
\\
& & \nonumber \\
z^{i}_{\infty}
& = &
\frac{\tilde{H}^{i}(A)+iA^{i}}{\tilde{H}^{0}(A) +iA^{0}}\, ,
\end{eqnarray}

\noindent 
but can always be solved, together with (\ref{eq:AB=0}) as follows: if we
write $X$ as

\begin{equation}
\label{eq:alphadef}
X = \tfrac{1}{\sqrt{2}} e^{U+i\alpha}\, ,  
\end{equation}

\noindent
then, from the definition of $\mathcal{I}^{M}$ (\ref{eq:RandIdef}) we get 

\begin{equation}
H^{M} = \sqrt{2} e^{-U}\, \Im\mathfrak{m}(e^{-i\alpha} \mathcal{V}^{M})\, ,  
\end{equation}

\noindent
and, at spatial infinity $\tau=0$, using asymptotic flatness
(\ref{eq:asympflat})

\begin{equation}
A^{M} = \sqrt{2} \, \Im\mathfrak{m}(e^{-i\alpha_{\infty}} \mathcal{V}^{M}_{\infty})\, .  
\end{equation}

\noindent
Now, to determine $\alpha_{\infty}$ we can use (\ref{eq:AB=0}) and the
definition of fake central charge (\ref{eq:fakecentralchargedef}).  Observe
that

\begin{equation}
A_{M}B^{M} 
=  
\langle\, H\mid B\, \rangle 
% = 
% \langle\, \mathcal{I}\mid B\,\rangle
=
\Im\mathfrak{m} \langle\, \mathcal{V}/X \mid B\,\rangle
=
\Im\mathfrak{m} (\tilde{\mathcal{Z}}/X ) =  
e^{-U}\Im\mathfrak{m} (e^{-i\alpha} \tilde{\mathcal{Z}} )
=0\, ,
\end{equation}

\noindent
from which it follows first that 

\begin{equation}
\label{eq:alpha}
e^{i\alpha}= \pm \tilde{\mathcal{Z}}/|\tilde{\mathcal{Z}}|\, ,   
\end{equation}

\noindent
and is then the general expression for the $A^{M}$ as a function of the
$B^{M}$ and the $z^{i}_{\infty}$:

\begin{equation}
\label{eq:AMextremal}
A^{M} = \pm \sqrt{2} \, 
\Im\mathfrak{m} \left(\frac{\tilde{\mathcal{Z}}^{*}_{\infty}}{|\tilde{\mathcal{Z}}_{\infty}|}
\mathcal{V}^{M}_{\infty}\right)\, .  
\end{equation}

\noindent
In general, the sign of $A^{M}$ should be chosen to make $H^{M}$ finite (and,
generically, the metric non-singular) in the range $\tau\in (-\infty,0)$. The
positivity of the mass is a physical condition that eliminates some
singularities of the metric. As we are going to see in
Eq.~(\ref{eq:massformulad4-1}), this requirement singles out the upper sign in
the above formula.

This reduces the problem of finding a complete solution to the determination
of the constants $B^{M}$ as functions of the physical parameters
$\mathcal{Q}^{M},z^{i}_{\infty}$, which must solve equations (\ref{eq:dV=dV})
and (\ref{eq:V=V}).

It is useful to analyze the near-horizon and spatial-infinity limits of these
two equations.  The near-horizon limit of (\ref{eq:V=V}) plus the definition
of the fake central charge lead to the following chain of
relations\footnote{In this and other equations, the expression $V_{\rm
    bh}\left(B,\mathcal{Q}\right)$ stands for standard the black-hole
  potential with the functions $H^{M}(\tau)$ replaced by the constants
  $B^{M}$.}

\begin{equation}
\label{eq:AvsVh4d}
S/\pi
=
\tfrac{1}{2}\mathsf{W}(B)
=
-V_{\rm bh}\left(B,\mathcal{Q}\right)
=
|\tilde{\cal Z}(B,B)|^{2}\, ,  
\end{equation}

\noindent
where $\tilde{\cal Z}(B,B)$ is the near-horizon value of the fake central
charge. The last of these relations, together with the condition
(\ref{eq:VandfakeZ}) imply that, on the horizon, the fake central charge
reaches an extremum

\begin{equation}
\partial_{i}|\tilde{\cal Z}(\phi_{h},B)|=0\, .  
\end{equation}

The near-horizon limit of  (\ref{eq:dV=dV}) leads to 

\begin{equation}
\label{eq:dVh=0}
\partial_{M}V_{\rm bh}(B,\mathcal{Q})=0\, ,  
\end{equation}

\noindent
which says that the $B^{M}$ extremize the value of the black-hole
potential on the horizon. Since the black-hole potential is invariant
under a global rescaling of the $H^{M}$, the solutions (that we will
call generically \textit{attractors} $B^{M}$) of these equations are
determined up to a global rescaling which can be fixed by imposing
Eq.~(\ref{eq:V=V}). 

The $B^{M}$ must transform under the duality group of the theory
(embedded in $Sp(2n+2,\mathbb{R})$) in the same representation as the
$H^{M}$, the charges $\mathcal{Q}^{M}$ and the constants $A^{M}$. In
certain cases this poses strong constraints on the possible solutions
since, building an object that transforms in the right representation
of the duality group and has dimensions of length squared from
$\mathcal{Q}^{M}$ and $z^{i}_{\infty}$ can be far from trivial. A
possibility which is always available is the Freudenthal dual defined
in Ref.~\cite{Ferrara:2011gv}, generalizing the definition made in
Ref.~\cite{Borsten:2009zy}, and further explored in \cite{Galli:2012ji}, where it was shown that the lagrangian (\ref{eq:effectiveaction2}) has indeed a gauge symmetry which is a generalization of a Frehudental transformation on $H^M$: Freudenthal duality in $\mathcal{N}=2,d=4$
theories can be understood as the transformation from the $H^{M}$ to
the $\tilde{H}_{M}(H)$ variables. The same transformation can be
applied to any symplectic vector, such as the charge vector. Then, in
our notation and conventions, the Freudenthal dual of the charge
vector, $\tilde{\cal Q}_{M}$, is defined by

\begin{equation}
\tilde{\cal Q}_{M} = 
\tfrac{1}{2}\frac{\partial \mathsf{W}(\mathcal{Q})}{\partial \mathcal{Q}^{M}}\, .
\end{equation}

\noindent
It is not hard to prove that this duality transformation is an
antiinvolution

\begin{equation}
\tilde{\tilde{\cal Q}}_{M} = -\mathcal{Q}_{M}\, ,
\end{equation}

\noindent
and using Eq.~(\ref{eq:Wdef}) to show that 

\begin{equation}
\mathsf{W}(\tilde{\cal Q}) = \mathsf{W}(\mathcal{Q})\, .  
\end{equation}

\noindent
It is harder to show that the critical points of the black-hole
potential are invariant under Freudenthal duality
\cite{Ferrara:2011gv}. Therefore, since $B^{M}=\mathcal{Q}^{M}$ is
always an attractor (the supersymmetric one),

\begin{equation}
B^{M} = \tilde{Q}^{M}\, ,  
\end{equation}

\noindent
will always be another attractor.

Let us now consider the spatial infinity limit taking into account the
definition of the mass in these spacetimes and the definition of the fake
central charge

\begin{equation}
\label{eq:massformulad4-1}
M=\dot{U}(0) 
= \tfrac{1}{\sqrt{2}} \langle\, \mathcal{R}(A) \mid B\rangle
= \pm  |\tilde{\cal Z}(A,B)| \, .
\end{equation}

\noindent
As mentioned before, to have a positive mass we must use exclusively the upper
sign in (\ref{eq:alpha}) and (\ref{eq:AMextremal}) and we do so from now
onwards.  In the supersymmetric case, when $B^{M}=\mathcal{Q}^{M}$ and the
fake central charge is the true one, this is the supersymmetric BPS relation.

The asymptotic limit of (\ref{eq:V=V}) plus (\ref{eq:VandfakeZ}) and the above
relation give

\begin{equation}
M^{2} + \left[\mathcal{G}^{ij^{*}}\mathcal{D}_{i}\tilde{\cal Z}
\mathcal{D}_{j^{*}}\tilde{\cal Z}^{*}\right]_{\infty}
+V_{\rm bh\, \infty} =0\, ,
\end{equation}

\noindent
which, when compared with the general BPS bound \cite{Ferrara:1997tw}, lead to
the identification of the scalar charges $\Sigma_{i}$ with the values of the
covariant derivatives of the fake central charges at spatial infinity

\begin{equation}
\Sigma_{i} = \left. \mathcal{D}_{i}\tilde{\cal Z}  \right|_{\infty}\, .
\end{equation}

%%%%%%%%%%%%%%%%%%%%%%%%%%%%%%%%%%%%%%%%%%%%%%%%%%%%%%%%%%%%%%%%%%%%%%
%%%%%%%%%%%%%%%%%%%%%%%%%%%%%%%%%%%%%%%%%%%%%%%%%%%%%%%%%%%%%%%%%%%%%%

\subsubsection{First-order formalism}
\label{First_order_ext}

%%%%%%%%%%%%%%%%%%%%%%%%%%%%%%%%%%%%%%%%%%%%%%%%%%%%%%%%%%%%%%%%%%%%%%
%%%%%%%%%%%%%%%%%%%%%%%%%%%%%%%%%%%%%%%%%%%%%%%%%%%%%%%%%%%%%%%%%%%%%%

First-order flow equations for extremal BPS and non-BPS black holes can be
easily found following \cite{Ortin:2011vm} but using the generic harmonic
functions (\ref{eq:harmonicfunctions}): let us consider the K\"ahler-covariant
derivative of the inverse of the auxiliary function

\begin{equation}
  \begin{array}{rcl}
\mathcal{D}X^{-1} 
& = &  
i \langle\, \mathcal{V} \mid \mathcal{V}^{*}\rangle \mathcal{D}X^{-1}
=
i \langle\, \mathcal{D}(\mathcal{V}/X) \mid \mathcal{V}^{*}\rangle 
=
i \langle\, d(\mathcal{V}/X) \mid \mathcal{V}^{*}\rangle 
\\
& & \\
& = &
i \langle\, d(\mathcal{V}/X) - d(\mathcal{V}/X)^{*} \mid
\mathcal{V}^{*}\rangle 
= 
-2 \langle\, d H \mid \mathcal{V}^{*}\rangle 
\\
& & \\
& = &
+\sqrt{2}\, \tilde{\cal Z}^{*}(\phi,B)\, d\tau\, ,
\end{array}
\end{equation}

\noindent
where we have used the normalization of the symplectic section in the first
step , the property $\langle\, \mathcal{D}\mathcal{V} \mid
\mathcal{V}^{*}\rangle =0$ in the second, the K\"ahler-neutrality of
$\mathcal{V}/X$ in the third, $\langle\, \mathcal{D}\mathcal{V}^{*} \mid
\mathcal{V}^{*}\rangle = \langle\, \mathcal{V}^{*} \mid \mathcal{V}^{*}\rangle
= 0$ in the fourth, the definition of $\mathcal{I}=H$ in the fifth, and the
ansatz (\ref{eq:harmonicfunctions}) and the definition of the fake central
charge (\ref{eq:fakecentralchargedef}) in the sixth. 

From this equation and (\ref{eq:alphadef}) and (\ref{eq:alpha}) we find the
standard first-order equation for the metric function $U$:

\begin{equation}\label{1st_order_Uext}
\frac{de^{-U}}{d\tau } = |\tilde{\cal Z}(\phi,B)|\, .  
\end{equation}

Let us now consider the differential of the complex scalar fields:

\begin{equation}
  \begin{array}{rcl}
dz^{i} 
& = &
i \mathcal{G}^{ij^{*}} 
\langle\, \mathcal{D}_{j^{*}}\mathcal{V}^{*} \mid
\mathcal{D}_{k}\mathcal{V}\, \rangle   
dz^{k}
=
iX \mathcal{G}^{ij^{*}} 
\langle\, \mathcal{D}_{j^{*}}\mathcal{V}^{*} \mid
\mathcal{D}_{k}(\mathcal{V}/X)\, \rangle   
dz^{k}
\\
& & \\
& = &
iX \mathcal{G}^{ij^{*}} 
\langle\, \mathcal{D}_{j^{*}}\mathcal{V}^{*} \mid
\partial_{k}(\mathcal{V}/X)\, \rangle   
dz^{k}
=
iX \mathcal{G}^{ij^{*}} 
\langle\, \mathcal{D}_{j^{*}}\mathcal{V}^{*} \mid
d(\mathcal{V}/X)\, \rangle   
\\
& & \\
& = &
iX \mathcal{G}^{ij^{*}} 
\langle\, \mathcal{D}_{j^{*}}\mathcal{V}^{*} \mid
d(\mathcal{V}/X)-d(\mathcal{V}/X)^{*}\, \rangle   
=
-2X \mathcal{G}^{ij^{*}} 
\langle\, \mathcal{D}_{j^{*}}\mathcal{V}^{*} \mid
dH\, \rangle   
\\
& & \\
& = &
+\sqrt{2}\, X \mathcal{G}^{ij^{*}} 
\langle\, \mathcal{D}_{j^{*}}\mathcal{V}^{*} \mid
B \, \rangle \, d\tau
=
\sqrt{2}\, X \mathcal{G}^{ij^{*}} 
\mathcal{D}_{j^{*}}\tilde{\cal  Z}^{*}(\phi,B)\, 
 d\tau\, ,  
\end{array}
\end{equation}

\noindent
where we have used the same properties as before. To put this expression in a
more conventional form we can use the covariant holomorphicity of $\tilde{\cal
  Z}$ writing

\begin{equation}
\mathcal{D}_{j^{*}}\tilde{\cal  Z}^{*} 
=  
\mathcal{D}_{j^{*}}\frac{|\tilde{\cal  Z}|^{2}}{\tilde{\cal  Z}} 
=  
\frac{2 |\tilde{\cal  Z}| \partial_{j^{*}}|\tilde{\cal  Z}|}{\tilde{\cal  Z}}
=2 e^{-i\alpha}\partial_{j^{*}}|\tilde{\cal  Z}|\, ,
\end{equation}

\noindent
and plugging this result in the expression above:

\begin{equation}\label{1st_order_Zext}
\frac{dz^{i}}{d\tau} 
= 
2e^{U} \mathcal{G}^{ij^{*}} 
\partial_{j^{*}}|\tilde{\cal  Z}|\, .
\end{equation}

It is easy to check that these first order equations imply the second-order
equations of motion 

\begin{eqnarray}
\label{2nd_eq_U}\ddot{U}
+e^{2U}V_{\rm bh}(\phi,B)
& = & 0\, ,\\ 
& & \nonumber \\
\label{2nd_eq_Z}\ddot{Z}^{i}
+\Gamma_{jk}{}^{i} \dot{Z}^{j} \dot{Z}^{k}
+e^{2U}\partial^{i}V_{\rm bh}(\phi,B)
& = & 0\, ,
\end{eqnarray}

\noindent
which coincide with the original ones if

\begin{equation}
V_{\rm bh}(\phi,B) = V_{\rm bh}(\phi,\mathcal{Q})\, ,  
\end{equation}

\noindent
for any $\phi$ (not just for ths solution; see the remark in footnote
\ref{foot:esa}).

%%%%%%%%%%%%%%%%%%%%%%%%%%%%%%%%%%%%%%%%%%%%%%%%%%%%%%%%%%%%%%%%%%%%%%
%%%%%%%%%%%%%%%%%%%%%%%%%%%%%%%%%%%%%%%%%%%%%%%%%%%%%%%%%%%%%%%%%%%%%%

\subsection{Non-extremal black holes: the generic Ansatz}
\label{sec-d4ansatz}

%%%%%%%%%%%%%%%%%%%%%%%%%%%%%%%%%%%%%%%%%%%%%%%%%%%%%%%%%%%%%%%%%%%%%%
%%%%%%%%%%%%%%%%%%%%%%%%%%%%%%%%%%%%%%%%%%%%%%%%%%%%%%%%%%%%%%%%%%%%%%

Precious experience \cite{Galli:2011fq} (see also \cite{Mohaupt:2011aa} and,
further, \cite{Meessen:2011bd,Meessen:2012su} for 5-dimensional examples)
suggests that a quite general ansatz for the variables $H^{M}$ for non-extremal
black holes of $\mathcal{N}=2,d=4$ supergravity is\footnote{See the caveats in
footnote~\ref{foot:laotra}.}

\begin{equation}
\label{eq:Ansatz}
H^{M}(\tau)
=
A^{M}\cosh\, r_{0}\tau+\frac{B^{M}}{r_{0}}\sinh\, r_{0}\tau\, ,   
\end{equation}

\noindent
for some integration constants $A^{M}$ and $B^{M}$ that, as in the extremal
case, have to be determined by solving the equations of motion and by imposing
the standard normalization of the physical fields at spatial infinity. 

Using this ansatz, the equations of motion
(\ref{eq:hamiltonianconstraint2})-(\ref{eq:Urewriten}) take the form

\begin{eqnarray}
\tfrac{1}{2}\partial_{P}\partial_{M}\partial_{N}\log \mathsf{W}\, 
\left[B^{M}B^{N} -r_{0}^{2}A^{M}A^{N}
\right]  
-\partial_{P}\left(V_{\rm bh}(\phi,\mathcal{Q})/\mathsf{W} \right)
 & = & 
0\, ,
\label{eq:equationsofmotion2nonext}
\\
& & \nonumber \\
-\tfrac{1}{2}\partial_{M}\partial_{N}\log\mathsf{W} 
\left[B^{M}B^{N} -r_{0}^{2}A^{M}A^{N}\right] 
-V_{\rm bh}(\phi,\mathcal{Q})/\mathsf{W}
& = &
0\, ,  
\label{eq:hamiltonianconstraint2nonext}
\\
& & \nonumber \\
A^{M} B_{M}
& = & 
0\, ,  
\label{eq:noNUTconstraintnonext}  
\end{eqnarray}
\noindent
where we have used the third equation and the homogeneity properties of the
Hessian potential $\mathsf{W}$ in order to simplify the first two.

In the non-extremal case we can define several fake central charges:

\begin{equation}
\tilde{\cal Z}(\phi,B)
\equiv
\langle\, \mathcal{V}\mid B \,  \rangle 
 \, ,
\hspace{1cm}  
\tilde{\cal Z}(\phi,B_{\pm})
\equiv
\langle\, \mathcal{V}\mid B_{\pm} \,  \rangle \, ,
\end{equation}

\noindent
where we have defined the shifted coefficients

\begin{equation}
\label{eq:shiftedB}
B^{M}_{\pm} 
\equiv 
\lim_{\tau\rightarrow \mp\infty}\frac{r_{0}H^{M}(\tau)}{\sinh r_{0}\tau} = 
B^{M}\mp r_{0} A^{M}\, .
\end{equation}

\noindent
Imposing now the same asymptotic conditions on the fields as in the extremal
case and the condition (\ref{eq:noNUTconstraintnonext}), we arrive again to
(\ref{eq:AMextremal}). Therefore, we only have to determine the $B^{M}$s
plus the non-extremality parameter $r_{0}$ by imposing the equations of
motion.

The mass is given again by Eqs.~(\ref{eq:massformulad4-1}) and the
expression for the event horizon area ($+$) and the Cauchy horizon
area ($-$) are given by

\begin{eqnarray}
\label{eq:AWBnonext}
\frac{A_{\rm h\pm}}{4\pi}
= 
\mathsf{W}(B_{\pm})\, .
\end{eqnarray}

\noindent
In the near-horizon limit, the equations of motion, upon use of the
above formulae for the area of the event horizon, lead to the
following relations

\begin{eqnarray}
\label{eq:AvsVhnonext}
\frac{A_{\rm h\pm}}{4\pi}
& = & 
-V_{\rm bh}(B_{\pm})\pm 2r_{0}\mathcal{M}_{MN}[\mathcal{F}(B_{\pm})] A^{M} B^{N}_{\pm}
=
\mathsf{W}(B_{\pm})\, ,
\\
& & \nonumber \\
\label{eq:dV}
\partial_{P}V_{\rm bh}(B_{\pm})
& = & 
\pm 2r_{0}\partial_{P}\mathcal{M}_{MN}[\mathcal{F}(B)] A^{M} B^{N}_{\pm}
=
- 2r^{2}_{0}\partial_{P}\mathcal{M}_{MN}[\mathcal{F}(B)] A^{M} A^{N}
\, ,  
\end{eqnarray}

\noindent
which generalize Eqs.~(\ref{eq:AvsVh4d}) and (\ref{eq:dVh=0}) to the
non-extremal case. In the last identity we have used the expression

\begin{equation}
H^{M}\partial_{P}\mathcal{M}_{MN}(\mathcal{F})=0\, .
\end{equation}

\noindent
The right-hand side of Eq.~(\ref{eq:dV}) would, then, vanish if $A^{M} \propto
B^{M}$. This is a special case that we will study in
Section~\ref{sec-d4nonextremaldoublyextremal}. Another possibility is that
$\mathcal{F}_{\Lambda\Sigma}$ and, henceforth, $\mathcal{M}_{MN}(\mathcal{F})$
are constant, as it happens in quadratic models but, in the general case
$\partial_{P}V_{\rm bh}(B_{\pm}) \neq 0$ for non-extremal black holes and we
conclude that, in general, the values of the scalars on the horizon do not
extremize the black-hole potential.

%%%%%%%%%%%%%%%%%%%%%%%%%%%%%%%%%%%%%%%%%%%%%%%%%%%%%%%%%%%%%%%%%%%%%%
%%%%%%%%%%%%%%%%%%%%%%%%%%%%%%%%%%%%%%%%%%%%%%%%%%%%%%%%%%%%%%%%%%%%%%

\subsubsection{First-order formalism}
\label{First_order_NnExt}

%%%%%%%%%%%%%%%%%%%%%%%%%%%%%%%%%%%%%%%%%%%%%%%%%%%%%%%%%%%%%%%%%%%%%%
%%%%%%%%%%%%%%%%%%%%%%%%%%%%%%%%%%%%%%%%%%%%%%%%%%%%%%%%%%%%%%%%%%%%%%

The derivation carried out for extremal black holes in
Section~\ref{First_order_ext} can be straightforwardly extended to the
non-extremal case. As in the 5-dimensional case studied in
Ref.~\cite{Meessen:2012su}, the trick is to define a new coordinate $\rho$ and
a function $f(\rho)$

\begin{equation}
  \rho\equiv\frac{\sinh r_{0}\tau}{r_{0}\cosh r_{0}\tau}
  \hspace{1.5cm}
  f(\rho)\equiv\frac{1}{\sqrt{1-r_{0}^{2}\rho^{2}}}=\cosh r_{0}\tau\,,
\end{equation} 

\noindent
so that the hyperbolic ansatz \eqref{eq:Ansatz} for $H^{M}$ can be
rewritten in the ``almost extremal form'':

\begin{equation}
H^{M}=f(\rho)(A^{M}+B^{M}\rho)\equiv f(\rho)\hat H^{M}\,.
\end{equation} 

\noindent
Then, following the same steps that lead to Eqs.~\eqref{1st_order_Uext} and
\eqref{1st_Zhat} one can obtain the first-order flow equations:

\begin{eqnarray}
 \label{1st_Uhat}
\frac{de^{-\hat{U}}}{d \rho}
& = &
\sqrt{2} |\tilde{\mathcal{\mathcal{Z}}}(\phi,B)|\,,\\
\nonumber\\
\label{1st_Zhat}
\frac{d z^{i}}{d \rho}
& = &
-2\sqrt{2}\,e^{\hat{U}}\mathcal{G}^{ij^{*}}
\partial_{j^{*}}|\tilde{\mathcal{\mathcal{Z}}}(\phi,B)|\,.
\end{eqnarray}

\noindent
where we ave introduced the hatted warp factor $\hat{U}=U+\log{f}$.

Similarly to the extremal case, it is not difficult to show that these
first-order equations imply the second order ones:

\begin{eqnarray}
\frac{d^{2} \hat{U}}{d\rho^{2}}
+
e^{2\hat{U}}V_{\text{bh}}(\phi,\sqrt{2}B)
& = & 
0\,,\\
\nonumber\\
\frac{d^{2} z^{i}}{d\rho^{2}}
+
\Gamma_{kl}{}^{i}\frac{d z^{k}}{d\rho}\frac{d z^{l}}{d\rho}
+
e^{2\hat{U}}\mathcal{G}^{ij^{*}}\partial_{j^{*}}V_{\text{bh}}(\phi,\sqrt{2}B)
& = &
0\, ,
\end{eqnarray}   

\noindent
plus the constraint\footnote{Observe that the right-hand side of this
  equationis not $r_{0}^{2}.$}

\begin{equation}
\label{eq:jamiltonian}
\left(\frac{d \hat{U}}{d\rho}  \right)^{2}
+
\mathcal{G}_{ij^{*}}\frac{d z^{i}}{d \rho}\frac{d z^{*\, j^{*}}}{d \rho}
+
e^{2\hat{U}}V_{\text{bh}}(\phi,\sqrt{2}B)
=
0\, ,
\end{equation}

\noindent
but now with respect to the new variable $\rho$ and the new function $\hat U$.
In order to compare these equations with the actual second-order equations of
the warp factor and the scalars we have to rewrite them in terms of the
variable $\tau$ and rescale $\hat{U}$ to $U$. For the former, by using
$d/drho=f^{2}d/d\tau$ and Eq.~\eqref{1st_Uhat} one finds:

\begin{equation}
\ddot{U} -\frac{2\sqrt{2} \rho}{f} e^{U} |\mathcal{Z}(\phi,\sqrt{2}B)| 
+\frac{r_{0}^{2}}{f^{2}}
+\frac{e^{2U}}{f^{2}} 
V_{\text{bh}}(\phi,\sqrt{2}B)\, ,
\end{equation} 

\noindent
from which it follows the relation between the true black hole potential and
the fake one that must hold for the above second-order equations to imply the
true ones:

\begin{equation}
\label{U_VBH}
e^{2U}V_{\text{bh}}(\phi,\mathcal{Q})
=
\frac{e^{2U}}{f^{2}} 
V_{\text{bh}}(\phi,\sqrt{2}B)
-\frac{2\sqrt{2} r_{0}^{2}\rho}{f} e^{U} |\mathcal{Z}(\phi,\sqrt{2}B)| 
+\frac{r_{0}^{2}}{f^{2}}\, .
\end{equation} 

The same condition ensures that the constraint Eq.~(\ref{eq:jamiltonian})
implies the standard Hamiltonian constraint. For the scalar equations we find 
the condition

\begin{equation}
\partial_{i}
\left\{
e^{2U}V_{\text{bh}}(\phi,\mathcal{Q})
-
\frac{e^{2U}}{f^{2}} 
V_{\text{bh}}(\phi,\sqrt{2}B)
+\frac{4\sqrt{2} r_{0}^{2}\rho}{f} e^{U} |\mathcal{Z}(\phi,\sqrt{2}B)| 
\right\}
=0\, .  
\end{equation}

There no other conditions to be satisfied for the firs-order equations to
imply all the second order ones.  Taking the derivative with respecto to
$\rho$ of Eq.~(\ref{U_VBH}) we find that, if we assume that this relation is
satisfied for any $\phi$ (or any $H^{M}$), then the last equation is also
satisfied and all the second-order equations are satisfied.

Evaluating Eq.~(\ref{U_VBH}) at spatial infinity, ($\tau=0$, which corresponds
to $\rho=0$) we find the following relation between the charges, the fake
charges, the moduli at infinity and the non-extremality parameter: 

\begin{equation}
\label{VBH_relation}
V_{\text{bh}}(\phi_{\infty},\mathcal{Q})
-V_{\text{bh}}(\phi_{\infty},\sqrt{2}B)
=
r_{0}^{2}\, .
\end{equation}

%%%%%%%%%%%%%%%%%%%%%%%%%%%%%%%%%%%%%%%%%%%%%%%%%%%%%%%%%%%%%%%%%%%%%%
%%%%%%%%%%%%%%%%%%%%%%%%%%%%%%%%%%%%%%%%%%%%%%%%%%%%%%%%%%%%%%%%%%%%%%

\subsection{Non-extremal generalizations of doubly-extremal black holes}
\label{sec-d4nonextremaldoublyextremal}

%%%%%%%%%%%%%%%%%%%%%%%%%%%%%%%%%%%%%%%%%%%%%%%%%%%%%%%%%%%%%%%%%%%%%%
%%%%%%%%%%%%%%%%%%%%%%%%%%%%%%%%%%%%%%%%%%%%%%%%%%%%%%%%%%%%%%%%%%%%%%

In this section we are going to solve the equations of the H-FGK system for
the non-extremal black holes whose scalars are constant over the whole
spacetime using the hyperbolic ansatz Eq.~(\ref{eq:Ansatz}) for any theory
of $\mathcal{N}=2,d=4$ supergravity. Thus, we assume

\begin{equation}
\label{eq:scalarsiguales}
z^{i}_{\infty} = z^{i}_{\rm h}\, ,  
\end{equation}

\noindent
which requires 

\begin{equation}
B^{M} \propto A^{M}\, ,
\end{equation}

\noindent
where the constants $A^{M}$ are given by Eq.~(\ref{eq:AMextremal}).

Using the proportionality of the $B^{M}$s and $A^{M}$s in the
$\tau\rightarrow 0^{-}$ or $\tau\rightarrow \pm \infty$ limit of
\eqref{eq:equationsofmotion2nonext} we get

\begin{equation}
\label{eq:attractordoubly}
\partial_{K}V_{\text{bh}}(\phi_{\infty},\mathcal{Q}) = 0\, , 
\end{equation}

\noindent
which proves that the scalars must assume a value
$\phi_{\infty}=\phi_{\rm h}$ which extremizes the black hole potential
just as in the extremal case (this is something that has to be taken into
account when one applies Eq.~(\ref{eq:AMextremal}) to find the
$A^{M}$). We can, therefore, use Eq.~(\ref{eq:AvsVh4d}) that gives the
value of the black-hole potential at the horizons in terms of the fake
central charge there $\tilde{\cal Z}(B,B)$ (not $\tilde{\cal
  Z}(\phi,B_{\pm})$)

\begin{equation}
\label{eq:VbhvsBdoubly}
-V_{\rm bh}(\phi_{\infty},\mathcal{Q}) =   |\tilde{\cal Z}(B,B)|^{2}\, .
\end{equation}

The proportionality constant between $B^{M}$ and $A^{M}$ is easily
determined to be $-\mathsf{W}^{1/2}(B)$ by using the normalization at
infinity $\mathsf{W}(A)=1$ and choosing the sign so as to make the
functions $H^{M}\neq 0$ for $\tau \in (-\infty,0)$. Then we can write

\begin{equation}
\label{eq:Bdoubly2}
H^{M}(\tau)
=
A^{M}
\left[
\cosh{r_{0}\tau} 
-\mathsf{W}^{1/2}(B)\frac{\sinh{r_{0}\tau}}{r_{0}} 
\right]\, .  
\end{equation}

\noindent
and the values of $B^{M}_{\pm}$ are

\begin{equation}
B^{M}_{\pm} = - [\mathsf{W}^{1/2}(B) \pm r_{0}]A^{M}\, ,  
\end{equation}

\noindent
and 

\begin{equation}
\mathsf{W}(B_{\pm}) 
= 
[\mathsf{W}^{1/2}(B) \pm r_{0}]^{2}\, .  
\end{equation}

A relation between the value of $\mathsf{W}^{1/2}(B)$ and physical
parameters and $r_{0}$ can be found by taking the $\tau\rightarrow
0^{-}$ of Eq.~\eqref{eq:hamiltonianconstraint2nonext}

\begin{equation}
\label{eq:Wdoubly}
\mathsf{W}(B) = r_{0}^{2} -V_{\text{bh}}(\phi_{\infty},\mathcal{Q})\, .  
\end{equation}

Another relation comes from the definition of mass $M= \dot{U}(0)$
which gives $M= \tilde{H}_{M}(A)B^{M}$. Using the proportionality
between $A^{M}$ and $B^{M}$ we find that 

\begin{equation}
M = \mathsf{W}^{1/2}(B)\, .  
\end{equation}
 
The final expression for the functions $H^{M}(\tau)$ and the entropies
of all these solutions, for any theory,  is

\begin{eqnarray}
H^{M}(\tau)
& = &
A^{M}
\left[
\cosh{r_{0}\tau} 
-M\frac{\sinh{r_{0}\tau}}{r_{0}} 
\right]\, ,
\\
& & \nonumber \\
\label{eq:entropy_doubly}
S_{\pm}
& = & 
\pi [M \pm r_{0}]^{2}\, ,\\  
\end{eqnarray}

\noindent
where the non-extremality parameter is, upon use of
Eq.~(\ref{eq:VbhvsBdoubly}) given by

\begin{equation}
r_{0} = \sqrt{M^{2}-|\tilde{\cal Z}(B,B)|^{2}}\, .  
\end{equation}
%\cleardoublepage

%%%%%%%%%%%%%%%%%%%%%%%%%%%%%%%%%%%%%%%%%%%%%%%%%%%%%%%%%%%%%%%%%%%%%%%
%%% PART II: APPLICATIONS OF THESE TECHNIQUES TO STRING THEORY
%%%%%%%%%%%%%%%%%%%%%%%%%%%%%%%%%%%%%%%%%%%%%%%%%%%%%%%%%%%%%%%%%%%%%%%
%\addtocontents{toc}{\contentsline{chapter}{\numberline {}}{}{}}
%\addtocontents{toc}{\newpage}
%\addtocontents{toc}{\contentsline{chapter}{\numberline {\small{PART II}}}{}{}}
%\renewcommand{\chaptername}{Part II\\ Chapter}
%%%%%%%%%%%%%%%%%%%%%%%%%%%%%%%%%%%%%%%%%%%%%%%%%%%%%%%%%%%%%%%%%%%%%%%
%%% CHAPTER 7: QUANTUM BLACK HOLES
%%%%%%%%%%%%%%%%%%%%%%%%%%%%%%%%%%%%%%%%%%%%%%%%%%%%%%%%%%%%%%%%%%%%%%%
%\addtocontents{toc}{\newpage}

\renewcommand{\leftmark}{\MakeUppercase{Chapter \thechapter. String Theory Quantum Black Holes}}
\chapter{Quantum black holes in String Theory}
\label{chapter:quantumbhs}

In this chapter we use the H-F.G.K. formalism (see section \ref{chapter:HFGK}) to define a new class of black hole solutions in Type-IIA String Theory compactified on a Calabi-Yau (C.Y.) three-fold, in the presence of perturbative and non-perturbative corrections. We have chosen to call them \emph{quantum} black holes, since they only exist when the \emph{quantum} corrections to the prepotential are present, and no classical limit can be assigned to them. We will also obtain the first explicit and complete black hole solution in the presence of non-perturbative corrections to the prepotential. Supersymmetric black hole solutions to $\mathcal{N}=2$ four-dimensional ungauged Supergravity in the presence quantum corrections have been previously considered in \cite{Gaida:1997id,Galli:2012pt}. The entropy of supersymmetric black holes in the presence of perturbative and non-perturbative corrections has been investigated in \cite{Behrndt:1997ny}.

%%%%%%%%%%%%%%%%%%%%%%%%%%%%%%%%%%%%%%%%%%%%%%%%%%%%%%%%%%%%%%%%%%%%%%
%%%%%%%%%%%%%%%%%%%%%%%%%%%%%%%%%%%%%%%%%%%%%%%%%%%%%%%%%%%%%%%%%%%%%%
%%%%%%%%%%%%%%%%%%%%%%%%%%%%%%%%%%%%%%%%%%%%%%%%%%%%%%%%%%%%%%%%%%%%%%
%%%%%%%%%%%%%%%%%%%%%%%%%%%%%%%%%%%%%%%%%%%%%%%%%%%%%%%%%%%%%%%%%%%%%%

\section{Type-IIA String Theory on a Calabi-Yau manifold}
\label{sec:perturbative}

%%%%%%%%%%%%%%%%%%%%%%%%%%%%%%%%%%%%%%%%%%%%%%%%%%%%%%%%%%%%%%%%%%%%%%
%%%%%%%%%%%%%%%%%%%%%%%%%%%%%%%%%%%%%%%%%%%%%%%%%%%%%%%%%%%%%%%%%%%%%%
%%%%%%%%%%%%%%%%%%%%%%%%%%%%%%%%%%%%%%%%%%%%%%%%%%%%%%%%%%%%%%%%%%%%%%
%%%%%%%%%%%%%%%%%%%%%%%%%%%%%%%%%%%%%%%%%%%%%%%%%%%%%%%%%%%%%%%%%%%%%%

Type-IIA String Theory compactified to four dimensions on a C.Y. three-fold, with Hodge numbers $(h^{1,1},h^{2,1})$, is described, up to two derivatives, by a $\mathcal{N}=2, d=4$ ungauged Supergravity whose prepotential is given in terms of an infinite series around $\Im{\rm m}z^i\rightarrow \infty$\footnote{Actually, the prepotential obtained in a Type-IIA C.Y. compactification is \emph{symplectically equivalent} to the prepotential (\ref{eq:IIaprepotential}).} \cite{Candelas:1990pi,Candelas:1990rm,Candelas:1990qd}
\begin{equation}
\label{eq:IIaprepotential}
\mathcal{F} =  -\frac{1}{3!}\kappa^{0}_{ijk} z^i z^j z^k +\frac{ic}{2}+\frac{i}{(2\pi)^3}\sum_{\{d_{i}\}} n_{\{d_{i}\}} Li_{3}\left(e^{2\pi i d_{i} z^{i}}\right) \ \, ,
\end{equation}

\noindent
where $z^i,~~ i =1,...,n_v+1=h^{1,1}$, are the scalars in the vector multiplets,\footnote{There are also $h^{2,1}+1$ hypermultiplets in the theory. However, they can be consistently set to a constant value.} $c=\frac{\chi\zeta(3)}{ (2\pi)^3}$ is a model dependent number\footnote{$\chi$ is the Euler characteristic, which for C.Y. three-folds is given by $\chi=2(h^{1,1}-h^{2,1})$.}, $\kappa^{0}_{ijk}$ are the classical intersection numbers, $d_{i}\in\mathbb{Z}^{+}$ is a $h^{1,1}$-dimensional summation index and $Li_{3}(x)$ is the third polylogarithmic function, defined in section \ref{sec:polylog} together with some of its properties. The first two addends in the prepotential correspond to tree level and perturbative contributions in the $\alpha^{\prime}$ expansion, respectively

\begin{equation}
\label{eq:IIapert}
\mathcal{F}_{\text{P}} =  -\frac{1}{3!}\kappa^{0}_{ijk} z^i z^j z^k +\frac{ic}{2}\ \, ,
\end{equation}

\noindent
whereas the third term accounts for non-perturbative corrections produced by world-sheet instantons. 

\begin{equation}
\label{eq:IIanonpert}
\mathcal{F}_{\text{NP}} = \frac{i}{(2\pi)^3}\sum_{\{d_{i}\}} n_{\{d_{i}\}} Li_{3}\left(e^{2\pi i d_{i} z^{i}}\right) \ \,.
\end{equation}

\noindent
These configurations get produced by (non-trivial) embeddings of the world-sheet into the C.Y. three-fold. The holomorphic mappings of the (genus 0) string world sheet onto the $h^{1,1}$ two-cycles of the C.Y. three-fold are classified by the nubers $d_i$, which count the number of wrappings of the world sheet around the $i-$th generator of the integer homology group $H_2(\text{C.Y.},\mathbb{Z})$. The number of different mappings for each set of $\{d_i\}$ $\left( \equiv \{   d_1,...,d_{h^{1,1}}\} \right)$ or, in other words, the number of genus $0$ instantons is denoted by $n_{\{d_i\}}$\footnote{See, e.g. \cite{Mohaupt:2000mj} for more details on the stringy origin of the prepotential.}

The full prepotential can be rewritten in homogeneous coordinates $\mathcal{X}^{\Lambda}$, $\Lambda=(0,i)$ as
\begin{align}
\label{eq:IIaprepotentialX}
F(\mathcal{X}) =  -\frac{1}{3!}\kappa^{0}_{ijk} \frac{\mathcal{X}^i \mathcal{X}^j \mathcal{X}^k}{\mathcal{X}^0} +\frac{ic(\mathcal{X}^0)^2}{2} +\frac{i(\mathcal{X}^0)^2}{(2\pi)^3}\sum_{\{d_{i}\}} n_{\{d_{i}\}} Li_{3}\left(e^{2\pi i d_{i} \frac{\mathcal{X}^{i}}{\mathcal{X}^0}}\right) \ \, ,
\end{align}

\noindent
with the scalars $z^i$ given by\footnote{This coordinate system is therefore only valid away from the locus $\mathcal{X}^{0}=0$. }
\begin{equation}
\label{eq:scalarsgeneral}
z^i=\frac{\mathcal{X}^{i}}{\mathcal{X}^0} \ \, .
\end{equation}

\noindent
We are interested in studying spherically symmetric, static, black hole solutions of the theory defined by Eq. (\ref{eq:IIaprepotential}). In order to do so we are going to use the H-F.G.K. formalism \cite{Galli:2011fq,Meessen:2011aa,Meessen:2012su}, based on the use of a new set of variables $H^M,~~ M=(\Lambda,\Lambda)$, that transform linearly under duality and reduce to harmonic functions on the transverse space $\mathbb{R}^{3}$ in the supersymmetric case\footnote{For more details, see section \ref{chapter:HFGK}.}. However, the theory defined by Eq. (\ref{eq:IIaprepotential}) is extremely complicated and therefore the task of obtaining explicit black hole solutions is almost hopeless. Therefore, we are going to consider a particular approximation, namely, the large volume limit $\Im {\rm m} z^{i}\rightarrow\infty$: in section \ref{sec:perturbative} we will discard the non-perturbative corrections and consider only the perturbative ones, and in section \ref{sec:nonperturbative} we will consider a self-mirror C.Y., and therefore the perturbative correction exactly vanishes, and the first non-trivial non-perturbative correction to the prepotential.

%%%%%%%%%%%%%%%%%%%%%%%%%%%%%%%%%%%%%%%%%%%%%%%%%%%%%%%%%%%%%%%%%%%%%%
%%%%%%%%%%%%%%%%%%%%%%%%%%%%%%%%%%%%%%%%%%%%%%%%%%%%%%%%%%%%%%%%%%%%%%
%%%%%%%%%%%%%%%%%%%%%%%%%%%%%%%%%%%%%%%%%%%%%%%%%%%%%%%%%%%%%%%%%%%%%%
%%%%%%%%%%%%%%%%%%%%%%%%%%%%%%%%%%%%%%%%%%%%%%%%%%%%%%%%%%%%%%%%%%%%%%

\section{Perturbative quantum black holes}
\label{sec:perturbative}

%%%%%%%%%%%%%%%%%%%%%%%%%%%%%%%%%%%%%%%%%%%%%%%%%%%%%%%%%%%%%%%%%%%%%%
%%%%%%%%%%%%%%%%%%%%%%%%%%%%%%%%%%%%%%%%%%%%%%%%%%%%%%%%%%%%%%%%%%%%%%
%%%%%%%%%%%%%%%%%%%%%%%%%%%%%%%%%%%%%%%%%%%%%%%%%%%%%%%%%%%%%%%%%%%%%%
%%%%%%%%%%%%%%%%%%%%%%%%%%%%%%%%%%%%%%%%%%%%%%%%%%%%%%%%%%%%%%%%%%%%%%

The non-perturbative corrections (\ref{eq:IIanonpert}) are exponentially suppressed and therefore can be safely ignored going to the large volume limit. Therefore our starting point is going to be Eq. (\ref{eq:IIapert}), which in homogeneous coordinates $\mathcal{X}^{\Lambda}, ~~\Lambda=(0,i)$, can be written as

\begin{equation}
\label{eq:prepotential}
F(\mathcal{X})=  - \frac{1}{3!}\kappa^{0}_{ijk}  \frac{\mathcal{X}^{i}\mathcal{X}^{j}\mathcal{X}^{k}}{\mathcal{X}^0} +\frac{ic}{2} \left(\mathcal{X}^0\right)^2 \ \, .
\end{equation}

\noindent
The scalars $z^i$ are given by

\begin{equation}
\label{eq:scalarsgeneral}
z^i=\frac{\mathcal{X}^{i}}{\mathcal{X}^0} \ \, .
\end{equation}

\noindent
The scalar geometry defined by (\ref{eq:prepotential}) is the so called \emph{quantum corrected} $d$-SK geometry \cite{deWit:1991nm,deWit:1992wf}.The attractor points of this class of models have been extensively studied in \cite{Bellucci:2007eh,Bellucci:2008tx,Bellucci:2009pg}. In this scenario, the classical case is modified and the scalar manifold, due to the correction encoded in $c$,  is no longer homogeneous, and therefore, the geometry has been \emph{corrected} by quantum effects.

%%%%%%%%%%%%%%%%%%%%%%%%%%%%%%%%%%%%%%%%%%%%%%%%%%%%%%%%%%%%%%%%%%%%%%
%%%%%%%%%%%%%%%%%%%%%%%%%%%%%%%%%%%%%%%%%%%%%%%%%%%%%%%%%%%%%%%%%%%%%%

\subsection{A \emph{quantum} class of black holes}
\label{sec:blackholes}

%%%%%%%%%%%%%%%%%%%%%%%%%%%%%%%%%%%%%%%%%%%%%%%%%%%%%%%%%%%%%%%%%%%%%%
%%%%%%%%%%%%%%%%%%%%%%%%%%%%%%%%%%%%%%%%%%%%%%%%%%%%%%%%%%%%%%%%%%%%%%

In chapter \ref{chapter:HFGK}, thanks to the H-F.G.K. formalism, we reduced the task of obtaining black hole solutions of four-dimensional ungauged $\mathcal{N}=2$ Supergravity to solving the following set of equations 

\begin{equation}
\label{eq:Eqsofmotion}
\mathcal{E}_{P}=\tfrac{1}{2}\partial_{P}\partial_{M}\partial_{N}\log \mathsf{W}\,
\left[\dot{H}^{M}\dot{H}^{N} -\tfrac{1}{2}\mathcal{Q}^{M}\mathcal{Q}^{N}
\right]
+\partial_{P}\partial_{M}\log \mathsf{W}\, \ddot{H}^{M}
-\frac{d}{d\tau}\left(\frac{\partial \Lambda}{\partial \dot{H}^{P}}\right)
+\frac{\partial \Lambda}{\partial H^{P}}=0\, ,
\end{equation}

\noindent
together with the \emph{Hamiltonian constraint}

\begin{equation}
\label{eq:Hamiltonianconstraint}
\mathcal{H}\equiv
-\tfrac{1}{2}\partial_{M}\partial_{N}\log\mathsf{W}
\left(\dot{H}^{M}\dot{H}^{N}-\tfrac{1}{2}\mathcal{Q}^{M}\mathcal{Q}^{N} \right)
+\left(\frac{\dot{H}^{M}H_{M}}{ \mathsf{W}}\right)^{2}
-\left(\frac{\mathcal{Q}^{M}H_{M}}{ \mathsf{W}}\right)^{2}
-r_{0}^{2}=0\, ,
\end{equation}

\noindent
where

\begin{equation}
\Lambda \equiv \left(\frac{\dot{H}^{M}H_{M}}{ \mathsf{W}}\right)^{2}
+\left(\frac{\mathcal{Q}^{M}H_{M}}{ \mathsf{W}}\right)^{2}\, ,
\end{equation}

\noindent
and

\begin{equation}
\label{eq:W(H)}
\mathsf{W}(H) \equiv \tilde{H}_{M}(H)H^{M} = e^{-2U}\, .
\end{equation}

\noindent
The theory is now expressed in terms of $2\left(n_{v}+1\right)$ variables $H^M$ and depends on $2\left(n_{v}+1\right)+1$ parameters: $2\left(n_{v}+1\right)$ charges $\mathcal{Q}^M$ and the non-extremality parameter $r_0$, from which one can reconstruct the solution in terms of the original fields of the theory (that is it, the space-time metric, scalars and vector fields).

For Eq. (\ref{eq:IIapert}), the general $\mathsf{W}(H)$ is an extremely involved function, and one cannot expect to solve in full generality the corresponding differential equations of motion, or even the associated algebraic equations of motion obtained by making use of the hyperbolic Ansatz for the $H^M$. Therefore, we are going to consider a particular truncation, which will give us the desired \emph{quantum} black holes

\begin{equation}
\label{eq:truncation}
H^0 = H_0 = H_i = 0,~~p^{0} = p_{0} = q_{i} = 0 \, .
\end{equation}

\noindent
Eq. (\ref{eq:truncation}) implies

\begin{equation}
\label{eq:hessian}
\mathsf{W}(H)=\alpha\left|\kappa^{0}_{ijk} H^i H^j H^k\right|^{2/3}\, ,
\end{equation}

\noindent
where $\alpha=\frac{\left( 3! c\right)^{1/3}}{2}$ must be positive in order to have a non-singular metric. Hence $c>0$ is a necessary condition in order to obtain  a regular solution and a consistent truncation. The corresponding \emph{black hole potential} reads

\begin{equation}
\label{eq:vbh}
V_{\rm bh} = \frac{\mathsf{W}(H)}{4}\partial_{ij}\log\mathsf{W}(H)\mathcal{Q}^{i}\mathcal{Q}^{j}\, ,
\end{equation}

\noindent
The scalar fields, purely imaginary, are given by

\begin{equation}
\label{eq:scalars}
z^i= i \left(3! c\right)^{1/3} \frac{H^i}{\left(\kappa^{0}_{ijk} H^i H^j H^k\right)^{1/3}} \, ,
\end{equation}

\noindent
and are subject to the following constraint, which ensures the regularity of the K\"ahler potential ($\mathcal{X}^0 = 1$ gauge)

\begin{equation}
\label{eq:kahlercondition}
\kappa^{0}_{ijk}\Im{\rm m}z^{i}\Im{\rm m}z^{j}\Im{\rm m}z^{k} > \frac{3c}{2}\, .
\end{equation}

\noindent
Substituting Eq. (\ref{eq:scalars}) into Eq. (\ref{eq:kahlercondition}), we obtain

\begin{equation}
\label{eq:kahlerconditionII}
c>\frac{c}{4}\ \, ,
\end{equation}

\noindent
which is an identity (assuming $c>0$) and therefore imposes no constraints on the scalars. This phenomenon can be traced back to the fact that the the K\"ahler potential is constant when evaluated on the solution, and given by

\begin{equation}
\label{eq:kahlerpotential}
e^{-\mathcal{K}}=6c\ \, ,
\end{equation}

\noindent
which is well defined, again, if $c>0$. Since the volume of the C.Y. manifold is proportional to $e^{-\mathcal{K}}$, Eq. (\ref{eq:kahlerpotential}) implies that such volume remains constant and, in particular, that the limit $\Im{\rm m}z^{i}\rightarrow \infty$ does not imply a large volume limit of the compactification C.Y. manifold, a remarkable fact that can be seen as a purely quantum characteristic of our solution\footnote{Notice that in order to consistently discard the non-perturbative terms in Eq. (\ref{eq:IIaprepotential}) we only need to take the limit $\Im{\rm m}z^{i}\rightarrow \infty$. Therefore, the behavior of the C.Y. volume in such limit plays no role.}. Notice that it is also possible to obtain the classical limit $\Im{\rm m}z^{i}\gg1$ taking $c\gg1$, that is, choosing a Calabi-Yau manifold with large enough $c$. In this case we would have also a truly large volume limit.

We have seen that, in order to obtain a consistent truncation, a necessary condition is $c>0$, which implies that $\mathsf{W}\left(H\right)$ is well defined. We can go even further and argue that this is a sufficient condition by studying the equations of motion $\mathcal{E}_{P}$: 

A consistent truncation requires that the equation of motion of the truncated field is identically solved for the \emph{truncation value} of the field. First, notice that the set of solutions of Eqs. (\ref{eq:Eqsofmotion}) and (\ref{eq:Hamiltonianconstraint}), taking into account (\ref{eq:truncation}), is non-empty, since there is a \emph{model-independent} solution, given by

\begin{equation}
\label{eq:universalsusy}
H^{i} = a^{i}-\frac{p^{i}}{\sqrt{2}}\tau,~~~~r_0=0\, ,
\end{equation}

\noindent
which corresponds to a supersymmetric black hole. However, the equations of motion $\mathcal{E}_{P}$ don't know about supersymmetry: it is system of differential equations whose solution can be written as

\begin{equation}
\label{eq:universalsolution}
H^{M} = H^M\left(a,b\right)\, ,
\end{equation}

\noindent
where we have made explicit the dependence in $2n_{v}+2$ integration constants. When the solution (\ref{eq:universalsolution}) is plugged into (\ref{eq:Hamiltonianconstraint}) is when we impose, through $r_0$, a particular condition about the extremality of the black hole. If $r_0 = 0$ the integration constants are fixed such as the solution is extremal. In general there is not a unique way of doing it, one of the possibilities being always the supersymmetric one. Therefore, given that for our particular truncation the supersymmetric solution always exists, we can expect the existence also of the corresponding solution (\ref{eq:universalsolution}) of the equations of motion, from which the supersymmetric solution may be obtained through a particular choice of the integration constants that make (\ref{eq:universalsolution}) fulfilling (\ref{eq:Hamiltonianconstraint}) for $r_{0} = 0$. 

\noindent
We conclude, hence, that

\begin{equation}
\label{eq:consistentproof}
\left\{ H^{P} = 0,\mathcal{Q}^P=0 \right\} ~~\Rightarrow \mathcal{E}_{P}=0\, ,
\end{equation}

\noindent
and therefore the truncation of as many $H$'s as we want, together with the correspondet $\mathcal{Q}$'s, is consistent as long as $\mathsf{W}\left(H\right)$ remains well defined, something that in our case is assured if $c>0$. From Eq. (\ref{eq:IIaprepotential}) it can be checked that the case $c=0$, that is $h^{1,1}=h^{2,1}$,  can be cured by non-perturbative effects. 

%The existence of regular non-extremal black holes is expected but has to be checked in a model-dependent way.

It is easy to see that the truncation is not consistent in the classical limit, and therefore, we can conclude that the corresponding solutions are \emph{genuinely} quantum solutions, which only exist when perturbative quantum effects are incorporated into the action. 

Hence, we can conclude that if we require our theory to contain regular \emph{quantum} black holes there is a topological restriction on the Calabi-Yau manifolds that we can choose to compactify Type-IIA String Theory. The condition can be expressed as

\begin{equation}
\label{eq:conditionc}
c>0 ~~ \Rightarrow ~~ h^{11}>h^{21} \, .
\end{equation}

\noindent
For the supersymmetric solution (\ref{eq:universalsusy}) it is possible to compute the entropy in a model independent way. The result reads

\begin{equation}
\label{eq:susyentropy}
S_{susy}=\pi\alpha\left|\kappa^{0}_{ijk} p^i p^j p^k\right|^{2/3}\, .
\end{equation}

\noindent
The class of supersymmetric black holes described here, with entropy (\ref{eq:susyentropy}), have no microscopic-String-Theory description, not even at the leading order, and therefore illustrate how the microscopic description of the entropy in String Theory is not well understood even for the simplest class of black holes, namely, the supersymmetric one.

%%%%%%%%%%%%%%%%%%%%%%%%%%%%%%%%%%%%%%%%%%%%%%%%%%%%%%%%%%%%%%%%%%%%%%
%%%%%%%%%%%%%%%%%%%%%%%%%%%%%%%%%%%%%%%%%%%%%%%%%%%%%%%%%%%%%%%%%%%%%%

\subsection{Quantum corrected \texorpdfstring{$STU$}{STU} model}
\label{sec:STUmodel}

%%%%%%%%%%%%%%%%%%%%%%%%%%%%%%%%%%%%%%%%%%%%%%%%%%%%%%%%%%%%%%%%%%%%%%
%%%%%%%%%%%%%%%%%%%%%%%%%%%%%%%%%%%%%%%%%%%%%%%%%%%%%%%%%%%%%%%%%%%%%%

In this section we consider a very special case, the so-called $STU$ model, obtaining the first non-extremal solution with non-constant scalars in the presence of perturbative quantum corrections. In order to do so, we set $n_v=3,~ \kappa^{0}_{123}=1$. From (\ref{eq:hessian}) we obtain

\begin{equation}
\label{eq:truncationSTU}
\mathsf{W}(H)=\alpha\left|H^1 H^2 H^3\right|^{2/3}\ \ ,
\end{equation}

\noindent
where $\alpha=3 c^{1/3}$. The scalar fields are given by

\begin{equation}
\label{eq:scalarsSTU}
z^i= i c^{1/3} \frac{H^i}{\left(H^1 H^2 H^3\right)^{1/3}} \ \ ,
\end{equation}

\noindent
The $\tau$-dependence of the $H^M$ can be found by solving Eqs. (\ref{eq:Eqsofmotion}) and (\ref{eq:Hamiltonianconstraint}), and the solution is given by 

\begin{equation}
\label{eq:solutionHSTU}
H^i = a^i \cosh \left(r_0\tau\right) + \frac{b^i}{r_0} \sinh \left(r_0\tau\right), ~~~~ b^i=s^i_b\sqrt{r^2_0 (a^i)^2+\frac{(p^i)^2}{2}}\ \ .
\end{equation}

\noindent
The three constants $a^i$ can be fixed relating them to the value of the scalars at infinity and imposing asymptotic flatness. We have, hence, four conditions for three parameters and therefore one would expect a relation among the $\Im{\rm m}z^i_{\infty}$, leaving $c$ undetermined. However, the explicit calculation shows that the fourth relation is compatible with the others, and therefore no extra constraint is necessary. The $a^{i}$ are given by

\begin{equation}
\label{eq:askSTU}
a^i=-s^i_b\frac{\Im{\rm m}z_{\infty}^i}{\sqrt{3 c}}\ \ .
\end{equation}

\noindent
The mass and the entropy, in turn, read

\begin{equation}
\label{eq:massSTU}
M=\frac{r_0}{3}\sum_i \sqrt{1+\frac{3c(p^i)^2}{2r_0^2(\Im{\rm m} z_{\infty}^i)^2}}\ \ ,
\end{equation}

\begin{equation}
\label{eq:entropySTU}
S_{\pm} = r_0^2\pi\prod_i \left(\sqrt{1+\frac{3c(p^i)^2}{2r_0^2 (\Im{\rm m} z_{\infty}^i)^2}}\pm 1 \right)^{2/3}\ \ ,
\end{equation}

\noindent
and therefore the product of the inner and outer entropy only depends on the charges

\begin{equation}
\label{eq:entropySTUproduct}
S_{+} S_{-} = \frac{\pi^2\alpha^2}{4}\prod_i\left(p^i\right)^{4/3}\ \ ,
\end{equation}

\noindent
In the extremal limit we obtain the supersymmetric as well as the non-supersymmetric extremal solutions, depending on the sign chosen for the charges.

%%%%%%%%%%%%%%%%%%%%%%%%%%%%%%%%%%%%%%%%%%%%%%%%%%%%%%%%%%%%%%%%%%%%%%
%%%%%%%%%%%%%%%%%%%%%%%%%%%%%%%%%%%%%%%%%%%%%%%%%%%%%%%%%%%%%%%%%%%%%%
%%%%%%%%%%%%%%%%%%%%%%%%%%%%%%%%%%%%%%%%%%%%%%%%%%%%%%%%%%%%%%%%%%%%%%
%%%%%%%%%%%%%%%%%%%%%%%%%%%%%%%%%%%%%%%%%%%%%%%%%%%%%%%%%%%%%%%%%%%%%%

\section{Non-perturbative quantum black holes}
\label{sec:nonperturbative}

%%%%%%%%%%%%%%%%%%%%%%%%%%%%%%%%%%%%%%%%%%%%%%%%%%%%%%%%%%%%%%%%%%%%%%
%%%%%%%%%%%%%%%%%%%%%%%%%%%%%%%%%%%%%%%%%%%%%%%%%%%%%%%%%%%%%%%%%%%%%%
%%%%%%%%%%%%%%%%%%%%%%%%%%%%%%%%%%%%%%%%%%%%%%%%%%%%%%%%%%%%%%%%%%%%%%
%%%%%%%%%%%%%%%%%%%%%%%%%%%%%%%%%%%%%%%%%%%%%%%%%%%%%%%%%%%%%%%%%%%%%%

In this section we are going to include non-perturbative corrections into the game. As we saw in section \ref{sec:perturbative}, the perturbative quantum black holes become singular when $c=0$, \emph{i.e.}, when the compactification C.Y. manifold is self-mirror. However, as we will see later, it is possible to make the formal limit $c\rightarrow 0$ regular by including non-perturbative corrections in the prepotential.  At the same time, we will construct the first explicit black hole solution in the presence of non-perturbative quantum corrections. 

Therefore we are going to consider, for the time being, the complete  (\ref{eq:IIaprepotential}), and impose the same particular truncation as in section \ref{sec:perturbative}

\begin{equation}
\label{eq:truncation}
H^0 = H_0 = H_i = 0,~~p^{0} = q_{0} = q_{i} = 0 \, .
\end{equation}

\noindent
Under this assumption, the stabilization equations, which can be directly read off from (\ref{eq:W(H)}) take the form

\begin{equation}
\label{eq:stab}
\left( \begin{array}{c} iH^i \\\mathcal{R}_i  \end{array} \right)=\frac{e^{\mathcal{K}/2}}{X}\left( \begin{array}{c} \mathcal{X}^i \\ \frac{\partial F(\mathcal{X})}{\partial\mathcal{X}^i}  \end{array} \right)
 \, ,
\left( \begin{array}{c} \mathcal{R}^0\\0  \end{array}\right)=\frac{e^{\mathcal{K}/2}}{X}\left( \begin{array}{c} \mathcal{X}^0 \\ \frac{\partial F(\mathcal{X})}{\partial\mathcal{X}^0}  \end{array} \right),
\end{equation}

\noindent
and the physical fields can be obtained in terms of the $H^i$ as
\begin{equation}
\label{eq:phy}
e^{-2U}=\mathcal{R}_i(H) H^i \, ,~~z^i=i\frac{H^i}{\mathcal{R}^0(H)}\, ,
\end{equation}

\noindent
as soon as $\mathcal{R}^0$ and $\mathcal{R}^i$ are determined. In order to obtain $\mathcal{R}^0$ as a function of $H^i$, we need to solve the highly involved equation

\begin{equation}
\label{eqR0}
\frac{\partial{F(H)}}{\partial \mathcal{R}^0}=0\, ,
\end{equation}

\noindent
where $F(H)$ stands for the prepotential expressed in terms of the $H^i$
\begin{align}
\label{eq:IIaprepotentialH}
F(H) =  \frac{i}{3!}\kappa^{0}_{ijk} \frac{H^i H^j H^k}{\mathcal{R}^0} +\frac{ic(\mathcal{R}^0)^2}{2} + \frac{i(\mathcal{R}^0)^2}{(2\pi)^3}\sum_{\{d_{i}\}} n_{\{d_{i}\}} Li_{3}\left(e^{-2\pi  d_{i} \frac{H^{i}}{\mathcal{R}^0}}\right)  \, .
\end{align}

\noindent
Once this is done, it is not difficult to express $\mathcal{R}^i$ in terms of $H^i$. Indeed, from (\ref{eq:stab}) we simply have

\begin{equation}
\label{eq:ri}
\mathcal{R}_i=-i\frac{\partial{F(H)}}{\partial H^i}\, ,
\end{equation}

\noindent
where $\mathcal{R}^{0}=\mathcal{R}^{0}(H)$ must be substituted only after we perform the derivative, and must be taken to be independent before. Let's see how involved is the equation for $\mathcal{R}^{0}$: if we expand (\ref{eqR0}), we find

\begin{align}
\label{d} \notag
  -\frac{1}{3!}\kappa^{0}_{ijk} \frac{H^i H^j H^k}{({\mathcal{R}^0})^3} +c+\frac{1}{4\pi^3}\sum_{\{d_{i}\}} n_{\{d_{i}\}} \left[Li_{3}\left(e^{-2\pi  d_{i} \frac{H^{i}}{\mathcal{R}^0}}\right) \right. & \\ \left. +Li_{2}\left(e^{-2\pi  d_{i} \frac{H^{i}}{\mathcal{R}^0}}\right)\left[\frac{\pi d_iH^i}{\mathcal{R}^0} \right] \right]=0\, .
\end{align}

\noindent
Solving (\ref{d}) for $\mathcal{R}^0$ in full generality seems to be an extremely difficult task. However, if we go to the large volume compactification limit ($\Im{\rm m} z^i >> 1$), we can make use of the following property for polylogarithmic functions
\begin{equation}
\label{eq:lilimit}
\lim_{|w|\rightarrow 0}Li_s(w)=w\, , \forall s\in \mathbb{N}\, ,
\end{equation}

\noindent
since, in our case, $w=e^{-2\pi d_{i}\Im m z^{i}}\, , \forall \left\{d_i\right\}\in\left(\mathbb{Z}^{+}\right)^{h^{1,1}}$. Eq. (\ref{eq:lilimit}) enables us to rewrite (\ref{d}) as

\begin{align}
\label{di}
  -\frac{1}{3!}\kappa^{0}_{ijk} \frac{H^i H^j H^k}{({\mathcal{R}^0})^3} +c+\frac{1}{4\pi^3}\sum_{\{d_{i}\}} n_{\{d_{i}\}} \left[e^{-2\pi  d_{i} \frac{H^{i}}{\mathcal{R}^0}}+e^{-2\pi  d_{i} \frac{H^{i}}{\mathcal{R}^0}}\left[\frac{\pi d_iH^i}{\mathcal{R}^0} \right] \right]= 0\, , 
\end{align}

\noindent
keeping in mind that we are assuming $\Im{\rm m}z^{i} >> 1$. The dominant contribution in this regime, aside from the cubic one, is given by $c$. In \cite{Bueno:2012jc}, the first non-extremal black hole solutions (with constant and non-constant scalars) of (\ref{eq:IIaprepotential}) were obtained ignoring the non-perturbative corrections. These solutions turned out to be purely \textit{quantum}\footnote{It is worth pointing out that, in this context, the term \textit{quantum} does not refer to space-time but to world-sheet properties \cite{Mohaupt:2000mj}. In this respect, although such denomination is widely spread in the literature, the adjective \textit{stringy} might result more acqurate.}, in the sense that not only the classical limit $c\rightarrow 0$ was ill-defined, but also the truncated theory became inconsistent and therefore no classical limit could be assigned to such solutions. An interesting question to ask now is whether the non-perturbative contributions could actually be able to cure or at least improve this behaviour. On the other hand, it is also interesting \textit{per se} to explore the existence of black hole solutions when the subleading contribution to the prepotential is not given by $c$, but has a non-perturbative origin. In order to tackle these two questions, let us restrict ourselves to C.Y. three-folds with vanishing Euler characteristic ($c=0$), the so-called self-mirror C.Y. three-folds. Under this assumption, and considering only the subleading contribution in (\ref{d}), which is now given by the fourth addend in (\ref{di}), such equation becomes\footnote{$e^{2\pi i d_{i}z^i}<< \pi |d_{i}\Im{\rm m}z^i| e^{2\pi i d_{i}z^i}$ for $\Im{\rm m}z^{i} >> 1$.}

\begin{equation}
\label{dii}
  -\frac{1}{3!}\kappa^{0}_{ijk} \frac{H^i H^j H^k}{({\mathcal{R}^0})^3}+\frac{1}{4\pi^3}\sum_{\{d_{i}\}} n_{\{d_{i}\}} e^{-2\pi  d_{i} \frac{H^{i}}{\mathcal{R}^0}}\left[\frac{\pi d_iH^i}{\mathcal{R}^0} \right] = 0\, .
\end{equation}

\noindent
The sum over $\{d_i\}$ in (\ref{dii}) will be dominated in each case by a certain term corresponding to a particular vector $\left\{\hat{d}_i\right\}$ %\footnote{In principle, it could be possible that several terms contributed with an exponential of the same weight to the series, but will omit this possibility.} 
(and, as a consequence, to a particular $n_{\hat{d}_{i}}\equiv \hat{n}$), which, consistently with the assumption $\Im{\rm m}z^{i}>>1$, is the only one that we are going to consider. Therefore, (\ref{dii}) becomes

\begin{equation}
\label{diii}
  -\frac{1}{3!}\kappa^{0}_{ijk} \frac{H^i H^j H^k}{({\mathcal{R}^0})^3}+\frac{\hat{n}}{4\pi^3}  e^{-2\pi  \hat{d}_{i} \frac{H^{i}}{\mathcal{R}^0}}\left[\frac{\pi \hat{d}_iH^i}{\mathcal{R}^0} \right] = 0\, ,~~ \Im{\rm m}z^{i} >> 1\, .
\end{equation}

\noindent
This equation is solved by\footnote{Henceforth we will be using $\mathsf{W}$ for the Hessian potential, and $W$ for the Lambert function. We hope this is not a source of confusion.}
\begin{equation}
\label{roW}
\mathcal{R}^0=\frac{\pi \hat{d}_lH^l}{ W_a\left(s_a \sqrt{\frac{3\hat{n}(\hat{d}_nH^n)^3}{2\kappa^0_{ijk}H^i H^j H^k}}\right)} \, ,
\end{equation}

\noindent
where $W_a(x), ~(a=0,-1)$ stands for (any of the two real branches of) the \textit{Lambert $W$ function}\footnote{See section \ref{sec:lambert} for more details.} (also known as \textit{product logarithm}), and $s_a=\pm1$. Using now Eqs. (\ref{roW}) and (\ref{eq:ri}) we can obtain $\mathcal{R}^i$. The result is

%\begin{equation}
%\mathcal{R}_i=\frac{1}{2}\kappa^{0}_{ijk} \frac{H^j H^k}{\mathcal{R}^0} %-\frac{\hat{n}\mathcal{R}^0}{(2\pi)^2}\hat{d}_i e^{-2\pi  \hat{d}_{l} \frac{H^{l}}{\mathcal{R}^0}}.
%\end{equation}

\begin{equation}
\mathcal{R}_i=\frac{1}{2}\kappa^{0}_{ijk} \frac{H^j H^k}{\pi \hat{d}_lH^l}W_a\left(s_a \sqrt{\frac{3\hat{n}(\hat{d}_mH^m)^3}{2\kappa^0_{pqr}H^p H^q H^r}}\right)\, .
\end{equation}

\noindent
The physical fields can now be written as a function of the $H^i$ as

\begin{equation}\label{hessian}
e^{-2U}=\mathsf{W}(H)=\frac{\kappa^{0}_{ijk} H^i H^j H^k}{2\pi \hat{d}_m H^m} W_a\left(s_a\sqrt{\frac{3\hat{n}(\hat{d}_lH^l)^3}{2\kappa^0_{pqr}H^p H^q H^r}}\right)\, ,
\end{equation}
\begin{equation}\label{scalar}
z^i=i\frac{H^i}{\pi \hat{d}_mH^m}W_a\left(s_a \sqrt{\frac{3\hat{n}(\hat{d}_lH^l)^3}{2\kappa^0_{pqr}H^p H^q H^r}}\right)\, .
\end{equation}

\noindent
In order to have a regular solution, we need to have a positive definite metric warp factor $e^{-2U}$. Since, as explained in section \ref{sec:lambert}, ${\rm sign} \left[W_a(x)\right]={\rm sign}\left[x\right],~a=0,-1,~x\in D^a_{\mathbb{R}}$,  we have to require that
\begin{equation}
s_0\equiv sign\left[\kappa^{0}_{ijk} \frac{H^i H^j H^k}{ \hat{d}_m H^m} \right]\, ,
\end{equation}

\noindent
for the real branch $W_{0}$ and

\begin{equation}
s_{-1}\equiv -1\, .
\end{equation}

\noindent
for the real branch $W_{-1}$. On the other hand, since $W_{0}(x)=0$ for $x=0$ and $W_{-1}(x)$ is a real function only when $x\in\left[-\frac{1}{e},0\right)$, we have to impose that the argument ${\rm Arg}\left[W_{a}\right]$ of $W_{a}(x)$ lies entirely either in $\left[-\frac{1}{e},0\right)$ or in $\left(0,\infty\right)$ for all $\tau\in\left(-\infty,0\right)$, since $e^{2U}$ cannot be zero in a regular black hole solution for any $\tau\in\left(-\infty,0\right)$. This condition must be imposed in a case by case basis, since it depends on the specific form of the symplectic vector $H^M=H^M (\tau)$ as a function of $\tau$. Notice that if ${\rm Arg}\left[W_{a}\right]\in \left[-\frac{1}{e},0\right)~~\forall~~\tau\in\left(-\infty,0\right)$ we can in principle\footnote{As we will see in section \ref{sec:susysolution}, the possibility $s_0=s_{-1}=-1$ will not be consistent with the large volume approximation we are considering.} choose either $W_{0}$ or $W_{-1}$ to build the solution, whereas if ${\rm Arg}\left[W_{a}\right]\in \left(0,+\infty \right)~~\forall~~\tau\in\left(-\infty,0\right)$, only $W_{0}$ is available.

Needless to say, in order to construct actual solutions, we have to solve the H-FGK equations of motion (\ref{eq:Eqsofmotion}) (plus hamiltonian constraint (\ref{eq:Hamiltonianconstraint})) using the Hessian potential given by (\ref{hessian}). Fortunately, such equations admit a \emph{model-independent} solution which is obtained choosing the $H^{i}$ to be harmonic functions in the flat transverse space, with one of the poles given in terms of the corresponding charge
\begin{equation}
\label{eq:universalsusy}
H^{i} = a^{i}-\frac{p^{i}}{\sqrt{2}}\tau,~~~~r_0=0\, .
\end{equation}

\noindent
This corresponds to a supersymmetric black hole.

%%%%%%%%%%%%%%%%%%%%%%%%%%%%%%%%%%%%%%%%%%%%%%%%%%%%%%%%%%%%%%%%%%%%%%
%%%%%%%%%%%%%%%%%%%%%%%%%%%%%%%%%%%%%%%%%%%%%%%%%%%%%%%%%%%%%%%%%%%%%%

\subsection{The general supersymmetric solution}
\label{sec:susysolution}

%%%%%%%%%%%%%%%%%%%%%%%%%%%%%%%%%%%%%%%%%%%%%%%%%%%%%%%%%%%%%%%%%%%%%%
%%%%%%%%%%%%%%%%%%%%%%%%%%%%%%%%%%%%%%%%%%%%%%%%%%%%%%%%%%%%%%%%%%%%%%

As we have said, plugging (\ref{eq:universalsusy}) into (\ref{scalar}) and (\ref{hessian}) provides us with a supersymmetric solution without solving any further equation. The entropy of such solution reads
\begin{equation}
S=\frac{1}{2}\kappa^{0}_{ijk} \frac{p^i p^j p^k}{ \hat{d}_m p^m} W_a\left(s_a \beta\right)\, ,
\end{equation}
\begin{equation}\notag
\beta=\sqrt{\frac{3\hat{n}(\hat{d}_l p^l)^3}{2\kappa^0_{pqr}p^p p^q p^r}}\, ,
\end{equation}

\noindent
and the mass is given by
\begin{align}
\label{mass}
M =\dot{U}(0)=\frac{1}{2\sqrt{2}}\left[\frac{3\kappa^0_{ijk}p^i a^j a^k}{\kappa^0_{pqr}a^p a^q a^r}\left[1-\frac{1}{1+W_a(s_a\alpha)} \right] \right.&\\ \notag \left.-\frac{d_l p^l}{d_n a^n}\left[1-\frac{3}{2\left(1+W_a(s_a\alpha)\right)} \right] \right] \, ,
\end{align}
\begin{equation}
\alpha= \sqrt{\frac{3\hat{n}(d_l a^l)^3}{2\kappa^0_{pqr}a^p a^q a^r}} \, .
\end{equation}

In the approximation under consideration, we are neglecting terms $\sim e^{-2\pi d_i \Im m z^i}$ with respect to those going as $\sim |d_i \Im m z^i| e^{-2\pi d_i \Im m z^i}$. Taking into account (\ref{scalar}), this assumption is translated into the condition

\begin{equation}
W_{a} (x) >> 1\, ,
\end{equation}

\noindent
where $x$ stands for the argument of the Lambert function (see (\ref{roW})). It is clear that this condition is satisfied for $a=0$ if Arg$[W_0]\in [\alpha,\beta]$ for positive and suficiently large values of $\alpha$ and $\beta$. However, it is not satisfied at all for Arg$[W_a]\in [-\frac{1}{e},0)$, which is the range for which both branches of the Lambert function are available. 

If we assume ${\rm Arg}\left[W_{a}\right] \in \left[\alpha,\beta \right]$ for suficiently large $\alpha,\beta\in\mathbb{R}^+$, $a=0$ and $W_0$ is the only real branch of the Lambert function. In that case, $s=s_0=1$, and we have
\begin{equation}\label{hessian0}
e^{-2U}=\frac{\kappa^{0}_{ijk} H^i H^j H^k}{2\pi \hat{d}_m H^m} W_0\left(\sqrt{\frac{3\hat{n}(\hat{d}_lH^l)^3}{2\kappa^0_{pqr}H^p H^q H^r}}\right)\, ,
\end{equation}
\begin{equation}\label{scalar0}
z^i=i\frac{H^i}{\pi \hat{d}_mH^m}W_0\left( \sqrt{\frac{3\hat{n}(\hat{d}_lH^l)^3}{2\kappa^0_{pqr}H^p H^q H^r}}\right)\, .
\end{equation}

\noindent
In the conformastatic coordinates we are working with, the metric warp factor $e^{-2U}$ is expected to diverge at the event horizon ($\tau\rightarrow -\infty$) as $\tau^2$. In addition, we have to require $e^{-2U}>0$ $\forall \tau \in (-\infty,0]$, and impose asymptotic flatness $e^{-2U(\tau=0)}=1$. The last two conditions read

\begin{equation}\label{posimetri}
\frac{\kappa^0_{ijk} H^i H^j H^k}{2\pi \hat{d}_n H^n}>0 ~~\forall \tau\in (-\infty,0] \, ,
\end{equation}
\begin{equation}\label{asyfla}
 \frac{\kappa^0_{ijk}a^i a^j a^k}{2\pi \hat{d}_m a^m}W_0\left(\alpha \right)=1\, ,
\end{equation}

\noindent
whereas the first one turns out to hold, since
\begin{equation}
e^{-2U}\overset{\tau \rightarrow -\infty}{\longrightarrow} \frac{\kappa^0_{ijk}p^i p^j p^k}{8\pi \hat{d}_m p^m}W_0\left(\beta\right)\tau^2  \,.
\end{equation}

\noindent
(\ref{posimetri}) and (\ref{asyfla}) can in general be safely imposed in any particular model we consider. Finally, the condition for a well-defined and positive mass $M>0$ can be read off from (\ref{mass}).

%%%%%%%%%%%%%%%%%%%%%%%%%%%%%%%%%%%%%%%%%%%%%%%%%%%%%%%%%%%%%%%%%%%%%%
%%%%%%%%%%%%%%%%%%%%%%%%%%%%%%%%%%%%%%%%%%%%%%%%%%%%%%%%%%%%%%%%%%%%%%

\subsection{Multivalued functions and black hole uniqueness theorems}

%%%%%%%%%%%%%%%%%%%%%%%%%%%%%%%%%%%%%%%%%%%%%%%%%%%%%%%%%%%%%%%%%%%%%%
%%%%%%%%%%%%%%%%%%%%%%%%%%%%%%%%%%%%%%%%%%%%%%%%%%%%%%%%%%%%%%%%%%%%%%

As we explained in the previous section, our approximation is not consistent with a solution such that Arg$[W_a]\in [-\frac{1}{e},0)$. This forbids the domain in which $W(x)$ is a multivalued function (both $W_0$ and $W_{-1}$ are real there). However, it seems legitimate to ask what the consequences of having two different branches would have been, if this constraint had not been present.
In principle, we could have tried to assign the asymptotic ($\tau \rightarrow 0$) and near horizon ($\tau \rightarrow -\infty$) limits to any particular pair of values of the arguments of $W_0$ and $W_{-1}$ through a suitable election of the parameters available in the solution. In particular, if we had chosen ${\rm Arg}\left[W_{0}\right]|_{\tau=0}={\rm Arg}\left[W_{-1}\right]|_{\tau=0}=-1/e$ and ${\rm Arg}\left[W_{0}\right]|_{\tau \rightarrow -\infty}={\rm Arg}\left[W_{-1}\right]|_{\tau \rightarrow -\infty}=\beta$, $\beta \in (-1/e,0)$, both solutions would have had exactly the same asymptotic behavior (and therefore the scalars of both solutions would have coincided at spatial infinity), and we would have been dealing with two completely different regular solutions with the same mass\footnote{Although $W^{\prime}_{0,-1}(x)$ are divergent at $x=-1/e$ (as explained in section \ref{sec:lambert}), and the definition of $M$ would involve derivatives of the Lambert function at that point, it would not be difficult to cure this behaviour and get a positive (and finite) mass by imposing $\dot{x}(\tau)\overset{\tau \rightarrow 0 }{\longrightarrow 0}$ faster than $|W^{\prime}_{0,-1}(x)|\overset{x \rightarrow -1/e}{\longrightarrow \infty}$.}, charges and asymptotic values of the scalar fields, in flagrant contradiction\footnote{Up to possible stability issues, which should be carefully studied.} with the corresponding black hole uniqueness theorem (conjecture). At this point, and provided that our approximation is not consistent with such presumable two-branched solution (therefore, we could say that ST forbids such possibility), the feasibility of this reasoning in a different context can only be catalogued as \textit{speculative} at the very least. However, as a matter of fact, a violation of the black hole uniqueness theorem (and, in turn, of the No-Hair conjecture) in four-dimensional ungauged Supergravity would have far-reaching consequences independently of whether the solution is embedded in ST or not. In this regard, the very possibility that the stabilization equations may admit (for certain more or less complicated prepotentials) solutions depending on multivalued functions seems to open up a window for possible violations of the black hole uniqueness theorems in the context of $\mathcal{N}=2$ $d=4$ ungauged Supergravity. The question (whose answer is widely assumed to be "no") is now: is it possible to find a four-dimensional (Super)gravity theory with a physically-admisible matter content admitting more than one stable black hole solution with the same mass, electric, magnetic and scalar charges? This question will be addressed in \cite{POS}.

%%%%%%%%%%%%%%%%%%%%%%%%%%%%%%%%%%%%%%%%%%%%%%%%%%%%%%%%%%%%%%%%%%%%%%
%%%%%%%%%%%%%%%%%%%%%%%%%%%%%%%%%%%%%%%%%%%%%%%%%%%%%%%%%%%%%%%%%%%%%%

\subsubsection{The polylogarithm}
\label{sec:polylog}

%%%%%%%%%%%%%%%%%%%%%%%%%%%%%%%%%%%%%%%%%%%%%%%%%%%%%%%%%%%%%%%%%%%%%%
%%%%%%%%%%%%%%%%%%%%%%%%%%%%%%%%%%%%%%%%%%%%%%%%%%%%%%%%%%%%%%%%%%%%%%

The \textit{polylogarithmic function} or \textit{polylogarithm} $Li_w(z)$ (see e.g. \cite{Lewin} for an exhaustive study) is a special function defined through the power series

\begin{equation}
Li_w(z)=\sum_{j=1}^{\infty} \frac{z^j}{j^w}\, ,~~z,w\in \mathbb{C}\, .
\end{equation}

\noindent
This definition is valid for arbitrary complex numbers $w$ and $z$ for $|z|<1$, but can be extended to $z's$ with $|z|\geq 1$ by analytic continuation. From its definition, it is easy to find the recurrence relation

\begin{equation}
Li_{w-1}(z)=z\frac{\partial Li_w(z)}{\partial z}\, .
\end{equation}
\noindent
The case $w=1$ corresponds to

\begin{equation}
Li_1(z)=-\log(1-z) \, ,
\end{equation}

\noindent
and from this it is easy to see that for $w=-n\in \mathbb{Z}^- \cup \left\{0 \right\}$, the polylogarithm is an elementary function given by

\begin{equation}
Li_0(z)=\frac{z}{1-z}\, ,~~ Li_{-n}(z)=\left(z\frac{\partial}{\partial z} \right)^n \frac{z}{1-z}\, .
\end{equation}

\noindent
The special cases $w=2,3$ are called \textit{dilogarithm} and \textit{trilogarithm} respectively, and their integral representations can be obtained from $Li_1(z)$ making use of

\begin{equation}
Li_{w}(z)=\int^{z}_0 \frac{Li_{w-1}(s)}{s}ds\, .
\end{equation}

%%%%%%%%%%%%%%%%%%%%%%%%%%%%%%%%%%%%%%%%%%%%%%%%%%%%%%%%%%%%%%%%%%%%%%
%%%%%%%%%%%%%%%%%%%%%%%%%%%%%%%%%%%%%%%%%%%%%%%%%%%%%%%%%%%%%%%%%%%%%%

\subsubsection{The Lambert W function}
\label{sec:lambert}

%%%%%%%%%%%%%%%%%%%%%%%%%%%%%%%%%%%%%%%%%%%%%%%%%%%%%%%%%%%%%%%%%%%%%%
%%%%%%%%%%%%%%%%%%%%%%%%%%%%%%%%%%%%%%%%%%%%%%%%%%%%%%%%%%%%%%%%%%%%%%

The \textit{Lambert W function} $W(z)$ (also known as \textit{product logarithm}) is named after Johann Heinrich Lambert (1728-1777), who was the first to introduce it in 1758 \cite{Lambert}. During its more than two hundred years of history, it has found numerous applications in different areas of physics (mainly during the 20th century) such as electrostatics, thermodynamics (e.g. \cite{Valluri}), statistical physics (e.g. \cite{JM}), QCD (e.g. \cite{Gardi:1998qr}, \cite{Magradze:1998ng}, \cite{Nesterenko:2003xb}, \cite{Cvetic:2011vy}, \cite{Sonoda:2013kia}), cosmology (e.g. \cite{Ashoorioon:2004vm}), quantum mechanics (e.g. \cite{Mann}) and general relativity (e.g. \cite{Mann:1996cb}). 

$W(z)$ is defined implicitly through the equation 

\begin{equation}
\label{Wf}
z=W(z)e^{W(z)}\, ,~~ \forall z\in \mathbb{C}\,.
\end{equation}

\noindent
Since $f(z)=ze^{z}$ is not an injective mapping, $W(z)$ is not uniquely defined, and $W(z)$ generically stands for the whole set of branches solving (\ref{Wf}). For $W:\mathbb{R}\rightarrow \mathbb{R}$, $W(x)$ has two branches $W_0(x)$ and $W_{-1}(x)$ defined in the intervals $x\in [-1/e,+\infty)$ and $x\in [-1/e,0)$ respectively (See Figure 1). Both functions coincide in the branching point $x=-1/e$, where $W_0(-1/e)=W_{-1}(-1/e)=-1$. As a consequence, the defining equation $x=W(x)e^{W(x)}$ admits two different solutions in the interval $x\in [-1/e,0)$.\\
\begin{figure}[h]
 \label{fig:u}
  \centering
    \includegraphics[scale=0.55]{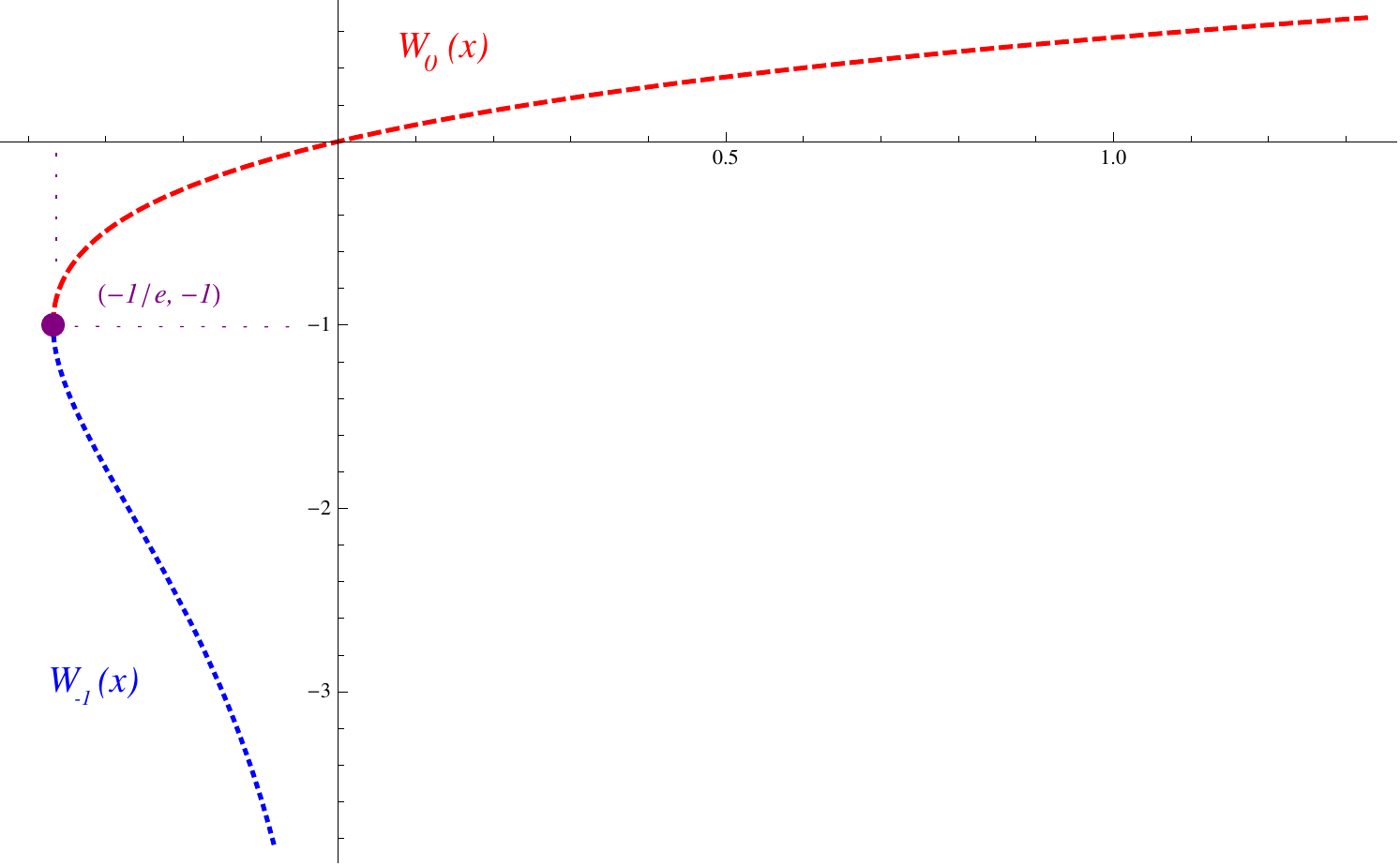}
 \caption{\small{The two real branches of $W(x)$.}}
\end{figure}

\noindent
The derivative of $W(z)$ reads

\begin{equation}
\frac{dW(z)}{dz}=\frac{W(z)}{z(1+W(z))},~~\forall z \notin \left\{0,-1/e \right\}; ~~ \frac{dW(z)}{dz}\bigg|_{z=0}=1 \, ,
\end{equation}

\noindent
and is not defined for $z=-1/e$ (the function is not differentiable there). At that point we have

\begin{equation}
\lim_{x\rightarrow -1/e}\frac{dW_0(x)}{dx}=\infty,~~\lim_{x\rightarrow -1/e}\frac{dW_{-1}(x)}{dx}=-\infty \, .
\end{equation}

%\cleardoublepage

%%%%%%%%%%%%%%%%%%%%%%%%%%%%%%%%%%%%%%%%%%%%%%%%%%%%%%%%%%%%%%%%%%%%%%%
%%% ANNEXE
%%%%%%%%%%%%%%%%%%%%%%%%%%%%%%%%%%%%%%%%%%%%%%%%%%%%%%%%%%%%%%%%%%%%%%%
%\addtocontents{toc}{\newpage}
%\addtocontents{toc}{\vspace{15pt}}
%\addtocontents{toc}{\contentsline{chapter}{\numberline {\small{ANNEXE}}}{}{}}
%\renewcommand{\chaptername}{Anexo}
%\renewcommand{\thechapter}{I}
%\input{Spanish}
%\cleardoublepage

%%%%%%%%%%%%%%%%%%%%%%%%%%%%%%%%%%%%%%%%%%%%%%%%%%%%%%%%%%%%%%%%%%%%%%%
%%% APPENDICES
%%%%%%%%%%%%%%%%%%%%%%%%%%%%%%%%%%%%%%%%%%%%%%%%%%%%%%%%%%%%%%%%%%%%%%%
%\addtocontents{toc}{\vspace{15pt}}
%\addtocontents{toc}{\contentsline{chapter}{\numberline {\small{APPENDICES}}}{}{}}
%\renewcommand{\chaptername}{Appendix}
\appendix

%%%%%%%%%%%%%%%%%%%%%%%%%%%%%%%%%%%%%%%%%%%%%%%%%%%%%%%%%%%%%%%%%%%%%%%

%\renewcommand{\leftmark}{\MakeUppercase{\chaptername\ \thechapter. Introducci\'on}}
%\input{./chapters/resumen}
%\cleardoublepage

%%%%%%%%%%%%%%%%%%%%%%%%%%%%%%%%%%%%%%%%%%%%%%%%%%%%%%%%%%%%%%%%%%%%%%%

%\renewcommand{\leftmark}{\MakeUppercase{\chaptername\ \thechapter. Resumen}}
%\input{./chapters/conclusiones}
%\cleardoublepage

%%%%%%%%%%%%%%%%%%%%%%%%%%%%%%%%%%%%%%%%%%%%%%%%%%%%%%%%%%%%%%%%%%%%%%%
%%% BIBLIOGRAPHY
%%%%%%%%%%%%%%%%%%%%%%%%%%%%%%%%%%%%%%%%%%%%%%%%%%%%%%%%%%%%%%%%%%%%%%%
\renewcommand{\leftmark}{\MakeUppercase{Bibliography}}
\phantomsection
\addcontentsline{toc}{chapter}{References}
\bibliographystyle{ThesisStyle}
\bibliography{/home/mains/Dropbox/Referencias/References.bib}
\label{biblio}
\clearpage

\end{document}